\documentclass[12pt]{article}

\usepackage[T1]{fontenc}
\usepackage[utf8]{inputenc}

\usepackage{amsmath,amssymb,slashed,hyperref}
\usepackage{amsfonts}
\usepackage{upgreek}
\usepackage{comment}

\numberwithin{equation}{section}

\usepackage[english]{babel}

\usepackage[mathscr]{eucal}
\usepackage{epsfig}
\usepackage{booktabs}
\usepackage{bbold}
\usepackage{slashed}

\DeclareMathAlphabet{\pxitfont}{OML}{pxmi}{m}{it}
\DeclareMathAlphabet{\pxitfontn}{U}{pxmia}{m}{it}
\def\Qb{\pxitfont Q}
\newcommand{\cAb}{{\mathcal{A}_{\pxitfont{b}}}}
\newcommand{\cAf}{{\mathcal{A}_{\pxitfont{f}}}}
\newcommand{\cFb}{{\mathcal{F}_{\pxitfont{b}}}}
\newcommand{\cDb}{{\mathcal{D}_{\pxitfont{b}}}}
\def\Nb{{\pxitfontn N}}

\DeclareFontFamily{U}{euc}{}%
\DeclareFontShape{U}{euc}{m}{n}{<-6>eurm5<6-8>eurm7<8->eurm10}{}%
\DeclareSymbolFont{AMSc}{U}{euc}{m}{n} %
\DeclareMathSymbol{\psitt}{\mathord}{AMSc}{"20}    
\DeclareMathSymbol{\chitt}{\mathord}{AMSc}{"1F}    
\def\psit{{\psitt_t}}

\def\chit{{\chitt_t}}
\def\U{{\mathrm U}}
\def\ang{\vartheta}

\newcommand{\bea}{\begin{array}}
\newcommand{\eea}{\end{array}}
\newcommand{\beq}{\begin{equation}}
\newcommand{\eeq}{\end{equation}}
\newcommand{\beqn}{\begin{eqnarray}}
\newcommand{\eeqn}{\end{eqnarray}}
\newcommand{\tr}{{\rm tr}}
\newcommand{\Tr}{{\rm Tr}}
\newcommand{\Str}{{\rm STr}}
\newcommand{\nnr}{\nonumber\\}
\newcommand{\ex}{{\rm e}}

\newcommand{\eps}{\epsilon}
\newcommand{\veps}{\varepsilon}

\newcommand{\bzeta}{\bar{\zeta}}
\newcommand{\CC}{{\mathbb C}}
\newcommand{\RR}{{\mathbb R}}
\newcommand{\ZZ}{{\mathbb Z}}
\newcommand{\Z}{\ZZ}
\newcommand{\fr}{\frac}
\newcommand{\pt}{\partial}
\newcommand{\cN}{\mathcal{N}}
\newcommand{\cA}{\mathcal{A}}
\newcommand{\cB}{\mathcal{B}}
\newcommand{\rmd}{{\rm d}}
\newcommand{\dA}{{\dot{A}}}
\newcommand{\dB}{{\dot{B}}}
\newcommand{\dC}{{\dot{C}}}
\newcommand{\dD}{{\dot{D}}}
\newcommand{\da}{{\dot{a}}}

\newcommand{\CS}{{\mathrm {CS}}}

\newcommand{\calK}{\mathcal{K}}
\newcommand{\SG}{SG}
\newcommand{\sg}{\mathfrak{sg}}
\newcommand{\gbos}{\mathfrak{g}_{\bar{0}}}
\newcommand{\gferm}{\mathfrak{g}_{\bar{1}}}
\newcommand{\gleft}{\mathfrak{g}_\ell}
\newcommand{\gright}{\mathfrak{g}_r}
\newcommand{\Dbos}{\Delta^+_{\bar{0}}}
\newcommand{\Dferm}{\Delta^+_{\bar{1}}}
\newcommand{\DfermI}{\bar{\Delta}^+_{\bar{1}}}
\newcommand{\bos}{{\bar{0}}}
\newcommand{\ferm}{{\bar{1}}}
\def\osp{\mathfrak{osp}}
\def\sp{\mathfrak{sp}}
\def\so{\mathfrak{so}}
\def\hyp{{\mathrm{hyp}}}
\def\mj{{\mathrm{j}}}
\def\sign{{\rm sign}}
\def\SO{{\rm SO}}
\def\Sp{{\rm Sp}}
\def\SU{{\mathrm {SU}}}
\def\frak{\mathfrak}
\def\cL{\mathcal{L}}
\def\cO{\mathcal{O}}

\font\teneusm=eusm10 
\font\seveneusm=eusm7 
\font\fiveeusm=eusm5
\newfam\eusmfam
\textfont\eusmfam=\teneusm \scriptfont\eusmfam=\seveneusm
\scriptscriptfont\eusmfam=\fiveeusm

\font\tencmmib=cmmib10 \skewchar\tencmmib='177
\font\sevencmmib=cmmib7 \skewchar\sevencmmib='177
\font\fivecmmib=cmmib5 \skewchar\fivecmmib='177
\newfam\cmmibfam
\textfont\cmmibfam=\tencmmib \scriptfont\cmmibfam=\sevencmmib
\scriptscriptfont\cmmibfam=\fivecmmib
\def\cmmib#1{{\fam\cmmibfam\relax#1}}

\font\teneurm=eurm10 
\font\seveneurm=eurm7 
\font\fiveeurm=eurm5
\newfam\eurmfam
\textfont\eurmfam=\teneurm \scriptfont\eurmfam=\seveneurm
\scriptscriptfont\eurmfam=\fiveeurm

\font\teneufm=eufm10 
\font\seveneufm=eufm7 
\font\fiveeufm=eufm5
\newfam\eufmfam
\textfont\eufmfam=\teneufm \scriptfont\eufmfam=\seveneufm
\scriptscriptfont\eufmfam=\fiveeufm

\def\H{{\mathcal H}}
\def\gw{{\gamma}}
\def\W{{\mathcal W}}
\def\bar{\overline}
\def\OSp{{\mathrm{OSp}}}
\def\Spin{{\mathrm{Spin}}}
\def\hat{\widehat}
\def\S{{\mathcal S}}
\def\Bbb{\mathbb}
\def\Z{{\Bbb Z}}
\def\p{{\cmmib p}}
\def\q{{\cmmib q}}
\def\AA{{\mathcal A}}
\def\AAb{\cAb}
\def\AAf{\cAf}

\def\d{{\mathrm d}}
\def\N{{\mathcal N}}
\def\YM{{\mathrm{YM}}}
\def\Str{{\mathrm{Str}}}
\def\tilde{\widetilde}
\def\t{\tilde}
\def\DD{{\mathcal D}}
\def\u{{\frak u}}
\def\R{{\mathcal R}}
\def\2{{\bf 2}}
\def\1{{\bf 1}}
\def\0{{\bf 0}}
\def\s{{\bf s}}
\def\bar{\overline}
\def\u{{\frak u}}
\def\su{{\frak {su}}}
\def\V{{\mathcal V}}
\def\O{{\mathcal O}}

\def\Q{Q}                     
\def\cQ{\mathcal{Q}}    
\def\fsigma{\xi}
\def\Pexp{{P\hspace{-0.7mm}\exp}}
\def\ta{\widetilde{a}}

\begin{document}

\thispagestyle{empty}
\begin{flushright}\footnotesize
~

\vspace{2.1cm}
\end{flushright}

\begin{center}
{\Large\textbf{\mathversion{bold} Branes And Supergroups}\par}

\vspace{2.1cm}

\textrm{Victor Mikhaylov$^{\flat}$ and Edward Witten$^{*}$}
\vspace{0.7cm}

\textit{$^{\flat}$Department of Physics, Princeton University, Princeton, NJ 08540 } \\
\vspace{4mm}
\textit{$^{*}$ School
of Natural Sciences, Institute for Advanced Study, Princeton, NJ 08540}\\
 \vspace{3mm}

\par\vspace{2cm}

\textbf{Abstract}
\end{center}
\noindent  
Extending previous work that involved D3-branes ending on a fivebrane with $\theta_{\mathrm{YM}}\not=0$, we consider a
 similar two-sided problem.  This construction, in case the fivebrane is of NS type, is associated
to the three-dimensional Chern-Simons theory of a supergroup $\U(m|n)$ or $\OSp(m|2n)$ rather than an ordinary Lie group as in the one-sided case.
By $S$-duality, we deduce a dual magnetic description of the supergroup Chern-Simons theory; a slightly different duality, in the orthosymplectic case,
 leads to a strong-weak coupling duality between certain supergroup Chern-Simons theories on $\RR^3$; and  a further $T$-duality leads to
a version of Khovanov homology for supergroups.  Some cases of these statements are known in 
 the literature.  We analyze how these dualities act on line and surface
 operators.  
\vspace*{\fill}

\begin{center}
February 11, 2015
\end{center}
\setcounter{page}{1}

\newpage

\tableofcontents

\section{Introduction}\label{intro}
$\N=4$ super Yang-Mills theory in four dimensions admits a wide variety of defects and boundary conditions that preserve some
of the supersymmetry.  Some familiar and much-studied examples arise from 
the interaction of D3-branes with fivebranes.  With a single fivebrane, one
 can consider either {\it (i)} a one-sided problem with D3-branes ending
on the fivebrane on one side,  or {\it (ii)} a
two-sided problem with D3-branes on both sides.  The present paper aims to generalize to case {\it (ii)} 
a recent analysis \cite{5knots} of certain aspects of case {\it (i)}.  We begin with a very short review of this previous 
work, simplifying some points. (Later in this paper, when relevant, we supply enough detail to make the paper reasonably self-contained.)

\subsection{A Mini-Review}\label{mini}

A system of $n$ parallel D3-branes supports a $\U(n)$ gauge theory with $\N=4$ supersymmetry. We write $g_\YM$ and $\theta_\YM$ for the coupling constant and theta-angle of the gauge theory. For $\theta_\YM=0$,
 D3-branes ending on an NS5-brane are described, in field
theory terms, by a simple half-BPS boundary condition -- Neumann boundary conditions for the gauge fields, extended to other fields
in a supersymmetric fashion. For $\theta_\YM\not=0$, a system of D3-branes
ending on an NS5-brane is still described by a half-BPS boundary condition, 
but the details are more subtle \cite{SuperBC}  and in particular  the
unbroken supersymmetries depend on $\theta_\YM$ and  $g_\YM$.  

Now let $M$ be a four-manifold with boundary $W$.  Consider 
$\N=4$ super Yang-Mills theory on $M$, with D3-NS5 boundary conditions along $W$.  A key point in \cite{5knots}
is that one can  pick one of the 
supercharges $\Qb$ such that $\Qb^2=0$ and the action $I$ is in a certain sense the sum of a $\Qb$-exact term and
a Chern-Simons action on $W$:
\begin{equation}\label{yetro}I =\int_M\{\Qb,V\}+\frac{i\calK}{4\pi} 
\int_{W}\Tr\,\left(\AA\wedge\d \AA+\frac{2}{3}\AA\wedge\AA\wedge \AA\right). \end{equation}
Here $\calK$ is a certain complex-valued 
function of $g_\YM$ and  $\theta_\YM$  that will be described later.\footnote{This function is denoted $\Psi$ in
\cite{5knots,Langlands}. In the present paper, we call it $\calK$ because of the analogy with the usual Chern-Simons
level $k$.}   Also, $\AA$ is a complexified version of the gauge field, roughly $\AA_\mu=A_\mu+i\phi_\mu$, where 
$A_\mu$ is the ordinary gauge field
and $\phi_\mu$ denotes some of the scalar fields of $\N=4$ super Yang-Mills 
theory (which scalar fields enter this formula depends on the choice of
$\Qb$).  The details of the functional $V$ are inessential.

Based on this formula, it is shown in \cite{5knots} that,
 if one specializes to $\Qb$-invariant observables, 
$\N=4$ super Yang-Mills theory on the four-manifold $M$ 
is closely related to  Chern-Simons gauge theory on the three-manifold $W$.    
The gauge group of the relevant Chern-Simons theory is
the same as the gauge group of the underlying $\N=4$ theory.    
In general, the theory obtained from this construction differs from the ordinary Chern-Simons theory on $W$ 
in the following unusual way: the integrand of the Feynman path integral is the same,  but the ``integration cycle'' in this
 path integral is not equivalent to the usual one \cite{Wittenold,Wittenoldone}.  However, 
for the important case of $W=\RR^3$, the  integration cycles are equivalent and the theory obtained
this way is equivalent to the  conventional Chern-Simons theory, or more precisely is an analytic continuation of it,
with the usual integer-valued coupling parameter $k$ of Chern-Simons theory generalized to the complex parameter $\calK$.

The main results in  \cite{5knots} came by studying this picture with the use of  standard dualities.  Applying $S$-duality
to the $\N=4$ gauge theory on $M$, one gets a dual ``magnetic'' description in terms of a D3-D5 system.  If we specialize to $\Qb$-invariant
observables -- such as Wilson loops in $W$ -- we get a magnetic dual description of Chern-Simons theory. Applied to knots in $\RR^3$, this dual description gives a new perspective on the invariants of knots -- such as the Jones polynomial
-- that can be derived from Chern-Simons theory.
After a  further $T$-duality to a D4-D6 system, the space of physical states of this system can be identified with 
what is known as the Khovanov
homology of a knot.  
Khovanov homology of a knot \cite{Kh} is a generalization of the Jones polynomial that is known to contain more information. For earlier
physics-based work on Khovanov homology, see \cite{GSV,OV}.

 An important detail here
is that although, for $\Qb$-invariant observables,  the ``electric'' description in the D3-NS5 system can be reduced to a three-dimensional Chern-Simons theory, the
dual ``magnetic'' description is essentially four-dimensional, even for those observables.  In section \ref{apps} below, we explain how to get a purely three-dimensional
duality for Chern-Simons theory of certain supergroups.

\subsection{The Two-Sided Problem And Supergroups}\label{content}

\begin{figure}
 \begin{center}
   \includegraphics[width=2.5in]{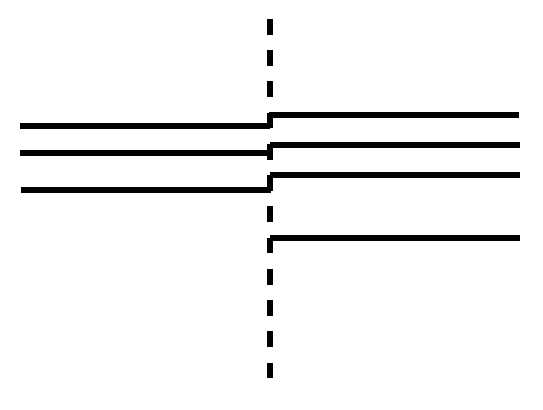}
 \end{center}
\caption{\small An NS5-brane (sketched as a vertical dotted line) with $m$ D3-branes ending on it from the left and $n$ from the right --
sketched here for $m=3$, $n=4$.  The D3-branes but not the NS5-brane extend in the $x_3$ direction, which is plotted horizontally, and
the NS5-brane but not the D3-branes extend in the $x_{4,5,6}$ directions, which are represented  symbolically by the vertical direction
in this figure.}
 \label{setup}
\end{figure}

In the present paper, we make a similar 
analysis of a problem (fig. \ref{setup}) with $m$ D3-branes on one side of an NS5-brane and $n$ D3-branes 
on the other side.  
Thus on one side of the NS5-brane, the gauge group is $\U(m)$ and on the other side it is $\U(n)$.
Let us write $M_\ell$ for the support of the $\U(m)$ gauge theory and $M_r$ for the support of the $\U(n)$ gauge theory.  
Thus $M_\ell$ and $M_r$ are four-manifolds that meet (from opposite sides) on a common boundary that we will call $W$.
The $\U(m)\times \U(n)$ gauge fields are coupled to a bifundamental hypermultiplet that lives on $W$.
In practice, the basic example we consider in this paper is simply that $M_\ell$ and $M_r$ are two half-spaces in
$\RR^4$, meeting along the codimension 1 linear subspace $W=\RR^3\subset \RR^4$.

Our main technical result is a formula with the same structure as eqn. (\ref{yetro}) but one crucial novelty. For a suitable choice of one of the supersymmetries $\Qb$,
such that $\Qb^2=0$, the action is the sum of a $\Qb$-exact term and a Chern-Simons interaction supported on $W$:
\begin{equation}\label{petro}I =\frac{i\calK}{4\pi} 
\int_{W}\Str\,\left(\AA\wedge\d \AA+\frac{2}{3}\AA\wedge\AA\wedge \AA\right)+\{\Qb,\dots\}. \end{equation}
(The $\Qb$-exact terms are a sum of terms supported on $M_\ell$, terms supported on $M_r$, and terms supported on $W$.)
Now, however, $\AA$ is a gauge field on $W$ whose structure group is the supergroup $\U(m|n)$, and the symbol $\Str$
represents a supertrace (with $\AA$ understood as a matrix-valued field acting on a $\Z_2$-graded vector space of dimension $m|n$).
The structure is clearer if we write $\AA$ in block-diagonal form:
\begin{equation}\label{blocks}\AA=\begin{pmatrix}\AA_{\U(n)} & \lambda \\ \tilde\lambda & \AA_{\U(m)}\end{pmatrix}.\end{equation}
Here $\AA_{\U(m)}$ is the complexified gauge connection on $M_\ell$, defined exactly as in eqn. (\ref{yetro}), ignoring the existence of branes and
fields on $M_r$.  And $\AA_{\U(n)}$ is defined in precisely the same way, now on $M_r$
rather than $M_\ell$.  In writing the Chern-Simons interaction in eqn. (\ref{petro}), we restrict $\AA_{\U(n)}$ and $\AA_{\U(m)}$
to $W$.  The off-diagonal blocks $\lambda$ and $\tilde\lambda$ are certain linear combinations of the fermionic fields contained
in the bifundamental hypermultiplet that lives on $W$, so in particular they are defined only on $W$.  Finally, the supertrace $\Str$ that appears in this formula is equivalent
to an ordinary trace when restricted to the Lie algebra of $\U(n)$, but to the negative of  a trace when restricted to the Lie algebra of $\U(m)$.  The minus
sign comes in because $M_\ell$ and $M_r$ end on $W$ with opposite orientations.  

The appearance of the supergroup $\U(m|n)$ in the  formula (\ref{petro}) may be surprising, but actually there were reasons to expect
this.  The field theory description of the D3-NS5 system at $\theta_\YM\not=0$ 
was analyzed in \cite{Janus} and shown to be closely related to a purely three-dimensional
theory with a Chern-Simons coupling and three-dimensional $\N=4$ supersymmetry (eight supercharges).  Moreover, it was shown that
such three-dimensional theories  are related to  supergroups.  The relationship with supergroups was somewhat mysterious, 
but was elucidated in \cite{KapustinSaulina}: a three-dimensional Chern-Simons theory with eight supercharges 
has a twisted version that is equivalent to Chern-Simons
theory for a supergroup.  This statement is proved via a three-dimensional  formula whose analog in  four dimensions is
our result (\ref{petro}).  Given the three-dimensional formula in \cite{KapustinSaulina}, it was natural to anticipate the
four-dimensional analog (\ref{petro}).

\subsection{Applications}\label{apps}

After we obtain the formula (\ref{petro}), the rest of this paper is devoted to applications, which 
 may be summarized as follows:

(1) Via the same $S$-duality and $T$-duality steps as in \cite{5knots}, we construct a magnetic dual of $\U(m|n)$ Chern-Simons
theory, and also an analog of Khovanov homology for this supergroup.  We analyze the behavior of line and surface operators
under these dualities.  We are able to get a reasonable picture, though the details are more involved than in the one-sided case and
a few details remain obscure.

(2) Our richest application, however, actually arises from an orientifold version of the whole construction.  In this
orientifold, the supergroup $\U(m|n)$ is replaced by an orthosymplectic group $\OSp(w|2s)$.  See
\cite{Baryons,HananyOrientifolds,FH,Evans,GWfour} for some basics concerning the relevant orientifolds.  One can again apply $S$-duality, possibly followed by $T$-duality, to find a magnetic dual description, and an analog
of Khovanov homology, for the orthosymplectic group.  

But an additional duality comes into play if $w$ is odd, say $w=2r+1$ for some $r\geq 0$. (This additional duality can also be
considered for $\U(m|n)$, but the result is not
 so interesting.)  Here we consider D3-branes interacting with a $(1,1)$-fivebrane
(or equivalently an NS5-brane with the theta-angle shifted by $2\pi$).  A $(1,1)$-fivebrane has the charges of
a composite of an NS5-brane and a D5-brane; it is invariant under a certain electric-magnetic duality operation.  
From this invariance, we deduce a duality for purely three-dimensional Chern-Simons theories (analytically
continued to non-integer values of the Chern-Simons level $k$).  This duality says that $\OSp(2r+1|2s)$ Chern-Simons
theory with coupling parameter $q$ is equivalent to $\OSp(2s+1|2r)$ Chern-Simons theory with coupling parameter $-q$.
The weak coupling region of Chern-Simons theory is $q\to 1$, while $q\to -1$ is a region of strong coupling.
So a duality that exchanges $q$ with $-q$ is a strong/weak coupling duality of Chern-Simons theory, not visible 
semiclassically.  Since the transformation $q\to -q$ is not compatible with the integrality of $k$, this duality has to be understood
as a statement about analytically-continued theories.

The novel duality mentioned in the last paragraph can also be combined with the more standard $S$- and $T$-dualities
in constructing an analog of Khovanov homology for $\OSp(2r+1|2s)$.  The upshot is that there are two closely related
theories that provide analogs of Khovanov homology for $\OSp(2r+1|2s)$.  Specialized to $s=0$, we suspect that these
theories correspond to what are usually called even and odd Khovanov homology for the odd orthogonal groups $\SO(2r+1)$
and their spin double covers (such as $ \mathrm{Spin}(3)\cong \SU(2)$, which is the most studied case). Odd Khovanov homology
was defined in \cite{ORS} and its relation to the orthosymplectic group was found from an algebraic point of view in \cite{Lauda}.

Everything we have said so far concerns a single NS5-brane interacting with D3-branes on the left and right.
What happens in the case of several parallel (and nonintersecting)
NS5-branes?  As we briefly explain in section \ref{quivers},  this case can be analyzed on the basis of the same ideas.
It leads to a certain analytic continuation of a product of supergroup Chern-Simons theories.

Knot and three-manifold invariants that are presumably related to those studied in the present paper have been previously
studied via quantum supergroups \cite{Zh,ZhTwo,ZhThree,Bluman}. Some previous work on supergroup Chern-Simons theories includes, in particular, \cite{Horne,VafaCS}.

This paper is organized as follows.
In section \ref{etheory}, we describe in detail the relation of the two-sided D3-NS5 system to supergroups.  Some complications are hard to avoid here, but
it  will be possible for the reader to understand the rest of the 
paper after only skimming section \ref{etheory}.  In section \ref{obs}, we analyze line and surface operators in this description.
In section \ref{magnetic}, we describe the $S$-dual D3-D5 system and describe line and surface operators in that description.
In section \ref{ortho}, we incorporate an O3 plane and discuss orthosymplectic groups and the novel duality that can arise in this case.
In section \ref{symbr}, we analyze a symmetry-breaking process that (for example) reduces $\U(m|n)$ to $\U(m-r|n-r)$.
And in section \ref{lifting}, we lift the magnetic description to a D4-D6 system and describe an analog of Khovanov homology
for supergroups. In particular, we describe candidates for odd and even Khovanov homology.  Some details are in appendices.

\section{Electric Theory}\label{etheory}
\subsection{Gauge Theory With An NS-Type Defect}\label{nstype}
As explained in the introduction, our starting point will be  
four-dimensional $\cN=4$ super Yang-Mills theory with a three-dimensional half-BPS defect. 
This theory can be defined in purely gauge-theoretic terms, but it will be 
useful to consider a brane construction, which gives a realization of the theory for  unitary and orthosymplectic gauge groups. 
We consider a familiar Type IIB setting \cite{HW}  of D3-branes interacting with an NS5-brane.  As sketched in 
 fig. \ref{setup} of the introduction, where we consider the horizontal direction to be parametrized by\footnote{Throughout the paper, notations $y$ and $x_3$ are used interchangeably for the same coordinate.} $y=x_3$,
 we assume 
that there are $m$ D3-branes and thus $\U(m)$ gauge symmetry for $y<0$ and $n$ D3-branes and thus $\U(n)$ gauge symmetry for $y>0$.
We take the NS5-brane to be 
 at $x_3=x_7=x_8=x_9=0$ and hence to be  parametrized by
$x_0,x_1,x_2$ and  $x_4,x_5,x_6$, while the semi-infinite D3-branes are 
parametrized by $x_0,x_1, x_2,x_3$.   With an orientifold projection, which we will introduce in section \ref{ortho}, the gauge groups become orthogonal and symplectic.  Purely from the point of view of four-dimensional field theory, there are other possibilities.

The theory in the bulk is $\cN=4$ super Yang-Mills, and it is coupled to some three-dimensional bifundamental hypermultiplets, which live on the defect at $y=0$ and come from the strings that join the two groups of D3-branes.   The bosonic fields of the theory are the gauge fields $A_i$, the scalars $\vec X$ that
describe motion of the D3-branes along the NS5-brane (that is, in the $x_4,x_5,x_6$ directions), and scalars $\vec Y$ that describe the motion of the D3-branes
normal to the NS5-brane (that is, in the $x_7,x_8,x_9$ directions).  

The relevant gauge theory action, including the effects of the defect at $y=0$, has been constructed in the paper~\cite{Janus}. In this section we recall some facts about this theory, mostly without derivation. More detailed explanations can be found in the original paper~\cite{Janus} or in the more technical Appendix \ref{technical1} below, which is, however, not necessary for understanding the main ideas of the present paper.

The half-BPS defect preserves $\cN=4$ superconformal supersymmetry in the three-dimensional sense; the corresponding superconformal group is $\OSp(4|4)$.
It is important that there exists a one-parameter family of inequivalent embeddings of this supergroup into the superconformal group $\mathrm{PSU}(2,2|4)$
of the bulk four-dimensional theory. 
For our purposes, it will suffice to describe the different embeddings just from the point of view of 
global supersymmetry (rather than the full superconformal
symmetry).  
The embeddings differ by which global supersymmetries are preserved by the defect.
The four-dimensional bulk theory is invariant under the product $\U_0=\SO(1,3)\times \SO(6)_R$ of the Lorentz group $\SO(1,3)$ and the $R$-symmetry group $SO(6)_R$  (or more
precisely, a double cover of this associated with spin); this is a subgroup of $\mathrm{PSU}(2,2|4)$. 
The three-dimensional half-BPS defect breaks $\U_0$ down to a subgroup $\U=
\SO(1,2)\times \SO(3)_X\times \SO(3)_Y$; this is a subgroup of $\OSp(4|4)$. Here in ten-dimensional terms, the two factors $\SO(3)_X$ and $\SO(3)_Y$
of the unbroken $R$-symmetry subgroup act by rotations in the $456$ and $789$ subspaces, respectively. ($\SO(6)_R$ is broken to $\SO(3)_X\times \SO(3)_Y$ because
the NS5-brane spans the 456 directions.) Under $\U_0$, the global  supersymmetries transform
 in a real representation $({\bf 2,1, 4})\oplus ({\bf 1,2, \bar{4}})$. Under $\U$ this becomes $V_8\otimes V_2$, where $V_8$ is a real eight-dimensional representation $({\bf 2,2,2})$ and $V_2$ is a two-dimensional real vector space with trivial action of $\U$.
 An embedding of $\OSp(4|4)$ in $\mathrm{PSU}(2,2|4)$ can be fixed by specifying which linear combination of the two copies of $V_8$ is left unbroken
 by the defect; these unbroken supersymmetries are of the form $V_8\otimes \varepsilon_0$, where $\varepsilon_0$ is a fixed vector
 in $V_2$.   
 Up to an irrelevant scaling, the choice of $\varepsilon_0$  is parametrized by an angle that we will call $\ang$.   This angle in turn is determined by
  the 
  string theory coupling parameter $\tau=i/g_{\mathrm{st}}+\theta/2\pi$, which in field theory terms is  $\tau=\fr{4\pi i}{g_{\rm YM}^2}+\fr{\theta_{\rm YM}}{2\pi}$. The relation can
  be found  in the brane description, as follows. Let $\veps_1$ and $\veps_2$ be the two ten-dimensional spinors that parametrize supersymmetry transformations in the underlying Type IIB theory. They transform in the ${\bf 16}$ of the ten-dimensional Lorentz group $\Spin(1,9)$, so
\beq
\Gamma_{012\dots 9}\veps_{i}=\veps_{i},~~i=1,2,
\eeq
where $\Gamma_{012\dots 9}$ is the product of the $\SO(1,9)$ gamma-matrices $\Gamma_I$, I=0,\dots, 9. The supersymmetry that is preserved by the D3-branes is defined by the condition
\beq
\veps_2=\Gamma_{0123}\veps_1\,,\label{d3susy}
\eeq
while the NS5-brane preserves supersymmetries that satisfy
\beq\label{bufog}
\veps_1=-\Gamma_{012456}(\sin\ang\,\veps_1-\cos\ang\,\veps_2)\,,
\eeq
where the angle $\ang$ is related to the coupling parameter $\tau$ by
\beq
\ang=\arg(\tau).\label{psi1}
\eeq
(When $\cos\ang=0$, (\ref{bufog}) must be supplemented by an additional condition on $\veps_2$.)
Altogether the above conditions imply
\beq
(B_2\sin\ang+B_1\cos\ang)\veps_1=\veps_1\,,\label{epseq}
\eeq
where $B_1=\Gamma_{3456}$ and $B_2=\Gamma_{3789}$ are operators that commute with the group $\U$ and thus act naturally in the two-dimensional space $V_2$. The
solutions of this condition are of the form  $\veps_1=\veps\otimes\veps_0$, where $\veps$ is any vector in $V_8$, and $\veps_0$ is a fixed, $\vartheta$-dependent
vector in $V_2$.  These are the generators of the unbroken supersymmetries.

It will be useful to introduce a new real parameter $\calK$ and to rewrite (\ref{psi1}) as
\beq
\tau=\calK\cos\ang\,\ex^{i\ang}.\label{canonical}
\eeq
The motivation for the notation is that $\calK$ generalizes the level $k$ of purely three-dimensional Chern-Simons theory.
For physical values of the coupling $\tau$, one has ${\mathrm {Im}}\,\tau>0$; this places a constraint on the
 variables $\calK$ and $\ang$. In the twisted topological field theory, $\calK$ will turn out to be what was called the canonical parameter $\Psi$
 in~\cite{Langlands}.

In general, let us write $G_\ell$ and $G_r$ for the gauge groups to the left or right of the defect.  From a purely field theory point of view,
$G_\ell$ and $G_r$ are completely arbitrary and moreover arbitrary hypermultiplets may be present at $x_3=0$
 as long as $\mathrm{Re}\,\tau=\theta_{\rm YM}/2\pi$ vanishes.\footnote{The gauge couplings $\tau_{\ell,r}$ and the angles $\ang_{\ell,r}$ can also be different at $y<0$ and $y>0$, as long as the canonical parameter $\mathcal{K}$ in eqn. (\ref{canonical}) is  the same \cite{Janus}.  
For our purposes, this generalization is not important.} 
However, as soon as $\theta_\YM\not=0$, $G_\ell $ and $G_r$ and the hypermultiplet representation are severely constrained; to maintain supersymmetry, 
the product $G_\ell\times G_r$ must
be a maximal bosonic subgroup of a supergroup whose odd part defines the hypermultiplet representation and whose Lie algebra
admits an invariant quadratic form with suitable properties.  These rather mysterious conditions \cite{Janus} have 
been given a more natural explanation in a closely
related three-dimensional problem \cite{KapustinSaulina}; as explained in the introduction, our initial task is to generalize that explanation
to four dimensions.  
We denote the Lie algebras of $G_\ell$ and $G_r$ as $\gleft$ and $\gright$, and denote the Killing forms on these
Lie algebras as  $\kappa_\ell$ and $\kappa_r$;
precise normalizations will be specified later. We will loosely write $-\tr(\dots)$ for $\kappa_\ell$ or $\kappa_r$. We also need a form $\kappa=-\kappa_\ell+\kappa_r$ on the direct sum of the two Lie algebras. This will be denoted by $-\Tr(\dots)$. The gauge indices for $\gleft\oplus\gright$ will be denoted by Latin letters $m,n,p$.

As already remarked, from a field theory point of view, as long as $\theta_{\rm YM}=0$,
  the defect at $y=0$ might support a system of $N$ hypermultiplets transforming in an arbitrary real symplectic representation of $G_\ell\times G_r$.   A real symplectic representation of $G_\ell\times G_r$
is a $4N$-dimensional real representation of $G_\ell\times G_r$, equipped with an action of $\SU(2)$ that commutes with $G_\ell\times G_r$. (In the context of the supersymmetric gauge theory, this $\SU(2)$ will become part of the R-symmetry group, as specified below.) This representation can be conveniently described as follows. Let $\R$ be a complex $2N$-dimensional symplectic representation of $G_\ell\times G_r$, with an invariant two-form $\omega_{IJ}$. We take the sum of two copies of this representation, with an $\SU(2)$ group acting on the two-dimensional multiplicity space, and impose a $G_\ell\times G_r\times\SU(2)$-invariant reality condition. This gives the desired $4N$-dimensional real representation. We denote indices valued in $\R$ as $I,J,K$, we write
$T^I_{mJ}$ for the $m^{th}$ generator of $G_\ell\times G_r$ acting in this representation, and we set  $\tau_{mIJ}= T^{S}_{mI}\omega_{SJ}$, which is symmetric in $I,J$ (and is related to the moment map for the action of
$G_\ell\times G_r$ on the hypermultiplets). As remarked above, for  $\theta_\YM\not=0$, the representation $\R$ is highly constrained. It turns out that a supersymmetric action for our system
with $\theta_\YM\not=0$  can be constructed if and only if 
\beq\label{urti}
\tau_{m(IJ}\tau_{K)Sn}\kappa^{mn}=0.
\eeq
This condition is equivalent \cite{Janus} to the fermionic Jacobi indentity for a superalgebra 
$\sg$, which has bosonic part $\gleft\oplus\gright$, with fermionic 
generators transforming in the representation $\mathcal R$ and with $\kappa\oplus\omega$ being an invariant and
nondegenerate graded-symmetric bilinear form on $\sg$; we will sometimes write this form
as $-{\rm Str}(\dots)$.  Concretely, if we denote the fermionic generators 
of $\sg$ as $f_I$, then the commutation relations of the superalgebra are 
\beqn
&&[T_m,T_n]=f_{mn}^s T_s\,,\nnr
&&[T_m,f_I]=T^K_{mI}f_K\,,\label{superalgebra}\\
&&\{f_I,f_J\}=\tau_{mIJ}\kappa^{mn}T_n.\nonumber
\eeqn
A short though admittedly mysterious calculation shows that the Jacobi identity for this algebra is precisely (\ref{urti}).
As already remarked, the closest to an intuitive explanation of this result has been provided
in \cite{KapustinSaulina}, in a related three-dimensional problem.   We will write $\SG$ for the supergroup with 
superalgebra $\sg$.

In more detail, the $\R$-valued hypermultiplet that lives on the defect consists of  scalar fields  
$\Q^{I\dA}$ and  fermions $\lambda_\alpha^{IA}$ that transform in the  representation $\R$ of the gauge group, and  transform respectively
as $({\bf 1,1,2})$ and $({\bf 2,2,1})$ under $\U=\SO(2,1)\times \SO(3)_X\times \SO(3)_Y$.   (Here $A,B=1,2$
are indices for the double cover $\SU(2)_X$ of $\SO(3)_X$, and $\dot A,\dot B$ are similarly related to $\SO(3)_Y$.) They are subject to a reality condition, which e.g. for the scalars reads  $\left(Q^I_\dA\right)^\dagger=\eps^{\dA\dB}\omega_{IJ}Q^J_\dB$.
To describe the coupling of the bulk fields to the defect theory, it is convenient to rewrite the bulk super 
Yang-Mills fields in three-dimensional language. 
The scalars $X^a$ and $Y^{\dot a}$, $a, \dot a=1,\dots,3$, transform
in the vector representations of $\SO(3)_X$ and $\SO(3)_Y$, respectively, and of course the gauge field $A_i$ is $\SO(3)_X\times \SO(3)_Y$ singlet.
The super Yang-Mills 
gaugino field $\Psi$  transforms in the representation $({\bf 2,1,4})\oplus({\bf 1,2,\bar{4}})$ of $\U_0$.  Under the subgroup $\U$, it splits into two spinors 
$\Psi_{1\alpha}^{A\dB}$ and $\Psi_{2\alpha}^{A\dB}$, which transform in the representation $(\bf{2,2,2})$, like the supersymmetry 
generator $\veps_\alpha^{A\dB}$. More precisely, we define
\beq
\Psi=-\Psi_2\otimes B_1\veps_0+\Psi_1\otimes B_2\veps_0.
\eeq
With this definition, it is straightforward to decompose the supersymmetry transformations of the 
four-dimensional super Yang-Mills to find the transformations that correspond~to~$\veps\otimes\veps_0$.  In particular, the bosons transform as 
\beqn
&&\delta A_i=\fr{1}{\sqrt{2}}\veps_{\alpha A\dB}\sigma_{i\beta}^\alpha\left(\Psi_1^{A\dB\beta}\sin\ang+\Psi_2^{A\dB\beta}\cos\ang\right)\,,\nnr
&&\delta X^a=-\fr{i}{\sqrt{2}}\veps^{A\alpha}_\dB\Psi_{1\alpha}^{B\dB}\sigma^a_{AB}\,,\nnr
&&\delta Y^{\da}=\fr{i}{\sqrt{2}}\veps_A^{\dA\alpha}\Psi_{2\alpha}^{A\dB}\sigma^{\da}_{\dA\dB}.\label{susy0}
\eeqn
Here $i,j,k$ and $\alpha,\beta$ are respectively vector and spinor indices of the three-dimensional Lorentz group $\SO(2,1)$, and $\sigma_i$ 
are the Pauli matrices. See Appendix A for some details on our conventions.

The action of the theory has the following form:
\beq
I_{\rm electric}=I_{\rm SYM}-\fr{\theta_{\rm YM}}{2\pi}\CS(A)+\calK I_{\mathrm{hyp}}.\label{act0}
\eeq
The terms on the right are as follows. $I_{\rm SYM}$ is the usual action of the $\cN=4$ super Yang-Mills in the bulk. 
The term proportional to $\theta_\YM$ reflects the bulk ``topological'' term of four-dimensional  Yang-Mills theory
\beq I_{\theta_\YM}=-\frac{\theta_\YM}{8\pi^2}\int_{x_3<0}\tr \,F\wedge F-\frac{\theta_\YM}{8\pi^2}\int_{x_3>0}\tr \,F\wedge F,\label{bulk}\eeq
which we have split into two contributions at $y<0$ and $y>0$ because in the present context the gauge field (and even the gauge
group)  jumps discontinuously at $y=0$.  Because of this discontinuity, even if we restrict ourselves to variations that are trivial
at infinity, $I_{\theta_\YM}$ has a nontrivial variation supported 
on the locus $W$ defined by $y=0$.  This variation is the same as that of $(\theta_\YM/2\pi)\CS(A)$, where $\CS(A)$
is the Chern-Simons interaction of $G_\ell\times G_r$:
\beq
\CS(A)=\fr{1}{4\pi}\int_W\Tr\left(A\wedge dA+\fr23 A\wedge A\wedge A\right).\label{ulk}
\eeq
(Recall that the symbol $\Tr$ includes the contributions of both $G_\ell$ and $G_r$, but with opposite signs.)
We lose some information when we replace $I_{\theta_\YM}$ by $(\theta_\YM/2\pi)\CS(A)$, since $I_{\theta_\YM}$ is gauge-invariant as a real number,
but $\CS(A)$ is only gauge-invariant modulo  an integer.  However, the replacement of $I_{\theta_\YM}$ by $(\theta_\YM/2\pi)\CS(A)$
is a convenient shorthand.   Finally, $I_{\mathrm{hyp}}$
 is the part of the action that involves the hypermultiplets.  
 More details concerning the action  are given in the Appendix \ref{technical1}.

We also need some facts about the boundary conditions and supersymmetry transformations in this theory. 
The bulk scalars $Y_{\dot a}$ obey a 
 Dirichlet type boundary condition.  In terms of $Y^m_{\dA\dB}=\sigma^{\dot a}_{\dA\dB}Y^{\dot a m}$, this boundary condition is
\beq
Y^m_{\dA\dB}=-\fr{1}{2\cos\ang} \tau^m_{IJ}\Q^I_{\dA}\Q^J_{\dB}.\label{Ybc1}
\eeq
In the brane picture, this boundary condition reflects the fact that  the fields $Y^\da$  describe displacement of the D3-branes from the NS5-brane in the 789 directions, and so vanish at $y=0$ if the hypermultiplets vanish.   Notice that, depending on whether $m$ labels a generator
of $G_\ell$ or $G_r$, the field $Y^m_{\dA\dB}$ is defined for $y\leq 0$ or for $y\geq 0$; but the boundary condition (\ref{Ybc1}) is valid
in both cases. A similar remark applies for other formulas below.
 Boundary conditions for other fields can be obtained from (\ref{Ybc1}) by  $\cN=4$ supersymmetry transformations, or 
by ensuring the vanishing of boundary contributions in the variation of the action. For the gauge fields, the relevant part of the action is
\beq
\fr{1}{2g_{\rm YM}^2}\int{\rm d}^4x\,\tr\,F_{\mu\nu}^2-\fr{\theta_{\rm YM}}{8\pi^2}\int\tr\, F\wedge F+\calK I_{\mathrm{hyp}}.\label{UNgauge}
\eeq
Taking the variation and reexpressing the coupling constant using (\ref{canonical}), one gets on the boundary
\beq
\sin\ang\,F^m_{k3}-\fr{1}{2}\cos\ang\,\eps_{kij}F^m_{ij}=\fr{2\pi}{\cos\ang} J^m_k\,,\label{Fphysbc}
\eeq
where $J_{mk}={\delta I_{\mathrm{hyp}}}/{\delta A^m_k}$ is the hypermultiplet current, and gauge indices are raised and lowered by the form $\kappa$.  There is a similar boundary condition for the $X^a$ scalar which we shall not write explicitly here. By making supersymmetry transformations (\ref{susy0}) of the equation (\ref{Ybc1}), one can also find the boundary condition for the bulk fermions,
\beq
\sqrt{2}\Psi^m_{2\alpha A\dB}=\fr{i}{\cos\ang} \tau^m_{IJ}\lambda^I_{\alpha A}Q^J_{\dB}.\label{bc0}
\eeq

It was shown in \cite{Janus} that this four-dimensional problem with a half-BPS defect is closely related to a purely three-dimensional Chern-Simons theory with three-dimensional $\N=4$ supersymmetry.  A three-dimensional Chern-Simons theory with $\N=3$ supersymmetry exists with arbitrary gauge group
and hypermultiplet representation, but with $\N=4$ supersymmetry, one needs precisely the constraints stated above: the gauge group $G$
is the bosonic part of a supergroup $\SG$, and the hypermultiplet representation corresponds to the odd part of the Lie algebra of $\SG$.
To compare the action of the 
four-dimensional model with the defect to the action of the purely three-dimensional model, we first decompose the hypermultiplet
action in (\ref{act0}) as
\beq  I_{\mathrm{hyp}} =I_{Q}(A)+I_{\mathrm{hyp}}',\label{decompose} \end{equation}
where $I_{Q}(A)$ is the part of the hypermultiplet action that contains couplings to no bulk fields except $A$, and $I'_{\mathrm{hyp}}$ contains
the couplings of hypermultipets to the bulk scalars and fermions.  (For details, see Appendix \ref{technical1}.)  
In these terms, the action of the purely three-dimensional theory  is
\beq -\calK\left(\CS(A)+I_{Q}(A)\right)   \label{threed}\eeq
while the contribution to the four-dimensional action at $y=0$ is
\beq
-\fr{\theta_{\rm YM}}{2\pi}\CS(A)-\calK\left(I_{Q}(A)+I'_\hyp\right).   \label{fogy}
\eeq
Thus, there are several  differences:  the defect part (\ref{fogy}) of the four-dimensional action contains the extra couplings in $I'_\hyp$,
and it has a different coefficient of the Chern-Simons term than that which appears in the purely three-dimensional action (\ref{threed}); also,
in (\ref{threed}), $A$ is a purely three-dimensional gauge field while in (\ref{fogy}), it is the restriction of a four-dimensional gauge field to $y=0$.
There also are differences in the supersymmetry transformations.
The supersymmetry transformations in the purely three-dimensional Chern-Simons theory are schematically
\beq
\delta A\sim \veps\lambda \Q\,,\label{zomb}
\eeq
In the four-dimensional theory with the defect, the transformation for the gauge field in (\ref{susy0}) is schematically 
\beq
\delta A\sim \veps(\Psi_1 +\Psi_2).\label{deltaAapprox}
\eeq
Clearly, the two formulas (\ref{zomb}) and (\ref{deltaAapprox}) do not coincide. 
With the help of the boundary condition (\ref{bc0}), we see that the $\Psi_2$ term in (\ref{deltaAapprox}),
when restricted to $y=0$, has the same form as the purely three-dimensional transformation law (\ref{zomb}).
The term involving $\Psi_1$ cannot be interpreted in that way; rather, before comparing
the four-dimensional theory with a defect to a purely three-dimensional theory, one must  redefine the connection $A$ in
a way that will eliminate the $\Psi_1$ term.    In section \ref{tt}, 
 generalizing the ideas in \cite{5knots} and in 
 \cite{KapustinSaulina}, we will explain how to reconcile the different formulas.

\subsection{Relation To Chern-Simons Theory Of A Supergroup}\label{tt}
\subsubsection{Topological Twisting}\label{tw}
After making a Wick rotation to  Euclidean signature on $\RR^4$, we want to  select a scalar supercharge $\Qb$, obeying $\Qb^2=0$,
in such a way that if we restrict to the cohomology of $\Qb$, we get a topological field theory.   As part of the mechanism to 
achieve topological invariance, we require $\Qb$
to be invariant under a twisted action of the rotations of $\RR^4$, that is, under rotations combined with suitable $R$-symmetries.
In Euclidean signature, the rotation and $R$-symmetry groups are the two factors of $\U_0^{E}=\SO(4)\times \SO(6)_R$, and the symmetries preserved by the defect are $\U^{E}=\SO(3)\times \SO(3)_X\times \SO(3)_Y$. The twisting relevant to our problem is the same procedure used in studying
 the geometric
Langlands correspondence  via gauge theory \cite{Langlands}. We pick a subgroup $\SO(4)_R\subset \SO(6)_R$, and define $\SO'(4)\subset U_0'$ to
be a diagonal subgroup of $\SO(4)\times \SO(4)_R$, such that from the ten-dimensional point of view, $\SO'(4)$ acts by simultaneous rotations in the 0123 and 4567 directions.     The space of ten-dimensional supersymmetries  transforms as 
 $({\bf 2,1,4})\oplus({\bf 1,2,\bar{4}})$ under $\U_0^E=\SO(4)\times \SO(6)_R\cong \SU(2)\times \SU(2)\times \SO(6)_R$.  Each summand has a one-dimensional  $\SO'(4)$-invariant subspace; this follows
 from the fact that the  representations ${\bf 4}$ and ${\bf \bar{4}}$ of $\SO(6)_R$ both decompose as $({\bf 2,1})\oplus ({\bf 1,2})$ under  $\SO'(4)$.
The two invariant vectors coming from $({\bf 2,1,4})$ and $({\bf 1,2,\bar{4}})$ give two supersymmetry parameters $\veps_\ell$ and $\veps_r$ with definite $\SO(4)$ chiralities.  Although there is no natural way to normalize $\veps_\ell$, there is a natural way\footnote{One sets $\veps_r=\sum_{\mu=0}^3\Gamma_{4+\mu,\mu}\veps_\ell/4$, as in eqn. (3.8) of \cite{Langlands}.} to define $\veps_r$ in terms of
$\veps_\ell$ and one can take $\Qb$ to be any linear combination $b\veps_\ell+a\veps_r$.  We only care about $\Qb$ up to scaling, so the relevant parameter
is $t=a/b$. 

In the bulk theory, we can make any choice of $t$, but in the 
 presence of the half-BPS defect, we must choose a supercharge that is preserved by the defect.
 As in section \ref{nstype}, the space of supersymmetries decomposes under $\U^E$ as $V_8\otimes V_2$, where $V_8$ transforms as
 $(\2,\2,\2)$, and $\U^E$ acts trivially on $V_2$.  (In Euclidean signature, the vector spaces $V_8$ and $V_2$ are not real.)
 The defect preserves supersymmetry generators of the form $\veps \otimes\veps_0$ with any $\veps\in V_8$ and with a fixed $\veps_0\in V_2$.
 Invariance under $\SO'(4)$ restricts to a 1-dimensional subspace of $V_8$, as explained in the next paragraph. 
So up to scaling, only one linear combination of $\veps_\ell$ and $\veps_r$ is preserved by the defect,
and $t$ is uniquely determined. 

To find the scalar supersymmetry generator in three-dimensional notation, we note that at $y=0$, $\SO'(4)$ can be naturally
restricted to $\SO'(3)$, which is a diagonal subgroup of  $\SO(3)\times \SO(3)_X\subset \SO(4)\times \SO(4)_R$. 
An $\SO'(3)$-invariant vector in $V_8$ must have the form
\beq
\veps_{\rm top}^{\alpha A\dA}=\eps^{\alpha A}v^\dA,\label{superparameter}
\eeq
where $\alpha, A,\dA=1,2$ label bases of the three factors of $V_8\sim \2\otimes \2\otimes \2$;
 $\eps^{\alpha A}$ is the antisymmetric symbol; and  $v^\dA$, which
 takes values in the $\2$ of $\SO(3)_Y$, 
 is not constrained by $\SO'(3)$ invariance.  However, $v^\dA$
  is determined up to scaling by $\SO'(4)$ invariance. In fact, for any particular $v^\dA$, the supersymmetry
parameter defined in eqn. (\ref{superparameter}) is invariant under a twisted rotation group  that  pairs the 0123 directions with 456${\bf v}$, where ${\bf{v}}^{\dot a}
\sim \bar v\sigma^{\dot a} v$ is some direction in the subspace 789 (here $\sigma^{\dot a}$ are the Pauli matrices). 
For $\SO'(4)$ invariance, we want to choose $v^\dA$ such that ${\bf v}$ is the direction $x_7$. A simple way to do that is to look at the $\U(1)_F$ symmetry subgroup of $\SO(3)_Y$ that rotates the 89 plane and commutes with
$\SO'(4)$; thus, $\U(1)_F$ rotates the last two components of $\vec Y=(Y_1,Y_2,Y_3)$.
We normalize the generator $F$ of $\U(1)_F$ so that the field $\sigma=\fr{Y_2-iY_3}{\sqrt{2}}$ has charge 2. Then using a standard representation of the
$\sigma^{\dot a}$, one has
\beq
Y^{\dA}_{\dB}\equiv Y^\da\sigma^{\dA}_{\da\dB}=i\left(
\bea{cc}
Y_1 & \sqrt{2}\,\sigma\\
\sqrt{2}\,\bar{\sigma}&-Y_1
\eea\right), \label{tork}
\eeq
and in this basis, the generator $F$ is
\beq
\left(\bea{cc}
1& 0\\
0&-1\eea\right).\label{urk}
\eeq
$\SO'(4)$ invariance implies that the supersymmetry parameter $\veps$ has charge $-1$ under $F$ (see eqn. (3.11) in \cite{Langlands}), so we can take
\beq
v^{\dA}= 2^{1/4}\,\ex^{-i\ang/2} \left(\bea{c} 0\\ 1 \eea\right).\label{v}
\eeq
The normalization factor here is to match the conventions of \cite{Langlands}. For future reference, we also define
\beq
u^\dA=2^{3/4}\,\ex^{i\ang/2}\left(\bea{c} 1\\0\eea\right).\label{u}
\eeq

We also will need the relation between the parameter $t$ and the angle $\ang$. For that, we use equation (2.26) from \cite{5knots} for the topological parameter $\veps_\ell+t\veps_r$. Comparing it to our eqn.~(\ref{epseq}), we find that 
\beq
t=\ex^{i(\pi-\ang)}.\label{tphys}
\eeq

In the  twisted theory, the  fields $\vec X$ and  $Y_1$  join into a one-form $\phi=\sum_{\mu=0}^3\phi_\mu \,d x^\mu$,
 with components $\phi_i=X_{i+1}$, $i=0,1,2$,  and $\phi_3=Y_1$.  $\Qb$-invariance (or more precisely the condition $\{\Qb,\zeta\}=0$ for
 any fermionic field $\zeta$) gives a system of equations for $A_\mu$ and $\phi_\mu$.  These equations, which have been extensively discussed in
 \cite{5knots}, take the form $\V^+=\V^-=\V^0=0$, with
 \begin{align}
\mathcal{V}^+&=\left(F-\phi\wedge\phi+t\,\d_A\phi\right)^+,\cr
\mathcal{V}^-&=\left(F-\phi\wedge\phi-t^{-1}\d_A\phi\right)^-, \label{localization}\cr
\mathcal{V}^0&=D_\mu\phi^\mu.
\end{align}
Here if $\Lambda$ is a two-form, we denote its selfdual and anti-selfdual projections as $\Lambda^+$ and $\Lambda^-$, respectively.

\subsubsection{Fields And Transformations}\label{relation}

If a four-dimensional gauge theory with a defect is related to a purely
three-dimensional theory on the defect, then what are the fields in the effective three-dimensional theory?
The hypermultiplets supported at $y=0$ give one obvious source of three-dimensional fields.   
So let us first discuss  these fields from the standpoint of the twisted theory.

The hypermultiplet contains 
scalar fields $\Q^{I\dA}$ that  transform as a doublet under $\SU(2)_Y$.  In the twisted theory,
$\SU(2)_Y$ is reduced to $\U(1)_F$, and accordingly we decompose the $\Q^{I\dA}$  in
multiplets $C^I$ and $\bar{C}^I$ with charges
 $\pm 1$ under $\U(1)_F$. (These are upper and lower components in the basis used in (\ref{urk}).)
 The fermionic part of the hypermultiplet $\lambda_\alpha^{AI}$ 
 has a more interesting decomposition in the twisted theory.  Under $\SO'(3)$, both the spinor index $\alpha$ and
 the $\SO(3)_X$ index $A$ carry spin $1/2$, so $\lambda_\alpha^{AI}$ is a sum of pieces of spin 1 and spin 0.
In other words, the fermionic part of the hypermultiplet decomposes into a vector $\cAf_{i}^I$ and a scalar $B^I$. 
 
 The supercharge $\Qb$ generates the following transformations of these fields:
 \begin{align}
\delta\cAf&=-\cDb C\,,   \cr
\delta C&=0\,,\cr
\delta \bar{C}&=B\,,\cr
\delta B&=\fr{1}{2}[\{C,C\},\bar{C}].\label{Q2}
\end{align}
Here for any field $\Phi$, we define $\delta\Phi=[\Qb,\Phi\}$, where $[~,~\}$ is a commutator or anticommutator for $\Phi$ bosonic
or fermionic; also, $\cDb$ is the coveriant derivative with a connection $\cAb$ that we define momentarily.

 The vector $\cA_{f}$ will become the fermionic part of the $\sg$-valued gauge field, which we will call $\cA$.
But where will we find $\cAb$, the bosonic part of $\cA$?  There is no candidate among the fields that are supported on the defect.
Rather,  $\cAb$  will be the restriction to the defect worldvolume  of a linear combination of bulk fields:
\beq
\cAb=A+i(\sin\ang) \phi.
\eeq
This formula defines both a $\mathfrak{g}_\ell$-valued part of $\cAb$ -- obtained by restricting $A+i(\sin\ang)\phi$ from $y\leq 0$ to $y=0$
-- and a $\mathfrak{g}_r$-valued part -- obtained by restricting $A+i(\sin\ang)\phi$ from $y\geq 0 $ to $y=0$.  (Here $\mathfrak g_\ell$ and $\mathfrak g_r$
are the Lie algebras of $G_\ell$ and $G_r$.)
The shift from $A$ to $\cAb$ 
removes the unwanted term with $\Psi_1$ in the topological supersymmetry variation~(\ref{deltaAapprox}), so that -- after restricting
to $y=0$ and using the boundary condition (\ref{bc0}) -- one gets 
\beq
\delta\cAb=\{C,\cAf\}.\label{Q1}
\eeq
Obviously, since $\cAf$ is only defined at $y=0$, $\delta\cAb$ can only be put in this form at $y=0$.

The interpretation of the formulas (\ref{Q2}) and (\ref{Q1}) was explained in \cite{KapustinSaulina} (where they arose in a purely three-dimensional context):
one can interpret $C$ as the ghost field for a partial gauge-fixing of the supergroup $\SG$ down to its maximal bosonic subgroup $G$,
and the supercharge $\Qb$ as the BRST operator for this partial gauge-fixing. 
Since $C$ has $\U(1)_F$ charge of 1, we should interpret $\U(1)_F$ as the ghost number.
Once we interpret $C$ as a ghost field, the transformation laws for $\cAb$ and $\cAf$ simply combine to say that acting on $\cA=\cAb+\cAf$,
$\Qb$ generates the BRST transformation $\delta\cA=-d_\cA C$ with gauge parameter $C$.  The gauge parameter $C$ has opposite
statistics from an ordinary gauge generator (it is a bosonic field but takes values in the odd part of the super Lie algebra $\sg$);
this is standard in BRST gauge-fixing of a gauge theory.   In such BRST gauge-fixing, one often introduces BRST-trivial multiplets
$(\bar C,B)$, where $\delta\bar C=B$ and $\delta B$ is whatever it must be to close the algebra.  In the most classical case, $\bar C$
is an antighost field, with $\U(1)_F$ charge $-1$, and $B$ is called a Lautrup-Nakanishi auxiliary field.  The multiplet $(\bar C,B)$ in (\ref{Q2})
has precisely this form.

If one finds a gauge transformation in which the gauge parameter has reversed statistics to be confusing,  one may wish to introduce a 
formal Grassman parameter $\eta$ and write $\delta'=\eta\delta$, so that for any field $\Phi$, $\delta'\Phi=[\eta \Qb,\Phi]$; $\eta \Qb$ is bosonic,
so there is an ordinary commutator here.  Then
\beq
\delta'\negthinspace \cA = -\mathcal{D}(\eta C),
\eeq
showing that the symmetry generated by $\eta \Qb$
 transforms the supergroup connection $\cA$ by a gauge transformation with the infinitesimal gauge parameter $\eta C$, which has normal statistics.

\subsubsection{The Action}\label{action}

After twisting, one can define the $\N=4$ super Yang-Mills theory on an arbitrary\footnote{If $M$ is not orientable, one
must interpret $\phi$ not as an ordinary 1-form but as a 1-form twisted by the orientation bundle of $M$.}
 four-manifold $M$, with the defect supported
on a three-dimensional oriented submanifold $W$.  Generically, in this generality, one preserves only the unique supercharge $\Qb$.

What is the form of the $\Qb$-invariant action of this twisted theory?   Any gauge-invariant expression $\{\Qb,\cdot\}$ is $\Qb$-invariant,
of course -- and also largely irrelevant as long as we calculate only $\Qb$-invariant observables, which are the natural observables
in the twisted theory.  But in addition, any gauge-invariant function of the complex connection $\cA$ is $\Qb$-invariant,
since $\Qb$ acts on $\cA$ as the generator of a gauge transformation.  $\cA$ is defined only on the oriented three-manifold $W$,
and as we are expecting to make  a topological field theory, the natural gauge-invariant function of $\cA$ is the 
Chern-Simons function.

Given this and previous results (concerning the case that there is no defect \cite{Langlands}, an analogous purely three-dimensional problem
\cite{KapustinSaulina}, and the case that the fields are nonzero only on one side of $W$ \cite{5knots}),  it is natural to suspect that the action of the twisted theory on $M$ may have the form
\beq
I=i\calK\,\CS(\cA)+\{\Qb,\dots\}=\fr{i\calK}{4\pi}\int_W \Str\left(\cA d\cA+\fr{2}{3}\cA^3\right)+\left\{\Qb,\dots\right\},\label{act2}
\eeq
where if there is a formula of this type, then the coefficient of $\CS(\cA)$ must be precisely $i\calK$, in view of what is already known
about the one-sided case.

This is indeed so.  Leaving some technical details for  Appendix \ref{technical1}, we simply make a few remarks here.
In the absence of a defect, and assuming that $M$ has no boundary, it was shown in \cite{Langlands} that the action of the twisted
super Yang-Mills theory is $\Qb$-exact modulo a topological term:
\beq
I_{\mathrm{SYM}}+\fr{i\theta_{\rm YM}}{8\pi^2}\int_M\tr\left(F\wedge F\right)=\fr{i\mathcal{K}}{4\pi}\int_M\tr\left(F\wedge F\right)+\left\{\Qb,\dots\right\}.
\label{zog}\eeq
(On the left, $I_{\mathrm{SYM}}$ is the part of the twisted super Yang-Mills
action that is proportional to $1/g_\YM^2$; the part proportional to $\theta_\YM$ is written out
explicitly.)  In \cite{5knots}, the case that $M$ has a boundary $W$ (and the D3-branes supported on $M$ end on an NS5-brane wrapping
$T^*W$) was analyzed.  It was shown that (\ref{zog}) remains valid, except that the topological term $\int_M \tr\, F\wedge F$ must be
replaced with a Chern-Simons function on $W=\partial M$, not of the real gauge field $A$ but of its complexification $\cAb$.
From the point of view of the present paper, this case means that $M$ intersects the NS5-brane worldvolume in a defect $W$,
and there are gauge fields only on one side of  $W$.  Part of the derivation of eqn. (\ref{act2}) is simply to use the identity (\ref{zog}) on both
$M_\ell$ and $M_r$, thinking of the integral of $\tr\,F\wedge F$ over $M_\ell$ or $M_r$ as a Chern-Simons coupling on the boundary.

To get the full desired result,
 we must include also the hypermultiplets $\Q$ that are supported on $W$. The full action of the theory was described in formulas (\ref{act0}) and (\ref{decompose}). In Euclidean signature it reads 
\beq
I_{\rm electric}=I_{\rm SYM}+\frac{i\theta_\YM}{2\pi}\CS(A)+\calK (I_Q(A)+I'_{\rm hyp}).
\eeq
The identity (\ref{zog}) has a generalization that includes the boundary terms:
\beq
I_{\rm electric}=i\calK\,\left( \CS_{G_r}(\AAb)-\CS_{G_\ell}(\AAb)\right)+\calK I_{Q}(\cAb) + \{\Qb,\dots\}.\label{zoob}
\eeq
Since the first three terms are defined purely on the three-manifold $W$, we can now invoke the result of \cite{KapustinSaulina}: 
this part of the action  is $i\calK \CS(\AA)+\{\Qb,\dots\}$, where now $\CS_{SG}(\AA)$ is the  Chern-Simons function for the full supergroup
gauge field $\cA=\cAb+\cAf$, and   the $\Qb$-exact terms  describe partial gauge-fixing from $\SG$ to $G$. 
This confirms the validity of eqn. (\ref{act2}).

We conclude by clarifying the meaning of the supergroup Chern-Simons function $\CS(\cA)$.  With $\cA=\cAb+\cAf$,
we have
\beq\CS(\cA)=\CS(\cAb)+\frac{1}{4\pi}\int_W \Str \, \cAf d_{\cAb} \cAf.  \label{hordo}\eeq
The term involving $\cAf$ is the integral over $W$ of a function with manifest 
gauge symmetry under $G_\ell\times G_r$ (and even its complexification).  It is not affected by the usual subtleties of the Chern-Simons function involving gauge transformations that are not
homotopic to the identity.  The reason for this is that the supergroup $\SG$ is contractible to its maximal bosonic subgroup $G$;
the topology is entirely contained in $G$.  Similarly, with 
$\AAb=A+i(\sin\ang)\phi$, we can expand the complex Chern-Simons function,
\beq \CS(\AAb)= \CS(A)+\frac{1}{4\pi}\int_W\Tr \,\left(i(\sin\ang) \phi\wedge  F-(\sin^2\ang)\phi\wedge\d_A\phi-i(\sin^3\ang)\phi\wedge\phi
\wedge\phi\right),  \label{zert}\eeq
and the topological subtleties affect only the first term $\CS(A)$ . 
Here, as in eqns. (\ref{ulk}) and (\ref{bulk}), to resolve the topological subtleties and put the action in a form that is well-defined for generic $\calK$, we should replace
$\CS(A)$ with the corresponding volume integral $(1/4\pi)\int_M \tr F\wedge F$.  There is no need for such a substitution in any of the other
terms, since they are all integrals over $W$ of gauge-invariant functions.  All this reflects the fact that a complex Lie group is contractible to a maximal
compact subgroup, so the topological subtlety in $\CS(\AAb)$ is entirely contained in $\CS(A)$.

It is convenient to simply write the action as $i\calK\, \CS(\AA)+\{\Q,\dots\}$, as we have done in eqn. (\ref{act2}), rather than always
explicitly replacing the term $\CS(A)$ in this action with a bulk integral.

\subsubsection{Analytic Continuation}\label{anacon}

To get the formula   (\ref{Q1}) along $W$, we have had to replace  $A$ by $A+i(\sin\ang)\phi$, with the result 
that the bosonic part of $\cA$ is complex-valued.  
This is related to an essential subtlety \cite{Wittenold,Wittenoldone} in the relation of the four-dimensional theory with a defect to a Chern-Simons
theory supported purely on the defect.  In general, four-dimensional $\N=4$ super Yang-Mills theory on a four-manifold $M$,
with a half-BPS defect of the type analyzed here on a three-manifold $W\subset M$, is not equivalent to standard Chern-Simons theory
on $W$ with gauge supergroup $\SG$, but to an analytic continuation of this theory.  The basic idea of this analytic continuation
is that localization on the space of solutions of the equations (\ref{localization}) defines an integration cycle in the complexified path
integral of the Chern-Simons theory.  This localization is justified using the fact that the $\Qb$-exact terms in (\ref{zoob}) can be scaled
up without affecting $\Qb$-invariant observables, so the path integral can be evaluated just on the locus where those terms vanish.  The condition
for these terms to vanish is the localization equations (\ref{localization}), which define the integration cycle.  (Thus, the integration cycle is characterized
by the fact that $\AAb$ is the restriction to $y=0$ of fields $A,\phi$ which obey the localization equations and have prescribed behavior for $y\to\pm\infty$.)

For generic $W$ and $M$, the integration cycle derived from $\N=4$ super Yang-Mills theory
differs from the standard one of three-dimensional Chern-Simons theory.
For the important case that $W=\RR^3$, there is essentially only one possible
integration cycle and therefore the two constructions are equivalent.  Thus,
after including Wilson loop operators (as we do in section \ref{obs}), the four-dimensional construction can be used to study the
usual knot invariants associated to three-dimensional Chern-Simons theory. 

Unfortunately, it turns out that for supergroups all the observables which can be defined using only closed Wilson loops in $\RR^3$ reduce to 
observables of an ordinary bosonic Chern-Simons theory.  We explain this in sections \ref{obs} and \ref{symbr}. 
To find novel observables, one needs to do
something more complicated.  All of the options seem to introduce some complications in the relation to four dimensions.  For example, one
can replace $\RR^3$ by $S^3$ and define  observables that appear to be genuinely new 
 by considering the path integral with insertion of a Wilson loop in a typical
representation (see section \ref{linerev}).   But the compactness of $S^3$ means that one encounters infrared questions in comparing to four dimensions.   Because of such complications, our results for  supergroup Chern-Simons theory are less complete then in the case of a bosonic Lie group.

A feature of the localization that is special to supergroups is that $\cAb$ is the boundary value of a four-dimensional field (which in the localization
procedure is constrained by the equations (\ref{localization})), but $\cAf$ is purely three-dimensional.  The reason that this happens is essentially
that the topology of the supergroup $\SG$ is contained entirely in its maximal  bosonic subgroup $G$.  Being fermionic, $\cAf$ is by nature
infinitesimal; the Berezin integral for fermions is an algebraic operation (a Gaussian integral in the case of Chern-Simons theory of a supergroup)
with no room for choosing different integration cycles.   By contrast, in the integration over the bosonic fields, it is possible to pick different
integration cycles and the relation to four-dimensional $\N=4$ super Yang-Mills theory does give a very particular one.

One important qualitative difference between purely three-dimensional Chern-Simons theory and what one gets by extension to four dimensions
is as follows.
 In the  three-dimensional theory, the ``level'' $k$ must be an integer, but in the analytically
continued version given by the relation to four-dimensional $\N=4$ super Yang-Mills, $k$ is generalized to a complex parameter
$\calK$.  Part of the mechanism for this is that although the Chern-Simons function $CS(A)$ is only gauge-invariant modulo 1, in the
four-dimensional context it can be replaced by a volume integral $\int_M \Tr\, F\wedge F$, which is entirely gauge-invariant, so there is no need to quantize the parameter.  

\subsubsection{Relation Among Parameters}\label{turogo} 

\def\sgn{{\mathrm{sign}}}
At first sight, eqn. (\ref{act2}) seems to tell us that the relation between the parameter $\calK$ in four dimensions
and the usual parameter $k$ of Chern-Simons theory, which appears in the purely three-dimensional action
\begin{equation}\label{threeac} i\frac{k}{4\pi}\int_W\Str\,\left(A\wedge \d A+\frac{2}{3}A\wedge A\wedge A\right),\end{equation}
would be $\calK=k$.  However, for the purely one-sided case, the relation, according to \cite{5knots}, is 
really\footnote{\label{rox} A careful reader will ask what precisely we mean by $k$ in the following formula.  In defining $k$ precisely,
we will assume that it is positive; if it is negative, one makes the same definitions after reversing orientations.
One precise definition is that $k$ is the level of a two-dimensional
current algebra theory that is related to the given Chern-Simons theory in three dimensions.  (The level is
defined as the coefficient of a $c$-number term appearing in the product of two currents.) 
Another precise definition  is that, for
integer $k$, the space of physical states of the Chern-Simons theory on a Riemann surface $\Sigma$
is $H^0(\mathcal M,\mathcal L^k)$, where
$\mathcal M$ is the moduli space of holomorphic $G$-bundles over $\Sigma$ and $\mathcal L$ generates the Picard group
of $\mathcal M$. (For simplicity, in this statement, we assume $G$ to be simply-connected.)}  \begin{equation}\label{pars}
\calK=k+h\,\,\mathrm{sign}\,(k). \end{equation}   An improved explanation of this is as follows.

The purely three-dimensional Chern-Simons theory for a compact gauge group $G$ 
involves a path integral over the space of real connections $A$.
This is an oscillatory integral and in particular, at one-loop level, in expanding 
around a classical solution, one has to perform an oscillatory Gaussian
integral.\footnote{The following is explained more fully on pp. 358-9 of \cite{WittenCS}, where however
a nonstandard normalization is used for $\eta$.  See also \cite{FreedGompf}.}  After diagonalizing the matrix that governs the fluctuations, the oscillatory Gaussian integral is a product of
one-dimensional integrals
\begin{equation}\label{oblo}\int_{-\infty}^\infty\frac{\d x}{\sqrt\pi}\exp(i\lambda x^2) =\frac{\exp(i(\pi/4)\sgn\lambda)}{|\lambda|},\end{equation}
where the phase comes from rotating the contour by $x=\exp(i(\pi/4)\sgn\lambda)x'$ to get a real convergent Gaussian
integral for $x'$.  In Chern-Simons gauge theory, the product of these phase factors over all modes of the gauge field
and the ghosts gives (after suitable regularization) a factor $\exp(i\pi \eta/4)$, where $\eta$ is the Atiyah-Patodi-Singer
$\eta$-invariant.  This factor has the effect
of shifting the effective value of $k$ in many observables to $k+h\,\,\sgn \,k$, where $h$ is the dual Coxeter number of $G$
(this formula is often written as $k\to k+h$, with $k$ assumed to be positive).  

One can think of the shift $k\to k+h\,\,\sgn \,k$ as arising in a Wick rotation in field space from the standard integration
cycle of Chern-Simons theory (real $A$) to an integration cycle on which the integral is convergent rather than
oscillatory.  But this is precisely the integration cycle that is used in the four-dimensional description (see \cite{Wittenold,Wittenoldone}).  Accordingly, in the four-dimensional
description, there is no one-loop shift in the effective value of $\calK$ and instead the shift must be absorbed in the relation
between parameters in the four- and three-dimensional descriptions by $\calK=k+h\,\,\sgn \,k$.

Up to a point, the same logic applies in our two-sided problem.  The four-dimensional path integral has no oscillatory phases
and hence no one-loop shift in the effective value of the Chern-Simons coupling.  So any such shift that would arise
in a purely three-dimensional description must be absorbed in the relationship between $\calK$ and a three-dimensional
parameter $k$.  We are therefore tempted to guess that the relationship between $\calK$ and the parameter $k$ of
a purely three-dimensional Chern-Simons theory of the supergroup $SG$ is
\begin{equation}\label{donox}\calK=k+h_{\sg}\,\sgn\,k, \end{equation}
where $h_\sg$ is the dual Coxeter number of the supergroup.  
The trouble with this formula is that it assumes that the effective Chern-Simons level for a supergroup has the same
one-loop renormalization as for a bosonic group. The validity of this claim is unclear for reasons explored in Appendix \ref{Anomalous}.
(In brief, the fact that the invariant quadratic
 form  on the bosonic part of the Lie superalgebra $\sg$ is typically not positive-definite means it is not clear what
 should be meant by $\sgn\,k$, and also means that a simple
 imitation of the standard one-loop computation of bosonic Chern-Simons theory does not give
the obvious shift $k\to k+h_\sg\,\sgn\,k$.)   We actually 
do not know the proper treatment of purely three-dimensional Chern-Simons
theory of a supergroup.  In this paper, we concentrate on the four-dimensional description, in which
the bosonic part of the path integral is convergent, not oscillatory, and accordingly there is no one-loop shift
in the effective value of $\calK$.  Thus we should just  think of $\calK$ as the effective parameter of the Chern-Simons
theory.

\def\g{{\frak g}}
Let us go back to the purely bosonic or one-sided case.  For $G$ simple and 
simply-laced, Chern-Simons theory is usually
parametrized in terms of
\begin{equation}\label{zeffo}q=\exp(2\pi i/(k+h\,\,\sgn \,k))=\exp(2\pi i/\calK). \end{equation}
If $G$ is not simply-laced, it is convenient to take $q=\exp(2\pi i/n_\g\calK)$, where $n_\g$ is the ratio of
length squared of long and short roots of $\g$.  Including the factor of $1/n_\g$ in the exponent ensures that $q$
is the instanton-counting parameter in a magnetic dual description.
Similarly, for a supergroup $SG$, we naturally parametrize
the theory in terms of
\begin{equation}\label{effo} q=\exp(2\pi i/n_\sg\calK), \end{equation}
where $n_\sg$ is the ratio of length squared of the longest and shortest roots of a maximal bosonic subgroup of $SG$,
computed using an invariant bilinear form on $\sg$ (for the supergroups we study in this paper, $n_\sg$ can be 1, 2, or 4).
To write this formula in terms of a purely three-dimensional parameter $k$, we would have to commit ourselves
to a precise definition of such a parameter.  Each of the definitions given for bosonic groups in footnote
\ref{rox} may generalize to supergroups, but in neither case is the proper generalization immediately clear.

\subsubsection{Quivers}\label{quivers}

\begin{figure}
 \begin{center}
   \includegraphics[width=4.5in]{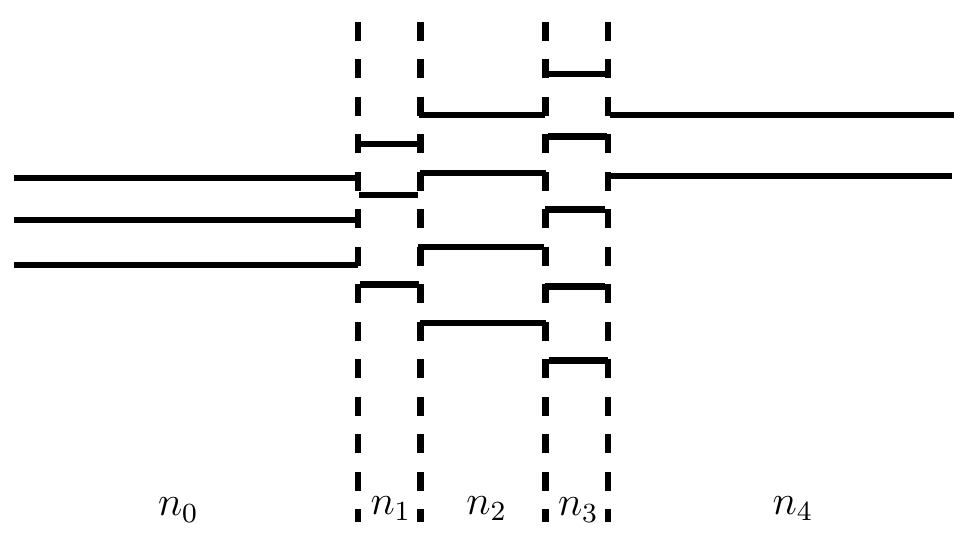}
 \end{center}
\caption{\small A ``quiver'' associated to a chain of $w$ NS5-branes, with $n_0,\dots,n_w$ D3-branes in the regions bounded by the NS5-branes; $w=4$ in
the example sketched here.}
 \label{quiver}
\end{figure}

This paper is devoted primarily to the study of D3-branes interacting on both sides of a single fivebrane.  But there is no difficulty in extending the basic ideas
to the case of any number of (nonintersecting) fivebranes, as we will now briefly indicate.

Consider $M=\RR^4$ with an arbitrary number $w$ of parallel codimension 1 defects $W_1,\dots, W_w$, dividing $M$ into $w+1$ regions (including two
semi-infinite regions at infinity) as indicated in fig \ref{quiver}.  We suppose that there are $n_i$ D3-branes, for $i=0,\dots,t$, in the $i^{th}$ region,
and that the defects represent intersections of $M$ with NS5-branes.  The same topologically twisted version of $\N=4$ super Yang-Mills theory that we have
studied in the presence of one defect makes sense in the presence of any number of defects, and the same arguments as before show that modulo $\{\Qb,\dots\}$,
it describes an analytic continuation of Chern-Simons gauge theory of the supergroup $\U(n_0|n_1)\times \U(n_1|n_2)\times \dots \times \U(n_{w-1}|n_w)$, with
a common coupling parameter $\calK$.  The integration cycle of this supergroup Chern-Simons theory is given by solving the equations (\ref{localization}) on the complement
of the defects.  This cycle is not simply a product of integration cycles for the individual factors of the gauge group.  The difference is unimportant as long as we
consider only knots in $\RR^3$, but it is significant in general.

Other constructions that will be presented later in this paper also have fairly evident generalizations to the case of any number of fivebranes.  For
instance, there is no difficulty in including any number of impurities in the dual magnetic description that is presented in section \ref{magnetic}.

A variant of this construction is possible if $n_0=n_w$, so that the number of D3-branes is the same at each end of the chain.
Then we can compactify the $x_3$ direction to a circle, replacing $W\times\RR$ with $W\times S^1$, and we get a framework
related to analytic continuation of $\U(n_1|n_2)\times \U(n_2|n_3)\times\dots\times \U(n_w|n_1)$ Chern-Simons theory.  In the special
case of only one NS5-brane, this reduces to $\U(n|n)$. 

\section{Observables In The Electric Theory}\label{obs}
The most important observables in ordinary  Chern-Simons gauge theory are Wilson line operators, labeled by representations of the gauge group.
To understand their analogs in supergroup Chern-Simons theory, we need to know something about representations of supergroups.
 The theory of Lie supergroups has some distinctive features, compared to the ordinary Lie group case, and these special features have  implications for Chern-Simons theory and its line observables. Accordingly, 
 we devote section \ref{superrev} to a brief review of Lie supergroups and superalgebras. Then in section \ref{throb}, we discuss the peculiarities of line observables in  three-dimensional supergroup Chern-Simons theory. In sections \ref{line4d} and \ref{surface}, we return to the four-dimensional construction, and explain, in fairly close parallel with \cite{5knots}, how  line operators of supergroup Chern-Simons theory are realized as line or surface operators in  $\cN=4$ super Yang-Mills theory. Finally, in section \ref{various} we summarize some unclear points.

In the four-dimensional construction, in addition to the line and surface operators considered here, it is possible to construct $\Qb$-invariant local
operators. They are described in Appendix \ref{finicky}.

\subsection{A Brief Review Of Lie Superalgebras}\label{superrev}
We begin with the basics of Lie superalgebras, Lie supergroups, and their representations. For a much more complete exposition see e.g. \cite{Kac2,Supersigma,Dictionary}.

A Lie superalgebra decomposes into its bosonic and fermionic parts,  $\sg=\gbos+\gferm$. We will assume that $ \gbos$ is a reductive Lie algebra (the sum
of a semi-simple Lie algebra and an abelian one). Moreover, to define the supergroup gauge theory action, we need the superalgebra  $\sg$ to possess a non-degenerate invariant bilinear form. (This also determines a superinvariant volume form on the $SG$ supergroup manifold.)
Finite-dimensional Lie superalgebras with these properties are  direct sums of some basic examples.  These include 
the unitary and the orthosymplectic superalgebras, as well as a one-parameter family of deformations of $\mathfrak{osp}(4|2)$, and two exceptional superalgebras, as specified in Table~\ref{superalgebras}. 
For the unitary Lie superalgebras, one can also restrict to the supertraceless matrices $\mathfrak{su}(m|n)$, and for $m=n$ further factor by the one-dimensional center down to $\mathfrak{psu}(n|n)$. 
In what follows, by a Lie superalgebra we mean a superalgebra from this list.\footnote{We avoid here using the term ``simple superalgebra,'' since, e.g., $\mathfrak{u}(1|1)$ is not simple (it is solvable), but is perfectly suitable for supergroup Chern-Simons theory. Let us mention that  Lie superalgebras with 
the properties we have required
which in addition are simple are called basic classical superalgebras.} 

Though we use real notation in denoting superalgebras, for instance in writing $\mathfrak{u}(m|n)$ and not $\mathfrak{gl}(m|n)$, we never
really are interested in choosing a real form on the full superalgebra. One  reason for this is that we will actually be  studying analytically-continued versions of supergroup
Chern-Simons theories.   If one considers all possible integration
cycles, then the real form is irrelevant. More fundamentally, as we have already explained in section \ref{anacon}, to define a path integral for supergroup Chern-Simons theory, one needs to pick a real integration cycle for the bosonic fields, but one does not need anything like this for the fermions. Correspondingly, we might need a real structure on $\gbos$ (and this will generally be the compact form) but not on the full supergroup or the superalgebra. So for our purposes,
 a three-dimensional Chern-Simons theory is naturally associated to a so-called $cs$-supergroup, which is a complex Lie supergroup together with a choice of  real form for its bosonic subgroup. 

If we choose the compact form of a maximal bosonic subgroup of a supergroup $SG$, 
then one can calculate the volume of $SG$ with respect to its superinvariant measure.   This volume has the following significance in Chern-Simons theory.
The starting point in Chern-Simons perturbation theory on a compact three-manifold is to expand around the trivial flat connection; in doing so one has
to divide by the volume of the gauge group.
But this volume is actually\footnote{\label{volumes} A quick proof is as follows.  Let $SG$ be a Lie supergroup whose maximal bosonic subgroup
is compact (this assumption ensures that there are no infrared subtleties in defining and computing the volume of $SG$). Suppose
that there is a fermionic generator $\mathcal C$ of $\frak{sg}$ with the property that $\{\mathcal C,\mathcal C\}=0$.  Such a $\mathcal C$ exists for every Lie supergroup except $\OSp(1|2n)$.  We view $\mathcal C$ as generating a supergroup ${\mathrm F}$ of dimension $0|1$, which we consider to act on $SG$ on (say) the left.
This gives a fibration $SG\to SG/{\mathrm F}$ with fibers ${\mathrm F}$.  The volume of $SG$ can be computed by first integrating over the fibers of
the fibration.  But the volume of the fibers is 0, so (given the existence of $\mathcal C$) the volume of $SG$ is 0.  The volume of the fibers
is 0 because, since $\{\mathcal C,\mathcal C\}=0$, there are local coordinates in which the fibers are parametrized by an odd variable $\psi$
and $\mathcal C=\partial/\partial\psi$.  $\mathcal C$-invariance of the volume then implies that the measure for integration over $\psi$
is invariant under adding a constant to $\psi$; the volume of the fiber is therefore $\int\d\psi\cdot 1=0$. } 0 for any Lie supergroup whose maximal bosonic subgroup is compact, 
with the exception of B$(0,n)=\OSp(1|2n)$.  This fact is certainly one reason that one cannot
expect to develop supergroup Chern-Simons theory by naively imitating the bosonic theory.

Another difference between ordinary groups and supergroups is that in the supergroup case, we have to distinguish between irreducible representations
and indecomposable ones.  A representation $R$ of $\sg$ is called irreducible if it does not contain a non-trivial $\sg$-invariant subspace $R_0$,
and it is called indecomposable if it cannot be decomposed as $R_0\oplus R_1$ where $R_0$ and $R_1$ are non-trivial $\sg$-invariant subspaces.  In
a reducible representation, the representation matrices are block triangular $\begin{pmatrix}*&*\cr 0&*\end{pmatrix}$, while in a decomposable representation,
they are block diagonal.
For ordinary reductive Lie algebras, these notions coincide (if the matrices are block triangular, there is a basis in which 
they are block diagonal), but for Lie superalgebras as defined above, they do not coincide, with the sole exception of B$(0,n)$.  It is not a coincidence
that B$(0,n)$ is an exception to both statements; a standard way to prove that a reducible representation of a compact Lie 
group  is also decomposable involves averaging over the
group, and this averaging only makes sense because the volume is nonzero.  For B$(0,n)$, taken with the compact form of its maximal bosonic subgroup,
the same proof works, since the volume is not zero.
  A physicist's explanation of the ``bosonic'' behavior of B$(0,n)$ might be that, as we argue later, the Chern-Simons theory 
  with this gauge supergroup is dual to an ordinary bosonic Chern-Simons theory with the gauge group $\SO(2n+1)$.  
  This forces B$(0,n)$ to behave somewhat like an ordinary bosonic group.

\begin{table}[t]
\beq
\bea{l|c|c|c}\toprule
{\rm superalgebra} & {\rm bosonic~part} & {\rm fermionic~part} & {\rm type}\\ \midrule
\mathfrak{u}(m|n)                        & \mathfrak{u}(m)\oplus\mathfrak{u}(n)       & (m,\bar{n})\oplus(\bar{m},n) &{\rm I}\\ \hline
B(m,n)\simeq \mathfrak{osp}(2m+1|2n)     & \mathfrak{so}(2m+1)\oplus\mathfrak{sp}(2n) & (2m+1,2n) &{\rm II}\\ \hline
C(n+1)\simeq \mathfrak{osp}(2|2n)        & \mathfrak{u}(1) \oplus \mathfrak{sp}(2n)   & (1,2n)\oplus(\bar{1},2n) & {\rm I}\\ \hline
D(m,n)\simeq \mathfrak{osp}(2m|2n), m>1  &\mathfrak{so}(2m)\oplus \mathfrak{sp}(2n)   & (2m,2n) &{\rm II}\\ \hline
D(2,1;\alpha), \alpha\in\CC\setminus\{0,-1\}                           &\mathfrak{so}(4)\oplus\mathfrak{sp}(2)\simeq \mathfrak{su}(2)\oplus\mathfrak{su}(2)\oplus\mathfrak{su}(2) & (2,2,2) &{\rm II} \\ \hline
G(3)                                     &\mathfrak{su}(2)\oplus \mathfrak{g}_2      & (2,7) &{\rm II}\\ \hline
 F(4)                                     &\mathfrak{su}(2)\oplus \mathfrak{so}(7)     & (2,8) &{\rm II}\\ \bottomrule
\eea\nonumber
\eeq
\caption{\label{superalgebras} \small Lie superalgebras suitable for the supergroup Chern-Simons theory. (We do not list explicitly the subquotients of the unitary superalgebra, which are mentioned in the text.)}
\end{table}

\begin{figure}
 \begin{center}
   \includegraphics[width=15cm]{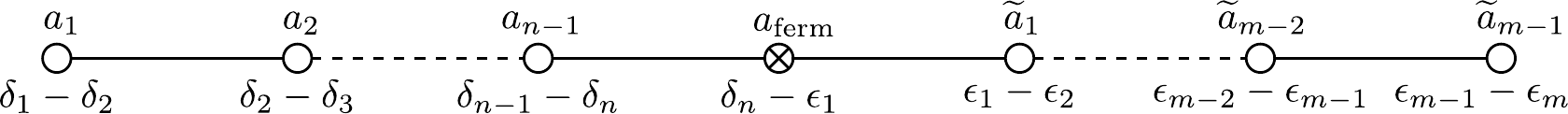}
 \end{center}
 \caption{\small Dynkin diagram for the $\mathfrak{su}(m|n)$ superalgebra. The subscripts are expressions for the roots in terms of the orthogonal basis $\delta_\bullet$, $\eps_\bullet$. The superscripts represent the Dynkin labels of a weight. The middle root denoted by a cross is fermionic.}
 \label{uDynkin}
\end{figure}

\begin{figure}
 \begin{center}
   \includegraphics[width=15cm]{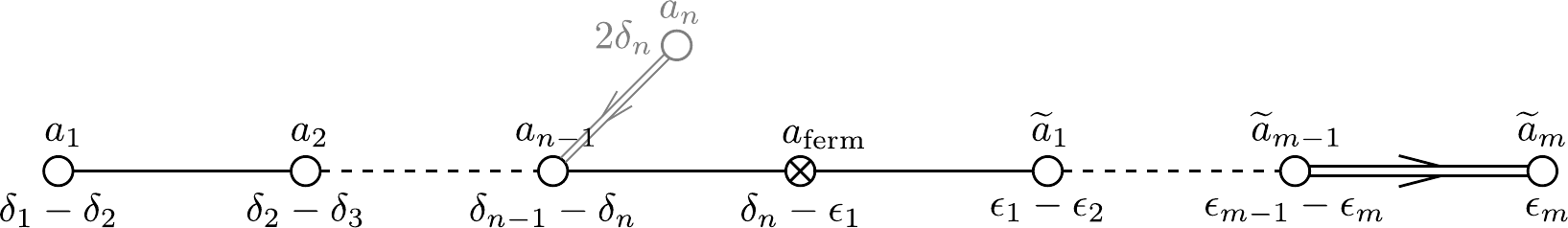}
 \end{center}
 \caption{\small Dynkin diagram for the $\mathfrak{osp}(2m+1|2n)$ superalgebra, $m\ge1$. The subscripts are expressions for the roots in terms of the orthogonal basis $\delta_\bullet$, $\eps_\bullet$. The superscripts represent the Dynkin labels of a weight. The arrows point in the direction of a shorter root. The middle root denoted by a cross is fermionic. Roots of the $\mathfrak{sp}(2n)$ and $\mathfrak{so}(2m+1)$ subalgebras are on the left and on the right of the fermionic root. The site shown in grey and labeled
 $a_n$ is the long simple root of the $\mathfrak{sp}(2n)$ subalgebra, which does not belong to the set of simple roots of the superalgebra.}
 \label{ospDynkin}
\end{figure}

The structure theory for a simple Lie superalgebra $\sg$ can be described similarly to the case of an ordinary Lie algebra. One starts by picking a Cartan subalgebra $\mathfrak{t}$, which for our superalgebras is just a Cartan subalgebra of the bosonic part. Then one decomposes $\sg$ into root subspaces. These subspaces lie either in $\gbos$ or in $\gferm$, and the roots are correspondingly called bosonic or fermionic. Then one makes a choice of positive roots, or, equivalently, of a Borel subalgebra $\mathfrak{b}\supset\mathfrak{t}$. Unlike in the bosonic case, different Borel subalgebras can be non-isomorphic. However, there is a distinguished Borel subalgebra -- the one which contains precisely one simple fermionic root. This is the choice that we shall make. For each choice of Borel subalgebra,
one can construct a Dynkin diagram. The distinguished Dynkin diagrams for the unitary and the odd orthosymplectic superalgebras are shown in fig.~\ref{uDynkin} and fig.~\ref{ospDynkin}. 

The fermionic $\ZZ_2$-grading of a Lie superalgebra can be lifted (in a way that is canonical up to conjugacy) to a $\ZZ$-grading, which can be defined as follows. The subalgebra of degree zero is generated by the Cartan subalgebra together with the bosonic simple roots of the superalgebra. The fermionic simple root of the distinguished Dynkin diagram is assigned degree one. The grading for the other elements of the superalgebra is then determined by the commutation relations. This $\ZZ$-grading is defined by a generator of $\frak g_{\overline 0}$.  

For example, for the unitary superalgebra this element can be taken to be the central generator of $\frak{u}(n)$. The degree zero subalgebra in this case is just the bosonic subalgebra, while the fermions decompose as $\gferm\simeq\mathfrak{g}_{-1}\oplus\mathfrak{g}_1$. Another example would be the odd orthosymplectic superalgebra $\osp(2m+1|2n)$, for which the situation is slightly different. There exists a simple root of the bosonic subalgebra, which is not a simple root of the superalgebra, but rather is a multiple of a fermionic simple root, and therefore will not have degree zero. It is shown in grey in fig.\ref{ospDynkin}. The degree zero subalgebra consists of a semisimple Lie algebra $\mathfrak{sl}(n)\oplus \mathfrak{o}(2m+1)$ with the Dynkin diagram obtained from fig.\ref{ospDynkin} by deleting the fermionic node, plus a central $\mathfrak{u}(1)$. This central element is the generator of the $\ZZ$-grading. The bosonic subalgebra decomposes into degrees $\pm2$ and $0$, while the fermions again live in degrees $\pm1$.

More generally, for any superalgebra, the distinguished $\ZZ$-grading takes values  from $-1$ to $1$ or from $-2$ to $2$, and the superalgebras are classified accordingly as type I or type II. In a type I superalgebra, the bosonic subalgebra lies completely in degree $0$. The representation of $\gbos$ on the fermionic subalgebra $\gferm$ is reducible, and $\gferm$ decomposes into subspaces of degree $-1$ and $1$.  The unitary superalgebra is an example of a type I superalgebra.  For the type II superalgebras, the action of $\gbos$ on $\gferm$ is irreducible. Under the $\ZZ$-grading, the bosonic subalgebra decomposes as $\gbos\simeq \mathfrak{g}_{-2}\oplus \mathfrak{g}_0\oplus\mathfrak{g}_2$, and the fermions decompose as $\gferm\simeq\mathfrak{g}_{-1}\oplus\mathfrak{g}_1$. The $\osp(2m+1|2n)$ superalgebra is an example of the type II case. The type of a superalgebra is important for representation theory, and we indicate it in Table~\ref{superalgebras}.

We need to introduce some further notation. Let $\Dbos$ and $\Dferm$ be the sets of positive bosonic and fermionic roots, respectively, and let $\DfermI$ be the set of positive fermionic roots with zero length.  The length is defined using the invariant quadratic form on $\sg$, which we normalize in a standard way so that the length squared of the longest
root is 2.  A root of zero length is called isotropic; isotropic roots
are always fermionic.   It is convenient to expand the roots and the weights in terms of a vector basis $\delta_\bullet$ and $\epsilon_\bullet$, orthogonal with respect to the invariant scalar product, with $\langle \delta_i,\delta_i\rangle=-\langle \eps_j,\eps_j\rangle>0$. For example, the positive roots for the unitary superalgebra $\frak{su}(m|n)$ are
\beqn
&&\Dbos=\bigl\{\delta_i-\delta_{i+p}, \eps_j-\eps_{j+p}\bigr\},\quad i=1\dots n, \quad j=1 \dots m,\quad p>0,\nnr
&&\Dferm=\DfermI=\bigl\{\delta_i-\eps_{j}\bigr\}.
\eeqn
The quadratic Casimir operator is defined  using the invariant form on $\sg$ (normalized in the standard way). In this paper, by the dual Coxeter number $h$ we mean one-half of the quadratic Casimir in the adjoint representation.\footnote{This definition is different from the definition of \cite{KacW}.} For future reference, in Table \ref{coxeters} we collect the superdimension (the difference between the dimension of $\gbos$ and that of $\gferm$) and the dual Coxeter number for the unitary and orthosymplectic superalgebras.

\begin{table}[t]
\beq
\bea{c|c|c|c}
\toprule
{~}           &   \mathfrak{su}(m|n),~n,m\ge 0& \mathfrak{osp}(m|2n),~m\ge0,~n\ge1 &\mathfrak{so}(n)  \\
\midrule
{h}           &  n-m              & n-m/2+1 &  n-2                    \\
\hline
{{\rm (s)dim}}        &  (n-m)^2-1        & (2n-m)(2n-m+1)/2& n(n-1)/2  \\
\bottomrule
\eea\nonumber
\eeq
\caption{\label{coxeters} \small (Super)dimensions and dual Coxeter numbers.}
\end{table}

For a given Borel subalgebra, one defines the bosonic and fermionic Weyl vectors as
\beq
\rho_\bos=\fr{1}{2}\sum_{\alpha\in\Dbos}\alpha\,,\quad \rho_\ferm=\fr{1}{2}\sum_{\alpha\in\Dferm}\alpha\,,
\eeq
and the superalgebra Weyl vector as $\rho=\rho_\bos-\rho_\ferm$. The Weyl group of a superalgebra, by definition, is generated by reflections with respect to the even (that is, bosonic) roots.

\subsubsection{Representations}\label{reps}
The finite-dimensional irreducible representations are labeled by their highest weights. The weights can be parametrized in terms of Dynkin labels. For a weight $\Lambda$, the Dynkin label associated to a simple root $\alpha_i$ is defined as $\displaystyle a_i=\fr{2\langle \Lambda,\alpha_i\rangle}{\langle \alpha_i,\alpha_i\rangle}$, if the length of the root $\alpha_i$ is non-zero, and $a_i=\langle \Lambda,\alpha_i\rangle$, if the length of the root is zero.

For a type I superalgebra, the Dynkin diagram coincides with the diagram for the semisimple part of the bosonic subalgebra $\gbos$, if one deletes the fermionic root. The finite-dimensional 
superalgebra representations are labeled  by the same data as the representations of the bosonic subalgebra. For example, for the dominant weights of  $\mathfrak{su}(m|n)$ all the Dynkin labels, except $a_{\rm ferm}$, must be non-negative integers. The fermionic label can be an arbitrary complex number, if we consider representations of the superalgebra, or an arbitrary integer, if we want the representation to be integrable to a representation of the compact form of
the bosonic subgroup.

For a type II superalgebra, if one deletes the fermionic node of the Dynkin diagram (and the links connecting to it), one gets a diagram for the semisimple part of the degree-zero subalgebra $\mathfrak{g}_0\subset \gbos$. The long simple root of the bosonic subalgebra $\gbos$ is ``hidden'' behind the fermionic simple root, and is no longer a simple root of the superalgebra. This is illustrated in fig.~\ref{ospDynkin} for the B$(m,n)$ case. For us it will be convenient to parametrize the dominant weights in terms of the Dynkin labels of the bosonic subalgebra, so, for type II, instead of $a_{\rm ferm}$ we will use the Dynkin label with respect to the long simple root of $\gbos$. For example, for B$(m,n)$ this label is\footnote{Our notation here is slightly unconventional: notation $a_n$ is usually used for what we call $a_{\rm ferm}$.} $a_n$, as shown on the figure, and the weights will be parametrized by $(a_1,\dots,a_n,\tilde{a}_1,\dots,\tilde{a}_m)$. Clearly, in this case for the superalgebra representation to be finite-dimensional, it is necessary for these Dynkin labels to be non-negative integers. It turns out that there is an additional supplementary condition. For example, for B$(m,n)$ this condition says that if $a_n<m$, then only the first $a_n$ of the labels $(\tilde{a}_1,\dots,\tilde{a}_m)$ can be non-zero. For the other type~II superalgebras the supplementary conditions can be found e.g. in Table 2 of \cite{Kac2}. The finite-dimensional irreducible representations are in one-to-one correspondence with  integral dominant weights that satisfy these extra conditions.

For a generic highest weight, the irreducible superalgebra representation can be constructed rather explicitly. For a type I superalgebra, one takes an arbitrary representation R$_\Lambda^0$ of the bosonic part $\gbos$, with highest weight $\Lambda$. A representation of the superalgebra can be induced from R$_\Lambda^0$ by setting the raising fermionic generators $\mathfrak{g}_1$ to act trivially on R$_\Lambda^0$, and the lowering fermionic generators $\mathfrak{g}_{-1}$ to act freely. The resulting representation in the vector space 
\beq
\mathcal{H}_\Lambda=\wedge^\bullet\mathfrak{g}_{-1}\times{\rm R}_\Lambda^0\label{tensor}
\eeq
is called the Kac module. For a generic highest weight, this gives the desired finite-dimensional irreducible representation. For a type II superalgebra, the representation can be similarly induced from a representation of the degree-zero subalgebra $\mathfrak{g}_0\subset\gbos$, but the answer is slightly more complicated than (\ref{tensor}), since the fermionic creation or annihilation  operators do not anticommute among themselves.

The Kac module, which one gets in this way, is irreducible only for a sufficiently generic highest weight. In this case, the highest weight $\Lambda$ and the representation are called typical. Typical representations share many properties of representations of  bosonic Lie algebras, e.g., a reducible representation
with a typical highest weight is always decomposable, and there exist simple analogs of the classical Weyl character formula for their characters and supercharacters.

However, if $\Lambda$ satisfies the equation 
\beq
\langle \Lambda+\rho, \alpha\rangle=0\,,\label{acondition}
\eeq
for some isotropic root $\alpha\in\DfermI$, then the Kac module acquires a null vector. The irreducible representation then is a quotient of the Kac module by a maximal submodule. Such weights and representations are called atypical. Let $\Delta(\Lambda)$ be the subset of $\DfermI$ for which (\ref{acondition}) is satisfied. The number of roots in $\Delta(\Lambda)$ is called the degree of atypicality of the weight and of the corresponding representation.

The maximal possible degree of atypicality of a dominant weight is called the defect of the superalgebra. For $\mathfrak{u}(m|n)$, 
for a dominant $\Lambda$ all the roots in $\Delta(\Lambda)$ are mutually orthogonal, and therefore the maximal number of such isotropic roots is min$(m,n)$. In the corresponding IIB brane configuration, this is the number of D3-branes which can be recombined and removed 
from the NS5-brane. (This symmetry breaking process is analyzed in
section \ref{symbr}.)

A Kac-Wakimoto conjecture \cite{KacW,SerganovaSD} states that the superdimension of a finite-dimensional irreducible representation is non-zero if and only if it has maximal atypicality. (For ordinary Lie algebras and for B$(0,n)$, the maximal atypicality is zero, and all  representations should be considered as both typical and maximally atypical.)

\subsubsection{The Casimir Operators And The Atypical Blocks}\label{ablocks}
The Casimir operators, by definition, are invariant polynomials in the generators of $\sg$; in a fancier language, they generate the center $\mathfrak{Z}$ of the universal enveloping algebra $\mathcal U(\sg)$. We introduce some facts about them, which will be useful for the discussion of Wilson lines.

There is a well-known formula for the value of the quadratic Casimir in a representation with highest weight $\Lambda$,
\beq
c_2(\Lambda)=\langle\Lambda+\rho,\Lambda+\rho\rangle-\langle\rho,\rho\rangle\,,\label{c2}
\eeq
which continues to hold in the superalgebra case. A remote analog of this formula for the higher 
Casimirs is known as the Harish-Chandra isomorphism (see e.g. \cite{Humphreys}), which we now briefly review.

By the Poincar\'e-Birkhoff-Witt theorem, a Casimir element $c\in\mathfrak{Z}$ can be brought to the normal-ordered form, 
where in the Chevalley basis, schematically, $c=\sum (E^-)^{k_1}H^{k_2}(E^+)^{k_1}$. When acting on the highest weight vector 
of some representation, the only non-zero contribution comes from the purely Cartan part. This gives a homomorphism 
$\hat{\xi}: \mathfrak{Z}\rightarrow S(\mathfrak{t})$, where $S(\mathfrak{t})$ are the symmetric polynomials in elements of $\mathfrak{t}$, 
and the value of the Casimir in a representation $R_\Lambda$ with highest weight $\Lambda$ is evaluated as $c(\Lambda)=(\hat{\xi}(c))[\Lambda]$. 
Here the square brackets mean the evaluation of a polynomial from $S(\mathfrak{t})$ on an element of $\mathfrak{t}^*$. By making appropriate shifts of the Lie algebra generators in the polynomial $\hat{\xi}(c)$, one can define a different polynomial $\xi(c)$, such that the formula becomes
\beq
c(\Lambda)=(\xi(c))[\Lambda+\rho]\,.
\eeq
This is a minor technical redefinition, which will be convenient.

For  ordinary Lie algebras, the Harish-Chandra theorem states that the image of the homomorphism $\xi$ consists of the Weyl-invariant polynomials 
$S_W(\mathfrak{t})\subset S(\mathfrak{t})$, and $\xi$ is actually an isomorphism of commutative algebras $\mathfrak{Z}\simeq S_W(\mathfrak{t})$. 
To summarize, the Casimirs can be represented by Weyl-invariant Cartan polynomials, and their values in a representation $R_\Lambda$ are 
obtained by evaluating these polynomials on $\Lambda+\rho$.

In the superalgebra case, the Harish-Chandra isomorphism \cite{KacHC} identifies $\mathfrak{Z}$ with a subalgebra 
$S^0_W(\mathfrak{t})\subset S_W(\mathfrak{t})$, consisting of Weyl-invariant polynomials $p$ with the following invariance property,
\beq
p[\Lambda+\rho+{\rm x}\alpha]=p[\Lambda+\rho]
\eeq
for any ${\rm x}\in\CC$ and $\alpha\in\Delta(\Lambda)$.

For a highest weight representation $R_\Lambda$, the corresponding set of eigenvalues of the Casimir operators (equivalently, a homomorphism from $\mathfrak{Z}$ into the complex numbers) is called the central character, denoted $\chi_\Lambda$. The Harish-Chandra isomorphism allows one to describe the sets of weights
which share the same central character. If the weight is typical, then the other weights with the same central character can be obtained by the shifted Weyl action $\Lambda\rightarrow w(\Lambda+\rho)-\rho$. The orbit of this transformation can contain no more than one dominant weight; therefore, two different typical finite-dimensional representations have different central characters. This is no longer the case for the atypical weights. Given an atypical dominant weight $\Lambda$, we can shift it by a linear combination of elements of $\Delta(\Lambda)$ to obtain new dominant weights with the same central character. More generally, we can
 apply a sequence of shifts and Weyl transformations without changing the central character. All the representations that are obtained in this way will have the same degree of atypicality, and they will share the same eigenvalues of the Casimir operators. The set of atypical finite-dimensional representations which have a common central character is called an atypical block. In this paper, we are interested mostly in the irreducible representations, and, somewhat imprecisely,\footnote{This
 phrasing is imprecise because it does not take account the difference between reducibility of a representation and decomposability.}  by an atypical block we will usually mean a set of irreducible representations (or, equivalently, dominant weights) with the same central character.

\begin{figure}
 \begin{center}
   \includegraphics[width=15cm]{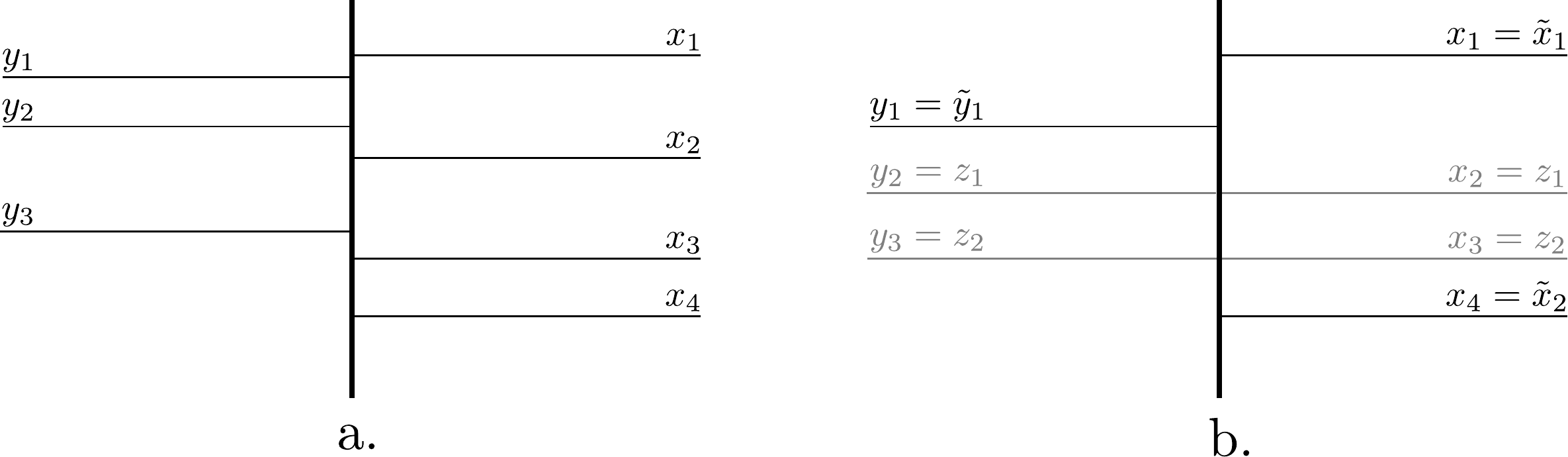}
 \end{center}
 \caption{\small Examples of dominant weights for $\mathfrak{u}(3|4)$. {\large a.} A typical weight. {\large b.} A weight of atypicality two, which
 is part of a block of atypical weights.
 The block is labeled by $\tilde{x}_1$, $\tilde{x}_2$, and  $\tilde{y}_1$, which correspond to a dominant weight of $\mathfrak{u}(1|2)$.
 The weights that make up  this block 
  are parametrized by $z_1$ and $z_2$, which can be thought of as labels of a maximally atypical weight of $\mathfrak{u}(2|2)$.}
 \label{uweights}
\end{figure}

As an example, let us describe the atypical blocks for the $\u(m|n)$ superalgebra. It is convenient to parametrize a weight $\Lambda$ as
\beq
\Lambda+\rho=\sum_{i=1}^n x_i \delta_i-\sum_{j=1}^m y_{j}\eps_{m+1-j}.
\eeq
For $\Lambda$ to be dominant, the two sequences $\{x_i\}$ and $\{y_j\}$ must be strictly increasing, and satisfy an appropriate integrality condition. A dominant weight can be represented graphically, as shown in  fig.~(\ref{uweights}a).  This is essentially the weight diagram of \cite{veracaro}.
The picture shows an obvious analogy between a dominant weight of $\u(m|n)$ and a vacuum of a brane system; we will develop this analogy in section \ref{gso}.  This description also confirms that dominant weights of $\u(m|n)$ correspond  to dominant
weights of the purely bosonic subalgebra $\u(m)\times \u(n)$.  In this correspondence, of the two central generators of $\u(m)\times \u(n)$, one
linear combination corresponds to the fermionic root $a_{\mathrm {ferm}}$ of $\frak{su}(m|n)$ and the other to the center of $\u(m|n)$.

For atypicality $r$, the set $\Delta(\Lambda)$ consists of $r$ isotropic roots $\delta_{i_l}-\eps_{j_l}$, $l=1\dots r$, which are mutually orthogonal, that is, each basis vector $\delta_\bullet$ or $\eps_\bullet$ can appear no more than once.\footnote{Suppose that in $\bigl\{\delta_1-\eps_1, \delta_1-\eps_2\bigr\}$ the vector $\delta_1$ appears more than once. Then, by taking a difference, we would get that $\langle \Lambda+\rho,\eps_1-\eps_2\rangle=0$, which contradicts the assumption that $\Lambda$ is dominant.} The atypicality condition (\ref{acondition}) then says that $r$ of the $x$-labels are ``aligned'' with (equal to) the $y$-labels. Let these labels be $x_{i_l}=y_{m+1-j_l}\equiv z_l$, $l=1\dots r$, and the others be $\tilde{x}_1,\dots,\tilde{x}_{n-r}$, $\tilde{y}_1,\dots,\tilde{y}_{m-r}$. Then the atypical blocks of atypicality $r$ are labeled by the numbers $\tilde{x}_\bullet$ and $\tilde{y}_\bullet$, which can be thought of as labels for a dominant weight of $\u(m-r|n-r)$, and the weights inside the same atypical block are parametrized by a sequence $z_\bullet$, which can be thought of as a dominant maximally atypical weight of $\u(r|r)$. An example is shown in fig.~(\ref{uweights}b). An atypical block is described by the following statement: the category of finite-dimensional representations (not necessarily irreducible) from the same atypical block of atypicality $r$ is equivalent to the category of maximally atypical representations of $\u(r|r)$ from the atypical block, which contains the trivial representation \cite{veracaro}. A similar statement holds for the orthosymplectic superalgebras; the role of $\u(r|r)$ is played by $\osp(2r|2r)$, $\osp(2r+1|2r)$ or $\osp(2r+2|2r)$.

\subsection{Line Observables In Three Dimensions}\label{throb}
\begin{figure}
 \begin{center}
   \includegraphics[width=7cm]{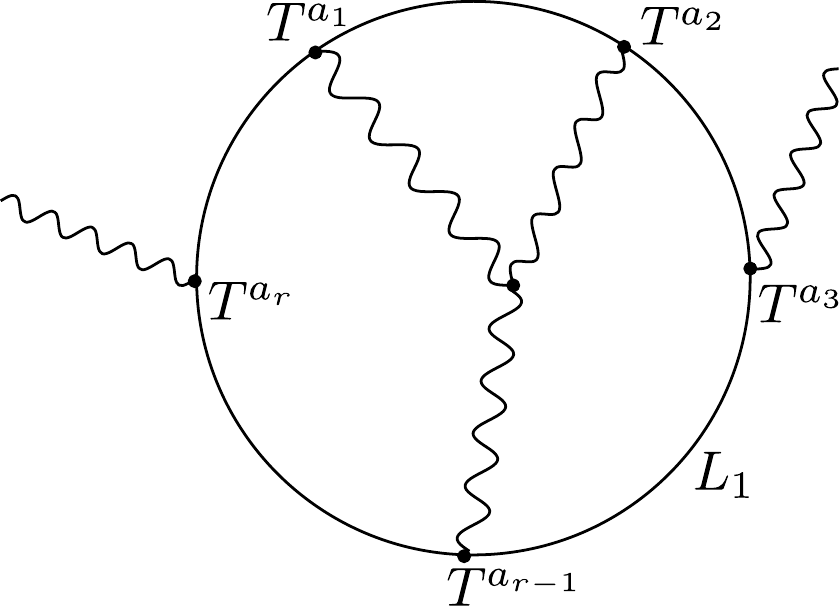}
 \end{center}
 \caption{\small A diagram contributing to the expectation value of a link. A component $L_1$ of the link is shown. The propagators running from $T^{a_3}$ and $T^{a_r}$ connect to the other components of the link.}
 \label{diagram}
\end{figure}
We begin the discussion of line operators by considering purely three-dimensional Chern-Simons theory of a supergroup. As we explain in Appendix~\ref{Anomalous}, there are some puzzles about this theory, but they do not really affect the following remarks.
In any event, these remarks are applicable to   the analytically-continued theory as defined in four dimensions, to which we return 
 in section~\ref{line4d}.

Consider a supergroup Chern-Simons theory on $\RR^3$ with a link $L$ which consists of $p$ closed Wilson loops $L_1,\dots,L_p$, labeled by representations $R_{\Lambda_1},\dots,R_{\Lambda_p}$ of the supergroup. Let us look at the perturbative expansion of this observable. On $\RR^3$, the trivial flat connection
is the only one, up to gauge transformation, and perturbation theory is an expansion about it.  The trivial flat connection is invariant under constant gauge
transformations, but as the generators of constant gauge transformations on $\RR^3$ are not normalizable, we do not need to divide by the volume of the group
of constant gauge transformations. This is just as well, as this volume is typically zero in the case of a supergroup.

A portion of a diagram that contributes to the expectation value is shown in fig.~\ref{diagram}. We focus on a single component of the link, say $L_1$, 
and sketch only the gluon lines that are attached to this component.  Let $r$ be the number of such lines.    Then in evaluating this diagram, we have to 
evaluate a trace
\beq
\Str_{R_{\Lambda_1}}(T^{a_1}\dots T^{a_r})\,d_{a_1\dots a_r}\,,
\eeq
where $T^{a_i}$ are bosonic or fermionic generators of the superalgebra, and everything except the group factor for the component $L_1$ is hidden inside the invariant tensor $d_{a_1\dots a_r}$ (whose construction depends on the rest of the diagram). By gauge invariance, the operator $T^{a_1}\dots T^{a_r}\,d_{a_1\dots a_r}$ is a Casimir operator $c_{L_1,p}\in\mathfrak{Z}$, acting in the representation $R_{\Lambda_1}$. The Casimir can be replaced simply by a number, and what then remains of the group factor is the supertrace of  the identity. So  this contribution to the expectation value can be written as $c_{L_1,p}({\Lambda_1})\,{\rm sdim} R_{\Lambda_1}$. From this we learn two things. First of all, up to a  constant factor, the expectation value for the link $L$ will not change if we replace any of the representations $R_{\Lambda_i}$ by a representation with the same values of the Casimirs. That is, for an irreducible atypical representation, the expectation value depends only on the atypical block, and not on the specific representative. Second, if the supertrace over any of the representations $R_{\Lambda_i}$ vanishes, the expectation value of the link in $\RR^3$ vanishes. Recall from the previous section that the superdimension can be non-zero only for a representation of maximal atypicality. We conclude that in $\RR^3$ for a non-trivial link observable, the components of the link should be labeled by maximally atypical blocks or else the expectation value will be zero. For example, for the unitary supergroup $\U(m|n)$, maximally atypical blocks correspond to irreducible representations of the ordinary Lie group $\U(|n-m|)$.

In section \ref{symbr}, we will argue that for knots on $\RR^3$ (and more generally on any space with enough non-compact directions)
one can give expectation values to the superghost fields $C$, without changing the expectation value of a product of loop
operators.  For instance, in this way, the $\U(m|n)$ theory can be Higgsed down to $\U(|n-m|)$.  Therefore, on $\RR^3$ the
supergroup theory does not give any new Wilson loop observables, beyond those that are familiar from $\U(|n-m|)$.
The symmetry breaking procedure shows that the expectation value of  a Wilson loop labeled by a maximally atypical representation
of $\U(m|n)$ is equal to the expectation value of the corresponding $\U(|n-m|)$ Wilson loop.

Yet it is known from the point of view of quantum supergroups \cite{Zh,ZhTwo} that knot invariants can be associated to arbitrary
highest weights of $\U(m|n)$, not necessarily maximally atypical.  It is believed that generically these invariants are new, that is,
they cannot be trivially reduced to invariants constructed using bosonic Lie groups.  
To make such a construction from the gauge theory point of view,  one needs to 
remove the supertrace which in the case of a representation that is not maximally 
atypical multiplies the expectation 
value by ${\rm sdim}R_\Lambda=0$. One strategy is to consider a Wilson operator supported not on a compact knot but on a non-compact 1-manifold that is asymptotic at infinity to a straight line in $\RR^3$
(but which may be knotted in the interior). The invariant of such a non-compact knot
 would be an operator acting on the representation $R_\Lambda$, 
rather than a number.  This approach may give invariants associated to arbitrary supergroup representations.  This strategy seems
plausible to us because it appears to make sense at least in perturbation theory, but we will not investigate it here.

The Higgsing argument suggests another approach that turns out to work well for typical representations.
(For representations that are neither typical not maximally atypical, the only strategy we see is the one mentioned in the last
paragraph.)  In this approach, we look at the loop observables on a manifold with less 
then three non-compact directions. We will focus on the case of $S^3$. Again perturbation theory is an expansion around the trivial flat 
connection. But now, unlike the $\RR^3$ case, the generators of constant
gauge transformations are normalizable and we do have to divide by the volume of the gauge group. As was mentioned in our superalgebra review, this volume is zero for any supergroup except OSp$(1|2n)$. Therefore, for almost all supergroups the partition function $Z(S^3)$ on $S^3$ is divergent,
\beq
Z(S^3)=\infty.\label{infinite}
\eeq
If we try to include a Wilson loop in a non-maximally atypical representation, we get an indeterminacy $0\cdot\infty$.  

There is a natural way to resolve this
indeterminacy in the case of typical representations, but it involves an additional tool.    In three-dimensional
Chern-Simons theory with a compact simple gauge group $G$,  Wilson line operators and line operators defined by a monodromy singularity are equivalent \cite{WittenCS,Elitzur,Beasley}. 
The proof involves using the Borel-Weil-Bott (BWB) theorem to ``de-quantize'' an irreducible representation of $G$, interpreting it as arising by quantizing
some auxiliary space (the flag manifold of $G$), in what we will call BWB quantum mechanics.  To resolve the indeterminacy that was just noted,
we need the analog of this for supergroups.

\subsubsection{BWB Quantum Mechanics}\label{bwbreview}
We first recall this story in the case of an ordinary bosonic group. Let $G$ be a compact reductive Lie group, $T\subset G$ a maximal torus, and $\uplambda\in \mathfrak{t}^*$ an integral weight. Assume in addition, that $\uplambda$ is regular, that is $\langle\uplambda,\alpha\rangle\ne 0$ for any root $\alpha$, or equivalently the coadjoint orbit of $\uplambda$ is $G/T$. 
(If this is not so, there is a similar story to what we will explain with $G/T$ replaced by $G/L$, where $L$ is a subgroup
of $G$ that contains $T$.  $L$ is called a Levi subgroup of $G$. Its Lie algebra is obtained by adjoining to $\mathfrak t$ the roots $\alpha$ that obey
$\langle\uplambda,\alpha\rangle=0$.) One can consider a quantum mechanics in phase space $G/T$ with the Kirillov-Kostant symplectic form corresponding to $\uplambda$. The functional integral for this theory can be written as 
\beq
\int{\rm D}h \exp\left(- i\int\uplambda\bigl(h^{-1}\partial_s h\,\bigr){\rm d}s\right)\,,\label{bwbqm}
\eeq
where we integrate over maps of a line (or a circle) to $G/T$.
The action here is defined using an arbitrary lift of the map $h(s)$ valued in $G/T$ into a map valued in $G$. The functional integral is independent of this lift, as long as the weight is integral. 

Let $V_\uplambda$ be a one-dimensional $T$-module, where $T$ acts with weight $\uplambda$. The prequantization line bundle over the phase space is defined as $\mathcal{L}_\uplambda\simeq G\times_T V_\uplambda$; thus, it is a line bundle associated to the principal $T$-bundle $G\rightarrow G/T$. To define an actual quantization, one needs to make a choice of polarization. For that we need a complex structure. To that end, pick a Borel subgroup $B\supset T$ in the complexified gauge group $G_\CC$. The complex K\"ahler manifold $\mathcal{M}\simeq G_\CC/B$ is isomorphic, as a real manifold, to our phase space, and this gives it a complex structure. The prequantum line bundle is likewise endowed with a holomorphic structure, $\mathcal{L}_\uplambda\simeq G_\CC\times_B V_\uplambda$. 

An accurate description of geometric quantization also involves the metaplectic correction. 
Instead of being just sections of the prequantum line bundle, the wave-functions are usually taken to be half-forms valued in this line bundle. For example, this is the source of the $1/2$ shift in the Bohr-Sommerfeld quantization condition. The metaplectic correction is important for showing the independence of the Hilbert space on the choice of polarization. In a holomorphic polarization, the bundle of half-densities is a square root of the canonical line bundle $K$. For the flag manifolds that we consider, $K$ is simply $\mathcal{L}_{-2\rho}$, where $\rho$ is the Weyl vector for the chosen Borel subgroup. So our wave-functions will live, roughly speaking, in $\mathcal{L}_\uplambda\otimes K^{1/2}\simeq \mathcal{L}_{\uplambda-\rho}$. 

The precise characterization of the Hilbert space is given by the Borel-Weil-Bott theorem. Let $w\in \mathcal{W}$ be the element of the Weyl group that conjugates $\uplambda$ into a weight that is dominant with respect to the chosen $B$. Since $\uplambda$ was assumed to be regular, the weight
\beq
\Lambda=w(\uplambda)-\rho\,,\label{Lambda}
\eeq
is also dominant. The BWB theorem states that the cohomology $H^\bullet(\mathcal{M},\mathcal{L}_{\uplambda-\rho})$ is non-vanishing precisely in one degree $\ell(w)$, which is the length of the element $w$ in terms of the simple reflections. The group $G_\CC$ acts naturally on the cohomology, and $H^{\ell(w)}(\mathcal{M},\mathcal{L}_{\uplambda-\rho})\simeq R_\Lambda$. This can be taken naturally as the Hilbert space $\mathcal{H}$ of our system. Clearly, $R_\Lambda$ depends only on $\uplambda$, and not on the choice of the Borel subgroup, that is, the polarization. If $B$ is taken such that $\uplambda$ is dominant, then this is the usual K\"ahler quantization, since $H^0(\mathcal{M},\mathcal{L}_{\uplambda-\rho})$ is the space of holomorphic sections.

The fact that the resulting Hilbert space   $H^{\ell(w)}(\mathcal{M},\mathcal{L}_{\uplambda-\rho})$ is independent of the choice of  complex structure (or equivalently
the choice of $B$) has a direct explanation. On a K\"ahler manifold, the bundle $\Omega^{0,\bullet}(\mathcal{M})\otimes K^{1/2}$ is isomorphic to the Dirac bundle $S\simeq S^+\oplus S^-$, where $S^+$ and $S^-$ are spinors of positive
or negative chirality. The Dirac operator is $\slashed{D}=\bar{\pt}+\bar{\pt}^*$, and the cohomology of $\bar{\pt}$ acting in $\Omega^{0,\bullet}(\mathcal{M})\otimes K^{1/2}$ is isomorphic to the space of zero-modes of the Dirac operator, by a standard Hodge argument. Therefore, the Hilbert space that we defined is simply the 
kernel of the Dirac operator acting on $S\otimes V_\uplambda$.

For the application to  Wilson operators, we need to decide if the particle  running in the loop is bosonic or fermionic. If the Hilbert space lies in the $\ell$-th cohomology group, it is natural to define the operator $(-1)^F$ that distinguishes bosons from fermions as $(-1)^F= (-1)^\ell$. In the Dirac operator terminology, the particle is a boson or a fermion depending on whether the zero-modes lie in $S^+$ or in $S^-$. In particular, the amplitude of propagation of the particle along a loop (with zero Hamiltonian) is naturally defined as the index of the Dirac operator ${\rm ind}\,\slashed{D}=\pm{\rm dim}\,R_\Lambda$, to account for the $-1$ factor for a fermion loop. Note that an elementary Weyl reflection of $\uplambda$ along a simple root reverses the orientation of $\mathcal{M}$, and therefore exchanges $S^+$ with $S^-$ and
exchanges bosons and fermions.

In what follows, we will always work in the Borel subalgebra in which $\uplambda$ is dominant, and therefore $\Lambda=\uplambda-\rho$.

Now we return to the supergroup case. We would like to write the same functional integral (\ref{bwbqm}), with matrices replaced by supermatrices.
A technical detail is as follows.  In the bosonic case, the integral goes over $G/T$, where $G$ is the real compact form of the group. 
In the supergroup case, we choose the compact real form of the bosonic subgroup $G_{\bar 0}$, since this is the only
choice that will lead to finite-dimensional representations of $SG$.  The compact form of $G_{\bar 0}$ may not extend to a real
form of $SG$ (for $\OSp(n|2m)$ it does not), so one has to develop the theory without assuming a real form of $SG$.
 Similarly to what we have said in the beginning of section ~\ref{superrev} for the Chern-Simons case, to make sense of the BWB path integral,
 a real form is needed only in the bosonic directions.  The path integral of the supergroup  BWB model goes over a sub-supermanifold in $SG_\CC/T_\CC$  whose reduced manifold is the bosonic phase space $G_{\bar{0}}/T$. For instance, in our analysis shortly of a type I supergroup, $h_0\in G_{\bar{0}}/T$, and $\theta$ and $\tilde{\theta}$ are independent variables valued in $\mathfrak g_1$ and $\mathfrak g_{-1}$, with no reality condition.
 
 We claim that a simple supergroup version of the BWB model 
 produces an irreducible representation of $SG$ as the Hilbert space. To exclude zero-modes, we assume that $\uplambda$ is regular, so that $\langle\uplambda,\alpha\rangle\ne0$ for any $\alpha\in\Delta^+$, bosonic or fermionic. It means in particular that the weight $\Lambda=\uplambda-\rho$ is typical. In this case, a direct analog of the BWB theorem exists \cite{Penkov}, and the same logic as for the bosonic group leads to the conclusion that the Hilbert space of the system is indeed the irreducible representation $R_\Lambda$.

For a type I superalgebra, this statement can be heuristically explained as follows. Take a parametrization of the supergroup element as $h=\ex^{{\theta}}h_0\ex^{\tilde{\theta}}$, with $h_0$ an element of the bosonic subgroup, and $\tilde{\theta}$ and $\theta$ belonging to $\mathfrak{g}_{-1}$ and $\mathfrak{g}_{1}$, respectively. The action of the theory is
\beq
-\int\uplambda(h^{-1}{\rm d} h)=\int\Str(\uplambda^\circ h^{-1}{\rm d} h)\,,
\eeq
where $\uplambda^\circ=\kappa^{mn}\uplambda_m T_n$ is the dual of $\uplambda$, defined using the superinvariant bilinear form.\footnote{The circle denotes the dual with respect to the bosonic part of the superinvariant bilinear form $\kappa=\kappa_r-\kappa_\ell$. The dual with respect to the positive definite form $\kappa_r+\kappa_\ell$ will be denoted by a star.\label{starcircle}}
Using the fact that $\{\mathfrak{g}_{-1},\mathfrak{g}_{-1}\}=0$, and the fact that the invariant bilinear form is even, one can rewrite this as
\beq
\int\Str\left(\uplambda^\circ h^{-1}{\rm d} h\,\right)= \int\Str\left(\uplambda^\circ h_0^{-1}{\rm d}h_0\right)+\int\Str\left(h_0[\tilde{\theta},\uplambda^\circ]h_0^{-1}{\rm d}{\theta}\right)\,,\label{factorize}
\eeq
If $\uplambda$ is regular, the commutation with it in $[\tilde{\theta},\uplambda^\circ]$ simply multiplies the different components of the fermion $\tilde{\theta}$ by non-zero numbers. Then we can set $\theta'=h_0[\tilde{\theta},\uplambda^\circ]h_0^{-1}$, with $\theta'$ a new fermionic variable. The resulting theory is a BWB quantum mechanics for the bosonic field $h_0$, together with the free fermions $\theta'$ and $\theta$. The Hilbert space is a tensor product (\ref{tensor}), as expected for a typical representation of a type I superalgebra.\footnote{There is a small caveat in this discussion. By our logic, the theory (\ref{factorize}) gives $\mathcal{H}\simeq  \wedge^\bullet\mathfrak{g}_{-1}\times R^{0}_{\uplambda-\rho_{\bar{0}}}$, which is the superalgebra representation with the highest weight $\uplambda-\rho_{\bar{0}}$, whereas the supergroup BWB predicts the highest weight to be $\uplambda-\rho$. Presumably, the discrepancy can be cured if one takes into account the Jacobian of the transformation from the superinvariant measure
in the full set of variables to the free measure in the $(\theta',\theta)$ variables.  In other words, that Jacobian gives the difference between the one-loop shift for $SG$ and for its maximal
bosonic subgroup $G$.}

What if $\uplambda$ is atypical, so that there exist isotropic fermionic roots $\alpha$ for which $\langle\uplambda,\alpha\rangle=0$? The usual BWB action (\ref{bwbqm})  is degenerate, as it is independent of some modes of $\theta$ and $\tilde\theta$. This is analogous to the problem that one has in the bosonic case if $\uplambda$
is non-regular, and one can proceed in a  similar way.  
We replace $SG/T$ with $SG/L$, where $L$ is a subgroup of $G$ whose Lie algebra includes the roots with 
 $\langle\uplambda,\alpha\rangle=0$.  ($L$ is a superanalog of a Levi subgroup of a simple bosonic Lie group.)
Then we quantize $SG/L$ instead of $SG/T$. This seems to give a well-defined quantum mechanics, but we will not try to analyze it. The BWB theory for  atypical representations is more complicated than a naive generalization from the bosonic case \cite{veracaro}. One expects the Hilbert space of the $SG/L$ model
to be  a finite-dimensional representation with highest weight $\Lambda$. However, rather than  the irreducible representation, it might be the Kac module, or some quotient of it, or some more complicated indecomposable representation.

\subsubsection{Monodromy Operators In The Three Dimensional Theory}\label{linerev}
By coupling the gauge field of  Chern-Simons theory to the currents of  BWB quantum mechanics, supported on a knot $K$, we can write a path integral
representation of a Wilson operator supported on $K$:
\beq
\Str_{R_\Lambda}P\exp\left(-\oint_K A\right)=\int{\rm D}h \exp\left(- i\oint_K\uplambda(h^{-1}\d_Ah)\right).\label{BWBline}
\eeq
Here $K$ is an arbitrary knot -- that is, a closed oriented 1-manifold in $W$.  As we have explained, this replacement is justified at least for typical representations. In the atypical case, we expect this replacement to be valid if $R_\Lambda$ is chosen correctly within its block.

 To establish the relation between Wilson lines and monodromy operators, we remove the BWB degrees of freedom by a gauge transformation.
 This is possible because $G$ acts transitively on $G/T$; thus, we can pick a gauge transformation along $K$ that maps $h$ to a constant
 element of $G/T$.   For a regular weight $\uplambda$, the choice of this constant element breaks the $G$ gauge symmetry along $K$
 down to $T$. What remains of the functional integral (\ref{BWBline}) is an insertion of an abelian Wilson line
\beq
\exp\left(- i\oint_K\uplambda(A)\right).\label{abWl}
\eeq
With this insertion, the classical equations derived from the Chern-Simons functional integral require the gauge field strength to have a delta-function singularity along the knot,
\beq
F=\fr{2\pi \uplambda^\circ}{\cal{K}}\delta_K.\label{clmonodromy}
\eeq
For example, if $r,\theta$ are polar coordinates in the normal plane to the knot, then this equation can be obeyed with
\beq A=\frac{\uplambda^\circ}{\cal K}\,\d\theta.\end{equation}
We note that $\d\theta$ is singular at $r=0$, that is, along $K$.
In quantum theory, the classical equations do not always hold.  However, to develop a sensible quantum theory, it is necessary to
work in a space of fields in which it is possible to obey the classical equations. One accomplishes this in the present case by quantizing
the theory in a space of fields characterized by
\beq A=\frac{\uplambda^\circ}{\cal K}\,\d\theta+\dots \label{zumbom}\eeq 
where the ellipses refer to terms less singular than $\d\theta$ at $r=0$.
This gives the definition of a monodromy operator.

Note that in (\ref{BWBline}), to rewrite a Wilson line for a dominant weight $\Lambda$, we used a weight $\uplambda=\Lambda+\rho$. The motivation for this shift was given in our review of the coadjoint orbit quantum mechanics, but this point requires more explanation. In the ordinary three-dimensional formulation of  Chern-Simons theory, it is known that such shifts of the weights should not be included in the definition of the monodromy operators, but rather they appear in the final answers as quantum corrections \cite{Elitzur}. This is analogous to the shift in the level $k\rightarrow k+h\,\,\sign(k)$.  However, in the analytically-continued theory, we have to put the shift of $k$ by hand into the classical action, and one expects that the same should be done with the shifts of the weights.\footnote{Both of these shifts arise from the phase of
an oscillatory Gaussian integral, as was explained in the case of $k$ in section \ref{turogo}.  In the 4d analytically-continued version
of the theory, the Gaussian integrals are real and will not generate shifts.} For example, let us look at the expectation value for the unknot, labeled with the spin $j$ representation, in the SU(2) Chern-Simons theory on $\RR^3$. This expectation value is
\beq
Z_{j}(\RR^3)=\fr{\sin\left(2\pi\,(j+1/2)/\calK\right)}{\sin\left(\pi/\calK\right)}.\label{unknotsu2}
\eeq
This formula is derived from the relation with conformal field theory, so $j$ here is a non-negative half-integer. In the analytically-continued theory, we want to replace the Wilson line of the spin $j$ representation with a monodromy operator, and assume that the answer is given by the same simple formula (\ref{unknotsu2}). The prescribed monodromy around the knot is defined by
\beq
F=2\pi i\fr{j'\,\sigma_3}{\cal{K}}\delta_K\,,\label{su2monodr}
\eeq
where $\sigma_3\in\mathfrak{su}(2)$ is the Pauli matrix. We need to choose between taking $j'=j$ or $j'=j+1/2$. Note that the Weyl transformation of the field in (\ref{su2monodr}) brings $j'$ to $-j'$. It should leave the expectation value invariant, up to sign.\footnote{For a knot in $\RR^3$ or $S^3$,
there is essentially only one integration cycle, so the Weyl reflection maps the integration cycle to an equivalent one.  But it may reverse the
orientation of the integration cycle, and that is the reason for the sign.  
A related explanation of the sign was given in section \ref{bwbreview}.} The symmetry of the formula (\ref{unknotsu2}) is consistent with this, if we take $j'=j+1/2$.

So we will assume that the monodromy operator in the analytically-continued theory, which corresponds to a representation with weight $\Lambda$, should be defined using the shifted weight $\uplambda=\Lambda+\rho$. However, let us comment on some possible issues related to these shifts.  For a type I superalgebra, the Weyl vector $\rho$ has integral Dynkin labels, so, if $\Lambda$ is an integral weight, then $\uplambda$ is also an integral weight of the superalgebra. But for the $\mathfrak{u}(m|n)$ case, it might not be an integral weight of the supergroup. This can be illustrated even in the purely bosonic case. For U(2), the quantum correction $\Lambda\rightarrow\Lambda+\rho$ shifts the SU(2) spin by one-half, and does not change the eigenvalue of the central generator ${\mathfrak u}(1)\subset{\mathfrak u}(2)$. The resulting weight is a well-defined weight of $\mathrm{SU}(2)\times {\mathrm U}(1)$, but not of ${\mathrm U}(2)\simeq (\mathrm{ SU}(2)\times {\mathrm U}(1))/\mathbb{Z}_2$. For a type II superalgebra, the problem can be worse. If $\Lambda$ is an integral weight of the superalgebra, $\uplambda$ might not be an integral weight of the superalgebra itself, because the Weyl vector $\rho$ can have non-integer Dynkin labels.\footnote{For any simple root $\alpha$ of the superalgebra, it is true that $2\langle\rho,\alpha\rangle=\langle \alpha,\alpha\rangle$. From this one infers that the Dynkin label of the Weyl vector  is either one or zero. However, for a type II superalgebra there exists a simple root of the bosonic subalgebra, which is not a simple root of the superalgebra, and for that root the Dynkin label of $\rho$ need not be integral.}

We will not try to resolve these puzzles, but will just note that in one approach to the line observables of the analytically-continued theory, one replaces a Chern-Simons monodromy operator by a surface operator in four dimensions. In that case, the fact that $\uplambda^\circ$ in eqn. (\ref{zumbom}) is defined using a non-integral weight presents no problem with gauge-invariance
for much the same reason that the non-integrality of $\calK$ presents no problem:  
the ``big'' gauge transformations that lead to integrality of the parameters
in purely three-dimensional Chern-Simons theory do not have analogs\footnote{In going from three to four dimensions, the support of a monodromy
operator is promoted from a knot $K$ to a two-manifold $C$ with boundary $K$.  If $C$ is compact, a homotopically non-trivial map from
$K$ to the maximal torus $T\subset G$ does not extend over $C$.  If $C=K\times \RR_+$, such a gauge transformation can be extended over $C$,
but the extension does not approach 1 at infinity.  In a noncompact setting, one only requires invariance under gauge transformations that are 1
at infinity.} in the four-dimensional setting.

Finally, we can return to the question of making sense of a path integral for a knot $K\subset S^3$ labeled by a typical representation.
As remarked following eqn. (\ref{infinite}), a direct attempt to do this in the language of Wilson operators
leads to a $0\cdot \infty$ degeneracy.  This degeneracy is naturally resolved by replacing the Wilson operator by a monodromy operator
with weight $\uplambda$.
In perturbation theory in the presence of a monodromy operator supported on a knot $K$, the functional integral is evaluated by expanding near classical flat connections on the complement of $K$ whose monodromy around $K$ has a prescribed conjugacy class.  
The group $H$ of unbroken gauge symmetries of
 any such flat connection, for $\uplambda $ typical, is a purely bosonic subgroup of $SG$, because the fermionic gauge symmetries have been
explicitly broken by the reduction of the gauge symmetry along $K$ from $SG$ to a bosonic subgroup (this subgroup is $T$ if $\uplambda$ is
regular as well as typical).\footnote{For any $K$, there is an abelian flat connection on the complement of $K$ with the prescribed monodromy around $K$,
unique up to gauge transformation.  Its automorphism group is $T$ if $\uplambda$ is regular as well as typical (and otherwise is a Levi subgroup $L$
that lies between $T$ and $G$).  In general, there may be nonabelian flat connections with the required monodromies;
their automorphism groups are smaller.}   To compute the functional integral expanded around a classical flat connection, one has to divide by the volume
of $H$, but this presents no problem: as $H$ is purely bosonic, its volume is not zero.  So in the monodromy operator approach, there is
no problem to define a path integral on $S^3$ with insertion of a knot labeled by a typical representation.

Now let us consider loop operators in $\RR^3$ rather than $S^3$.
We have claimed that a path integral on $\RR^3$ with a Wilson operator labeled by a  representation of non-maximal atypicality is 0. This
must remain true if the Wilson operator is replaced by a corresponding monodromy operator.
Let us see how this happens.  The difference between $\RR^3$ and $S^3$ is that in defining the path integral on $\RR^3$, we only
divide by gauge transformations that are trivial at infinity. If on $S^3$ a flat connection has an automorphism subgroup $H$, then on $\RR^3$ it will give rise to a family of irreducible connections, with moduli space $SG/H$. 
The volume of this moduli space will appear as a factor (in the numerator!) in evaluating the path integral.
If $H$ is purely
bosonic, then the quotient $SG/H$ has fermionic directions, and its volume generally vanishes.\footnote{As always, the exception is a supergroup from the series B$(0,n)$.} Therefore, the expectation value of a closed monodromy operator in $\RR^3$, for $\uplambda$ typical, vanishes (except for B$(0,n)$), in agreement with the corresponding statement for the Wilson loop. To analyze the case that the weight $\uplambda$ is not typical, 
we need to extend the BWB quantum mechanics for atypical weights, and presumably we will then 
 need to compute the invariant volume of a homogeneous space $SG/H$, where now $H$ will be a subsupergroup. It is plausible
that for $\uplambda$ of sufficient atypicality,  this volume can be non-zero,\footnote{In view of an argument explained in footnote \ref{volumes},
a necessary condition is that any fermionic generator ${\mathcal C}$ of $SG$ that obeys $\{{\mathcal C},{\mathcal C}\}=0$ must be conjugate to a generator
of $H$.  This ensures that the group $\mathrm F$ generated by $\mathcal C$ does not act freely on $SG/H$, so that the argument of footnote \ref{volumes} cannot
be used to show that the volume of $SG/H$ is 0.  For $\U(m|n)$, it follows from this criterion 
that $SG/H$ has zero volume except possibly
if $\uplambda$ is maximally atypical.} so that the monodromy operator can have a non-trivial expectation value.  But we have not performed this computation.

\subsection{Line Observables In Four Dimensions}\label{line4d}

Our next goal is to interpret the line operators that we have discussed in the full four-dimensional construction. First we consider  Wilson lines, and explore their symmetries in the physical 4d super Yang-Mills theory associated to the D3-NS5 system.

\subsubsection{Wilson  Operators}\label{loopoperators}
For generic values of $t$, $\N=4$ super Yang-Mills theory in bulk  does not
admit $\Qb$-invariant Wilson operators.  (They exist precisely if $t^2=-1$, a fact that is important in the geometric Langlands correspondence
\cite{Langlands}.)  
However, on the defect $W$ there always exist supersymmetric Wilson operators 
\beq
\W_R(K)=\Str_R \Pexp\left(-\oint_K \cA\right)\,,\label{Wilson}
\eeq
labeled by an arbitrary  representation $R$ of the supergroup $\SG$.  Here $\cA$ is the supergroup gauge field
and $\Qb$-invariance is clear since $\Qb$ acts on $\cA$ by gauge transformations. 

These are the most obvious $\Qb$-invariant line operators, but they have a drawback that makes them harder to study: as operators
in the physical $\N=4$ super Yang-Mills theory, they have less symmetry than one might expect. We will analyze the symmetry of these
operators in different situations.

The procedure by which we constructed a topological field theory involved twisting four of the six scalars of $\N=4$ super Yang-Mills
theory, leaving two untwisted scalars and hence an unbroken $R$-symmetry group $\U(1)_F=\SO(2)\subset \SO(6)_R$.  In the special case that
$M= \RR\times W$ with a product metric,  there is no need for twisting in the $\RR$ direction to maintain
supersymmetry, so three scalars remain untwisted and $\U(1)_F$ is enhanced to $\SU(2)_Y$.  The supercharge $\Qb$
that we chose in constructing a topological field theory was one component of an $\SU(2)_Y$ doublet.  For $M=\RR\times W$,
the twisted action is invariant under $\SU(2)_Y$ as well as $\Qb$, so it inevitably preserves two supercharges -- both components of the doublet
containing $\Qb$.
Likewise, the Wilson loop operators (\ref{Wilson}) are invariant under $\SU(2)_Y$ as well as $\Qb$, so on $M=\RR\times W$, they really preserve
two supersymmetries.

Now let us specialize further to the case that  $W=\RR^3$ is flat, with $M=\RR\times W=\RR^4$.  In this case, no topological twisting is necessary,
but the half-BPS defect supported on $W$ breaks the $R$-symmetry group to $\SO(3)_X\times \SO(3)_Y$.  In addition, there is an unbroken rotation
group $\SO(3)$, and, as explained in section \ref{nstype}, the unbroken supersymmetries transform as $(\2,\2,\2)$ under $\SO(3)\times \SO(3)_X\times
\SO(3)_Y$.  Let us consider a Wilson operator $\W_R(K)$ where $K$ is a straight 
line $\RR\subset W$, say the line $x_1=x_2=0$, parametrized by $x_0$.
$K$ is invariant under a  subgroup $\SO(2)\subset \SO(3)$ of rotations of $x_1$ and $x_2$.  To identify the global symmetry of  $\W_R(K)$  involves a crucial subtlety.
First let us consider the one-sided case studied in \cite{5knots}, in other words the case of an ordinary gauge group $G$ rather than a supergroup $\SG$.
In this case, the supergroup connection reduces to $\AAb=A+i(\sin\ang)\phi$, and the Wilson operator for a straight Wilson line depends on
one component $\phi_0$ of a triplet $(\phi_0,\phi_1,\phi_2)$ of $\SO(3)_X$.  This field is invariant under a subgroup $\SO(2)_X\subset \SO(3)_X$,
and hence a straight Wilson line in the case of an ordinary gauge group has global (bosonic) symmetry $\SO(2)\times \SO(2)_X\times \SO(3)_Y$.
In the supergroup case, we must remember that the supergroup connection also has a fermionic part $\AAf$ which began life
as part of a field that transforms as $(\2,\2)$ under $\SO(3)\times \SO(3)_X$.  As a result, the component of $\AAf$ in the $x_0$ direction
is not separately invariant under $\SO(2)$ and $\SO(2)_X$ but only under a diagonal combination $\SO'(2)\subset \SO(2)\times \SO(2)_X$.
Hence the bosonic global symmetry of a straight Wilson line in the supergroup case is $\SO'(2)\times \SO(3)_Y$, reduced from the corresponding
symmetry in the case of an ordinary Lie group.

The supersymmetry of a straight Wilson line $\W_R(K)$ is likewise reduced in the supergroup case from what it is in the case of an ordinary Lie group.
A supersymmetry has no chance to preserve the straight Wilson line
if its commutator with the complexified bosonic gauge field $\cAb$ has a contribution proportional to $\Psi_1$.
 Indeed, the boundary conditions do not tell us anything about the behaviour of $\Psi_1$ at $x_3=0$, so there would be no way to cancel
 such a term.   Inspection of the supersymmetry transformations (\ref{susy0}) reveals that,
 apart from the $\SO'(3)$-invariant supersymmetries  with generators
 \beq
\veps_{\rm }^{\alpha A\dA}=\eps^{\alpha A}w^\dA\label{zuperparameter}
\eeq
 (familiar from eqn. (\ref{superparameter})), with arbitrary two-component spinor $w^\dA$,
  the only supersymmetries that do not produce variations of $\AAb$ proportional to $\Psi_1$ are
  those with generators
  \beq
\veps^{\alpha A\dA} = \sigma_{0}^{\alpha A}\tilde{w}^\dA\,,\label{Qtilde}
\eeq
where again $\tilde{w}^\dA$ is an arbitrary spinor.  Since $\tilde w^\dA$ transforms as a spinor of $\SU(2)_Y$, an  $\SU(2)_Y$-invariant Wilson operator
is invariant under this transformation for all $\tilde w^\dA$ if and only if it is invariant for some particular nonzero $\tilde w^\dA$.
A choice that is convenient because it enables us to write simple formulas in the language of the twisted theory
is to set $\tilde{w}^\dA=v^\dA$ (where $v^\dA$ was defined in (\ref{u})).  Writing $\tilde\delta$ for the transformation generated
by the corresponding supersymmetry, one computes that
\beq
\tilde{\delta}\cA_0=-{i}[\bar{C},\cB\}\,,\label{QtildeA}
\eeq
where we define
\beq
\cB=\{C,\bar{C}\}+ B.
\eeq
Since (\ref{QtildeA}) is non-zero, our Wilson lines do not preserve supersymmetries (\ref{Qtilde}) for a generic representation. Therefore, they preserve only the two supersymmetries (\ref{zuperparameter}).
They  are 1/4-BPS objects from the standpoint of the defect theory (or $1/8$ BPS relative to the underlying $\N=4$ super Yang-Mills theory). This is an important difference from the case of a purely bosonic gauge group, in which Wilson lines preserve four supersymmetries (a fact that greatly simplifies the analysis of the dual 't Hooft operators \cite{5knots,Mikhaylov}).
In fact, if the representation $R$ that labels the Wilson line $\W_R(K)$ is such that the fermionic generators act trivially, then (\ref{QtildeA}) vanishes, and $\W_R(K)$ becomes 1/2-BPS (in the defect
theory), as in the bosonic or one-sided case. More generally, for (\ref{QtildeA}) to vanish it is enough that the anticommutators of the fermionic generators vanish in the representation $R$.  Of course, in the case
of a supergroup such as $\U(m|n)$, this is a very restrictive condition.

One can also construct other $\Qb$-invariant Wilson operators in the electric theory, by adding a polynomial of the Higgs field $\mathcal{B}$ to the connection in the Wilson line. The resulting operators preserve 1/4 or 1/8 of the three-dimensional supersymmetry. In the $\Qb$-cohomology, such operators are equivalent to the ordinary Wilson lines (\ref{Wilson}), and for this reason we will not say more about them.

Why do we care about the reduced supersymmetry of the supergroup Wilson loop operators?
One of our goals will be to understand what happens to  line operators under nonperturbative dualities.  For this purpose, 
the fact that the supergroup Wilson operators  are only 1/4 and not 1/2 BPS is rather inconvenient.  Possible
constructions of a dual operator that preserve 4 supercharges are much more restrictive than possible constructions that preserve only
2 supercharges.  We will obtain a reasonable duality picture for certain 
1/2 BPS Wilson-'t Hooft line operators that will be introduced in section \ref{defmonodr}.  
These Wilson-'t Hooft operators are labeled by weights of $SG$
and the way they are constructed suggests that from
the point of view of the twisted topological field theory -- the supergroup Chern-Simons theory -- they
are equivalent to Wilson operators.  But because of their enhanced supersymmetry, it is much easier to find their duals.  

About the Wilson operators, we make the following remarks.  We were not able to find a construction of 't Hooft-like disorder operators
-- characterized by a singularity of some kind -- with precisely the right global symmetries so that they might be dual to the Wilson operators
constructed above.  It may be that one has to supplement an 't Hooft-like construction by adding some quantum mechanical variables
that live along the line operators (analogous to the BWB variables that we discussed in section \ref{bwbreview}).  With only 2 supersymmetries
to be preserved, there are many possibilities and we do not know a good approach.  Also, the fact that the two-dimensional space of
supersymmetries preserved by a Wilson operator is not real suggests that it is difficult to realize such an object in string theory.
A string theory realization would probably have helped in understanding the action of duality.

\subsubsection{Wilson-'t~Hooft Operators}\label{defmonodr}
For all these reasons, we now move on to consider Wilson-'t Hooft operators.

$\cN=4$ super Yang-Mills theory supports BPS Wilson-'t~Hooft line operators in the bulk \cite{Kapustin}. 
Though they preserve 8 supersymmetries,  generically these do not include the specific supersymmetry $\Qb$.
The condition 
for  a Wilson-'t Hooft operator in bulk to be $\Qb$-invariant is that  its electric and magnetic charges must be proportional with a ratio $\calK$  \cite{Langlands}. Since both charges have to be integral, $\Qb$-invariant Wilson-'t~Hooft operators exist in the bulk only for rational values of the canonical parameter $\calK$.  In the present
paper, we generally assume $\calK$ to be generic.

However, we are interested in operators that  are supported not in the bulk but along the defect at $x^3=0$. The gauge theories with gauge groups $G_\ell$ and $G_r$ live in half-spaces, 
and the magnetic flux for each gauge group can escape through the boundary of the half-space and so is not quantized.
So a  Wilson-'t~Hooft operator that lives only at $y=0$ is no longer constrained to have an integral magnetic charge.  Such operators can exist for any (integer) electric charge and arbitrary $\calK$. To define them precisely, we work in the weak coupling regime, where $g_\YM$ is small, and therefore, according to (\ref{canonical}), $\calK$ is large. The weight of the representation is taken to scale with $\calK$, so that the monodromy of the gauge field, which is proportional to  $\uplambda/\calK$, is fixed.

\begin{figure}
 \begin{center}
   \includegraphics[width=9cm]{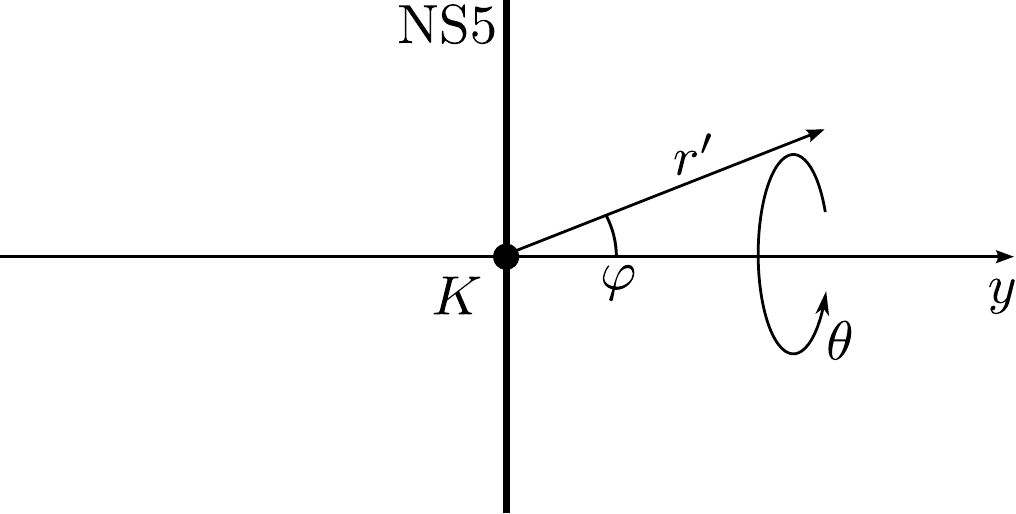}
 \end{center}
\caption{\small The hyperplane $x^0=0$ showing our notation for the coordinates. $y=x^3$ runs horizontally; the $x^0=0$ section of the knot $K$ is shown as a dot in the center.}
 \label{geom}
\end{figure}

Consider a line operator located at $y=0$ along the $x^0$ axis. (See fig. \ref{geom} for the notation.) We want to find a model solution of the BPS equations (\ref{localization}) that will define the singular asymptotics of the fields near the operator. For definiteness, consider the Yang-Mills theory on the right of the three-dimensional defect. We make a conformally-invariant abelian ansatz which preserves the $\SO(2)\times \SO(2)_X\times \SO(3)_Y$ symmetry:
\beqn\label{muff}
&&A=c_a\fr{{\rm d}x^0}{r'}+{\mathfrak{m}_r}(1-\cos\varphi){\rm d}\theta\,,\nnr
&&\phi=c_\phi\fr{{\rm d}x^0}{r'}.\label{tHW}
\eeqn
Here $\mathfrak{m}_r$ is the magnetic charge (which as noted above will not be quantized). The ray $\varphi=0$ points in the $y>0$ direction, and the signs were chosen such that there is no Dirac string along this ray. The localization equations (\ref{localization}) are satisfied if
\beq
c_a=i{\mathfrak{m}_r}\tan\ang\,,\quad c_\phi=-\fr{\mathfrak{m}_r}{\cos\ang}\,,\label{tHW2}
\eeq
where $\ang$ is the angle related to the twisting parameter $t$, as introduced in section \ref{etheory}. The hypermultiplet fields are
taken to vanish. The factor of $i$ in the Coulomb singularity of the gauge field $A$ is an artifact of the Euclidean continuation; in Lorentz signature, the solution would be real.
 Eqn. (\ref{muff}) fixes the behaviour of the bulk fields near a line operator. For a generic magnetic charge, the fields of the hypermultiplet do not commute with the singularity in (\ref{tHW}), and thus are required to vanish along the operator.

Let us check that our model solution satisfies also the boundary conditions at $y=0$. The boundary conditions can be derived from (\ref{Fphysbc}) and an analogous expression for the scalar $X^a$. This is done in Appendix~\ref{bcsection}. However, in the topological theory one can understand the relevant features by a more simple argument. The boundary condition should require vanishing of the boundary part of the variation of the action of the theory. Suppose that we consider a configuration in which all the fermions vanish, and the bosonic fields satisfy the localization equations. The variation of the non-$\cQ$-exact Chern-Simons term (equivalently, the topological term) gives the gauge field strength $\cFb$. The $\cQ$-exact terms in the action come in two different sorts. There is a bulk contribution, whose bosonic part is proportional to the sum of squares of the localization equations (\ref{localization}). The variation of these terms vanishes when the equations are satisfied. There are also $\cQ$-exact terms supported on the defect; they furnish gauge fixing of the fermionic gauge symmetry of the supergroup Chern-Simons. Their variation is proportional to the hypermultiplet fields. Therefore, we conclude that if the fields satisfy the localization equations, and the three-dimensional hypermultiplet vanishes, the boundary condition reduces to
\beq
\imath^*\left(\cFb\right)=0\,,\label{fiszero}
\eeq
where $\imath:\,W\hookrightarrow M$ is the natural embedding of the three-manifold into the bulk manifold. This boundary condition is indeed satisfied by the model solution (\ref{tHW}), (\ref{tHW2}), because in the complexified gauge field $\cAb=A+i(\sin\ang)\phi$, the Coulomb parts of $A$ and $\phi$ cancel. (The magnetic part is annihilated by $\imath^*$.)  In fact, at $y=0$, the complexified field $\cAb$ reduces to the field of a Chern-Simons monodromy operator (\ref{clmonodromy}), if we identify $\mathfrak{m}=\uplambda^\circ/\calK$, where now $\mathfrak{m}$ includes both the part in $\mathfrak{g}_\ell$ and in $\mathfrak{g}_r$.

In Chern-Simons theory, in the  presence of a monodromy defect, the bulk action is supplemented with an abelian Wilson line (\ref{abWl}) along
the defect; in our derivation in section (\ref{linerev}), this is what remained after gauge-fixing the BWB action. The Chern-Simons action with an insertion of an abelian Wilson line is characterized by the fact that its variation near the background singular field (\ref{clmonodromy}) does not have a delta function
term supported on the knot (a delta function term that would come from the variation of Chern-Simons in the presence of the monodromy
singularity is canceled by the variation of the abelian Wilson operator). In four dimensions, in the presence of a singularity along a knot $K$, the topological action (\ref{act2}) should be integrated over the four-manifold with a neighborhood of $K$ cut out, and taking into account
the singularity along $K$ of the Wilson-'t Hooft operator, this produces a term in the variation with delta-function support along $K$:
\beq
\delta\left(\fr{i\calK}{4\pi}\int_{M\setminus K} \tr\left(\mathcal{F}\wedge\mathcal{F}\right)\right)=i\oint_K \Str\left(\uplambda^\circ \delta\cA\right).
\eeq
To cancel this variation, just like in three dimensions, one inserts an abelian Wilson line (\ref{abWl}). 

But now we learn something fundamental.  Although the Wilson-'t Hooft operators that we have constructed do not have a 
quantized magnetic charge, they have a quantized electric charge.
The abelian Wilson line is only gauge-invariant if $\uplambda$ is an integral weight of $G_\ell\times G_r$.  For a type I superalgebra such
as $\frak{u}(m|n)$, an integral weight of $G_\ell\times G_r$ corresponds to an integral weight of the supergroup $\U(m|n)$ and therefore,
these Wilson operators are classified by integral weights of the supergroup.  The Weyl group of $\U(m|n)$ is the same as that of its
bosonic subgroup $\U(m)\times \U(n)$, so an equivalent statement is that Wilson operators of the supergroup (for irreducible typical representations, or some particular atypical representations) are in correspondence with this class of Wilson-'t Hooft operators.
The advantage of the Wilson-'t Hooft operators is that they have more symmetry: in addition to $\Qb$-invariance, they are half-BPS operators with
 the full $\SO(2)\times \SO(2)_X\times \SO(3)_Y$ symmetry, just like a Wilson line in the one-sided problem.
 
 For a type II superalgebra,
such as $\osp(2m+1|2n)$, there is a slight complication.  For such algebras, 
some  ``small'' dominant weights do not correspond to
representations.  (These are the weights that do not satisfy the ``supplementary condition,'' as defined in section~\ref{reps}. See also section \ref{osptheory} for details in the case of $\OSp(2m+1|2n)$.)  Our construction gives a half-BPS line operator
for every dominant weight whether or not this weight corresponds to a representation. 
 It is hard to study explicitly why some Wilson-'t Hooft operators with small weights do not correspond to representations, 
because  the semiclassical description  of a Wilson-'t Hooft operator is valid for large weights.\footnote{Given 
this, one may wonder if the
half-BPS property is lost when the weights are too small.  We doubt that this is the right interpretation because  the construction
of half-BPS line operators on the magnetic side, discussed in section \ref{magline}, appears to be valid for all weights.}

\subsubsection{Twisted Line Operators}\label{twisted}
In section \ref{ortho}, we will discuss a non-perturbative duality for  Chern-Simons theory with orthosymplectic supergroup $\OSp(r|2n)$. 
It will turn out  that line operators labeled by dominant weights of the supergroup are not a closed set of operators under that duality. To get
a duality-invariant picture, one needs to include what we will call twisted line operators.   

The clearest explanation seems to be also the most naive one.  We consider 4d super Yang-Mills theory on $W\times \RR_y$, where $\RR_y$
is parametrized by $y$.  For $y<0$, the gauge group is $\SO(r)$; for $y>0$, it is $\Sp(2n)$.  Along $W\times \{y=0\}$ is a bifundamental
hypermultiplet.

Now we pick a knot $K\subset W$, and define a line operator supported on $K$ by saying that the hypermultiplet fermions change sign under
monodromy around $K$.  Locally, this makes perfect sense.  Globally, to make sense of it, we have to say essentially that the hypermultiplets
are not just bifundamentals, but are twisted by a $\Z_2$ bundle defined on $W\times \{y=0\}$ that has monodromy around $K$.  If such
a flat bundle does not exist, we say that the path integral with insertion of the given line operator is 0.  If there are inequivalent choices for
this flat bundle, we sum over the choices.  

This procedure actually defines not just a single new line operator, but a whole class of them, which we will call twisted line operators.
The reason is that the monodromy around $K$ forces the hypermultiplets to vanish along $K$, and therefore there is no problem
to include arbitrary $\SO(r)\times \Sp(2n)$ Wilson operators along $K$.  This class of operators will be important in section \ref{ortho}.

Can we do something similar for $\U(m|n)$?  In this case, we can pick an arbitrary nonzero complex number $e^{ic}$, embedded
as an element of the  center of $\U(n)$ (or of its complexification if $c$ is not real), and twist the hypermultiplet fields by $e^{ic}$ under
monodromy around $K$.  Then we proceed as just explained, and get a family of line operators that depend on the parameter $c$.   Again, from a global point of view,
this means the hypermultiplets are bifundamentals twisted by a flat line bundle with monodromy $e^{ic}$ around $K$, and we define the path integral 
by summing over the possible flat bundles that obey this condition.
And again, we can generalize this definition by including Wilson operators of $\U(m)\times \U(n)$. 

\subsection{Surface Operators}\label{surface}

In the relation of 3d Chern-Simons theory to 4d gauge theory, there are two possible strategies for finding a 4d construction
related to a line operator in the Chern-Simons theory.

In one picture, the 3d line operator is promoted to a 4d line operator that lives on the defect that supports the Chern-Simons
gauge fields.  In the second picture, a line operator in 3d is considered to have its support in codimension 2,
and it is promoted to a surface operator in 4d, whose support is in codimension 2.

So if Chern-Simons theory on a three-manifold $W$ is related to 4d super Yang-Mills on $W\times \RR_y$,
where $\RR_y$ is a copy of $\RR$ parametrized by $y$ with a defect at $y=0$,  then in the first approach,
a 3d line operator supported on $K\subset W$ is promoted in 4d to a line operator supported on $K\times \{y=0\}$.
In the second approach, a 3d line operator supported on $K$ is promoted to a 4d surface operator supported on a two-manifold
$C$ such that $C\cap \{y=0\}=K$.  For example, $C$ might be simply $K\times \RR_y$.

Both of these viewpoints were explored in \cite{5knots} for the one-sided problem, 
although the first one based on 4d line operators was developed
in more detail.  In the two-sided case, we have followed the first viewpoint so far but now we turn to the second
one and consider surface operators.

We focus on the  simplest half-BPS surface operators, which were described in the bulk in \cite{Surface}.
Our problem is to understand what happens when one of these operators 
intersects a fivebrane.  In the present section, we answer this question on the electric side (that is, for an NS5-brane).
In section \ref{magnetic}, we answer the question on the magnetic side (that is, for a D5-brane).

One advantage to the formulation via surface operators in four dimensions rather than line operators 
 is that the behavior under $S$-duality is simple to understand.    That is
because, in the 4d bulk,  one already knows the behavior under $S$-duality of the  surface operators we will be studying.
Given a surface operator intersecting an NS5-brane, the $S$-dual of this configuration will have to consist of the $S$-dual
surface operator intersecting a D5-brane.  So all we have to do is to determine what happens when a surface operator
intersects an NS5-brane or a D5-brane.  $S$-duality will then take care of itself.

\subsubsection{Surface Operators In The Bulk}\label{sbulk}
The simplest half-BPS surface operators in $\N=4$ super Yang-Mills theory   are 
labeled by a set of four parameters $(\alpha,\beta,\gamma,\eta)$. The first three define the singular behavior of the 
fields near the support of the operator, which will be a two-manifold $C$. If $r$ and $\theta$ are polar coordinates in the normal plane to  $C$, 
we require the fields near $C$  to behave like
\beqn
&&A=\alpha\,{\rm d}\theta+\dots\,,\nnr
&&\phi=\beta\,\frac{{\rm d}r}{r}-\gamma\, {\rm d}\theta+\dots\,,\label{singty}
\eeqn
where the ellipses represent less singular terms. The parameters $\alpha$, $\beta$ and $\gamma$ take values in a Cartan subalgebra $\mathfrak{t}\subset\mathfrak{g}$. More precisely, one can make big gauge transformations on the complement of $C$ that shift $\alpha$ by an arbitrary cocharacter; therefore, $\alpha$ should be considered as an element of the maximal torus\footnote{In this section we discuss only 
the bulk $\cN=4$ Yang-Mills theory, and all our notation refers to its bosonic gauge group, and not to a supergroup.} $T\simeq \mathfrak{t}/\Gamma_{\rm cochar}$. 

The meaning of the fourth parameter $\eta$ is the following. Assume that the triple $(\alpha,\beta,\gamma)$ is regular, that is, it commutes only with $\mathfrak t$. In this case the singular behavior (\ref{singty}) reduces the gauge group along the surface operator  to its maximal torus ${T}$, and it makes sense to speak of the first Chern class of the resulting ${T}$-bundle on $C$. One can define the $\mathfrak{t}^*$-valued theta-angle $\eta$ coupled to this Chern class, and introduce a factor
\beq
\exp\left(i\int_C \eta(F)\right)\label{etaterm}
\eeq
in the functional integral. By integrality of the first Chern class, this expression is invariant under a shift of the theta-angle by an element of the character lattice $\Gamma_{\rm char}\subset \mathfrak{t}^*$, so $\eta$ really takes values in the maximal torus of the Langlands-dual group, $\eta\in {T}^\vee\simeq \mathfrak{t}^*/\Gamma_{\rm char}$.
Dividing by the action of the Weyl group $\mathcal{W}$, which is a remnant of the non-abelian gauge symmetry, we get that the parameters $(\alpha,\beta,\gamma,\eta)$ take values in $({T},\mathfrak{t},\mathfrak{t},{T}^\vee)/\mathcal{W}$.

The singular asymptotics of the fields (\ref{singty}) satisfy the localization equations (\ref{localization}) for any value of $t$, if supplemented with appropriate sources,
\begin{align}\label{locsurf}   F-\phi\wedge\phi&=2\pi\alpha\,\delta_C\,,\cr
\d_A\phi&=-2\pi\gamma\,\delta_C\cr
\d_A\star\phi&=2\pi\beta\,{\rm d}x^0\wedge{\rm d}y\wedge\delta_C \end{align}
where $\delta_C={\rm d}({\rm d}\theta)/2\pi$ is the $\delta$-function 2-form that is Poincar\'e dual to the surface $C$, and $x^0$ and $y$ are  coordinates along the surface. 

The prescribed singularities (\ref{singty}) define the space of fields over which one integrates to define $\N=4$ super Yang-Mills theory in the
presence of the surface operator. Let us also define more precisely what functional we are integrating over this space. The action of the bulk topological theory consists of the topological term and some $\cQ$-exact terms (\ref{zog}). In the presence of the surface operator, the topological term is defined as
\beq
\fr{i\calK}{4\pi}\int'_M\tr(F\wedge F)\,,
\eeq
where the symbol $\int'_M$ denotes an integral over $M\setminus C$, not including a delta function contribution along $C$. Alternatively, we can write this as an integral over the whole four-manifold, and explicitly subtract the contribution which comes from the delta-function singularity of the curvature:
\begin{equation}
\fr{i\calK}{4\pi}\int'_M\tr(F\wedge F)=\fr{i\calK}{4\pi}\int_M\tr(F\wedge F)-i\calK\int_C\tr(\alpha F)-i\pi\calK\,\tr(\alpha^2)\, C\cap C.\label{topdelta}
\end{equation}
The c-number contribution proportional to the self-intersection number $C\cap C$ appears here from the square of the delta-function.

In the absence of the surface operator, the $\cQ$-exact part of the action has the form
\beq
-\fr{1}{g^2_\YM}\int\tr\left(\frac{2t^{-1}}{t+t^{-1}}\,\mathcal{V}^+\hspace{-2pt}\wedge\mathcal{V}^+-\frac{2t}{t+t^{-1}}\,\mathcal{V}^-\hspace{-2pt}\wedge\mathcal{V}^-+\mathcal{V}^0\wedge\star\mathcal{V}^0\right)\,,\label{Qpart}
\eeq
where $\mathcal{V}^+$, $\mathcal{V}^-$ and $\mathcal{V}^0$ are the left hand sides of the supersymmetric localization equations, as defined in (\ref{localization}). In the presence of the surface operator, the localization equations acquire delta-function sources, as in (\ref{locsurf}). The action (\ref{Qpart}) is modified accordingly, {\it e.g.}, the first term becomes
\beq
-\fr{1}{g^2_\YM}\int\tr\left(\frac{2t^{-1}}{t+t^{-1}}\left(\mathcal{V}^+-2\pi(\alpha-t\gamma)\delta^+_C\right)\wedge\left(\mathcal{V}^+-2\pi(\alpha-t\gamma)\delta^+_C\right)\right).\label{QexactDeltas}
\eeq
Because it contains the square of a delta function, this expression is at risk of being divergent.  To make the action finite,
one works in a class of fields in which the localization equations (\ref{locsurf}) are satisfied, modulo smooth terms.  In other
words, the left hand side of the localization equations must contain the same delta functions as the right hand side.

In the definition of the surface operator, it was assumed that the singularity defined by $(\alpha,\beta,\gamma)$ is regular, that is, the gauge group along the operator is broken down to the maximal torus. This is the case for which the theta-angles $\eta$ can be
defined classically. But it can be argued that the surface operator is actually well-defined quantum mechanically as long as the
full collection of couplings $(\alpha,\beta,\gamma,\eta)$ is regular.  One approach to showing this involves
 a different construction of these surface operators with additional degrees of freedom along the surface as described in section 3 of \cite{Rigid}. In this paper, we will try to avoid these issues.

\subsubsection{Surface Operators In The Electric Theory}\label{surel}
Let us specialize to  a four-manifold $M=W\times\RR_y$, with an NS5-type defect along $W\times\{y=0\}$. To incorporate a loop operator along the knot $K$ in the Chern-Simons theory, we insert surface operators in the left and right Yang-Mills theories along a two-surface $C=C_\ell\cup C_r$ that intersects the $y=0$ hyperplane along $K$. We could
simply take $C$ to be an infinite cylinder $K\times \RR_y$, or we could take an arbitrary finite 2-surface.  The orientations are taken to be such that $\partial C_r=-\partial C_\ell=K$. The parameters of the surface operators on the right and on the left will be denoted by letters with a subscript $r$ or $\ell$. Sometimes we will use notation without subscript to denote the combined set of parameters on the right and on the left (e.g., $\alpha=(\alpha_r,\alpha_\ell)$).

We would like to understand the meaning of the parameters of a surface operator in the Chern-Simons theory. It is clear that a surface operator with $\beta=\gamma=\eta=0$ and non-zero $\alpha$ is equivalent to a monodromy operator in Chern-Simons, with weight $\uplambda^\circ=\calK\alpha$. Such a surface operator can be obtained e.g. as a Dirac string, which is produced by moving a Wilson-'t~Hooft line operator in the four-dimensional theory into the bulk.

The parameter $\beta$ has no direct interpretation in Chern-Simons, and defines just a deformation of the integration contour, without changing the path integral. As noted in \cite{5knots}, sometimes it might not be possible to turn on $\beta$. For example, let the bosonic gauge group be abelian, and let the three-manifold $W$ be compact (e.g., $W\simeq S^3$). If we have a link with components labeled by $\beta_1,\dots,\beta_p$, then, integrating the third equation in (\ref{locsurf}) over $W$, we get that $\sum \beta_i l_i=0$, where $l_i$ is the length of the $i$-th component of the link. We see that if there is only one component, then $\beta$ has to be zero.

The case of a surface operator with non-zero $\gamma$ is a little subtle. It is not clear to us whether such an operator in the physical theory\footnote{By 
the ``physical theory'' we mean the theory that in flat space describes the D3-NS5 intersection.  In this theory, $t$ is given by (\ref{tphys}) and lies on the unit circle, and $\calK$ is real. By the ``topological theory,'' we mean the theory which arises naturally from the Morse theory construction \cite{Wittenold,Wittenoldone}, with $t$ being real, and $\calK$ in general complex. In this paper, we focus on the physical theory.} can intersect (or end on) the three-dimensional defect in a $\Qb$-invariant way, and if it can, then to what line operator in Chern-Simons theory it would correspond. In topological theory, when one takes the parameter $t$ to be real, such an operator makes perfect sense and has a natural Morse theory interpretation \cite{5knots, Wittenold}. In that case, the bosonic part of the action, modulo $\Qb$-exact terms, is defined in presence of a surface operator by an integral of the local density $\tr(\cFb\wedge\cFb)$ over the four-manifold $M\setminus C$. Up to some field-independent constants, we have, analogously to (\ref{topdelta}),
\beq
\fr{i\calK}{4\pi}\int_{M_r}'\tr(\cFb\wedge\cFb)=\fr{i\calK}{4\pi}\int_{M_r}\tr(\cFb\wedge\cFb)-i\calK\int_{C_r}\tr((\alpha_r-w\gamma_r)\cFb).\label{elinsertion}
\eeq
(Here we focus on the integral on the right hand side of the defect.) The combination $\alpha_r-w\gamma_r$ under the trace came from the monodromy of the complexified gauge field $\cAb=A+w\phi$, where $w$ is some complex number with non-zero imaginary part. (In physical theory, $w=i\sin\ang$.) Such an operator clearly corresponds to a Chern-Simons monodromy operator of weight $\uplambda^\circ=\calK(\alpha-w\gamma)$, which generically is complex. Now, the problem with such an operator in the physical theory is that the right hand side of (\ref{elinsertion}) contains an integral of $i\calK w\tr(\gamma F)$ over $C$, which cannot have any interpretation in the bulk physical theory, since $w$ is not real. (Comparing e.g. to (\ref{etaterm}), we could say that this insertion corresponds to $\eta=w\calK\gamma$, which is not an element of the real Lie algebra.) What one should really do in the physical theory is to write the action as a four-dimensional integral of the density $\tr(F\wedge F)$, with gauge field non-complexified, plus the three-dimensional integral of a three-form which can be found on the right hand side of equation (\ref{zert}). In the presence of a surface operator, one should omit $C$ from the four-dimensional integral of $\tr(F\wedge F)$, and the knot $K$ from the boundary integral of the just-mentioned three-form. In the bulk, this gives an ordinary surface operator of the sort reviewed in section \ref{sbulk}. However, it is not completely clear  whether with this definition the intersection of the operator with the defect at $y=0$ can be made $\Qb$-invariant, and to what Chern-Simons weight it would correspond.  In the $S$-dual description of the theory in section \ref{magnetic}, we will find natural half-BPS surface operators with non-zero $\gamma^\vee$, and the Chern-Simons weight will not depend on this parameter. So we would expect that in the physical theory, $\Qb$-invariant surface operators with $\gamma\ne 0$, intersecting the boundary, do exist, and that $\gamma$ plays much the same role as $\beta$~--~that is, it only deforms the integration contour. But this point is not completely clear.

Finally, turning on the parameter $\eta$ of the surface operator corresponds to adding an abelian Wilson insertion along the line $K$, where the surface operator crosses the $y=0$ hyperplane. Naively, this
 happens because of the ``identity'' $\exp(i\eta\int_{C\cap M_r}F)=\exp(i\eta\oint_{K}A)$ where
$A$ is an abelian gauge field with curvature $F$.   We cannot take this formula literally, since $\oint_{K}A$ is only gauge-invariant mod $2\pi\Z$.
But the ``identity'' is correct for computing classical equations of motion, and thus 
shifting $\eta_{\ell,r}$ has the same effect on the equations of motion as shifting the electric charges that live on $K=C\cap W$. Note that in presence of the three-dimensional defect the parameter $\eta$ is lifted from the maximal torus $T^\vee$, and takes values in the dual Cartan subalgebra $\mathfrak{t}^*$.

Let us briefly summarize. A surface operator with parameters $(\alpha,\beta,0,\eta)$, supported on a surface $C=C_\ell\cup C_r$, corresponds in the analytically-continued three-dimensional Chern-Simons theory to a monodromy operator with weight $\uplambda^\circ=\calK\alpha-\eta^*$. (Recall that a circle denotes the dual with respect to the superinvariant bilinear form $\kappa=\kappa_r-\kappa_\ell$, and a star represents the dual with respect to the positive definite form $\kappa_r+\kappa_\ell$.) Let $\uplambda_\ell$ and $\uplambda_r$ be the parts of the weight, lying in the Cartan of the left and right bosonic gauge groups, respectively. 
Then, more explicitly,
\beqn
&&\uplambda_{\ell}=-\calK\alpha_\ell^*+\eta_\ell\,,\nnr
&&\uplambda_r=\calK\alpha_r^*-\eta_r.\label{surfweight}
\eeqn
We have set $\gamma$ to zero, since its role is not completely clear. For a given weight $\uplambda$, we have a freedom to change $\alpha$ and $\eta$, while preserving $\uplambda_{\ell,r}$. So a given line operator in the Chern-Simons theory can be represented by a family of surface operators in the four-dimensional theory.

Now let us specialize for a moment to the operators of type $(\alpha,0,0,0)$. The action of the Weyl group on $\alpha$, together with the large gauge transformations which shift $\alpha$ by an element of the coroot lattice\footnote{For simplicity, here we restrict to a simply-connected gauge group, where the cocharacter lattice is the coroot lattice.} $\Gamma_w^*$ of the bosonic subalgebra, generate the action of the affine Weyl group $\hat{\mathcal{W}}_1=\mathcal{W}\ltimes \Gamma_w^*$ at level 1. Equivalently, on the quantum-corrected weights $\uplambda$ these transformations act as the affine Weyl group $\hat{\mathcal{W}}_\calK=\mathcal{W}\ltimes\calK\Gamma^*_w$ at level\footnote{By the affine Weyl group at some level $\p$ we mean the group which acts on the Cartan subalgebra by ordinary Weyl transformations together with shifts by $\p$ times a coroot. Our terminology is slightly imprecise, since as an abstract group, the affine Weyl group does not depend on the level.} $\calK$. Though the description by  surface operators makes sense for arbitrary $\uplambda$, let us look specifically at the integral weights $\uplambda\in\Gamma_w$. For generic $\calK$, the subgroup of $\hat{\mathcal{W}}_\calK$ which maps the weight lattice to itself consists only of the ordinary Weyl transformations. Therefore, the space of integral weights modulo the action of $\hat{\mathcal{W}}_\calK$ in this case is the space $\Gamma_w/\mathcal{W}$ of dominant weights  of the superalgebra, and the Chern-Simons observables corresponding to these weights are generically all inequivalent.  Of course, this is a statement about the analytically-continued theory, which is the only theory
that makes sense for generic $\calK$.  If however $\calK$ is a rational number $\p/\q$, then there are infinitely many elements of the affine Weyl group, which preserve the integral weight lattice $\Gamma_w$. (For example, such are all the transformations from $\hat{\mathcal{W}}_\p\subset\hat{\mathcal{W}}_\calK$.) Modulo these transformations, there is only a finite set of inequivalent integral weights. 

For an ordinary bosonic Chern-Simons theory and integer leve,l this can be compared to the well-known three-dimensional result according to which
 the inequivalent Chern-Simons line operators are labeled by the integrable weights $\Lambda\in\Gamma_w/\hat{\mathcal{W}}_k$. The connection between the two descriptions is that the weight $\Lambda$ is integrable at level $k$ if and only if the corresponding quantum corrected weight $\uplambda=\Lambda+\rho$ belongs to the {\it interior} of the fundamental Weyl chamber $\Gamma_w/\hat{\mathcal{W}}_\calK$, while the operators with $\uplambda$ belonging to the boundary of the fundamental Weyl chamber decouple in the Chern-Simons. This explains how the four-dimensional description by codimension-two operators with quantum-corrected level $\calK$ and weight $\uplambda$ can be equivalent (for integer $\calK$ and if the four-dimensional
theory is specialized to an appropriate class of observables) to the analogous  three-dimensional description by operators defined with ordinary $k$ and $\Lambda$. For the case of a supergroup, where the purely three-dimensional description is not completely clear (see Appendix \ref{Anomalous}), this discussion supports the view that, similarly to the bosonic case, at integer level there is a distinguished theory with only a finite set of inequivalent line operators. One detail to mention is that  in the four-dimensional construction, we did not show that the operators with $\uplambda$ lying on the boundary of the affine Weyl chamber decouple from the theory. We do not know for sure if this is true for supergroups in the context of a hypothetical
theory with only the distinguished set of line operators. Another caveat is that we worked with the half-BPS surface operators, and therefore our conclusion might not hold for the atypical supergroup representations.

\subsection{Various Problems}\label{various}
We conclude by emphasizing a few unclear points.

In the four-dimensional construction, we have separately defined Wilson line operators and Wilson-'t Hooft line operators in the
3d defect $W\subset M$.  They are parametrized
by the same data -- at least in the case of typical weights.  The Wilson line operators generically have less symmetry.  Is it conceivable
that they flow in the infrared to Wilson-'t Hooft line operators with enhanced symmetry?

For an atypical weight, there are many possible Wilson operators but only one half-BPS Wilson-'t Hooft operator.  This in itself is no contradiction.
But in the $S$-dual description of section \ref{magnetic} (see in particular section \ref{redsolutions}), 
we will find several half-BPS line operators for a given atypical weight.
The counterparts of this on the electric side seem to be missing.

One more technical puzzle arises  for  type II superalgebras. The half-BPS Wilson-'t Hooft operators seem to be well-defined for an arbitrary integral weight $\uplambda$, at least if it is typical, even though in some cases there is no corresponding representation.
(For a weight to correspond to a finite-dimensional representation, the weight should satisfy an extra
constraint, as was recalled in section \ref{reps}.)  There is no contradiction, but it is perhaps a surprise to apparently
find half-BPS Wilson-'t Hooft line operators that do not correspond to representations.

Additional line operators can presumably be constructed by coupling the bulk fields to some quantum mechanical degrees of freedom
that live only along the line operator.  This may help in constructing additional half-BPS line operators.  Perhaps it is important to understand better the BWB quantum mechanics for atypical weights.

\section{Magnetic Theory}\label{magnetic}
\subsection{Preliminaries}\label{magpre}
In this section we explore the $S$-dual description of our theory. Throughout this section 
the reader may assume that the theory considered corresponds to the supergroup $SG=\U(m|n)$.  This means in 
particular that the maximal bosonic subgroup $SG_{\bar{0}}=\U(m)\times \U(n)$ is
simply-laced. Some minor modifications that arise for other supergroups will be discussed in section \ref{ortho}.

We would like to recall how the supersymmetries and various parameters transform under  $S$-duality. It is convenient to look again on the Type IIB picture. Under the element
\beq\label{helpme}
\mathcal{M}=\left(\begin{array}{cc} a&b\\c&d\end{array}\right)
\eeq
of the $S$-duality group $SL(2,\ZZ)$,  the coupling constant of the theory transforms as 
\beq
\tau\rightarrow \fr{a\tau+b}{c\tau+d}.
\eeq
The supersymmetries of the Type IIB theory transform according to 
\beq
\veps_1+i\veps_2\rightarrow \ex^{i\alpha/2}(\veps_1+i\veps_2)\,,\label{iibs}
\eeq
where $\alpha=-\arg(c\tau+d)$. In particular, for the supersymmetries that are preserved by the D3-brane we can use the relation (\ref{d3susy}) to rewrite this as
\beq
\veps_1\rightarrow \exp\left(-\fr{1}{2}\alpha\,\Gamma_{0123}\right)\veps_1\,,
\eeq
in Lorentz signature. In \cite{Langlands} this relation was derived from the field theory point of view.

Under the duality transformation $\mathcal M$,  the charges of the fivebranes transform as
\beq\label{polk}
(p\,\,\,q)\rightarrow (p\,\,\,q)\mathcal{M}^{-1}\,,
\eeq
where $(p,q)=(1,0)$ for the NS5-brane and $(p,q)=(0,\pm1)$ for the D5- or $\bar{\rm D5}$-brane. For future reference we describe the supersymmetries that are preserved by a defect consisting of a general $(p,q)$-fivebrane. The supersymmetries preserved by such a brane, stretched in the 012456 directions, are given by the same formula as in (\ref{bufog}), where now
\beq\label{zumbo}
\ang=\arg(p\tau+q).
\eeq
Equation (\ref{bufog}) can be rewritten in a more convenient form
\beq
\veps_1+i\veps_2=i\ex^{i\ang}\,\Gamma_{012456}(\veps_1-i\veps_2).\label{pq}
\eeq
Under the $S$-duality, $\ang$ is shifted by angle $\alpha=-\arg(c\tau+d)$, so one can see that equation (\ref{pq}) indeed transforms covariantly, if the supersymmetries are mapped as in equation (\ref{iibs}). The twisting parameter $t=-\ex^{-i\ang}$ is multiplied by $\ex^{-i\alpha}$, that is,
\beq\label{zolk}
t\rightarrow t\fr{c\tau+d}{|c\tau+d|}.
\eeq

The canonical parameter $\calK$ of the bulk theory was defined in equation (\ref{zog}). In terms of the gauge coupling and the parameter $t$,
\beq
\calK=\fr{\theta_\YM}{2\pi}+\fr{4\pi i}{g^2_\YM}\fr{t-t^{-1}}{t+t^{-1}}.\label{canonical4d}
\eeq
For the special case that $t$ corresponds to the supersymmetry preserved by the D3-NS5 system, this reduces to eqn. (\ref{canonical}). Under $S$-duality, the canonical parameter transforms \cite{Langlands} in the same way as the gauge coupling,
\beq
\calK\rightarrow \fr{a\,\calK+b}{c\,\calK+d}.\label{cKtransform}
\eeq

Let us specialize to the case of  interest. The basic $S$-duality transformation  that exchanges electric and magnetic 
fields is usually described (for simply-laced groups) 
as $\tau\to -1/\tau$, but this does not specify it uniquely, since it does not determine the sign of the matrix 
$\mathcal M$  of eqn. (\ref{helpme}).  
We fix the sign by taking
\begin{equation}\mathcal M=\begin{pmatrix} 0 & 1 \cr -1 & 0 \end{pmatrix}.\end{equation}
This means, according to eqn. (\ref{polk}) that an NS5-brane,
with $(p,q)=(1,0)$, transforms to a $\bar{\mathrm D5}$-brane, with $(p,q)=(0,-1)$, so that according to eqn. (\ref{zumbo}), $\ang^\vee=\pi$ and $t^\vee=1$. 
Then from the definition (\ref{canonical4d}) of the canonical parameter, 
it follows that $\calK^\vee=\fr{\theta^\vee_\YM}{2\pi}$.

Unlike in the electric theory, the twisted action is very simple
on the dual magnetic side.  As in the purely bosonic case \cite{5knots}, the action is $\Qb$-exact except for a multiple of the instanton
number (see Appendix \ref{tech2} for a detailed explanation). In Euclidean signature, we have 
\beq
I_{\mathrm{magnetic}}=\fr{i\theta_\YM^\vee}{8\pi^2}\int\tr\left(F\wedge F\right)+\{\Qb,\dots\}.\label{Imagnetic}
\eeq
 If we set
\beq
q=\exp(-i\theta^\vee_\YM)\,,
\eeq
then the dependence of the theory on $q$ is easily described:  a solution of the localization equations of instanton number $n$
makes a contribution $\pm q^n$ to the path integral. (The sign is given by the sign of the fermion determinant.)
This simple result arises in the usual way because of cancellation between bosonic and fermionic fluctuations around 
a solution of the localization equations.  
  If therefore the instanton number is integer-valued and  is bounded above and below in all solutions of the localization equations,\footnote{\label{noninteger}One expects  the instanton number to be bounded in any solution, though this has not been proved.  However, the claim that the instanton number is integer-valued is oversimplified;
for example, if the gauge group is simply-connected or $M$ is contractible, the instanton number takes values in $\Z+c$
where $c$ is a constant determined by the boundary conditions.  In such a situation, the partition function is $q^c$ times
a Laurent polynomial in $q$.} then the path integral is a Laurent
polynomial in $q$ with integer coefficients, namely
\begin{equation}\label{omex}Z=\sum_n a_n q^n, \end{equation}
where $a_n$ is the number of solutions (weighted by sign) of instanton number $n$.

It is straightforward to express $q$ in terms of the parameters of the electric theory.  As explained above,
in the magnetic theory $\calK^\vee=\theta^\vee_\YM/2\pi$; also, according to (\ref{cKtransform}), $\calK^\vee=-1/\calK$.
So 
\beq
\theta^\vee_\YM=-2\pi/\calK\,,\label{thetaunitary}
\eeq
and hence
\beq
q=\exp\left(\frac{2\pi i}{\calK}\right).\label{qmagn}
\eeq
For an ordinary (simple, compact, and simply-laced) bosonic group, this is the standard variable
in which the quantum knot invariants are conveniently expressed, and for a supergroup it is the closest analog.
These matters were described in section \ref{turogo}.

 We now proceed to describing the localization equations and the boundary conditions in the magnetic theory, leaving
 many technical details for  Appendix~\ref{tech2}. Some relevant aspects of the gauge
 theory have been studied in \cite{SuperBC}. The details depend on the difference of the numbers of  D3-branes on the two sides of the D5-brane. We describe different cases in the subsequent sections.

\subsection{Gauge Groups Of Equal Rank}\label{eqrank}
In the case of an equal number of D3-branes on the two sides, the effective theory is a $\U(n)$ super Yang-Mills theory in the whole four-dimensional space, with an additional three-dimensional matter hypermultiplet localized on the defect, at $x_3=0$. This hypermultiplet comes from the strings that join the D5-brane and the D3-branes, and therefore it transforms in the fundamental of the $\U(n)$ gauge group. Under the global bosonic symmetries $\U=\SO(1,2)\times \SO(3)_X\times \SO(3)_Y$, the scalars $Z^A$ of the hypermultiplet  transform as a doublet $({\bf1, 2,1})$, and the fermions $\zeta^{\alpha\dA}$ transform as $({\bf 2,1,2})$.  The bulk fields have discontinuities at $x_3=0$ as a result of their
interaction with the defect.   For example, the
equations of motion of the gauge field, in Euclidean signature, can be deduced from the action
\beq
-\fr{1}{2(g^\vee_\YM)^2}\int {\rm d}^4 x\,\tr\,F^2_{\mu\nu}+\fr{1}{(g^\vee_{\YM})^2}I^\vee_{\rm hyp}.
\eeq
(In the magnetic description, the topological term $\int \tr \,F\wedge F$ is integrated over all of $\Bbb R^4$ and so
does not affect the equations of motion.) The equations of motion that come from the variation of 
this action have a delta-term supported on the defect,
\beq
D_3F^m_{3i} - \fr{1}{2}\delta(x_3)J^m_i\,=0\,,
\eeq
where $J^i_m=\delta I^\vee_{\rm hyp}/\delta A_i^m$ is the current.\footnote{Indices $m,n$ continue to denote
gauge indices, although now the gauge group is just one copy of $\U(n)$ throughout $\Bbb R^4$. Gauge indices are raised and lowered with the positive-definite Killing form $\delta_{mn}=-\tr(T_mT_n)$.} The delta-term in this equation means that the gauge field has a cusp at $x_3=0$, so that $F_{3i}$ has a discontinuity:
\beq
\left.F^m_{3i}\right|^\pm=\fr{1}{2}J^m_i.
\eeq
Here and in what follows we use the notation $\varphi|^\pm= \varphi(x_3+0)-\varphi(x_3-0)$ for the jump of a field across the defect. By supersymmetry, this discontinuity equation can be extended to a full three-dimensional current supermultiplet. The most important for us will be the lowest component of the current multiplet, 
which governs the discontinuity of the bulk scalar fields $X^a$:
\beq
X^{am}\bigr|^+_-=\fr{1}{2}\mu^{am}\,,\label{discX}
\eeq
where  the hyperkahler moment map for the defect hypermultiplets is
\beq
\mu^a_m=\bar{Z}_A\sigma^{aA}_BT_m Z^B.\label{momentmap}
\eeq
(The other bulk scalar fields $Y^{\dot a}$ are continuous at $x_3=0$.)

Now we turn to the twisted theory. Recall, that for twisting we use an $\SO(4)$ 
subgroup of the $R$-symmetry, which on the defect naturally reduces to $\SO(3)_X$. Thus, the 
hypermultiplet scalars $Z^A$ become spinors $Z^\alpha$ under the twisted Lorentz group. They are 
invariant under $\SU(2)_Y$, and therefore have ghost number zero. The hypermultiplet fermions 
$\zeta^{\alpha\dA}$ remain spinors.  Since they also transform as a doublet of $\SU(2)_Y$, 
we can expand them in the basis given by the vectors $u$ and $v$ of eqns. (\ref{u}) and (\ref{v}) (where
now we take $\ang^\vee=\pi$):
\beqn \label{momo}
&&\zeta^\dA=iu^\dA\zeta_u+iv^\dA\zeta_v\,,\nnr
&&\bzeta^\dA=iu^\dA\bzeta_u+iv^\dA\bzeta_v.
\eeqn
 The $u$- and $v$-components of $\zeta$ and $\bzeta$ have ghost number plus or minus one, respectively.

As usual, the path integral can be localized on the solutions of the BPS equations 
$\{\Qb,\xi\}=0$, where $\xi$ is any fermionic field. The resulting equations for the bulk 
fermions were partly described in eqn. (\ref{localization}). At $t^\vee=1$, they have a particularly simple form,
\beqn\label{lono}
&&F-\phi\wedge\phi+\star \d_A\phi=\fr{1}{2}\star \left(\delta_W\wedge\mu\right)\,,\nnr
&&D_\mu\phi^\mu=0.
\eeqn
Here $\delta_W=\delta(x_3)\d x_3$ is Poincar\'e dual to the three-manifold $W$ on which the defect
is supported.  The delta function term on the right hand side of the first equation in (\ref{lono})
is related to  the discontinuity (\ref{discX}) of the 1-form field $\phi$. 
There is no such term in the second equation, because the only field whose $x_3$ derivative appears
in this equation is $\phi_3$; this field originates as a component of $Y^{\dot a}$, and is continuous at $x_3=0$.
The condition that $\{\Qb,\xi\}=0$ for all $\xi$ also leads to  conditions
on  the ghost field $\sigma$:
\beq\label{melb}
D_\mu\sigma=[\phi_\mu,\sigma]=[\bar{\sigma},\sigma]=0.
\eeq
These equations say that the infinitesimal gauge transformation generated by $\sigma$ is a 
symmetry of the solution. In this paper we generally do not consider reducible solutions, so we generally can
set $\sigma$ to 0.

We also should consider the condition $\{\Qb,\xi\}=0$ where $\xi$ is one of the defect fermions. 
For the $u$-component of the fermions that are defined in eqn. (\ref{momo}), $\{\Qb,\xi\}$ equals the variation
of the defect fields under the gauge transformation generated by $\sigma$, so the condition for it to vanish, when
combined with (\ref{melb}) says that the full configuration including the fields on the defect is $\sigma$-invariant.  
More important for us
will be the condition $\{\Qb,\xi\}=0$ for the $v$-components:
\beq\label{bono}
\slashed{D}Z+\phi_3 Z=0\,,\quad \slashed{D}\bar{Z}-\bar{Z}\phi_3=0.
\eeq
Eqns. (\ref{lono}) and (\ref{bono}) together give the condition for a supersymmetric configuration.

\subsection{Gauge Groups Of Unequal Rank}\label{unequalrk}
Now consider the case that the number of D3 branes jumps from $n$ to $n+r$, $r>0$, upon 
crossing the D5-brane. The gauge groups on the left and on the right are $\U(n)$ and $\U(n+r)$, and will 
be denoted by $G_\ell$ and $G_r$, respectively. The behavior along the
defect has been described in \cite{SuperBC}.    In contrast to the case $r=0$, there are no
hypermultiplets supported along the defect at $y=0$.  
What does happen is different according to whether $r=1$ or $r>1$.  We first describe the behavior for $r>1$.

\def\frak{\mathfrak}
The main feature of this problem is that some of the bulk fields have a singular behavior (known as a Nahm pole
singularity) near $y=0$.  
Assuming that $r$ is positive, the singular behavior arises as one approaches $y=0$ from above.  
To describe the singularity, we first pick a subgroup $H=\U(n)\times \U(r)\subset \U(n+r)$, and we set
$H'=\U(n)\times \U(1)$, where $\U(1)$ is the center of the second factor in $H$.  The singularity will
break $G_r=\U(n+r)$ to $H'$.
The fields with a singular behavior are the scalar fields that we have called $X^a$ in the untwisted
theory or as $\phi_i$ in the twisted theory.  
The behavior of $\phi$ as $y$ approaches 0 from above is
\beq\label{mexo}
\phi_i=\frac{t_i}{y}+\dots\,,
\eeq
where the ellipsis represent less singular terms, and the matrices $t_i$ represent an irreducible
embedding of  $\mathfrak{su}(2)$ into the Lie algebra $\frak{u}(r)$ of the second factor
of $H=\U(n)\times \U(r)$.     Thus the matrices $t_i$  are $(n+r)\times (n+r)$ matrices that vanish except
for a single $r\times r$ block, as shown here for $n=2,\,r=3$:
\begin{equation}\label{forz}\begin{pmatrix} 0 & 0 &0 &0 &0 \cr
    0 & 0 & 0& 0& 0 \cr
    0&0&*&*&*\cr 0 &0&*&*&*\cr 0&0&*&*&* \end{pmatrix}.\end{equation}
    These matrices are traceless, so their nonzero blocks are actually valued in $\frak{su}(r)\subset \frak{u}(r)$.

The Nahm pole singularity breaks the gauge symmetry for $y>0$ from $\U(n+r)$ to $H'=\U(n)\times \U(1)$,
and there is to begin with a $G_\ell=\U(n)$ gauge symmetry for $y<0$.  There is therefore a $\U(n)$ gauge symmetry on
both sides of the defect, and the condition obeyed by the $\U(n)$ gauge fields is just that they are continuous at $y=0$,
making  a $\U(n)$ gauge symmetry throughout the whole spacetime.  On the other hand, the fields supported
at $y>0$ that do not commute with the Nahm pole singularity acquire very large masses near $y=0$, and they vanish for $y\to 0$.   (This statement applies to fields in the adjoint representation of $\frak{su}(r)$
and also to fields in the bifundamental of $\U(n)\times \U(r)$.)
To finish describing the gauge theory of the defect, we must explain the behavior at $y=0$ of the fields
in the second factor of $H'=\U(n)\times \U(1)$.  These fields make up a single vector multiplet, which obeys
what we might call Dirichlet boundary conditions (the gauge fields $A_i$ and scalars $Y^{\dot a}$ in this multiplet
obey Dirichlet boundary conditions, while the scalars $X^a$ obey Neumann boundary conditions; these conditions
are extended to the fermions in a fashion determined by supersymmetry).

For $r=1$, this description requires some modification, because $\frak{su}(1)=0$ and accordingly the matrices $t_i$ vanish.
Still, the defect breaks the $G_r=\U(n+1)$ gauge symmetry for $y>0$ to a subgroup $H'=\U(n)\times \U(1)\subset \U(n+1)$. Just
as at $r>1$, the
$\frak{u}(n)$-valued gauge fields on the two sides of the defect fit smoothly into continuous $\frak{u}(n)$-valued fields
throughout the whole spacetime.  For $y>0$, the gauge fields valued in the orthocomplement of $\frak{u}(n)$ obey
the same Dirichlet boundary conditions described at the end of the last paragraph.

So far, we have described this construction as if the matrices $t_i$ in eqn. (\ref{mexo}) are just constant matrices.
This makes sense if $W=\RR^3$, but in general, we must recall that in the twisted theory on $M=W\times\RR$,
 $\phi=\sum_i\phi_i\d x^i$ transforms as a 1-form along $W$.  The proper interpretation of the Nahm pole
 singularity in this general setting is as follows (see section 3.4 of \cite{5knots};
 the considerations there carry over to the present case without essential change).  The $\frak{u}(r)$
 bundle along $W$ must be derived from a spin bundle  $S_W$ via a homomorphism $\varrho:\frak{su}(2)\to \frak{u}(k)$
 defined by the $t_i$.  The restriction to $W\times \{y=0\}$ of the $\frak{u}(r)$-valued part of the gauge field is the Levi-Civita
 connection $\omega$ of $S_W$, embedded in $\frak{su}(r)$ via $\varrho$.  We describe this by saying that
 when restricted to $y=0$, the $\frak{u}(r)$-valued part of the gauge field $A$ is $A_{\frak u(r)}=\varrho(\omega)$.

\subsubsection{The Framing Anomaly}\label{franom}

It is now possible to make an interesting check of the relationship between Chern-Simons theory of $\U(n|n+r)$ and
the defect theory just described. Here we will be rather brief, assuming that the reader is familiar with the description
of the one-sided case in section 3.5.3 of \cite{5knots}. 
Recall that in general the partition function of  Chern-Simons 
theory on a three-manifold $W$ is not quite a topological invariant of $W$;
$W$ must be endowed with a framing (or more precisely a two-framing \cite{Atiyah}) to define
this partition function.  A framing is a trivialization of the tangent bundle of $W$.
 Under a unit change of framing, the partition function 
acquires a factor \cite{WittenCS}
\beq
\exp(2\pi ic\,\sign(k))/24)\,,\label{globalframing}
\eeq
where $c$ is the central charge of the relevant current algebra at level $k$.  For a compact simple gauge group $G$
this is $c=k \,\mathrm{dim}\,G/(k+h\,\,\sign(k))$, where $h$ is the dual Coxeter number of $G$.  We will
assume that the same formula for $c$
applies, at least modulo an integer, for a simple supergroup $SG$, which in our case will be $\SU(n|n+r)$:
\beq\label{ovely}
c=\frac{k\,{\rm sdim}\,SG}{k+h_{\mathfrak{sg}}\sign(k)}~~{\rm mod}\,\Bbb Z.
\eeq
This is a non-trivial assumption, since some of the standard arguments do not
apply for supergroups, as we describe in Appendix \ref{Anomalous}.  
(Replacing $\SU(n|n+r)$ by $\U(n|n+r)$, which is isomorphic locally to 
$\SU(n|n+r)\times \U(1)$, shifts $c$ by 1, which will not be important as we will only study $c$ mod $\Bbb Z$.
So the following discussion will be phrased for the simple supergroup $\SU(n|n+r)$, rather than $\U(n|n+r)$.) 
It is useful to factor
(\ref{globalframing}) as follows:
\beq\label{pingo}
\exp\left(2\pi i\,\sign(k)\,{\rm sdim}\,SG/24 \right)\cdot q^{-h_\sg\,\rm{sdim}SG/24}.
\eeq
Perturbation theory is an expansion in powers of $1/\calK$, with an $\ell$-loop diagram making
a contribution of order $\calK^{1-\ell}$.  Accordingly, the exponent $2\pi i\,\sign(k)\,{\rm sdim}\,SG/24$ in the first
factor in (\ref{pingo}), being invariant under scaling of $k$, is a 1-loop effect.  Since it is not analytic in $\calK$,
 we cannot hope to reproduce it from 
four dimensions.  If this factor -- or a similar one that arises if $c$ is shifted by an integer;
see Appendix \ref{Anomalous} -- appears in a purely three-dimensional construction, then it must appear in a comparison
between the relevant measures in three and four dimensions, as discussed in section \ref{turogo} above and in section 3.5.3
of \cite{5knots}.  
However, the second factor in (\ref{pingo}), which is a simple power of $q$, comes from diagrams with $\geq 2$ loops
and can be reproduced from four dimensions.

As in 
\cite{5knots}, this factor arises from a subtlety in the definition of  
instanton number in the presence of the Nahm pole. 
The condition that along $W\times \{y=0\}$, $A_{\frak{u}(r)}=\varrho(\omega)$ means that the instanton number, defined in the obvious
way from the integral $\int_{M_\ell}\Tr\, F\wedge F+\int_{M_r}\Tr\,F\wedge F$, is not a topological invariant.
If one varies the metric of $W$, the second term picks up a variation from the change in $\omega$.
To compensate for this, one must add to the instanton number a multiple of the Chern-Simons invariant of $\omega$, but this is only gauge-invariant (as a real number) once we pick a framing on $W$.
From the viewpoint of the dual magnetic description, that is why Chern-Simons theory on $W$ requires a framing
of $W$.  
To adapt the analysis of \cite{5knots} to the present problem, we simply proceed as follows.  In the $\U(n|n+r)$ case, the Nahm pole is embedded in a
$\mathfrak{u}(r)$ subalgebra, and therefore the framing-dependence that is introduced when we define the instanton
number for this problem is independent of $n$ and is the same as it is for the one-sided problem with $n=0$
and gauge group $\U(r)$.  Hence,  to obtain in the magnetic
description the expected factor $q^{-h_{\sg}{\mathrm{sdim}}\, G/24}$ in the framing dependence, we need the identity
\beq\label{needid}
h_{\mathfrak{su}(n|n+r)}\,{\rm sdim}\,SU(n|n+r)=h_{\mathfrak{su}(r)}\,{\rm dim}\,SU(r).
\eeq
This is true because ${\rm sdim}\,SU(n|n+r)$ is independent of $n$ and likewise $h_{\frak{su}(n|n+r)}$ is independent of
$n$.  
See Table \ref{coxeters}.

\subsection{Line And Surface Operators In The Magnetic Theory}\label{magline}
Our next goal is to identify the $S$-duals of the line and surface operators that we have found on the electric side. We use the fact that we know how 
$S$-duality acts on the bulk surface operators. For an ``electric'' surface operator, the magnetic dual \cite{Surface} has parameters $(\alpha^\vee,\beta^\vee,\gamma^\vee,\eta^\vee)=(\eta,|\tau|\beta^*,|\tau|\gamma^*,-\alpha)$, where $\tau$ is the gauge coupling constant. This determines the singularity of the fields along the operator in the bulk, away from the three-dimensional defect. We still have to find the model solution which describes the behavior of the fields near the end of the surface operator at $y=0$. This will be the main subject of the present section. 

In bulk, for a surface operator with parameters $(\alpha,\beta,\gamma,\eta)$, the parameters $\alpha$ and $\eta$ are both periodic.  In the presence
of a defect, this is no longer the case.  In the electric description, $\eta$ is not a periodic variable on a D3-brane that ends on (or intersects) an
NS5-brane.  Shifting $\eta$ by an integral character would add a unit of charge along the defect.  Dually to this, for the D3-D5 system,
in the case of a surface operator with parameters $(\alpha^\vee,\beta^\vee,\gamma^\vee,\eta^\vee)$, $\alpha^\vee $ is not a periodic variable.
In the model solutions that we construct below, if $\alpha^\vee$ is shifted by an integral cocharacter (of $G^\vee$), then the solution is unchanged
in the bulk up to a gauge transformation, but is modified along the defect.

It follows from this that once we construct model solutions for surface operators
with parameters $(\alpha^\vee,\beta^\vee,\gamma^\vee)$, we can trivially construct
magnetic line operators.  We return to this in section \ref{lineop}.

\subsubsection{Reduction Of The Equations}
We  focus first on the case of gauge groups of equal rank, as described in section \ref{eqrank}. The discussion can be transferred to the unequal rank case in a straightforward way, and we shall comment on this later. 

To give a definition of a surface operator whose support intersects the three-dimensional defect, we have to find a model solution of the localization equations (\ref{lono}) and (\ref{bono}) for the fields near the surface $C$ and near the hyperplane $y=0$. The classical solution does not depend on the two-dimensional theta-angles $\eta^\vee$, so we label it by three parameters $(\alpha^\vee,\beta^\vee,\gamma^\vee)$. We consider a surface operator stretched along $C=\RR_{x^0}\times \RR_y$ in $\RR^4$,and look for a time-independent, scale-invariant solution. We aim to construct a model solution that is 1/2-BPS, that is, it preserves the four supersymmetries (\ref{zuperparameter}) and (\ref{Qtilde}). It should also be invariant under the SO$(3)_Y$ subgroup of the R-symmetry groups. The symmetries allow us to considerably reduce the localization equations. An analogous problem in the one-sided theory was considered in section 3.6 of \cite{5knots}, where the reader can find many details which we do not repeat here.

First of all, for an irreducible solution the field $\sigma$ is zero, and therefore, by SO$(3)_Y$ symmetry, $\phi_3$  should also vanish. The $\Qb$-invariance together with SO$(3)_Y$ symmetry makes the solution invariant under the first pair of supersymmetries (\ref{zuperparameter}). Using the explicit formulas for the transformations (\ref{N4formulas}), one can also impose invariance under the second pair of supersymmetries (\ref{Qtilde}). For $t^\vee=1$, which is the case in the magnetic theory, this fixes $A_0$ to be zero. The reduced localization equations can be written in a concise form, after introducing some convenient notation. Following \cite{5knots}, we define three operators
\beqn
&&\mathcal{D}_1=2D_{\bar{z}}\,,\nnr
&&\mathcal{D}_2=D_3-i\phi_0\,,\nnr
&&\mathcal{D}_3=2\phi_z\,,
\eeqn
where $z=x_1+ix_2$ is a complex coordinate, $\phi_z=(\phi_1-i\phi_2)/2$ is the $z$-component of $\phi$, and $D_{\bar{z}}$ and $D_3$ are covariant derivatives. We also denote the components of the bosonic spinor field $Z^\alpha$ as $Z\equiv Z^1$ and $\tilde{Z}\equiv(Z^2)^\dagger$. For simplicity, we assume the gauge group $G^\vee$ to be U$(n)$. Then the components of the moment map (\ref{momentmap}) can be written as
\beq
\mu_0=i(\tilde{Z}^\dagger\otimes \tilde{Z}-Z\otimes Z^\dagger)\,,\quad \mu_z=-iZ\otimes\tilde{Z}.
\eeq

With this notation, the reduced localization equations are
\beqn
&&[\mathcal{D}_1,\mathcal{D}_2]=0\,,\quad [\mathcal{D}_3,\mathcal{D}_1]=0\,,\quad [\mathcal{D}_2,\mathcal{D}_3]-\mu_z\delta(y)=0\,,\nnr
&&\mathcal{D}_1Z=\mathcal{D}_1\tilde{Z}=0\,,\label{grone}
\eeqn
together with
\beq
\sum_i[\mathcal{D}_i,\mathcal{D}^\dagger_i]+i\mu_0\delta(y)=0.\label{grtwo}
\eeq
The space of fields in which we look for the solution is the space of continuous connections on $\RR^4\setminus C$, and Higgs fields with an arbitrary discontinuity across the hyperplane $y=0$. (The fields should also be vanishing at infinity.) The correct discontinuity (\ref{discX}) is enforced by the delta-terms in the localization  equations. To put the real and imaginary parts $A_3$ and $\phi_0$ of the connection in $\mathcal{D}_2$ on equal footing, let us also allow $A_3$ to have an arbitrary discontinuity across $y=0$, and to compensate for this, we divide the space of solutions by the gauge transformations, which are allowed to have a cusp across the defect hyperplane.

The analysis of these equations in the one-sided case in \cite{5knots} was based on the fact that the equations (\ref{grone}) are actually invariant under
complex-valued gauge transformations, not just real-valued ones.  One can try to solve the equations in a two-step procedure in which one first
solves eqn. (\ref{grone}) and then tries to find a complex-valued gauge transformation to a set of fields that obeys (\ref{grtwo}) as well.

Though we could follow that strategy here as well, we will instead follow a more direct approach.  We are motivated by the fact that the basic
surface operator in the absence of any defect or boundary is described by a  trivial abelian solution.    In the one-sided problem, one requires
a Nahm pole along the boundary and therefore the full solution is always irreducible.  However, in the two-sided case with equal ranks,
there is no Nahm pole.  Is it too much to hope that we can find something interesting by taking simple abelian solutions for $y<0$ and $y>0$,
somehow glued together along $y=0$?

\subsubsection{Some ``Abelian'' Solutions}\label{absol}
We look for a model solution for a surface operator with parameters $(\alpha^\vee,0,0)$, and initially we assume $\alpha^\vee$ regular.
Since we take $\beta^\vee=\gamma^\vee=0$, we look
for a model solution  invariant under the SO$(2)$ group of rotations in the 12-plane, and under the SO$(2)_X$ subgroup of the R-symmetry. 
Accordingly, the field $\phi_z$ should vanish. Indeed, the SO$(2)_X$ acts on $\phi_z$ by multiplication by a phase. 
In a fully non-abelian solution, this phase could possibly
 be undone by a gauge transformation, but in a solution that is abelian away from $y=0$ -- as we will assume here -- that is not possible and
 $\phi_z$ must vanish.  Therefore, from the discontinuity equation for $\phi_z$ it follows that either $Z$ or $\tilde{Z}$ should vanish. 
 So for definiteness, assume that $\tilde{Z}=0$ and $Z\not=0$.   
 
 For now we focus on irreducible solutions, for which the gauge group along $K$ is broken completely. We postpone the discussion of
 reducible solutions.  
 
  A simple abelian solution of the localization equations would be $A=\alpha^\vee\cos\varphi\d\theta$, $\phi=\alpha^\vee \d x^0/r'$.
 For $y\to\infty$, $\phi$ vanishes, and $A$ approaches the simple surface operator solution $\alpha^\vee \d\theta$ for $y\to+\infty$ ($\theta=0$)
 or $-\alpha^\vee\d\theta$ for $y\to -\infty$ ($\theta=\pi$).  However, we want a solution in which $A$ will approach independent limits $\alpha_\ell^\vee\d\theta$ and $\alpha_r^\vee\d\theta$
 for $y\to-\infty$ and $y\to+\infty$.  Also we want to allow for the possibility that a gauge transformation by a constant matrix $g$ has to be made to
 match the solutions for $y<0$ and $y>0$.  So we try
\begin{align}\label{abansatz}
y>0:&~~~ A=\alpha_r^\vee \cos\varphi\,\d\theta\,,   ~~~~~~~~~~~~~~~ \phi=\alpha_r^\vee \frac{\d x^0}{r'}\,, \cr
y<0:&~~~~ A=-g\alpha_\ell^\vee g^{-1}\cos\varphi\,\d\theta\,, ~~~~~\phi=-g\alpha_\ell^\vee g^{-1}\frac{\d x^0}{r'}. 
\end{align}

We also have to impose the discontinuity equation $\phi_0\bigr|^\pm=\fr{i}{2}(\tilde{Z}^\dagger\otimes\tilde{Z}-Z\otimes Z^\dagger)$. Note first of all that taking the trace of this gives $i(\tr(\alpha_r^\vee)+\,\tr(\alpha_\ell^\vee))=r'(|Z|^2-|\tilde{Z}|^2)/2$. Therefore, the choice of whether $Z$ or $\tilde{Z}$ is non-zero is determined by the sign of the combination of parameters on the left hand side of this equation. We assume this combination to be positive, and take 
\begin{equation}\label{telbo}Z=\frac{\mathtt{v}}{\sqrt{z}},\end{equation}
 where $\mathtt{v}$ is some constant vector. We have taken $Z$ to be holomorphic to satisfy $\mathcal D_1Z=0$ (this is one of the localization
 equations, eqn. (\ref{grone})).  Note that
 $A$ does not appear in this equation, since it vanishes at $y=0$, so the formula for $Z$ does not depend on $\alpha^\vee_\ell$ or $\alpha^\vee_r$.  Also, (\ref{telbo}) 
means that $Z$ has a monodromy $-1$ around the knot, which in this description is located at $z=0$.  So we have to assume that this
monodromy of $Z$ is part of the definition of the surface operator in this magnetic description.

The discontinuity equation now becomes
\beq
i\alpha^\vee_r+ig\alpha^\vee_\ell g^{-1}=\fr{1}{2}\mathtt{v}\otimes \mathtt{v}^\dagger.\label{ghjk}
\eeq
This is a set of $n^2$ equations for a unitary matrix $g$ and a vector $\mathtt{v}$, which are together $n^2+n$ variables. 
The equations are invariant under the diagonal unitary gauge transformations, which remove $n$ parameters. Therefore, 
generically one expects to have a finite number of solutions. 

The equations can be conveniently formulated as follows. For a given hermitian matrix $N=i\alpha^\vee_r$, 
find a vector $\mathtt{v}$, such that the hermitian matrix $N'=N-\fr{1}{2}\mathtt{v}\otimes \mathtt{v}^\dagger$ has the same eigenvalues as $M=-i\alpha^\vee_\ell$. 
Using the identity $\det(X+\mathtt{v}\otimes \mathtt{v}^\dagger)=(1+\mathtt{v}^\dagger X^{-1} \mathtt{v})\det(X)$, 
the characteristic polynomial for $N'$ can be written as
\beq
\det\left(\mathbb{1}\cdot \lambda-N+\fr{1}{2}\mathtt{v}\otimes \mathtt{v}^\dagger\right)=
\det(\mathbb{1}\cdot \lambda-N)\left(1+\fr{1}{2}\sum_{i=1}^n\fr{|\mathtt{u}_i^\dagger \mathtt{v}|^2}{\lambda-\lambda_i}\right)\,,
\eeq 
where $\mathtt{u}_i$ are the eigenvectors of $N$ with eigenvalues $\lambda_i$. First let us assume
that $\mathtt{u}_i^\dagger \mathtt{v}\ne0$ for all $i$. Then the eigenvalues of $N'$ are solutions of the equation
\beq\
1+\fr{1}{2}\sum_{i=1}^n\fr{|\mathtt{u}_i^\dagger \mathtt{v}|^2}{\lambda-\lambda_i}=0.\label{asdf}
\eeq
Note that all the eigenvalues of $N$ are distinct~--~this is the regularity condition for the weight, which says that 
$\langle\uplambda,\alpha_\bos\rangle\equiv\langle\Lambda+\rho,\alpha_\bos\rangle\ne0$ for all the superalgebra bosonic 
roots $\alpha_\bos$. By sketching a plot of the function in the left hand side of (\ref{asdf}), it is easy to observe that 
the equation has $n$ solutions $\lambda=\lambda_i', $ $i=1,\dots,n$.  These solutions
interlace the eigenvalues $\lambda_i$, in the sense that if the $\lambda_i$ and $\lambda'_i$ are arranged in increasing order then $\lambda'_1<\lambda_1<\lambda_2'<
\dots<\lambda_n$.  Had we assumed $\t Z$ rather than $Z$ to be non-zero, we would have obtained the opposed interlacing 
condition $\lambda_1<\lambda_1'<\lambda_2<\dots<\lambda_n'$.
Moreover, by tuning the $n$ coefficients $|\mathtt{u}_i^\dagger \mathtt{v}|^2$ of the equation, one can in a unique way put these 
solutions to arbitrary points inside the intervals $(-\infty,\lambda_1)$, $(\lambda_1,\lambda_2)$, \dots, $(\lambda_{n-1},\lambda_n)$, 
to which they belong.  To do this, we simply view eqn. (\ref{asdf}) as a system of linear equations for the constants
$|\mathtt{u}_i^\dagger \mathtt{v}|^2$. The interlacing condition ensures that there is no problem with the positivity of those constants. An important special case is that  $|\mathtt{u}_i^\dagger \mathtt{v}|^2\to 0$ precisely when $\lambda'_j$ (for $j=i$ or $i\pm 1$) approaches $\lambda_i$. 
The facts we have just stated are used in some applications of random matrix theory; for example, see p. 16 of \cite{sriv}.

We conclude that the equation (\ref{ghjk}) has a solution, which moreover is unique (modulo diagonal gauge transformations), if and only if the eigenvalues of $i\alpha^\vee_r$ and $-i\alpha^\vee_\ell$ are interlaced. Since the eigenvalues of $i\alpha^\vee_r$ and $i\alpha^\vee_\ell$ should be the weights of a dual
Wilson-'t Hooft operator on the electric side, we have a reasonable candidate for the dual of such operators when certain inequalities
are satisfied.  If some of the eigenvalues of $i\alpha^\vee_r$ coincide with eigenvalues of $-i\alpha^\vee_\ell$, then the corresponding components of $Z=\mathtt{v}/\sqrt{z}$ vanish.  (We return to this point in section \ref{redsolutions}.)

If the eigenvalues are not interlaced, the abelian ansatz fails.  As a motivation to understand what to do in this case, we will describe a possibly more
familiar problem that leads to the same equations and conditions that we have just encountered.  We look at the system of N D3-branes intersecting a D5-brane
from a different point of view.  Instead of studying a surface operator, we look for a supersymmetric vacuum state in which the fields $\vec X$ have one asymptotic
limit $\vec X_\ell$ for $y\to-\infty$ and another limit $\vec X_r$ for $y\to+\infty$.  Such a vacuum exists for any choice of $\vec X_\ell$, $\vec X_r$, and is
unique up to a gauge transformation.  Macroscopically, this vacuum is often just understood  by saying that a D3-brane can end on a D5-brane so the value of $\vec X$
can jump from $\vec X_\ell$ to $\vec X_r$ in crossing the D5-brane.  Thus, one represents the vacuum by the simple picture of fig. \ref{setup} of section \ref{content},
but now with the fivebrane in the picture understood as a D5-brane.  

Although this picture is correct macroscopically, from a more microscopic point of view, the vacuum of the D3-D5 system is found by solving Nahm's equations
for the D3 system, with the D3-D5 intersection contributing a hypermultiplet that appears as an impurity. This has been analyzed in detail in \cite{SuperBC}.
Let us just consider the case that the branes are separated at $y\to \pm \infty$ only in the $X_4$ direction, where $X_4$ corresponds to $\phi_0$ in our notation here. 
A natural ansatz would then be to assume that $X_5=X_6=0$ everywhere.  That leads to simple equations.
 Nahm's equations
with  $X_5=X_6=0$ just reduce to $\d X_4/\d y=0$ (for $y\not=0$), so $X_4$ is one constant matrix for $y>0$ and a second constant matrix for $y<0$.
After diagonalizing $X_4$ for $y>0$, we can write $X_4=\alpha_r^\vee$ for $y>0$, $X_4=-g\alpha_\ell^\vee g^{-1}$ for $y<0$, with $\alpha_\ell^\vee,\alpha_r^\vee\in
\frak t$, $g\in \U(n)$.  Finally, in the construction of the vacuum, the jump condition at the location of the hypermultiplet is precisely (\ref{discX}).

\begin{figure}
 \begin{center}
   \includegraphics[width=15cm]{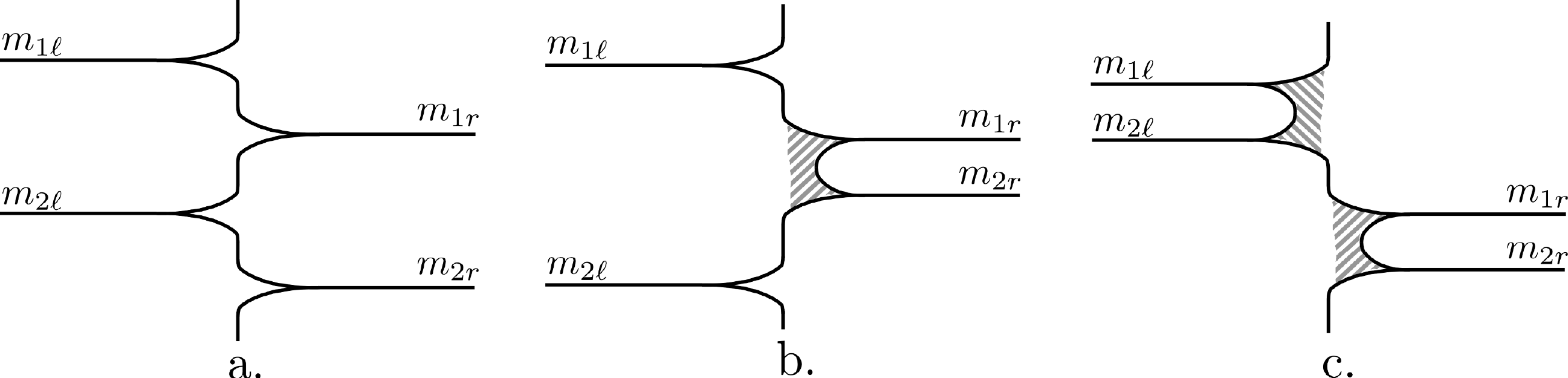}
 \end{center}
\caption{\small D3-branes ending on the two sides of a D5-brane. If the branes are not interlaced, they can form a fuzzy funnel.}
 \label{interlaced}
\end{figure}

So in constructing the vacuum assuming that $X_5=X_6=0$ identically, the solution exists if and only if the eigenvalues of $X_4$ are interlaced, so that the branes are placed as shown in  fig. \ref{interlaced}(a). What if they are not interlaced? A unique vacuum solution still exists, but the assumption that $X_5$ and $X_6$ 
are identically 0 is no longer valid.  For example, if two of the eigenvalues of $X_4$ for $y\to -\infty$ or for $y\to +\infty$ are very close -- in other words
if two of the $\lambda_i$ or two of the $\lambda'_i$ are very close -- then the neighboring branes form a fuzzy funnel, as in fig. \ref{interlaced}(b,c).
The fuzzy funnel  is described  \cite{FMT} by a nonabelian solution of Nahm's equations, with $X_4,X_5\not=0$.    If $X_4,X_5\to 0$ for  $y\to \pm\infty$,
then in the appropriate solution of Nahm's equations, $X_4\pm i X_5$ is nilpotent, but not zero \cite{SuperBC}.    This suggests that we should try a new ansatz with $\phi_z$ nilpotent but not zero in order
to find the missing solutions when the  weights are not interlaced.    For now, we present this as heuristic motivation for
a more general ansatz, but later we will explain a precise map between the problem of finding half-BPS surface operators 
and Nahm's equations for a D3-D5 vacuum.

\subsubsection{General Solution For U(2)}\label{gensol}
We consider the first non-trivial example of this problem, which is for gauge group $\U(2)$, corresponding to $\U(2|2)$ on the electric side. 
We  focus on the configuration shown in fig. \ref{interlaced}(b). The positions of the branes in that figure  should be interpreted as the eigenvalues of the matrices
which appear in the $1/r'$ singularity of the field $\phi_0$. If the weights are $\alpha^\vee_r=i\,{\rm diag}(m_{1r},m_{2r})$ and $\alpha^\vee_\ell=-i\,{\rm diag}(m_{1\ell},m_{2\ell})$, then $m_{1r,\ell}$ and $m_{2r,\ell}$ label the positions of the horizontal lines in fig. \ref{interlaced}. We assume that, by a Weyl conjugation, $\alpha^\vee$ was brought to the form with $m_{1r}>m_{2r}$ and $m_{1\ell}>m_{2\ell}$. 

We introduce a convenient variable $\varsigma$ defined as $\sinh\,\varsigma=\cot\varphi$ (or $\tanh \varsigma=\cos\varphi$). It runs from $-\infty$ to $0$ on the left of the defect, and from $0$ to $+\infty$ on the right. For the fields on the left of the defect, we use the same abelian ansatz (\ref{abansatz}). For the fields on the right, we want to find a conformally- and SO$(2)_X$-invariant solution with $\phi_z$ belonging to the non-trivial nilpotent conjugacy class. A family of such solutions, which actually contains all the solutions with these symmetries, was found in \cite{5knots}, and has the following form,
\beqn \label{zoff}
&&A=\fr{i}{2}\left(\begin{array}{cc}m_{1r}+m_{2r}+\partial_\varsigma V_r&0\\0&m_{1r}+m_{2r}-\partial_\varsigma V_r\end{array}\right)\cos\varphi\,{\rm d}\theta\,,\nnr
&&\phi_0=\fr{i}{2r'}\left(\begin{array}{cc}m_{1r}+m_{2r}+\partial_\varsigma V_r&0\\0&m_{1r}+m_{2r}-\partial_\varsigma V_r\end{array}\right)\,,\nnr
&&\phi_z=\fr{1}{2z}\left(\begin{array}{cc}0&1\\0&0\end{array}\right)\exp(-V_r)\,,\label{fright}
\eeqn
where the function $V_r(\varsigma)$ is found from the localization equations to be
\beq
V_r=\log\left(\fr{\sinh(a_r\varsigma+b_r)}{a_r}\right).
\eeq
The ansatz is $\SO(2)_X$-invariant up to a diagonal gauge transformation.
In (\ref{zoff}),  $a_r$ and $b_r$ are some unknown constants. We choose $a_r$ to be positive. Then $b_r$ should also be positive, so that no singularity appears\footnote{The singularity that the solution has at $a_r\varsigma+b_r=0$ is the Nahm pole. In the one-sided problem, one chooses $b_r=0$ to have this pole precisely at $\varsigma=0$.} in the interval $\varsigma\in(0,\infty)$. The requirement that the behavior of the gauge field at $\varsigma\rightarrow\infty$ should agree with the surface operator $A=\alpha^\vee_r{\rm d}\theta$ fixes $a=m_{1r}-m_{2r}$. (Had we chosen the opposite Weyl chamber for $\alpha^\vee$, we would have to make a Weyl transformation on the ansatz (\ref{fright}), making $\phi_z$ lower-triangular.) Note that, due to the $\cos\varphi$ factor, the gauge field at $y=0$ vanishes; this  agrees with our requirement that  $Z^\alpha\sim 1/\sqrt z$ should have monodromy $-1$. The next step is to impose the discontinuity equations at $\varsigma=0$, and to hope that they will have a solution for some positive real $b_r$. The $z$-component of the discontinuity equations tells us that the hypermultiplet fields should have the form
\beq
Z=\fr{1}{\sqrt{z}}\left(\begin{array}{c}\mathtt{s}\\0\end{array}\right)\,,\quad \tilde{Z}=\fr{1}{\sqrt{z}}(0\,\,i\mathtt{w}).\label{hypfields}
\eeq
Unlike the interlaced case, here there is no freedom to include a general non-abelian  gauge transformation in gluing the left and the right side. Such a gauge transformation would not be consistent with the symmetry, since generically it would not commute with the U$(1)$ subgroup of the gauge group which is used to undo the SO$(2)_X$ rotations. The only possible non-abelian gluing gauge transformation is the Weyl conjugation. The equations will tell us that in this case it is not needed. The $\phi_0$ and $\phi_z$ discontinuity conditions give
\begin{align}
\frac{{a_r}}{\sinh b_r}&=\mathtt{s}\mathtt{w}\,,\nnr \cr
m_{1r}+m_{2r}-2m_{1\ell}+a_r\coth b_r&=-|\mathtt{s}|^2\,,\nnr\cr
m_{1r}+m_{2r}-2m_{2\ell}-a_r\coth b_r&=|\mathtt{w}|^2.\label{algeqs}
\end{align}
Subtracting the last two equations, we see that a solution with positive $b$ cannot exist unless $m_{1\ell}-m_{2\ell}>0$. This is consistent with our choice of the Weyl chamber, so no gluing gauge transformation is needed. Eliminating $\mathtt{s}$ and $\mathtt{w}$ from (\ref{algeqs}), we obtain
\beq
\fr{m_{1\ell}-m_{2\ell}}{m_{1r}-m_{2r}}=\coth b_r+\sqrt{\left(\fr{m_{1r}+m_{2r}-m_{1\ell}-m_{2\ell}}{m_{1r}-m_{2r}}\right)^2+\fr{1}{\sinh^2b_r}}.
\eeq
The function on the right is monotonically decreasing. It is easy to see that the equation has a solution $b_r>0$ if and only if the eigenvalues are arranged as in fig. \ref{interlaced}b.

The last case to consider for the U$(2)$ group is that of fig. \ref{interlaced}c. Here fields on both sides of the defect should have a non-zero nilpotent $\phi_z$. The fields on the right are given by the same ansatz (\ref{fright}). The fields on the left are given by the same ansatz, but with $V_r$ replaced by
\beq
V_\ell=\log\left(\fr{\sinh(-a_\ell\varsigma+b_\ell)}{a_\ell}\right).
\eeq
Again, we assume $a_\ell$ to be positive, and then $b_\ell$ should also be positive to avoid the singularity on the interval $\varsigma\in(-\infty,0)$. We fix $a_\ell$ from the asymptotics at $\varsigma\rightarrow-\infty$ to be $a_\ell=m_{1\ell}-m_{2\ell}$, though in this case the gauge field $A$ asymptotically is proportional to ${\rm diag}(m_{2\ell},m_{1\ell})$. We could make a Weyl gauge transformation to bring it to the other Weyl chamber.

In gluing left and right, we cannot make any non-diagonal gauge transformations, as follows again from the $\SO(2)_X$ symmetry. There are two separate cases to consider. First assume that $\phi_z$ has a non-trivial jump at $y=0$. This forces the hypermultiplet fields $Z$ and $\tilde{Z}$ to have the form (\ref{hypfields}). The discontinuity equations give 
\begin{align}
\fr{a_r}{\sinh b_r}\pm\fr{a_\ell}{\sinh b_\ell}&=\mathtt{s}\mathtt{w}\,,\nnr\cr
a_r\coth b_r+a_\ell\coth b_\ell&=-\fr{|\mathtt{w}|^2+|\mathtt{s}|^2}{2}\,,\label{uio} \cr \\
m_{1r}+m_{2r}-m_{1\ell}-m_{2\ell}&=\fr{|\mathtt{w}|^2-|\mathtt{s}|^2}{2}.\nonumber
\end{align}
(The sign in the first equation can be exchanged by an abelian gluing gauge transformation.) The second equation clearly has no positive solutions for $b_{r,\ell}$.

Therefore, the field $\phi_z$ has to be continuous at $y=0$. In this case, either $Z$ or $\tilde{Z}$ should be zero. Assume that it is $\tilde{Z}$, and 
\beq
Z=\fr{1}{\sqrt{z}}\left(\begin{array}{c}\mathtt{s}\\\mathtt{w}\end{array}\right).
\eeq
Since the field $\phi_0$ is diagonal, the matrix $Z\otimes Z^\dagger$ should be also diagonal, so either $\mathtt{s}$ or $\mathtt{w}$ is zero. We have to choose $\mathtt{s}=0$ to avoid the same sign problem which caused trouble in the second equation in (\ref{uio}). The discontinuity equations become
\begin{align}
\fr{a_r}{\sinh b_r}-\fr{a_\ell}{\sinh b_\ell}&=0\,,\nnr  \cr
a_r\coth b_r+a_\ell\coth b_\ell&=|\mathtt{w}|^2/2\,,\nnr \cr
m_{1r}+m_{2r}-m_{1\ell}-m_{2\ell}&=-|\mathtt{w}|^2/2.
\end{align}
The last equation here implies that $m_{1r}+m_{2r}<m_{1\ell}+m_{2\ell}$. In the opposite case, we would have to take $Z$ and not $\tilde{Z}$ to be zero. Eliminating $|\mathtt{w}|$ and $b_\ell$, we get
\beq
\fr{m_{1\ell}+m_{2\ell}-m_{1r}-m_{2r}}{m_{1r}-m_{2r}}=\coth b_r+\sqrt{\left(\fr{m_{1\ell}-m_{2\ell}}{m_{1r}-m_{2r}}\right)^2+\fr{1}{\sinh^2b_r}}.
\eeq
This equation has a solution precisely when the eigenvalues are arranged as in fig. \ref{interlaced}c.

\subsubsection{General Surface Operators}\label{gso}
We have described the abelian solutions for the U$(n|n)$ case, and some more general solutions for U$(2|2)$ for surface operators of type $(\alpha^\vee,0,0)$. In this section we look at the general singularities of type $(\alpha^\vee,\beta^\vee,\gamma^\vee)$, aiming to make a precise statement about the correspondence between surface operators and supersymmetric vacua of the theory.

Let us go from the coordinates $(t,x_1,x_2,y)$ to $(t,\varsigma,\theta,r')$, in which the rotational and scaling symmetries act in the most simple way. The flat metric in these coordinates is conformally equivalent to $\cosh^2\varsigma({{\rm d}t^2+{\rm d}r'^2})/{r'^2}+{\rm d}\varsigma^2+{\rm d}\theta^2$, which is $AdS_2\times \RR_\varsigma\times S^1_\theta$, up to a warping factor $\cosh^2\varsigma$. In conformal field theory, finding a model solution for a surface operator is equivalent to finding a vacuum configuration in this space, with the asymptotics of the scalar fields at $\varsigma\rightarrow\pm\infty$ defined by the charges of the surface operator. To make this intuition precise, let us rewrite our localization equations (\ref{grone}), (\ref{grtwo}) in terms of these coordinates. We make a general scale-invariant and rotationally-invariant ansatz for the fields,
\beq
\phi_0=\fr{1}{r'}M(\varsigma)\,,\quad \phi_z=\fr{1}{z}N(\varsigma)\,,\quad A=M_1(\varsigma)\cos\varphi\,{\rm d}\theta.\label{redvac}
\eeq
(We could have absorbed $\cos\varphi=\tanh\,\varsigma$ into $M_1$, but it is more convenient to write it this way.) The equations reduce to
\beq
[\partial_\varsigma-iM,N]=0\,,\quad [\partial_\varsigma-iM_1,N]=0\,,\quad [\partial_\varsigma-iM,\partial_\varsigma-iM_1]+\fr{2i}{\sinh 2\varsigma}(M-M_1)=0\,,\label{nahm1}
\eeq
together with 
\beq
\fr{\sinh^2\varsigma \partial_\varsigma M_1+\partial_\varsigma M}{\cosh^2\varsigma}+2i[N,N^\dagger]+\fr{\sinh\,\varsigma}{\cosh^3\varsigma}(M_1-M)=0.
\eeq
The first set of equations almost implies that $M_1=M$. In fact, there is a class of reducible solutions for which this equality is not true. They will be described in the next section, but for now we take $M_1=M$ as an ansatz. Then the equations reduce simply to  Nahm's equations $[\partial_\varsigma-iM,N]=0$ and $\partial_\varsigma M+2i[N,N^\dagger]=0$ for the scalar fields $M$, Re$(N)$ and Im$(N)$. At $\varsigma\rightarrow\pm\infty$, these fields should approach limiting values given by the parameters of the surface operator $\alpha^\vee$, $\beta^\vee$ and $\gamma^\vee$. At $\varsigma=0$, assuming the regularity of $M(\varsigma)$, the conformally invariant solution for $Z$ and $\tilde{Z}$ is given by $1/\sqrt{z}$ times some constant vectors, which should be found from the discontinuity equations.

In this way, the problem of finding the model solution for a surface operator is indeed reduced to the problem of finding the supersymmetric vacuum of the D3-D5 system for given asymptotic values of the scalar fields. To actually find the solutions, one needs to find the solutions of the Nahm's equations on a half-line, with asymptotics of the fields given by the regular triple $(\alpha^\vee,\beta^\vee,\gamma^\vee)$, and then glue them at $y=0$, according to the discontinuity equations. The relevant solutions of the Nahm's equations can be found e.g. in \cite{Mikhaylov}. The problem reduces to solving a set of algebraic equations for the integration constants of the solutions and the components of the hypermultiplet field $Z^\alpha$. Solving these equations seems like a tedious problem even for the U$(2)$ case, and we will not attempt to do it here. The relation to the supersymmetric vacua guarantees that for any values of the parameters a model solution exists, unique
up to gauge invariance.

The reduction that we have just described works for the unequal rank case as well. The gluing conditions of section \ref{unequalrk} for the conformally-invariant solution (\ref{redvac}) at $y=0$ reduce to the gluing conditions for the scalar fields $M$ and $N$. In particular, a $1/y$ Nahm pole boundary condition translates into a $1/\varsigma$ Nahm pole for the vacuum scalar fields.

\subsubsection{Reducible Solutions}\label{redsolutions}
So far we have concentrated on irreducible solutions, but there are  reducible solutions as well. 

Returning to eqn. (\ref{nahm1}), instead of setting $M_1=M$, we write $M_1=M+S$.  We find that the equations are obeyed if $M$ and $N$ obey the same
conditions as before, while
\begin{equation}\label{urtz} [S,N]=[S,M]=\partial_\varsigma S+\frac{2}{\sinh 2\varsigma}S=0. \end{equation}
The last equation means that 
\begin{equation}\label{murtz}S=\coth\,\zeta S_0=\frac{1}{\cos\varphi}S_0 \end{equation}
with a constant matrix $S_0$.  

The interpretation is very simple. First we describe the equal rank case.  In $\U(n)$, we pick a subgroup $\U(n-m)\times \U(m)$.
In $\U(n-m)$, we pick matrices $M,N$ and defect fields $Z,\t Z$ that satisfy  Nahm's equations and the jump conditions at $y=0$,
giving an irreducible solution (in $\U(n-m)$) as described in section \ref{gso}.  In $\U(m)$, we embed a trivial
abelian solution with $A=\alpha^\vee\d\theta$, $\phi_z=(\beta^\vee+i\gamma^\vee)/(2z),$ $\phi_0=0$.  (This trivial solution is obtained
by taking $S=\alpha^\vee$, and taking the $\frak{u}(m)$-valued part of $N$ to be the constant matrix $(\beta^\vee+i\gamma^\vee)/2$.)  
This describes a solution that can exist if $m$ eigenvalues of $\vec\zeta_\ell^\vee=(\alpha_\ell^\vee,\beta_\ell^\vee,\gamma_\ell^\vee)$ coincide
with $m$ eigenvalues of $\vec\zeta_r^\vee=(\alpha_r^\vee,\beta_r^\vee,\gamma_r^\vee)$.  
For left and right eigenvalues to coincide is the condition for an atypical weight,
so these solutions govern atypical weights.  

For the same atypical weight, however, we could have simply used the irreducible $\U(n)$-valued solution with $S=0$ constructed in section \ref{gso}.
After all, this solution exists for any weights. More generally, consider an atypical weight of $\U(n|n)$ with $s$ eigenvalues of $\vec\zeta_\ell^\vee$ 
equal to corresponding eigenvalues of $\vec\zeta_r^\vee$.   
For any $m\leq s$, we can obtain a surface operator solution with this weight, based on a subgroup $\U(n-m)\times \U(m)\subset\U(n)$.  We simply
take a trivial abelian solution in $\U(m)$ based on $m$ of the $s$ common weights, and combine this
with an irreducible solution in $\U(n-m)$ for all the other weights.  For each $m$, there are $\begin{pmatrix}s\cr m\end{pmatrix}$ such solutions,
since we had to pick $m$ of the $s$ common weights.  Considering all values of $m$ from 0 to $s$,
this gives $2^s$ surface operator solutions for a weight of $\U(n|n)$ of atypicality $s$.  Qualitatively, this is in agreement with what one finds on the electric side,
where a finite-dimensional representation with a given highest weight is unique only if the weight is typical.  In the case that the weights $\alpha_\ell^\vee$
and $\alpha_r^\vee$ are integral and $\beta_\ell^\vee,\gamma_\ell^\vee$ and $\beta_r^\vee,\gamma_r^\vee$ all vanish, so that the model solutions that we have constructed 
are related to line operators (see section \ref{lineop}), this leads to $2^s$ line operators associated to a weight of atypicality $s$; we suspect
that they  are dual to $2^s$ distinguished representations with the given highest weight.

The story is similar for unequal ranks.  The gauge group is $\U(n)$ for $y<0$ and $\U(n+r)$ for $y>0$.  
We pick subgroups $\U(n-m)\times \U(m)\subset \U(n)$
and $\U(n+r-m)\times \U(m)\subset \U(n+r)$.  We combine a trivial abelian $\U(m)$-valued solution on the whole 
$y$ line with an irreducible solution based on
$\U(n-m)$ for $y<0$ and $\U(n+r-m)$ for $y>0$.  Just as in the last paragraph, we get $2^s$ solutions for a weight of $\U(n|n+r)$ of atypicality $s$.

Another type of reducible solution was found in section \ref{absol}.  If one of the eigenvalues of $\alpha_r^\vee$ 
is equal to an eigenvalue of $-\alpha_\ell^\vee$,
then the corresponding matrix elements of $Z$ and $\t Z$ vanish and a $\U(1)$ subgroup of the gauge group is unbroken.  The basic phenomenon
occurs actually for the gauge group $\U(1)$, corresponding to the supergroup $\U(1|1)$.  There is a surface operator described by a trivial
abelian solution with $A=\alpha^\vee \cos\varphi \,\d \theta$ and $\phi=\alpha^\vee\,\d x^0/r'$ everywhere and $Z=\t Z=0$.
(This solution has $\alpha_r^\vee=\alpha=-\alpha_\ell^\vee$ because $\cos\varphi=1$ on the positive
$y$ axis and $-1$ on the negative $y$ axis.)  Clearly since $Z$ and $\t Z$ vanish, the $\U(1)$ gauge
symmetry is unbroken.  This is a reducible solution that can occur for a typical weight, since $\alpha_r^\vee=-\alpha_\ell^\vee$ is not a condition for
atypicality. 
Such a surface operator does not seem to be well-defined. Since the gauge symmetry remains unbroken along the knot $K$, the gauge field near $K$ is free to fluctuate. In particular, it follows that the variation of the topological term in the presence of this model singularity is not zero, but is proportional to $\int_K\alpha\delta A$, and therefore, the action is not $\Qb$-invariant. We do not know how to interpret the singularity that seems to arise
when an eigenvalue of $\alpha_\ell^\vee$ approaches one of $-\alpha_r^\vee$, or how to describe a half-BPS surface operator in this
case.  A possibly similar
problem arises in the bulk in $\N=4$ super Yang-Mills theory with any nonabelian gauge group if one tries to define a surface operator with parameters
$(0,0,0,\eta^\vee)$.   Classically, it is hard to see how to do this, since the definition of $\eta^\vee$ requires a reduction of the gauge symmetry to
the maximal torus along the support of the surface operator, and this is lacking classically if $\alpha^\vee=\beta^\vee=\gamma^\vee=0$.  Yet the
surface operator in question certainly exists; it is $S$-dual to a surface operator with parameters $(\alpha,0,0,0)$ that can be constructed
semiclassically.  One approach to defining it involves adding additional variables along the surface (see section 3 of \cite{Rigid}).

\subsection{Line Operators And Their Dualities}\label{lineop}
We have constructed surface operators, but there is an easy way to construct line operators from them.  We simply observe that
if we set $\beta^\vee=\gamma^\vee=0$, and also take $\alpha^\vee$ to be integral, then the bulk solution $A=\alpha^\vee\,\d\theta$ defining a surface operator in the absence of any D5-brane can
be gauged away.  So for those parameters, the surface operators that we have constructed are trivial far away from the D5-brane defect.
That means that those surface operators reduce macroscopically to line operators supported on the defect.  

Saying that $\alpha^\vee$ is ``integral'' means that it is a cocharacter of the maximal torus of the dual group $G^\vee$, or in other words a 
character of the maximal torus of $G$.  Up to the action of the  Weyl group, this character corresponds to a dominant weight of $G$.  In other words, 
we have found line operators of the magnetic description by $G^\vee$ gauge theory
that are classified by dominant weights (or representations)  of the electric group $G$.

In all these statements, $G$ is either $G_\ell$ or $G_r$, the gauge group to the left or right of the D5-brane defect.  Taking account of the
behavior on both sides,  these line operators are really
classified by dominant weights of $G_\ell\times G_r$.
(In our main example of $\U(m|n)$, $G$ is $\U(m)$ or $\U(n)$ and the distinction between $G$ and its dual group $G^\vee$ is not important.
However, this part of the analysis is more general and carries over also to the orthosymplectic case that we discuss in section \ref{ortho}.)

Wilson-'t Hooft operators of the ``electric'' description involving an NS5-brane are also classified by dominant weights of $G_\ell\times G_r$
(or equivalently by dominant weights of the supergroup $SG$),
as we learned in section \ref{defmonodr}.
Thus an obvious duality conjecture presents itself: the line operator associated to a given weight of $G_\ell\times G_r$ in one description
is dual to the line operator associated to the same weight in the other description.  

This statement is a natural 
analog of the usual duality between
Wilson and 't Hooft operators, adapted to the present situation.  But  a detail remains to be clarified.   In the standard mapping
between Wilson operators of $G$  and 't Hooft operators of $G^\vee$, there is a minus sign that to some extent is a matter of convention.  That is because electric-magnetic
duality could be composed with charge conjugation for either $G$ or $G^\vee$.  Charge conjugation acts by reversing the sign of a weight, up to a Weyl
transformation.  

In the supergroup case, let $(\uplambda_\ell,\uplambda_r)$ be a weight of $G_\ell\times G_r$, and let $(\alpha^\vee_\ell,\alpha^\vee_r)$ be a magnetic
weight of $G_\ell^\vee\times G_r^\vee$.  If we specify that we want a duality transformation
that maps $\uplambda_\ell$ to $+\alpha^\vee_\ell$, then it becomes a well-defined question whether $\uplambda_r$ maps to $+\alpha^\vee_r$
or to $-\alpha_r^\vee$.  The correct answer is the one with a minus sign:
\begin{equation}\label{zomo}(\uplambda_\ell,\uplambda_r)\leftrightarrow (\alpha_\ell^\vee,-\alpha_r^\vee). \end{equation}
   To see this, we observe that there is a symmetry of the problem that
exchanges the left and right of the defect and exchanges $\uplambda_\ell$ with $\uplambda_r$ but $\alpha_\ell^\vee$ with $-\alpha_r^\vee$.
For a defect at $x^3=0$ and a line operator supported on the line $L:x^1=x^2=x^3=0$, we can take this symmetry to 
be $x^2\to -x^2, \, x^3\to -x^3$, with $x^0,x^1$ fixed.  This has been chosen to exchange the left and right sides of the defect, while  mapping
the line $L$ to itself and
preserving the orientation of spacetime, so as to leave $\calK$ fixed.   It does not affect electric charge, but it reverses the sign of $\alpha^\vee$
because it reverses the orientation of the $x^1x^2$ plane.

As was already remarked in section \ref{various}, in the case of an atypical weight,
 our pictures on the magnetic and  electric sides  do not quite match. On the magnetic side, for a given atypical weight,  we have
 found multiple possible 1/2 BPS surface and line operators, as  explained in section~\ref{redsolutions}. On the electric side, for any weight, even atypical, we found only  a single 1/2 BPS surface or Wilson-'t~Hooft line operator.

\subsection{A Magnetic Formula For  Knot And Link Invariants}\label{gettingpolynomials}

The $\Qb$-invariant line and surface operators that we have constructed can be used to get magnetic formulas
for knot and link invariants.  In the case of line operators, we have little to add  to what was stated in eqn. (\ref{omex}). Here
we will elaborate on the construction of knot and link invariants using surface operators.  After some general observations,
we will  comment on what happens for atypical weights.

We start on the electric side with a knot invariant defined by including a surface operator with parameters 
$(\alpha,\beta,\gamma,\eta)$ supported on a two-surface $C$ that intersects the hyperplane $y=0$ along a knot $K$.  One can take simply 
$C=K\times \RR_y$ (where $\RR_y$ is parametrized by $y$) or one can choose $C$ to be compact.   The dual magnetic description involves 
 a surface operator wrapped on $C$ with parameters $(\alpha^\vee,\beta^\vee,\gamma^\vee,\eta^\vee)=(\eta,|\tau|\beta^*,|\tau|\gamma^*,-\alpha).$
 
  The parameters of the surface operator in the magnetic case define the singularities of the fields near $C$, but also they determine some insertions
  that must be made in the functional integral along $C$.  The action of the theory in the presence of the surface operator is
\beq
\fr{i\calK^\vee}{4\pi}\int_{M}\tr(F\wedge F)-i\int_{C}\tr\left((\calK^\vee\alpha^\vee-\eta^{\vee*}) F\right)\,,\label{magact}
\eeq
modulo $\Qb$-exact terms. We have used eqns. (\ref{topdelta}) and (\ref{etaterm}) for the terms proportional to $\alpha^\vee$
and $\eta^{\vee*}$.
The integral in the four-dimensional topological term is taken over $M$, but alternatively, we could take it over $M\setminus C$, 
and that would absorb the term proportional to $\alpha^\vee$. Note that the objects which appear in this 
formula are topological invariants, because the bundle is naturally trivialized both at infinity and in the vicinity 
of $K$, where the fields $Z^\alpha$ become large. (For now we consider the generic irreducible case, when  the gauge group is completely broken along $K$; 
we do not consider the problem mentioned at the end of section \ref{redsolutions}.) Using the relation (\ref{surfweight}) between weights and parameters of 
the surface operator, the action can be alternatively written as
\beq
\fr{i\calK^\vee}{4\pi}\int_{M}\tr(F\wedge F)+i\calK^\vee\int_{C_r}\tr(\uplambda_r F)-i\calK^\vee\int_{C_\ell}\tr(\uplambda_\ell F).\label{magninsertion}
\eeq
The insertion of the two-dimensional observable in this formula is essentially the $S$-dual of the analogous insertion 
in the electric theory. This statement can be justified explicitly if the gauge group is abelian. In that case, the  
two-observable $\int F$ is the second descendant of the $\Qb$-closed field $\sigma$. Under  $S$-duality, 
both the gauge-invariant polynomials of $\sigma$ and their descendants are mapped to each other. 
(See Appendix \ref{finicky} for details on the descent procedure in the presence of the three-dimensional defect.)  

The functional integral in the magnetic theory can be localized on the space of solutions to the localization equations (\ref{lono}), (\ref{bono}). The knot polynomial can be obtained by counting the solutions of the localization equations in the presence of a singularity of type type $(\alpha^\vee,\beta^\vee,\gamma^\vee)$, weighted by the combination (\ref{magninsertion}) of topological numbers of the solution, as well as the sign of the fermion determinant. 
(These statements hold for both the equal-rank and  unequal rank cases, though one uses different equations and model solutions in the two
cases.) For a given weight, there are different possible choices of surface operator. We can vary $\alpha^\vee$ and $\eta^\vee$, as long as their appropriate combination is equal to the weight. We can also turn on arbitrary $\gamma^\vee$ and $\beta^\vee$, as long as it is not forbidden for topological reasons. All this simply reflects the fact that  the problem of counting solutions of elliptic equations is formally invariant under continuous deformations of  parameters.
Note that, in particular, the operators with $\gamma^\vee\ne 0$ are well-defined and 1/2-BPS, and changing $\gamma^\vee$ does not change the weight in (\ref{magninsertion}), with which the solutions of the localization equations are counted. This supports the view, 
proposed in section \ref{surel}, that in the physical theory $\gamma$ plays much the same role, as $\beta$: it deforms the contour of integration in the functional integral, without changing the Chern-Simons observables.\footnote{All this is true for the physical theory, where both $\calK$ and the weights are real. We expect the situation to be different in the topological theory, where on the electric side the surface operators with $\gamma\ne 0$ are defined according to eq. (\ref{elinsertion}). In that case, $\gamma$ is related to the imaginary part of the weight. In particular, the insertion of $i\calK w\int\tr(\gamma\mathcal{F})$ in (\ref{elinsertion}) will lead on the magnetic side to a similar insertion, which will complexify the weight in eq. (\ref{magninsertion}).}

It is conceivable that 
 the counting of the solutions of the localization equations is only generically
 independent of the parameters $(\alpha^\vee,\beta^\vee,\gamma^\vee)$, and that wall-crossing phenomena can occur.
 (A prototype of what might happen has been seen for the three-dimensional Seiberg-Witten equations  \cite{MT}.)  We will not attempt to analyze this possibility
 here, and will simply assume that for any regular triple $(\alpha^\vee,\beta^\vee,\gamma^\vee)$, the counting of solutions is the same. Let $\S_0$ be the space of these solutions. 
It is convenient to introduce variables $\mathbf{t}_r=q^{-\uplambda_r^*}$ and $\mathbf{t}_\ell=q^{\uplambda_\ell^*}$, valued in the complexification of the maximal tori of the left and the right bosonic gauge groups of the electric theory. The knot polynomial is then given by
\beq
\sum_{s\in \S_0} (-1)^{f}q^{{\Nb^\vee}}\, \mathbf{t}_\ell^{c_{1\ell}}\,\label{knotpoly} \mathbf{t}_r^{c_{1r}}.
\eeq
Here $(-1)^f$ is the sign of the fermion determinant, evaluated in the background of the classical solution $s$, $\Nb^\vee=\fr{1}{8\pi^2}\int_{M\setminus K}\tr(F\wedge F)$ is the instanton number, and $c_{1\ell,r}=\fr{1}{2\pi}\int_{C_{r,\ell}}F$ are the $\mathfrak{t}^\vee$-valued relative first Chern classes for the abelian bundles on $C_r$ and $C_\ell$. One can consider (\ref{knotpoly}) as a polynomial in $q$, after expressing $\mathbf{t}_{\ell,r}$ in terms of $q$ for a particular weight $\uplambda$, but one can also treat $\mathbf{t}_{\ell,r}$ as independent formal variables.

What happens if the weight $\uplambda$ is atypical? By varying $\alpha^\vee$ and $\eta^\vee$, while preserving $\uplambda$, we can still make the model solution irreducible. So we can use the solutions from $\S_0$ to obtain the knot polynomial, and simply substitute our $\uplambda$ in eqn.(\ref{knotpoly}). We expect that this polynomial will correspond to the Kac module of highest weight $\uplambda$. This expectation follows from the fact that a typical representation can be continuously deformed into an atypical one by varying the fermionic Dynkin label $a_{\rm ferm}$. Since this label need not be integral, this variation makes sense, and the limit of this typical representation, when the weight becomes atypical, is the Kac module. In the magnetic theory, to take the limit of a knot invariant, we simply substitute the atypical weight into the universal polynomial (\ref{knotpoly}), evaluated on $\S_0$. So this type of polynomial indeed corresponds to the Kac module.

For an atypical weight, rather than an irreducible model solution, we can also use surface operators defined by reducible solutions.
For any weight of atypicality at least $p$, we can consider a surface operator whose irreducible part is associated to a surface operator of
$\U(m-p|n-p)$.   This surface operator breaks the bosonic group $\U(m)\times \U(n)$ to an subgroup $H$ that generically is $\U(1)^p$ (it
can be a nonabelian group containing $\U(1)^p$ if the reducible part of the solution is non-regular).    Let $T_H\cong \U(1)^p$ be
the maximal torus of $H$.  The group $H$ acts on the space of solutions
of the localization equations.  In such a situation, by standard localization arguments,\footnote{Generically, one expects that the
solutions consist of a finite set of points, and if so, these points are all invariant under the continuous group $T_H$.  However,
suppose that some of the solutions make up a manifold $U$ that has a non-trivial action of $T_H$.  Then by standard arguments of
cohomological field theory \cite{coho}, the contribution of the manifold $U$ to the counting of solutions is $(-1)^f\chi(U;V)$, where $(-1)^f$ is
the sign of the fermion determinant, $V\to U$ is a certain ``obstruction bundle''  (a real vector bundle of rank equal to the dimension of
$U$), and $\chi(U;V)$ is the Euler characteristic of $V\to U$.  Let $U'$ be the fixed point set of the action of $T_H$ on $U$
and let $V'\to U'$ be the $T_H$-invariant subbundle of $V|_{U'}$.  A standard topological argument shows that  $(-1)^f\chi(U;V)=(-1)^{f'}\chi(U';V')$ (if $U'$ is not connected, one must write a sum over components on the right hand side).  
In our problem, this means that we can consider only the $\U(m-p|n-p)$
solutions and count them just as we would for $\U(m-p|n-p)$, ignoring the embedding in $\U(m|n)$.}
the invariants can be computed
by just counting the $T_H$-invariant solutions.   The $T_H$-invariant subgroup of $\U(m)\times \U(n)$ is $T_H\times \U(m-p)\times \U(n-p)$.  There are no interesting solutions valued in the abelian group $T_H$, so in fact, the $\U(m|n)$ invariants with a surface operator of this
type can be computed by counting solutions for $\U(m-p|n-p)$.   Some simple group theory shows  that the signs of the two fermion determinants are the same
and hence the $\U(m|n)$ invariants for a weight of atypicality $\geq p$ coincide  with  $\U(m-p|n-p)$ invariants. 
In particular, $\U(m|n)$ invariants of maximal atypicality coincide with invariants of the bosonic group $\U(|n-m|)$.  (This reasoning also
makes it clear that the knot and link invariants constructed using a reducible model solution do not depend on the weights in the abelian
part of the solution.)

For a weight of atypicality $r$, we can take any $p\leq r$ in this construction.  We have argued that for $p=0$, we expect to get
invariants associated to the Kac module, while $p=r$ presumably corresponds to the irreducible atypical representation.  
The intermediate values of $p$ plausibly correspond to the reducible indecomposables, which are obtained by taking non-minimal
subquotients of the Kac module.   

In section \ref{symbr}, we give an alternative approach to comparing $\U(m|n)$ with $\U(m-p|n-p)$.  The key idea there is gauge symmetry
breaking.  This approach is very natural on the electric side.

In the rather formal discussion that we have given here, we have not taken into account some of the insight from section \ref{linerev}.
From that analysis, we know that for the knot invariants to be nonzero, we can consider a typical weight for a knot in $S^3$ or
a maximally atypical weight for a knot in $\RR^3$.  For other weights, a slightly different approach is needed.  We have not understood
the analogs of these statements on the magnetic side.

\subsection{A Possible Application}\label{possible}

Here we will briefly indicate a possible application of this work, for gauge group $\U(1|1)$.

Using the fact that the supergroup $\U(1|1)$ is solvable, the invariant for a knot $K\subset S^3$ labeled by a typical representation of $\U(1|1)$ can be explicitly
computed by repeated Gaussian integrals.   It turns out to equal the Alexander polynomial 
\cite{RSone,RStwo,RS}.   The usual variable $q$ on which the Alexander polynomial depends is a certain function
of the Chern-Simons coupling and the typical weight.

The Alexander polynomial of $K$ can also be computed \cite{MT} by counting solutions of a 3d version of the
Seiberg-Witten equations with a prescribed singularity along $K$.  Such solutions can be labeled by an integer-valued invariant $\Theta$
(a certain relative first Chern class), and if $b_n$ is the number of solutions with $\Theta=n$ (weighted as usual with the sign of
a certain fermion determinant), then the Alexander polynomial is  $Z(q)=\sum_n b_n q^n$.  The proof that  $Z(q)$ equals the Alexander
polynomial is made by showing that the two functions obey the same ``skein relations.''

The question arises of whether one could find a more direct explanation of this result, or perhaps a more direct link between $\U(1|1)$ Chern-Simons theory and the Seiberg-Witten
equations.  From the point of view of the present paper, $\U(1|1)$ Chern-Simons theory can be represented in terms of $\N=4$ super
Yang-Mills theory with gauge group $\U(1)_\ell\times \U(1)_r$ on $S^3\times \RR$, interacting with a bifundamental 
hypermultiplet that is supported on $S^3\times\{0\}$.
However, as was actually already remarked at the end of section \ref{quivers}, we can just as well replace $\RR$ here by $S^1$.  
If we do that, we get $\U(1|1)$ Chern-Simons theory with a different integration cycle.  However, as long as one considers only
Wilson operators on $\RR^3$ or $S^3$, all integration cycles are equivalent and so $\N=4$ super Yang-Mills theory on $S^3\times S^1$
with a bifundamental hypermultiplet on $S^3\times \{0\}$ should give another way to study the Alexander polynomial.\footnote{Once we replace
$S^3\times \RR$ with $S^3\times  S^1$, the left and right of the defect are connected.  So we now have a single $\U(1)$ vector multiplet
on $S^3\times S^1$, with the fields allowed to have different limits as $S^3\times \{0\}$ is approached from the left or right.  The  two limits
give two different sets of 3d fields to which the ``bifundamental'' hypermultiplet is coupled.}  

$S$-duality converts this to a ``magnetic'' problem on $S^3\times S^1$, now with $\U(1)$ gauge fields in bulk and a twisted hypermultiplet
supported on $S^3\times S^1$.  If one takes the radius of $S^1$ to be small compared to that of $S^3$, the four-dimensional localization
equations can be expected to reduce to three-dimensional effective equations.  These will be equations in which $\U(1)$ gauge fields are
coupled to a hypermultiplet, and one can argue that the relevant equations are the Seiberg-Witten equations.

Thus one can hope that, as in \cite{MT}, it will be possible to compute the Alexander polynomial by counting solutions of the Seiberg-Witten
equations.  Unfortunately, in trying to implement this program, we ran into a number of technical difficulties, which hopefully will be resolved
in the future.   Some of the more significant difficulties
involve the compactness of $S^3\times S^1$, as a result of which some of the standard arguments relating Chern-Simons theory to
$\N=4$ super Yang-Mills theory do not quite apply.  

\section{Orthosymplectic Chern-Simons Theory}\label{ortho}

In this section, we return to the D3-NS5 system of fig. \ref{setup}, but now we add an O3-plane parallel to the D3-branes. A D3-O3 system can have orthogonal or symplectic gauge symmetry, depending on which type of O3-plane is chosen.
The gauge symmetry jumps from orthogonal to symplectic in crossing an NS5-brane.  Accordingly, the construction of section
\ref{etheory}, with an O3-plane added, is related to Chern-Simons theory of an 
 orthosymplectic gauge group $\OSp(r|2n)$, where the integers $r$ and $n$ depend on the numbers of D3-branes on the two sides
 of the NS5-brane.    As in section \ref{magnetic}, an $S$-duality transformation that converts the D3-O3-NS5 system to a D3-O3-D5
 system gives a magnetic dual of three-dimensional $\OSp(r|2n)$ Chern-Simons theory.  This is a close analog of what
 we have already seen for unitary groups. 
 
  However, something novel happens if $r=2m+1$ is odd.  In this case, a slightly different procedure yields a duality
  between two ``electric'' descriptions.  In three-dimensional terms, we will  learn that  Chern-Simons theory of $\OSp(2m+1|2n)$, with coupling parameter $q$, is equivalent
  to  Chern-Simons
 theory of $\OSp(2n+1|2m)$, with coupling parameter $-q$.  (The Chern-Simons theories that appear in this statement are defined
via the brane constructions which as usual allow analytic continuation away from integer levels.)  Since weak coupling in Chern-Simons theory is $q\to 1$, while $q\to -1$ is
 a strongly-coupled limit, this duality exchanges strong and weak coupling.

\begin{figure}
 \begin{center}
   \includegraphics[width=5cm]{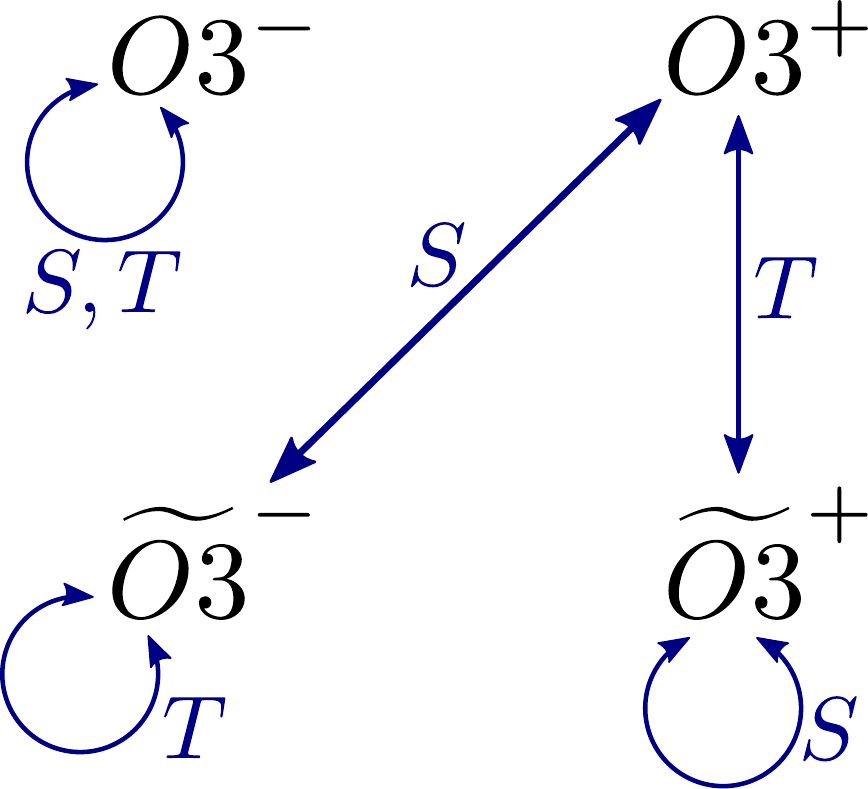}
 \end{center}
\caption{\small Action of the $S$-duality group on the orientifold planes.}
 \label{oplanes}
\end{figure}
\subsection{Review Of Orientifold Planes}\label{review}
 We start with a brief review of the orientifold 3-planes of Type IIB superstring 
 theory \cite{Baryons,HananyOrientifolds} (see also section 7 of \cite{GWfour}).

\def\\SO{{\mathrm{SO}}}
\def\O{{\mathrm{O}}}
\def\Sp{{\mathrm{Sp}}}

There are four kinds of  O3-plane, distinguished by  $\ZZ_2$-valued discrete fluxes of the NS and RR two-form fields of 
Type IIB supergravity. 
An O3-plane in which both fluxes vanish is denoted O3$^-$; in the presence of $m$ parallel D3-branes (and their
images) it gives $\O(2m)$ gauge symmetry (for some purposes, we consider only the connected component $\SO(2m)$).  Adding discrete RR flux gives an $\widetilde{{\rm O}3}^-$-plane, which with the addition of $m$ parallel
D3-branes gives $\O(2m+1)$ gauge symmetry.  An orientifold 3-plane with only NS flux is denoted $\O 3^+$ and gives
$\Sp(2m)$ gauge symmetry.  
Finally, the orientifold $\widetilde{{\rm O}3}^+$ with both kinds of flux gives again $\Sp(2m)$ gauge symmetry, but (as we recall shortly) with a shift
in the value of the theta-angle $\theta_\YM$, a fact that we abbreviate by saying that the gauge group is $\Sp'(2m)$.  The transformation properties of the orientifold 3-planes under the SL$(2,\ZZ)$ $S$-duality group are summarized in fig.~\ref{oplanes}.

When an O3-plane crosses an NS5-brane, its NS flux jumps; when it crosses a D5-brane, its RR flux jumps.  More generally, when an
O3-plane crosses a $(p,q)$-fivebrane its $(\mathrm{NS},\mathrm{RR})$ fluxes jump by $(p,q)$ mod 2.  

Regardless of the type of O3-plane, a D3-O3 system has the same supersymmetry as a system of D3-branes only. In particular, this supersymmetry is parametrized by the angle $\ang$, which is related to the string coupling in the usual way, as in eqn. (\ref{zumbo}). To find the classical effective action for the gauge theory that describes a D3-O3 system at low energies, we simply take the effective
action of a D3-brane system, restrict the fields to be invariant under the orientifold projection, and divide by 2.  The restriction
reduces a $\U(n)$ gauge symmetry to $\O(n)$ or $\Sp(n)$, depending on the type of O3-plane.  We divide by 2 because the orientifolding
operation is a sort of discrete gauge symmetry in string theory.  (As we explain shortly, there is a subtlety in dividing $\theta_\YM$ by 2.)
The same procedure of restricting to the invariant subspace and dividing by 2
enables us to deduce  the effective action of a D3-O3-NS5 or D3-O3-D5 system  from those of a D3-NS5 or D3-D5 system. 

\def\YM{{\mathrm{YM}}}
For the $\U(n)$ gauge fields along a system of $n$ parallel D3-branes, we write the gauge theory action as
\beq
\fr{1}{2g_{\rm YM}^2}\int{\rm d}^4x\,\tr\,F_{\mu\nu}^2-\fr{\theta_{\rm YM}}{8\pi^2}\int\tr\, F\wedge F\,,\label{SOaction}
\eeq
where $\tr$ is the trace in the fundamental representation of $\U(n)$, and the Yang-Mills parameters $g_\YM$ and $\theta_\YM$ are related
to the $\tau$ parameter of the underlying Type IIB superstring theory by the standard formula
\beq \label{helbo}\tau=\frac{\theta_\YM}{2\pi}+\frac{2\pi i}{g_\YM^2}. \end{equation}
The action (\ref{SOaction}) is defined so that $\theta_\YM$ couples precisely to the instanton number
\beq\label{elba}\Nb=\frac{1}{8\pi^2}\int \tr\, F\wedge F, \eeq
normalized to be an integer on a four-manifold without boundary.
This ensures that the theory is invariant under $\tau\to\tau+1$, which corresponds to $\theta_\YM\to\theta_\YM+2\pi$.

If we include an O3 plane that reduces the gauge symmetry from $\U(n)$ to $\O(n)$, then we write the action in the same way, with $\tr$ now
representing a trace in the fundamental representation of $\O(n)$.  But since we have to divide the action by 2, we express 
the gauge theory parameters in terms of $\tau$ not by (\ref{helbo}) but by
\beq \label{melba} \frac{\tau}{2}=\frac{\theta_\YM}{2\pi}+\frac{2\pi i}{g_\YM^2}. \end{equation}
We write
\beq\label{lba}\frac{\tau}{2}=\tau_\YM,\end{equation}
where  $\tau_\YM$ is expressed in terms of $g_\YM$ and $\theta_\YM$ in the usual way.
An important detail now is that the quantity $\Nb$,
which is $\Z$-valued in $\U(n)$ gauge theory, takes values in\footnote{For $n\geq 4$, an $\O(n)$ instanton of minimal instanton number
can be embedded in an $\SO(4)$ subgroup.  An $\SO(4)$ instanton of minimal instanton number (on $\RR^4$; we do not consider here effects
associated to the second Stieffel-Whitney class) is simply an $\SU(2)$ instanton of
instanton number 1, embedded in one of the two factors of $\Spin(4)\cong \SU(2)\times \SU(2)$.  Upon embedding $\O(n)$ in $\U(n)$,
the $\O(n)$ instanton constructed this way is a $\U(n)$ instanton of instanton number 2, explaining why the
instanton number normalized as in (\ref{elba}) is an even integer in $\O(n)$.  In the case of $\O(3)$, there is not room for
the construction just described, and the minimal instanton has $\Nb=4$.}    
$2\Z$ in $\O(n)$ gauge theory for $n\geq 4$.  Because of this, the $\O(n)$ gauge theory is invariant under $\tau\to\tau+1$, even though
$\theta_\YM$ couples to $\Nb/2$.

Next consider the orientifold plane to be O$3^+$, reducing the gauge symmetry from $\U(n)$ to $\Sp(n)$ (here $n$ must be even).
The action is still defined as in eqn. (\ref{SOaction}), now with $\tr$ representing the trace in the fundamental representation of $\Sp(n)$.
Furthermore, the coupling parameter $\tau$ of Type IIB superstring theory is still related to the gauge theory parameters as in (\ref{melba}).
Now, however, the quantity $\Nb$ is integer-valued (a minimal $\Sp(n)$ instanton is an $\SU(2)$ instanton of instanton number 1 embedded
in $\Sp(2)\cong \SU(2)$), so the operation $\tau\to \tau+1$ of the underlying string theory is not a symmetry of the gauge theory.  Instead,
this operation maps an $\O3^+$ orientifold plane to a $\widetilde{{\rm O}3}^+$-plane, in which the gauge group is still $\Sp(n)$ but the relation between
string theory and gauge theory parameters is shifted from (\ref{melba}) to 
\beq \label{melbax} \frac{\tau+1}{2}=\frac{\theta_\YM}{2\pi}+\frac{2\pi i}{g_\YM^2}. \end{equation}
The term $\Sp'(n)$ gauge theory is an abbreviation for $\Sp(n)$ gauge theory with coupling parameters related in this way to the underlying
string theory parameters.

\subsection{The Even Orthosymplectic Theory}\label{evenorth}
\begin{figure}
 \begin{center}
   \includegraphics[width=17cm]{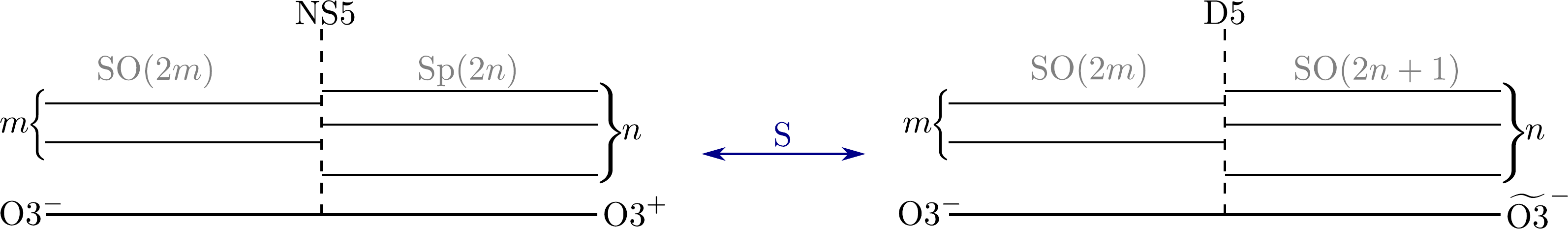}
 \end{center}
\caption{\small The brane configurations that realize the electric and magnetic theory for the four-dimensional construction of the OSp$(2m|2n)$ Chern-Simons theory.}
 \label{ospeven}
\end{figure}

Now we begin our study of the D3-O3 system interacting with a fivebrane.   On the left of  fig. \ref{ospeven}, we sketch an $\O3^-$-plane that converts to an $\O3^+$-plane in crossing
an NS5-brane.  The gauge group is therefore $\SO(2m)$ on the left and $\Sp(2n)$ on the right,
where $m$ and $n$ are the relevant numbers of D3-branes.  In the topologically twisted version of the theory, along the defect,
one sees a Chern-Simons theory of the supergroup $\OSp(2m|2n)$.    
After the orientifold projection, the action can be written just as in eqn. (\ref{act2}):
\beq
I=\fr{i\calK_\osp}{4\pi}\int_W \Str\left(\cA \d\cA+\fr{2}{3}\cA^3\right)+\left\{\Qb,\dots\right\},\label{nact}\eeq
Now $\Str$ denotes the supertrace in the fundamental representation of the orthosymplectic group.
This follows by simply projecting the effective action described in
section \ref{etheory} onto the part that is invariant under the orientifold projection. The expression for $\calK_\osp$ in terms of string
theory parameters $\tau,\ang$ is the same as in equation (\ref{canonical}) except for a factor of 2 associated to the orientifolding:
\beq
\frac{\tau}{2}=
\tau_\YM=\calK_\osp \cos\ang\,\ex^{i\ang}.\label{eq1even}
\eeq
Note that the bosonic part of the Chern-Simons action in (\ref{nact}) can be also expressed as
\beq
\fr{i\calK_\osp}{4\pi}\int_W \Tr\left(\cAb \d\cAb+\fr{2}{3}\cAb^3\right)=i\calK_\osp\Bigl(\CS(A_\sp)-2\CS(A_\so)\Bigr)\,,\label{ospnorm}
\eeq
where the Chern-Simons functionals $\CS(A_\sp)$ and $\CS(A_\so)$ are normalized to take values in $\RR/2\pi\ZZ$ for simply connected gauge groups and $m>1$.

Now we apply  the usual $S$-duality transformation $\tau\to\tau^\vee=-1/\tau$.  As indicated in the figure, this leaves the $\O3^-$-plane
invariant but converts the $\O3^+$-plane
to an $\t{\O3}^-$-plane; now the gauge group is $\SO(2m)$ on the left and $\SO(2n+1)$ on the right.   What we get this way is 
a magnetic dual of Chern-Simons theory of $\OSp(2m|2n)$.

The appropriate effective
action to describe this situation is found by simply projecting the effective action described  in section
\ref{unequalrk} onto the part invariant under the orientifold projection.  There is no analog  of the case $m=n$ that was important in section
\ref{unequalrk}, since $2m$ never coincides with $2n+1$.  The condition analogous to $|n-m|\geq 2$ is  $|2m-(2n+1)|\geq 3$. If this is the
case,  the appropriate
description involves a Nahm pole associated to an irreducible embedding $\frak{su}(2)\to \frak{so}(|2m-(2n+1)|)$. The Nahm pole
appears on the left or the right of the defect depending on the sign of $2m-(2n+1)$.   What commutes with the Nahm pole is an $\SO(w)$
gauge theory theory that fills all space; here $w$ is the smaller of $2m$ and $2n+1$.
 If $|2m-(2n+1)|=1$, then as in section
\ref{unequalrk}, there is no Nahm pole and the vector multiplets that transform in the fundamental representation of $\SO(w)$ obey Dirichlet
boundary conditions along the defect.

The action can still be expressed as in  (\ref{Imagnetic})
\beq
I_{\mathrm{magnetic}}=\fr{i\theta_\YM^\vee}{8\pi^2}\int\tr\left(F\wedge F\right)+\{\Qb,\dots\},\label{Ima}
\eeq
 where now $\tr$ is the trace in the fundamental representation of the orthogonal group, and  $\tau_\YM^\vee=\theta^\vee_\YM/2\pi
 +4\pi i/(g_\YM^\vee)^2$ is related to the underlying string theory parameters by
\beq 
\label{doppo} \tau_\YM^\vee=\fr{1}{2}\tau^\vee=-\frac{1}{2\tau}. \eeq
We recall from section \ref{review} that the instanton number $\Nb^\vee=(1/8\pi^2)\int\tr \,F\wedge F$ takes even integer values in the case of
an orthogonal gauge group.  
Hence the natural instanton-counting parameter is
\beq\label{icon} q=\exp(-2i\theta^\vee_\YM),\end{equation}
in the sense that a field of $\Nb^\vee=2r$ contributes $\pm q^r$ to the path integral (as usual the sign depends on the sign of the fermion
determinant).  

The variable $q$ can be expressed in terms of the canonical parameter $\calK_\osp$ of the electric description. 
In (\ref{thetaunitary}), we have obtained ${\rm Re}\,(\tau^\vee)=-1/\calK$, 
where $\calK$ is the canonical parameter for the theory with no orientifolds. In the orientifolded theory, 
the canonical parameter $\calK_\osp$  that appears in the action (\ref{nact}) is one-half of that. Hence, using equation (\ref{doppo}), we find that 
\beq\frac{\theta^\vee_\YM}{2\pi}=
{\rm Re}\,\tau_\YM^\vee=\frac{1}{2}{\rm Re}\,\tau^\vee=-\fr{1}{2\calK}=-\fr{1}{4\calK_\osp}\,,\label{thetaospdual}
\eeq
and therefore the definition (\ref{icon}) gives
\beq
q=\exp\left(\fr{\pi i}{\calK_{\osp}}\right).\label{qeven}
\eeq
By contrast, Chern-Simons theory or two-dimensional current algebra for a purely bosonic group $G$ with Lie algebra $\frak g$
is naturally parametrized by
\beq
q_\g=\exp\left(\fr{2\pi i}{n_\g\calK_{\g}}\right)\,,\label{usdef}
\eeq
where $n_\g$ is the ratio of length squared of long and short roots of $\g$.  (This is also the natural instanton-counting parameter in the magnetic
dual description of this theory \cite{5knots}.)  The parameter $q$ defined in eqn. (\ref{qeven}) is an analog of this, with $n_\g$ replaced by
the ratio of length squared of the longest and shortest bosonic roots; for 
 $\osp(2m|2n)$, this ratio is equal to 2.
\begin{figure}
 \begin{center}
   \includegraphics[width=17cm]{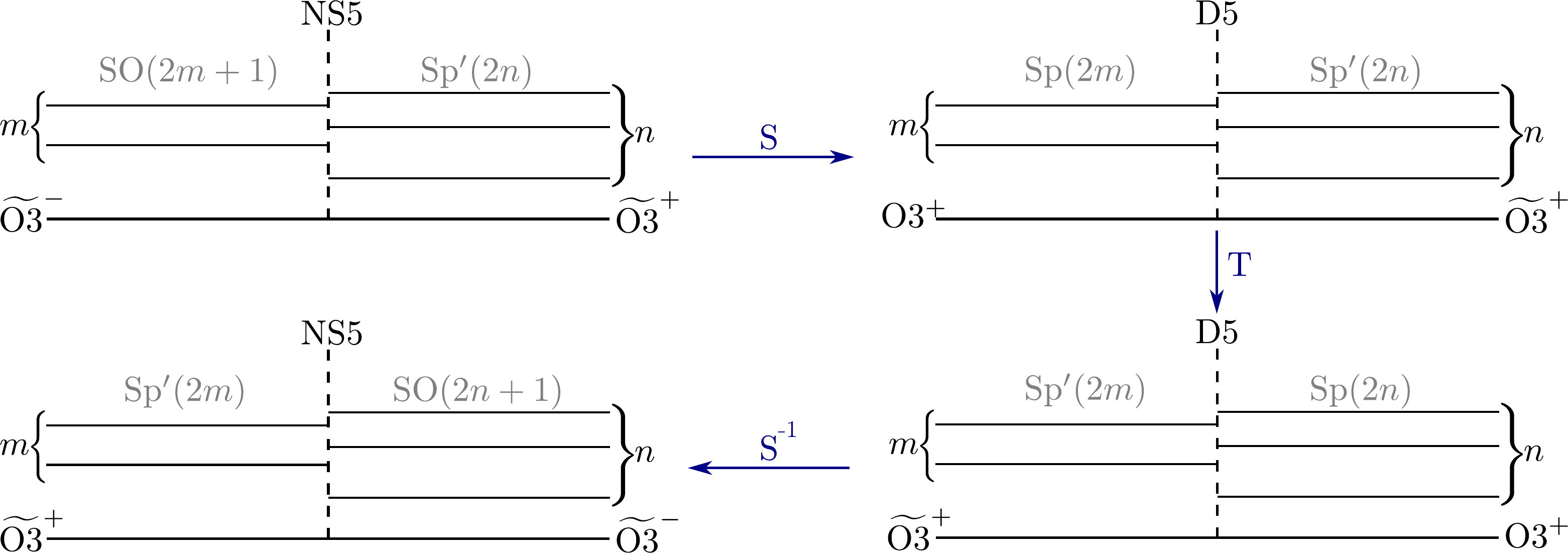}
 \end{center}
\caption{\small The figure in the upper left corner shows the brane configuration, which gives the four-dimensional construction for the OSp$(2m+1|2n)$ Chern-Simons theory. The other figures are obtained by acting with various elements of the SL$(2,\ZZ)$ $S$-duality group.  In particular, the
transformation $S^{-1}TS$ maps the configuration in the upper left to the one in the lower left.}
 \label{ospodd}
\end{figure}

\subsection{The Odd Orthosymplectic Theory}\label{oddorth}

\subsubsection{Preliminaries}\label{prelims}
Now we will repeat the analysis of the D3-O3-NS5 system, with just one important change:  we give the O3-planes a discrete 
 RR flux.  As depicted in the upper left
 of fig.~\ref{ospodd}, we take the O3-plane to be of type $\t{\O 3}^-$ to the left of the NS5-brane and (therefore) of type $\t{\O 3}^+$ to the right.
 The gauge groups realized on the D3-O3 system on the two sides of the defect are $\SO(2m+1)$ and $\Sp'(2n)$, so this configuration describes
 an analytically-continued version of $\OSp(2m+1|2n)$ Chern-Simons theory.   Up to a point, the four-dimensional gauge theory description of this
 system can be found just as in section \ref{evenorth}:  we restrict the fields of the familiar $\U(2m+1|2n)$ system to be invariant under the
orientifold projection, and divide the action by 2.  
 
However, there are some crucial subtleties that do not have a  close analog in the previous case:
 
(1)  The gauge theory theta-angle jumps by $\pi$ in crossing the defect, because the gauge theory on the right is of type $\Sp'(2m)$.
  By itself, this would spoil the supersymmetry of the defect system,
 since when one verifies supersymmetry at the classical level, one assumes that $\tau_\YM$ is continuous in crossing the defect.\footnote{Supersymmetry
 actually allows certain discontinuities \cite{Janus}, but not a jump in $\theta_\YM$ at fixed $\ang$.}
 
(2) This suggests that a quantum anomaly may be relevant, and in fact there is one: in three dimensions,
the bifundamental hypermultiplet of $\SO(2m+1)\times \Sp(2n)$ that is supported on the three-dimensional defect
suffers from a global anomaly.
 
 These two problems compensate each other, but this requires some explanation. 
 
 \subsubsection{The Anomaly In Three Dimensions}\label{thanom}
 
  \def\Pf{{\mathrm{Pf}}}
  \def\slD{{\slashed{D}}}
 In this section, we will describe an anomaly that arises in three dimensions (or in general in $8k+3$ dimensions)
 for fermions in the fundamental representation of an orthogonal or symplectic gauge group \cite{Redlich,AlvarezGaume}.  As we will see, the result
 for the symplectic group is really a special case of the result for the orthogonal group.
 
 First of all, in general, in any dimension, in a physically sensible theory like super Yang-Mills theory, fermions are always real
 in Lorentz signature.  After analytic continuation to Euclidean signature, they are in general not real (for example, Euclidean signature
 fermions  in the Standard Model of particle physics do not have a real structure).  In three dimensions, which we will focus on,
 there is up to isomorphism a unique irreducible spinor representation of the Lorentz group $\SO(2,1)$ or rather its double cover
 $\Spin(2,1)$;  it is two-dimensional.  
 Moreover, fermi statistics let us write an action for a field $\psi$ transforming in this representation; we write this action schematically
 \begin{equation}\label{faction} I_\psi=\int \d^3 x\,\,\left(\psi,\slashed{D}\psi\right) . \end{equation}
 The most general free theory of fermions in three dimensions is obtained by simply taking $k$ copies of this one for some positive
 integer $k$.  The global symmetry is $\O(k)$.  Classically, any subgroup of $\O(k)$ can be gauged.
 By ``fermions in the fundamental representation of $\O(k)$,'' we mean this theory of $k$ fermion doublets, the global
 symmetry being $\O(k)$.
 By ``fermions in the fundamental representation of $\Sp(2n)$,'' we actually mean a theory with $4n$ fermion doublets, in which
 we focus on an $\Sp(2n)$ subgroup of the full symmetry group $\O(4n)$, using the embedding $\Sp(2n)\times \SU(2)\subset \Spin(4n)$.
 (For fermions that come from a hypermultiplet in the fundamental representation of $\Sp(2n)$, the $\SU(2)$ factor here is part of the
 $R$-symmetry group.)  
 
 After analytic continuation to Euclidean signature, where we will work, the Lorentz group $\SO(2,1)$ is replaced by $\SO(3)$ and the
 spinor representation of $\Spin(3)$ is pseudoreal rather than real.  Hence, in a physically sensible theory in Euclidean signature, fermions
 always consist of $k$ copies of this pseudoreal representation for some $k$ (some subgroup of $\O(k)$ may be gauged) and so
 always have a pseudoreal structure.\footnote{\label{melo} In the case of $4n$ copies of the spinor representation, with 
  the gauge group being a subgroup $\Sp(2n)\subset\SO(4n)$, the fermions simply consist of two copies of a set of real
  fermions transforming as $(\bf{2},\bf{2n})$ under $\Spin(3)\times \Sp(2n)$.  In this case, the fermions can be given
  a real structure as well as a pseudoreal one. We will not emphasize this in what follows as it would make it difficult to
  describe the $\SO(k)$ and $\Sp(2n)$ cases together.}  This is enough to ensure that the fermion path integral is real.
 
 The fermion path integral is traditionally called  a ``determinant.'' However, it is more natural to think of this path integral as a Pfaffian.
 The idea here is that because of fermi statistics, the fermion kinetic energy (\ref{faction}) is naturally understood as an antisymmetric or skew
 form (in infinitely many variables). The fermion path integral can be regarded as a determinant if there is a suitable $\U(1)$ symmetry (electric charge in QED, baryon number in QCD, ghost number in the topologically twisted
 version of the theory studied in the present paper), as a result of which the skew form whose Pfaffian we want can be written
 \begin{equation}\label{fluff}\begin{pmatrix} 0 & M \cr -M^{\mathrm{tr}} & 0 \end{pmatrix}.\end{equation}
 In a suitable sense, the Pfaffian of such a form can be identified as the determinant of $M$, and this is the basis for the idea of a
 fermion ``determinant.''  But generically we will not be in that situation, so it is more useful to consider Pfaffians.
 
 We will write $\slashed{D}$ for the skew form associated to the fermion action, and $\Pf(\slashed{D})$ for
 the fermion path integral.  For reasons just explained, the general case we consider is that of
  $k$ copies of the spinor representation of $\SO(2,1)$, with a subgroup $G\subset \SO(k)$
 being gauged, on a three-manifold $W$.
 
 Any nondegenerate skew form is the direct sum of $2\times 2$ blocks
 \begin{equation}\label{mexic}\begin{pmatrix} 0 & \lambda_i \cr -\lambda_i & 0 \end{pmatrix}, \end{equation}
 with complex numbers $\lambda_i$.  In three dimensions, the pseudoreal structure means that the $\lambda_i$ are real.
 (The usual way to say this is that $i\slashed{D}$, regarded as a matrix rather than a form, is hermitian.  See also footnote
 \ref{zelo}.)
 The Pfaffian of such a form is formally
 \begin{equation}\label{mero}\Pf(\slashed{D})=\prod_i (\pm \lambda_i),  \end{equation}
 and in particular is naturally real.
 Of course, this infinite product needs to be regularized, for instance by $\zeta$-function regularization.  This regularization leads
 to a perfectly natural (and completely gauge-invariant)
 definition of the absolute value   $|\Pf(\slashed{D})|$ of the Pfaffian, or its square $\Pf(\slD)^2$.  However,
 there is no natural definition of the sign of $\Pf(\slashed{D})$, because there is no good way to decide, mod 2, how many of the $\lambda_i$
 are negative.
 
 The most simple idea of how to define the sign of $\Pf(\slD)$ is to start with some particular gauge field $A_0$, and
 declare $\Pf(\slD)$ to be positive at $A=A_0$.  (For example, this procedure was followed in \cite{GlobalAnom} in a four-dimensional problem
 similar to the present one.)  Then one lets $\Pf(\slD)$ evolve continuously as a function of $A$.  In this way, one can define $\Pf(\slD)$
 as a continuous function on a connected component of the space of gauge fields.  The components are classified by topological
 types of $G$-bundles over $W$, and possibly by operator insertions that may create singularities in $A$.
 
 The only problem with this definition, apart from the dependence on the choice of $A_0$, is that $\Pf(\slD)$ defined this way
 may not be gauge-invariant.   It is always gauge-invariant under gauge transformations that are continuously connected to the identity,
 but it is not necessarily invariant under ``big'' gauge transformations that are not continuously connected to the identity.  Let $A_0^g$
 be the transform of $A_0$ under a big gauge transformation $g$.  
 One can continuously evolve the gauge field from $A_0$ to $A_0^g$ (for example by the
 one-parameter family $tA_0+(1-t)A_0^g$, $0\leq t\leq 1$).  When one does this, $\Pf(\slD)$ comes back to itself up to sign, but
 in general only up to sign.  (Concretely, this happens because some of the $\lambda_i$ change sign during evolution from $t=0$ to $t=1$,
 causing a sign change of $\Pf(\slD)$.)  
 
 A simple source of ``big'' gauge transformations  is $\pi_3(G)$: for any simple Lie group $G$, $\pi_3(G)\cong \Z$.  If $G$ is
 the  $\SO(k)$ symmetry of $k$ spinors of $\SO(2,1)$ with\footnote{For $k=2,3$, there is no anomaly associated to $\pi_3(G)$, but there
 is still an anomaly in the sign of $\Pf(\slD)$ for suitable $W$ and suitable $G$-bundles over $W$.  For $k=2$, so that $\SO(k)\cong \U(1)$,
 this is important in the theory of topological insulators \cite{HK,QZ,HM,AB}.}  $k\geq 4$, or is an 
 $\Sp(2n)$ subgroup of $\SO(4n)$ acting on $4n$ such spinors, then $\Pf(\slD)$ changes sign
 under a generator of $\pi_3(G)$.
 
 A similar problem arises in 
 four dimensions for fermions transforming in a single copy of the fundamental representation of $\Sp(2n)$, and in this case the
 theory is actually inconsistent \cite{GlobalAnom}.  In three dimensions, however, one can cancel the anomaly at the cost of explicitly violating
 the parity and time-reversal symmetries ($P$ and $T$) of the classical fermion action.  The phenomenon is therefore traditionally called a parity
 anomaly \cite{Redlich,AlvarezGaume}.
 
 \def\CS{{\mathrm{CS}}}
A simple way to describe how to cancel the anomaly is adequate for many purposes.  One simply adds to the three-dimensional action a Chern-Simons term with a half-integral coefficient.  Thus if
 $\CS(A)$ is the Chern-Simons form, properly normalized to be defined mod $2\pi$, we include a term $is\CS(A)$ in the action,
 with $s\in \frac{1}{2}+\Z$.  Any $s$ will suffice for canceling the anomaly, but for our purposes it is convenient to take $s=-1/2$.
 Thus instead of $\Pf(\slD)$, we consider the product
 \begin{equation}\label{mixo} \Pf(\slD)\exp\left(\frac{i}{2}\CS(A)\right). \end{equation}
 This product is completely gauge-invariant, and for many purposes, it gives an adequate description of the physics associated to the parity
 anomaly.
 
 The limitation of this definition is that it does not give a clear definition of the overall sign of the path integral.  There is a dependence on the choice
 of the point $A=A_0$ at which we declared $\Pf(\slD)$ to be positive, and we also have to decide how to define the sign of
 $\exp\left(\frac{i}{2}\CS(A)\right)$ at $A=A_0$.  The main case that the definition (\ref{mixo}) is adequate is that $\pi_0(G)=\pi_1(G)=0$
 and we consider in the path integral only operators that are functions of $A$ (as opposed to monopole operators or monodromy
 operators that are defined by prescribed singularities in $A$).  The condition on $G$ ensures that any $G$-bundle is trivial, and the condition
 on the operators ensures that it makes sense to pick $A_0=0$.  At $A=A_0=0$, we define both factors in (\ref{mixo}) to be positive,
 and then the dependence on $A$ is uniquely determined.  For gauge groups $G$ that do not satisfy these conditions, such as $\O(k)$ or $\SO(k)$,
 and for any $G$ in the presence of monopole or monodromy operators, the choice $A_0=0$ is not possible in general and we are left
 with an ill-defined  overall sign of the path integral for each choice of $G$-bundle and/or set of insertions of monopole or monodromy operators.
 
 One can treat such questions piecemeal for each choice of $G$ and of the operators considered, but there is actually a fairly universal
 answer.  This is to replace the Chern-Simons function $\CS(A)$ in (\ref{mixo}) by the Atiyah-Patodi-Singer $\eta$-invariant.
 A theorem of Atiyah, Patodi, and Singer \cite{APS} 
 says that a certain multiple of $\eta$ coincides with $\CS(A)$ modulo a constant.  Replacing
 $\CS(A)$ by $\eta$ eliminates the constant indeterminacy in the more elementary approach to canceling the anomaly via the Chern-Simons function.
 
 \def\TT{{\mathcal T}}
To define $\eta$, we have to consider the eigenvalues of the hermitian Dirac operator $i\slD$.  The fact that the fermions are pseudoreal
means there is an antiunitary operator $\TT$ that commutes with $i\slD$ and obeys $\TT^2=-1$.  As is familiar in quantum mechanics
(where it leads to Kramers degeneracy for systems with half-integer spin), 
the existence of such a symmetry ensures that the eigenvalues $\lambda_i$ of $i\slD$
have even multiplicity.\footnote{ \label{zelo} If viewed as a skew form as in eqn. (\ref{mexic}), $\slD$ is the direct sum of $2\times 2$ blocks.  This decomposition
gives a pairing of states which corresponds to the Kramers doubling.  There is a subtlety here.  If viewed as a quadratic form, $\slD$
is antisymmetric (because of fermi statistics) and thus is a direct sum of antisymmetric blocks, as in (\ref{mexic}).  To view $\slD$ as an operator
(a matrix $M^i{}_j$ with one ``upper'' and one ``lower'' index) rather than a form (lower indices $M_{ij}$), one has to raise an index.  In the
context of a pseudoreal set of fermions, the raising is done using an antisymmetric tensor, and converts the $2\times 2$ block in (\ref{mexic})
into a $2\times 2$ matrix $\mathrm{diag}(\lambda_i,\lambda_i)$, with 2 equal eigenvalues (not equal and opposite) reflecting the Kramers degeneracy.}
For $G=\Sp(2n)\subset \SO(4n)$, there is a more elementary way to explain the  doubling. As described in footnote \ref{melo},
the fermions make up two copies of a certain real representation, and the operator $i\slD$ is the direct sum of two copies of an
operator $i\slD'$ acting on only one copy of this representation.

The $\eta$-invariant of the gauge field $A$ is defined as
\begin{equation}\label{etainv}\eta=\lim_{s\to 0}\sum_i |\lambda_i|^{-s}\mathrm{sign}(\lambda_i), \end{equation}
where $\lambda_i$ are the eigenvalues of the hermitian  operator $i\slD$.  ({\it A priori}, the sum on the right hand side
converges for sufficiently large $\mathrm{Re}\,s$.  One proves that in odd dimensions,
the function defined by this sum can be analytically continued to $s=0$, and $\eta$ is defined to be the value of the function at $s=0$.) For our purposes, it is convenient
to eliminate the Kramers doubling and define
\begin{equation}\label{tainv}\eta'=\frac{\eta}{2}. \end{equation}
For $G=\Sp(2n)$, $\eta'$ is the $\eta$-invariant of the operator $i\slD'$; in general, it is just a convenient definition.

$\eta'$ is real-valued for all $A$ and varies smoothly as a function of $A$ except that it jumps by 
$\pm 2$ whenever one of the $\lambda_i$ passes through 0.  
Accordingly, $\exp(i\pi \eta'/2)$ has modulus 1 in general, and jumps in sign when one of the $\lambda_i$ passes through 0.
Therefore the product
\begin{equation}\label{teflon}|\Pf(\slD)|\exp(i\pi\eta'/2) \end{equation}
varies smoothly as a function of $A$.  Indeed, whenever one of the $\lambda_i$ passes through 0, so $\Pf(\slD)\sim\lambda_i$ but
$|\Pf(\slD)|\sim |\lambda_i|$, $\exp(i\pi \eta'/2)$ jumps in sign so that the product in (\ref{teflon}) varies as $\lambda_i$.  Therefore,
this product is a smooth function of $A$.
The product in (\ref{teflon}) is  also completely gauge-invariant.  In fact, both factors in (\ref{teflon}) are separately gauge-invariant.
From the beginning, there was never a problem in defining the absolute value $|\Pf(\slD)|$, and the $\eta$-invariant is defined in eqn. (\ref{etainv})
as a function of the gauge-invariant eigenvalues of $i\slD$, so it is certainly gauge-invariant.    The Atiyah-Patodi-Singer theorem concerning
the variation of the $\eta$-invariant says that the ratio of (\ref{teflon}) and (\ref{mixo}) is a constant, invariant under variations
in $A$, so the two formulas differ
(in each sector of field space defined by a choice of $W$ and of a $G$-bundle and a set of monopole and/or monodromy operators)
by a multiplicative constant.  It is hard to be completely precise about this constant, beause (\ref{teflon}) is completely well-defined
and (\ref{mixo}) is not.   Eqn. (\ref{teflon}) is the precise formula, while (\ref{mixo}) is an informal description that is sufficient for many purposes.
We will use this informal description in the present paper where it is adequate.

Passing from (\ref{mixo}) to the more precise (\ref{teflon}) does not change the fact that in a purely three-dimensional theory,
the anomaly in the sign of the fermion path integral can only be canceled by explicitly violating the $P$ and $T$ symmetries of the classical
action.

\subsubsection{The Anomaly In Four Dimensions}\label{anomfour}

In the present paper, we are interested in fermions that are in the bifundamental representation of $\SO(2m+1)\times\Sp(2n)$.
From the point of view of $\SO(2m+1)$, these fermions consist of $4n$ copies of the fundamental representation. (Naively, there are $2n$ copies,
but an extra doubling occurs because the fundamental representation of $\Sp(2n)$ is pseudoreal rather than real.)  Although
there would be an anomaly in the sign of the fermion path integral for one copy of this representation, the anomaly cancels
when we consider $4n$ copies.  From the point of view of the $\Sp(2n)$ gauge group, the fermions form $2m+1$ copies
of the fundamental representation.  Since this is an odd number, the anomaly does not cancel.

In the informal description of eqn. (\ref{mixo}), the anomaly is canceled by adding a half-integral Chern-Simons interaction 
 \begin{equation}\label{toxo}-\frac{i}{8\pi}\int\,\tr_\sp\,\left(A\wedge \d A+\frac{2}{3}A\wedge A\wedge A\right). \end{equation}
 In an abstract three-dimensional theory, this coupling violates $P$ and $T$ symmetry and there is no way to restore
 those symmetries while maintaining gauge-invariance.
 
 However, in the present paper, we are really studying a four-dimensional theory.  The bifundamental fermions live on a three-dimensional
 submanifold $W$ of a four-manifold $M$.  $W$ divides $M$ into a region $M_\ell$ in which the gauge group is $\SO(2m+1)$ and a region
 $M_r$ in which it is $\Sp(2n)$.  
 
Adding the term (\ref{toxo}) to the defect action will, by itself, break supersymmetry in our problem: this term is certainly not supersymmetric
by itself. By adding this term, we have restored gauge-invariance but apparently at the cost of supersymmetry.  However, at this point we must
remember from section (\ref{prelims}) that we had a problem with supersymmetry even before we considered the anomaly,
because $\theta_\YM$ jumps by $\pi$ in crossing from $M_\ell$ to $M_r$.  The two problems precisely cancel.  The sum of the
two troublesome terms is
 \begin{equation}\label{moxo}-\frac{i}{8\pi}\int_{W}\tr_\sp\,\left(A\wedge \d A+\frac{2}{3}A\wedge A\wedge A\right)+\frac{i}{8\pi}\int_{M_r}\tr_\sp\,F\wedge F \end{equation}
 and this sum is completely supersymmetric: the variation of the bulk integral $\tr\,F\wedge F$ in (\ref{moxo}) precisely cancels
 the variation of the Chern-Simons function.  
 
At this point, however, the reader may be confused.  A standard definition of the Chern-Simons function on a three-manifold $W$ expresses it
via the integral $\int_{M_r} \tr \,F\wedge F$ on a manifold $M_r$ with boundary $W$.  Therefore, is not (\ref{moxo}) simply zero?  And
if it is zero, how could it have helped in canceling the anomaly in $\Pf(\slD)$?  And given the Chern-Simons interaction on $W$,
is there really in any meaningful sense a jump in $\theta_\YM$ by $\pi$ in crossing from $M_\ell$ to $M_r$?  Will local observables that
depend on $\theta_\YM$ see this jump, or is it canceled by the effects of the half-integral Chern-Simons term on the boundary?

It is difficult to clarify these points without using the more
precise description based on the $\eta$-invariant rather than the Chern-Simons function.  The root of the problem is that the Chern-Simons
function in (\ref{moxo}) is not really well-defined, since it is not properly normalized. To answer a delicate question, we need
to use well-defined expressions.  
To get a clear picture, we consider the contribution
of the interactions (\ref{moxo}) to the path integral, but we also include\footnote{In doing this, we consider only the fermions
supported on the defect, ignoring the fact that they are actually coupled by boundary conditions to the bulk fermions.  This coupling
does not affect the anomaly.  However, in section \ref{thedual}, it is important to include this coupling.} the fermion Pfaffian, and we use $\eta$ instead of Chern-Simons.  To avoid some details that could be confusing at first sight, we assume at first that $m=0$, meaning that the gauge
group is trivial on the left (or is the finite group $\O(2m+1)=\O(1)=\Z_2$, which does not much affect the discussion) 
and is $\Sp(2n)$ on the right.  So there is a single fundamental
of $\Sp(2n)$ along the defect, as assumed in the three-dimensional context in section \ref{thanom}.  Hence we study the product
\beq\label{orez} |\Pf(\slD)|\exp(i\pi\eta'/2)\exp\left(-\frac{i}{8\pi}\int_{M_r}\tr_\sp \,F\wedge F\right).
\eeq
There is no problem with gauge-invariance, since each of the three factors is separately gauge-invariant.  There is no problem with supersymmetry,
since the verification of supersymmetry only involves local questions, for which eqn. (\ref{moxo}) is sufficient.  Finally, in what sense
does $\theta_\YM$ jump by $\pi$ in crossing from $M_\ell$ to $M_r$?  To answer this question, let us evaluate eqn. (\ref{orez})
assuming that the gauge field is gauge-equivalent to 0 along $W$.  While maintaining these conditions, we can change $A$ so as to shift
the instanton number $\Nb $ on $M_r$ by an integer $w$.  When we do this, only the last factor in (\ref{orez}) changes
and (\ref{orez}) is multiplied by $(-1)^w$.  Of course,\footnote{We are jumping ahead of the story.  Since we set $m=0$, it is not possible to add an instanton
on $M_\ell$.  But after we generalize to $m>0$ by replacing eqn. (\ref{orez}) with eqn. (\ref{roboxx}) below, it remains true that adding an instanton on $M_\ell$
has no effect.}
(\ref{orez}) is unchanged if we similarly add an instanton on $M_\ell$.  This comparison gives a precise meaning to the assertion that $\theta_\YM$
jumps by $\pi$ in crossing from $M_\ell$ to $M_r$.  This difference will be reflected in the values of any local observables that depend on $\theta_\YM$.

So far in this discussion we have ignored the twisting that is used to maintain supersymmetry when $W$ is not flat. But actually, this twisting has no essential influence on the above analysis.  Twisting means that instead of $4n$ spinors transforming
as 2 copies of the fundamental representation of $\Sp(2n)$, we should consider $4n$ fermions transforming as a bifundamental of $\SU(2)\times\Sp(2n)$, where
the $\SU(2)$ bundle in question is the spin bundle of $W$. (The Dirac operator in this situation is known as a twisted
signature operator.)  If we consider the metric of $W$ to be fixed, although arbitrary, so that the spin bundle is fixed, then twisting 
really does not affect the above analysis.   When the metric of $W$ is varied, one refinement is needed: 
to maintain topological invariance of eqn. (\ref{orez}) and other formulas below, we have to subtract from the exponent a gravitational Chern-Simons term (related to the $R^2$ term in eqn. (\ref{merox}) below).  To keep
our formulas relatively simple, we will omit this.

The  system we have described actually has  a striking analogy with three-dimensional topological insulators, which have been much-studied recently
(see for example \cite{HK,QZ,HM,AB} for introductions).  The surface of 
such a material supports  electrically charged gapless modes whose relevant properties
are analogous to those of the bifundamental fermions in our problem: their path integral is real but from a purely three-dimensional point of view its
sign cannot be defined in a gauge-invariant and time-reversal symmetric fashion.    Associated to this, the electromagnetic
theta-angle jumps by $\pi$ in crossing from the vacuum into such a material, rather like the jump in $\theta_\YM$ in our problem.  
In the theory of these materials, it is very important that time-reversal symmetry can be maintained in this situation even though
in a purely $2+1$-dimensional material supporting the same gapless modes, this would not be the case.

\def\I{{\mathcal I}}
\def\h{\widehat}
The last statement is also true in our problem: as long as $\theta_\YM$ is 0 or $\pi$ in $M_\ell$ (and hence is $\pi$ or 0 in $M_r$), our system
is $P$- and $T$-invariant.  In showing this, we will assume that $M_r$ is compact, with boundary $W$. (We believe that the following
explanation is also illuminating for topological insulators.) We consider the Atiyah-Patodi-Singer theorem for the index of the four-dimensional Dirac operator\footnote{As we have explained, we really need a four-dimensional operator, which in three dimensions would reduce to what we called the twisted signature operator. The actual twisted signature operator in four dimensions would reduce to two copies of this~--~one acting on the even forms, and one acting on the odd forms. To get only one copy, in four dimensions we take the operator, which arises in the linearization of the self-duality equation. According to the index theorem for this operator, what we denote by $\int R^2$ in the formula below is the integral of a density, which would give $(\chi-\sigma)/2$ on a closed four-manifold. (Here $\chi$ and $\sigma$ are the Euler characteristic and the signature, respectively.)}, acting on a single multiplet of positive chirality fermions in the fundamental
representation of $\Sp(2n)$. Let $\I'$ be one-half of this index, which is an integer, because of the Kramers doubling. The theorem says that
\begin{equation}\label{merox}\I' =-\frac{\eta'}{2} +\frac{1}{8\pi^2}\int_{M_r}\tr_\sp \, F\wedge F +n\int_{M_r}R^2,\end{equation}
where $R^2$ is an appropriate quadratic expression in the Riemann tensor of $M_r$.  If $M_r$ admits an orientation-reversing symmetry (which could be interpreted physically
as $P$ or $T$), then $\int_{M_r} R^2=0$.  Since $\I'$ is an integer,  this vanishing implies that
the phase factor in (\ref{orez}) is $\pm 1$:
\beq\label{nevbo} \exp(i\pi\eta'/2)\exp\left(-\frac{i}{8\pi}\int_{M_r}\tr_\sp \,F\wedge F\right)=e^{-i\pi \I'}=\pm 1.\eeq
The left hand side of (\ref{nevbo}) is mapped to its own inverse by $P$ or $T$ symmetry, but equals its own inverse
by virtue of (\ref{nevbo}).  So the anomaly-canceling mechanism with the jump in $\theta_\YM$ by $\pi$ does preserve $P$ and $T$ symmetry.

\def\o{{\mathfrak o}}
All this is for $m=0$, that is for the orthosymplectic group $\OSp(1|2n)$.  
The generalization to $m>0$ involves a few inconvenient details.  We replace $\eta'$ by what we will call $\hat\eta$,
one-half of the $\eta$-invariant of the bifundamental fermions. Writing $A_\so$ and $A_\sp$ for the $\SO(2m+1)$ and $\Sp(2n)$ gauge
fields, the Atiyah-Patodi-Singer theorem says that the variation of $\eta'$ is the same as the variation of a linear combination of the corresponding
Chern-Simons invariants, namely
$(m+1/2)\CS(A_\sp)+2n\CS(A_\so)$.  One way to proceed is to replace $\exp(i\pi\eta'/2)$ by
\begin{equation}\label{robox}\exp(i\pi \h\eta/2-im\CS(A_\sp)-2in\CS(A_\so)) \end{equation}
(whose variation is the same as that of $\frac{i}{2}\CS(A_\sp)$).  
Everything then proceeds much as before.
The properly normalized Chern-Simons interactions $m\CS(A_\sp)+2n\CS(A_\so)$ can also be replaced by curvature
integrals, shifting $\theta_\YM$ by $-2\pi n$ on $M_\ell$ and by $2\pi m$ on $M_r$.   We will use a convenient
abbreviation
\begin{equation}\label{roboxx}\exp(i\pi\eta'/2)=\exp(i\pi \h\eta/2-im\CS(A_\sp)-2in\CS(A_\so)), \end{equation}
so that (\ref{orez}) holds for any $m$.

\subsubsection{The Dual Theory}\label{thedual}
We can find now a magnetic dual of $\OSp(2m+1|2n)$ Chern-Simons theory  by applying the $S$-duality transformation $\tau\to -1/\tau$. 
Its action on the brane configuration 
is shown in the upper part of fig.~\ref{ospodd}. The new string coupling is $\tau^\vee=-1/\tau$. The gauge groups are
now $\Sp(2m)$ in $M_\ell$ and $\Sp'(2n)$ in $M_r$.   We continue to use the notation $\tau^\vee_\YM=\fr{1}{2}\tau^\vee$ for 
the gauge coupling. The minimal instanton number 
for the symplectic group is 1, so the natural instanton-counting parameter analogous to (\ref{icon}) is  
$q=\exp(-i\theta^\vee_\YM)$. Using (\ref{thetaospdual}), this can be presented as
\beq
q=\exp\left(\fr{\pi i}{2\calK_{\osp}}\right).\label{qodd}
\eeq
This agrees with the general definition (\ref{usdef}), since the ratio of length squared of the longest and shortest bosonic
roots for the odd orthosymplectic algebras is $n_\g=4$.

In the ``magnetic'' description,  one of the orientifold planes is again of type $\widetilde{{\rm O}3}^+$, which means that the 
$\theta_\YM$  jumps by $\pi$ upon crossing the defect.  As in the electric description, this jump appears to violate supersymmetry.
The resolution is similar to what it was in the electric description.  First we consider the case that $m=n$.  For this case, the gauge group
is simply $\Sp(2n)$ filling all of spacetime.  There is a fundamental hypermultiplet supported on the defect.  Its Pfaffian has the sign anomaly
that was reviewed in section \ref{thanom}.  The anomaly is canceled roughly speaking via a half-integral Chern-Simons term supported on the defect,
or more accurately via an $\eta$-invariant.  The combined path integral involving the fermion Pfaffian, the $\eta$-invariant, and the jump in 
$\theta_\YM$ (as well as other factors) is gauge-invariant and supersymmetric.  The factors involved in the anomaly cancellation
are the familiar ones from eqn. (\ref{orez}):
\beq\label{nevbox}  |\Pf(\slD)|\exp(i\pi\eta'/2)\exp\left(-\frac{i}{8\pi}\int_{M_r}\tr_\sp \,F\wedge F\right).\eeq
As see in eqn. (\ref{nevbo}), the product of the last two factors equals $\pm 1$ (possibly multiplied by a factor that only depends on
$\int_{M_r}R^2$).  This factor of $\pm 1$ must be incorporated in the sum over instanton solutions.  We denote
it as
\beq\label{wevbox}\sgn_{y\geq 0}=\exp(i\pi\eta'/2)\exp\left(-\frac{i}{8\pi}\int_{M_r}\tr_\sp \,F\wedge F\right).  \eeq

What happens if $n\not=m$?  In this case, there are no hypermultiplets supported on the defect.  Instead, there is a jump in the gauge
group in crossing the defect.  Along the defect there is a Nahm pole, associated to an irreducible embedding of $\frak{su}(2)$ in 
$\frak{sp}(|2n-2m|)$.  As usual, the pole is on the side on which the gauge group is larger.
The gauge group that is unbroken throughout all space is $\Sp(2s)$, where $s$ is the smaller of $n$ and $m$.

At first sight, it is not clear how to generalize (\ref{nevbox}) to $n\not=m$.  If there are no fermions supported on the defect, how
can we possibly use an anomaly in a fermion determinant as part of a mechanism to compensate for a jump in $\theta_\YM$ by $\pi$?
To understand what must happen, recall that we can deform from $n=m$ to $n\not=m$ by Higgsing -- by moving some of the D3-branes
(on one side or the other of the defect) away from the rest of the system.  When we do this, the bifundamental hypermultiplet which
is responsible for some of the interesting factors in (\ref{nevbox}) does not simply vanish in a puff of smoke.  It mixes with some of the bulk
degrees of freedom and gains a large mass.  When this happens, whatever bulk degrees of freedom remain will carry whatever anomaly
existed before the Higgsing process.

So the resolution of the puzzle must involve a subtlety in the fermion path integral for $n\not=m$.   Going back to (\ref{nevbox}), naively
$\slD$ is the Dirac operator just of the defect fermions and $\eta'$ is one-half their $\eta$-invariant.  There are also bulk fermions,
but they have no anomaly and vanishing $\eta$-invariant, so it does not seem interesting to include them in (\ref{nevbox}).  However,
precisely because they have no anomaly and vanishing $\eta$-invariant, we could include them in (\ref{nevbox}) (and their coupling
to the defect fermions) without changing anything.  This is a better starting point to study the Higgsing process, since Higgsing disturbs
the decoupling.  

Upon Higgsing, the first two factors in eqn. (\ref{nevbox}) keep their form, but some modes become massive and -- in the limit
that $|2n-2m|$ D3-branes are removed on one side or the other -- the defect fermions disappear and we are left with an expression
of the same form as (\ref{nevbox}), but now the Pfaffian and the $\eta$-invariant are those of the bulk fermions in the presence of the Nahm pole.
The Dirac operator of the bulk fermions in the presence of the Nahm pole can be properly defined, with some subtlety, as an 
elliptic differential operator \cite{MazzeoWitten}.  This gives a framework in which one could investigate its Pfaffian and $\eta$-invariant.
For the theory that we are discussing here to make sense,  there must be an anomaly in the sign of the Pfaffian of this operator,
and it must also have a nontrivial $\eta$-invariant that compensates in the familiar way for the jump in $\theta_\YM$.  These points
have not yet been investigated, but there do not seem to be any general principles that exclude the required behavior.

\subsection{The Framing Anomalies}\label{ospfranom}
In section~\ref{franom} we have verified that our constructions predict the correct value for the global framing anomaly for the Chern-Simons theory of the unitary supergroup. Here we repeat the same analysis for the orthosymplectic gauge group.

In the non-simply-laced case, the analog of the formula (\ref{pingo}) for the framing factor is
\beq
\exp\left(2\pi i\,\sign(k)\,{\rm sdim}\,SG/24 \right)\cdot q^{-n_\g h_\sg\,\rm{sdim}\,SG/24}.
\eeq
The difference with the simply-laced case is the factor of $n_\g$ in the exponent, which compensates for the analogous factor in the definition (\ref{usdef}) of the $q$ variable. As usual in this paper, we will ignore the one-loop contribution to the anomaly, and focus only on the power of $q$. To compare the anomalies for different groups, it is convenient to express them in terms of the theta-angle of the magnetic theory. What we need to know is that for a theory with a bosonic gauge group the variable $q$ is defined as $q=\exp(-2i\theta^\vee_\YM)$, if the gauge group in the magnetic description is orthogonal, and as $q=\exp(-i\theta^\vee_\YM)$, if this group is symplectic. We have explained the reason behind this definition, when we discussed the magnetic theories for the orthosymplectic supergroups. 

Consider first the even orthosymplectic algebra $\osp(2m|2n)$. As we recalled in section \ref{franom}, the framing anomaly in the magnetic description comes from the peculiarities of the definition of the instanton number in the presence of the Nahm pole. We set $r=n-m$. For $r>0$, the Nahm pole in the magnetic theory is embedded into an  $\so(2r+1)$ subalgebra of $\so(2n+1)$. This means that the framing anomaly depends only on $r$ and not on $m$; setting $m=0$, we reduce to the magnetic dual of  Sp$(2r)$ Chern-Simons theory and we should
get the same framing anomaly. The anomaly factor for the orthosymplectic case is expected to be
\beq
q_\osp^{-n_{\osp(2m|2n)}h_\osp{\rm sdim OSp}/24}=\exp\left(-4i\theta^\vee_\YM h_\osp{\rm sdim\, OSp}/24\right).\label{evenospanomaly}
\eeq
For the symplectic gauge group this factor is
\beq
q_\sp^{-n_{\sp}h_\sp{\rm dim Sp}/24}=\exp\left(-4i\theta^\vee_\YM h_\sp{\rm dim \,Sp}/24\right).
\eeq
The two expressions agree, since
\beq\label{zelbo}
h_{\osp(2m|2n)}\,{\rm sdim}\,\OSp(2m|2n)=h_{\sp(2r)}\,{\rm dim}\,\Sp(2r)=2r(r+1/2)(r+1).
\eeq  
This identity is the analog of (\ref{needid}); see Table~\ref{coxeters} for the numerical values.

If $r<0$, the Nahm pole lives in the $\so(-2r-1)$ subalgebra on the other side of the defect. This is the same Nahm pole
that would arise in the magnetic dual of  SO$(-2r)$ Chern-Simons theory, so the framing anomaly should agree with that theory. For the bosonic theory with the even orthogonal gauge group we have
\beq
q_\so^{-h_\so{\rm dim\,SO}}=\exp\left(-2i\theta^\vee_\YM h_\so{\rm dim\, SO}/24\right).
\eeq
This agrees with (\ref{evenospanomaly}), since
\beq
h_{\osp(2m|2n)}\,{\rm sdim}\,\OSp(2m|2n)=-\fr{1}{2}h_{so(-2r)}{\rm dim\,SO(-2r)}=2r(r+1/2)(r+1).
\eeq
The minus sign appears here, because the Nahm pole for the orthosymplectic theory with $r<0$ is on the left side of the defect.

Alternatively, we could think of the $\so(-2r-1)$ Nahm pole as corresponding to the Sp$(-2r-2)$ electric theory. This would give the same result.

Let us repeat the same story for the odd orthosymplectic superalgebra $\osp(2m+1|2n)$. Again, we set $r=n-m$. The Nahm pole is embedded in the $\mathfrak{sp}(2|r|)$ subalgebra. In the purely bosonic case, the same embedding 
would arise for the $SO(2|n-m|+1)$ electric theory. Therefore, we would expect that the global framing anomaly for the superalgebra 
case is the same as for this purely bosonic Lie algebra, at least above one loop. 
The framing factor for the odd orthosymplectic case should be
\beq
q_\osp^{-n_{\osp(2m+1|2n)}h_\osp{\rm sdim OSp}/24}=\exp\left(-4i\theta^\vee_\YM h_\osp{\rm sdim\, OSp}/24\right).\label{oddospanomaly}
\eeq
In the SO$(2|r|+1)$ the answer is
\beq
q_\so^{-n_\so h_\so{\rm dim\,SO}}=\exp\left(-2i\theta^\vee_\YM h_\so{\rm dim\, SO}/24\right).\label{qwer}
\eeq
The two expressions (\ref{oddospanomaly}) and (\ref{qwer}) agree, since from Table~\ref{coxeters} we have
\beq
h_{\osp(2m+1|2n)}\,{\rm sdim}\,\OSp(2m+1|2n)=\fr{1}{2}h_{so(2|r|+1)}{\rm dim\,SO(2|r|+1)}=2r(r^2-1/4).
\eeq
The sign in the right hand side changes, depending on the sign of $r$, in accord with the fact that the Nahm pole is on the right or on the left of the defect. Note also, that up to this change of sign the formula is symmetric under the exchange of $m$ and $n$. This reason for this symmetry will become clear in section~\ref{another}.

\subsection{Another  Duality}\label{another}

\def\SL{\mathrm{SL}}
So far in this paper, we have just exploited the duality $S:\tau\to -1/\tau$, exchanging NS5-branes with D5-branes.
 The full $S$-duality group $\SL(2,\Z)$ of Type IIB superstring
theory contains much more.  In particular, it has a non-trivial subgroup that maps an NS5-brane to itself.
This  subgroup is generated by the element
\beq
S^{-1}TS=\left(\begin{array}{cc}1&0\\-1&1\end{array}\right).
\eeq
That this element maps an NS5-brane to itself follows from the action of duality on fivebrane charges given in eqn. (\ref{polk}).  (Concretely, $S$ converts an NS5-brane to a D5-brane, $T$ leaves fixed the D5-brane, and $S^{-1}$ maps back to an NS5-brane.)
This transformation will map a D3-NS5 system, possibly with an O3-plane, to a system of the same type.  In the approach to Chern-Simons
theories followed in the
present paper, this transformation will map an ``electric'' description to another ``electric'' description, and thus it will give a duality of Chern-Simons theories
(analytically continued away from integer levels).

Let us first see what this duality does to a D3-NS5 system, associated to the supergroup $\U(m|n)$.  The operation $S^{-1}TS$  maps D3-branes and NS5-branes
to themselves, so it maps the Chern-Simons theory of $\U(m|n)$ to itself, 
while transforming the canonical parameter according to (\ref{cKtransform}), which in this case gives
\beq
\frac{1}{\calK}\rightarrow \fr{1}{\calK}-1=\frac{1}{\calK'}.\label{cKtransform1}
\eeq
This transformation leaves fixed the variable $q=\exp(2\pi i/\calK)$ in terms of which the knot invariants are usually expressed. 
(In fact, the symmetry (\ref{cKtransform1}) can be viewed as the reason that the knot invariants can be expressed in terms of $q$ rather
than being more general functions of $\calK$.) This duality acts trivially on line operators of $\U(m|n)$.  To argue this, we just observe that $T$ can be understood classically -- as a $2\pi$
shift in $\theta_\YM$ -- and does not affect the model solution that is used to define a line operator. 

The action of $STS^{-1}$ on  a surface operator can be determined by looking at the behavior far away from the defect.
We have
\begin{equation}\label{longeqn}\begin{pmatrix}\alpha\cr \eta\end{pmatrix}\xrightarrow{S} \begin{pmatrix}\eta\cr -\alpha\end{pmatrix}\xrightarrow{T}
\begin{pmatrix}\eta\cr \eta-\alpha\end{pmatrix}\xrightarrow{S^{-1}}\begin{pmatrix}\alpha-\eta\cr \eta\end{pmatrix} . \end{equation}
Using the relation (\ref{surfweight}), the action on  the weight $\uplambda$ can be conveniently written
\begin{equation}\label{convenaction}\frac{\uplambda'}{\calK'}=\frac{\uplambda}{\calK}. \end{equation}
Since knot invariants computed using surface operators by the procedure explained in section \ref{gettingpolynomials}
only depend on the ratio $\uplambda/\calK$, this shows that they are invariant under $S^{-1}TS$.
Using the relation (\ref{cKtransform1}) between $\calK'$ and $\calK$, eqn. (\ref{convenaction}) is equivalent to
\begin{equation}\label{anotherone} \uplambda'=\uplambda+\calK'\uplambda. \end{equation}

Let us check whether these formulas are consistent with the idea that if $\uplambda$ is integral, the same knot and link invariants
can be computed  using either line operators or surface operators.  $S^{-1}TS$ acts trivially on the weight of a line operator,
but acts on the weight of a surface operator as in (\ref{anotherone}).  However, knot invariants computed from surface operators
are unchanged in shifting $\uplambda$ by $\calK$ times an integral cocharacter.  Since the groups $\U(n)$ and $\U(m)$ are selfdual,
if $\uplambda$ is an integral character, it is also an integral cocharacter.  

Now let us apply this duality to the configuration of fig.~\ref{ospeven}, which corresponds to an even orthosymplectic group $\OSp(2m|2n)$.
The transformation $S^{-1}TS$ maps the O3-planes that appear in this configuration to themselves, so again it maps 
Chern-Simons theory of $\OSp(2m|2n)$ to itself.   The canonical parameter $\calK_\osp$ of the orthosymplectic theory was defined as one-half of 
the object $\calK$ defined in section  \ref{magnetic}, so the transformation rule (\ref{cKtransform1}) can be written
\beq
\frac{1}{\calK_\osp}\rightarrow\frac{1}{\calK_\osp}-2=\fr{1}{\calK_\osp'}\,,\label{Koddtransform}
\eeq
Therefore, the natural Chern-Simons parameter $q=\exp(\pi i/\calK_\osp)$,  defined in eqn. (\ref{qeven}), is invariant, just as for the unitary case.
The Chern-Simons theory  again is simply mapped to itself. It takes a little more effort  to understand the duality action on line and surface operators. For this reason, the discussion of the operator mapping will be presented in a separate section \ref{operators}. There we will find that, unlike for the unitary superalgebra, the duality acts on the set of line operators by a non-trivial involution. 

For the odd orthosymplectic group $\OSp(2m+1|2n)$, matters are more interesting. The action of  $S^{-1}TS$ 
on the brane configuration associated to $\OSp(2m+1|2n)$ is described in  fig.~\ref{ospodd}. Chasing clockwise around the figure
from upper left to lower left, we see  that the duality maps a brane
configuration associated to $\OSp(2m+1|2n)$ to one associated to $\OSp(2n+1|2m)$.  Since the gauge group changes, this
 is definitely a non-trivial duality of 
(analytically-continued) Chern-Simons theories.  For example, setting $n=0$, we get a duality between Chern-Simons theory of the ordinary
bosonic group $\O(2m+1)$ and Chern-Simons theory of the supergroup $\OSp(1|2m)$.  How does this duality act on the natural
variable $q$ that parametrizes the knot invariants?  For the odd orthosymplectic group, the natural variable in terms of which the
knot invariants are expressed is $q=\exp(\pi i/2\calK_\osp)$, introduced in eqn. (\ref{qodd}).  The transformation (\ref{Koddtransform}) 
acts on this variable by\footnote{There is a subtlety here.  The Killing form for a superalgebra can be defined 
with either sign. Since the duality maps theories with, say, Sp group at $y>0$ to Sp group at $y<0$, it exchanges the two choices. 
If we want to define the sign of the Killing form to be always positive, say, for the $\sp$ subalgebra, we should rather say that $q$ maps to $-q^{-1}$.  What is written in the text assumes that the sign of the Killing form in $M_\ell$ or $M_r$ is unchanged in the duality.}
\begin{equation}\label{telme} q\to -q.\end{equation}
The minus sign means that the duality we have found exchanges weak and strong coupling.  Indeed, in three-dimensional Chern-Simons
theory, the weak coupling limit is $q\to 1$, and $q\to -1$ is a point of strong coupling.  

It is inevitable that the duality must map weak coupling
to strong coupling, since the classical representation theories of $\OSp(2m+1|2n)$ and $\OSp(2n+1|2m)$ are not equivalent.  A duality
mapping weak coupling to weak coupling would imply an equivalence between the two classical limits, but this does not hold.

 Some instances of the duality
predicted by the brane construction
have been discovered previously.
For $n=0$ and $m=1$, the relation between knot invariants has been discussed in \cite{osp12}; for $n=0$ and any $m$, this subject
has been discussed in \cite{Blumen} in a different language. For  related discussion from the standpoint of quantum groups 
see \cite{qOSp}, and see \cite{Rittenberg} for associated representation theory.  We will say more on some of these results in section \ref{operators}.

Now let us look at the same duality in the magnetic dual language.
Our two electric theories are sketched in the upper and lower left of fig.~\ref{ospodd}, and the corresponding magnetic duals, obtained by
acting with $S$,  are shown
in the upper and lower right of the same figure.   One involves an $\Sp(2m)\times \Sp'(2n)$ gauge theory, and the other involves
an $\Sp'(2m)\times \Sp(2n)$ gauge theory.   There is no change in the  gauge groups, the localization equations, or in 
 the hypermultiplet fermions if $n=m$ or in  the Nahm 
pole singularity if $n\not=m$.   The only difference is that in one case $\theta_\YM$ differs on the right  by $\pi$
from the underlying Type IIB theta-angle, and in the other case, it differs on the left by $\pi$ from the underlying Type IIB
theta-angle.   In the upper right of  fig. \ref{ospodd}, a solution of the localization equations with instanton number $\Nb^\vee$ 
is weighted by the product of $q^{\Nb^\vee}$ with the sign factor of eqn. (\ref{wevbox}).   There is an additional
sign that we will call $(-1)^f$; this is the sign of the determinant of the operator obtained by linearizing around a solution
of the localization equations. This factor is not affected by the duality.  The combination is
\begin{equation}\label{evbox}(-1)^fq^{\Nb^\vee}\sign_{y\geq 0}=(-1)^fq^{\Nb^\vee}\exp(i\pi\eta'/2)\exp\left(-\frac{i}{8\pi}\int_{M_r}\tr_\sp \,F\wedge F\right).
\end{equation}
On the lower left of the figure,  the sign factor $\sign_{y\geq0}$ is replaced with
\begin{equation}\label{yevbox}\sign_{y\leq 0}=\exp(i\pi \eta'/2)\exp\left(+\frac{i}{8\pi}\int_{M_\ell}\tr_\sp \,F\wedge F\right).\end{equation}
We also have to replace $q$ with $-q$.  So (\ref{evbox}) is replaced with
\begin{equation}\label{zevbox} (-1)^f(-q)^{\Nb^\vee}\exp(i\pi\eta'/2) \exp\left(+\frac{i}{8\pi}\int_{M_\ell}\tr_\sp \,F\wedge F\right). \end{equation}
The two expressions (\ref{evbox}) and (\ref{zevbox}) are equal, since 
\begin{equation}\label{vbox}\Nb^\vee=\Nb_\ell^\vee+\Nb_r^\vee,\end{equation}
with \begin{equation}\label{rbox}\Nb_\ell^\vee=\frac{1}{8\pi^2}\int_{M_\ell}\tr_\sp \,F\wedge F,~~~~\Nb_r^\vee=\frac{1}{8\pi^2}\int_{M_r}\tr_\sp \,F\wedge F. \end{equation}

The above formulas can be written more elegantly by using the Atiyah-Patodi-Singer  (APS) index theorem \cite{APS} for the
Dirac operator on a manifold with boundary.  This will also be useful later.
 We let $\nu_\ell$ (or $\nu_r$) be the index of the Dirac operator on $M_\ell$
(or $M_r$), acting on spinors with values in the fundamental representation of $\Sp(2n)$ (or $\Sp(2m)$).  This index is
defined by counting zero-modes of spinor fields that are required to be square-integrable at infinite ends of $M_\ell$ or $M_r$,
and to obey APS global boundary conditions along the finite boundary $W$.  The APS index theorem gives
\begin{align}\label{multifo}(-1)^{\nu_\ell}&=\exp(i\pi \eta'/2)\exp\left(+\frac{i}{8\pi}\int_{M_\ell}\tr_\sp \,F\wedge F\right)\cr
(-1)^{\nu_r}&= \exp(i\pi\eta'/2)\exp\left(-\frac{i}{8\pi}\int_{M_r}\tr_\sp \,F\wedge F\right).\end{align}
Thus the factors weighting a given solution in the dual constructions of fig. \ref{ospodd} are respectively
\begin{equation}\label{tifo} (-1)^f(-q)^{\Nb^\vee}(-1)^{\nu_\ell} \end{equation}
and
\begin{equation}\label{nifo}(-1)^f q^{\Nb^\vee}(-1)^{\nu_r}. \end{equation}
The most convenient way to compare these two formulas is as follows.  Let $\nu$ be the index of the Dirac operator on the whole
four-manifold $M=M_\ell\cup M_r$.  Additivity of the index under gluing gives
\begin{equation}\label{wifo}\nu=\nu_\ell+\nu_r. \end{equation}
But we also have
\begin{equation}\label{zifo} \nu=\Nb^\vee. \end{equation}
To obtain this formula, one can first deform the gauge field into an $\Sp(2s)$ subgroup, where $s=\mathrm{min}(n,m)$,
so as not to have to consider the jump from $n$ to $m$ (which is not present in standard formulations of index problems).
Then (\ref{zifo}) is a consequence of the ordinary Atiyah-Singer index theorem, or of the APS theorem on the noncompact
four-manifold $M=W\times\RR$ (with the contributions of the ends at infinity canceling).  
It follows from these statements that
\begin{equation}\label{wonno} (-1)^{\Nb^\vee} (-1)^{\nu_r}=(-1)^{\nu_\ell},\end{equation}
showing that the two descriptions do give the same result.

We now proceed to describe the action of the duality on line and surface operators of the orthosymplectic theory.

\subsection{Duality Transformation Of Orthosymplectic Line And Surface Operators}\label{operators}
\subsubsection{Magnetic Duals Of Twisted Line Operators}\label{mtws}
Before we can describe the action of the duality on  line operators, we need  some preparation. In section \ref{twisted}, we have introduced the twisted line operators in the electric description. One needs to include them in the story to get a consistent picture for the $S^{-1}TS$ duality of line operators in the orthosymplectic theory. For this reason, here we make a digression to describe their magnetic duals.

This question arises already for U$(m|n)$, so we start there. Consider a knot $K$ in a three-manifold $W$. $W$ is embedded in 
a four-manifold $M$, for example $W\times \RR$. The definition of twisted line operators on the electric side depended on the existence of a flat line bundle with some twist $c$ around the knot $K$. For a generic twist, such a bundle can only exist if the cycle $K$ is trivial in $H_1(M)$.  In addition to the twist, the line operator also supports a Wilson operator of the bosonic
subgroup with some weight $\Lambda$.
 In the magnetic theory, we propose the following definition for the dual of a twisted operator. Let $\uplambda=\Lambda+\rho_{\bar{0}}$ be the quantum-corrected weight. Note that here we use the bosonic Weyl vector for the quantum correction, since $\Lambda$ was the highest weight of a representation of the bosonic subgroup. For a twisted operator of quantum-corrected weight $\uplambda$, we define the dual magnetic operator, using the irreducible model solution of section \ref{magline}, corresponding to the weight $\uplambda$, but also make the following modification. For definiteness, let $n\ge m$. Then the U$(m)$-part of the gauge field is continuous across the three-dimensional defect.
Pick a surface $\Sigma$ bounded by $K$, or, more precisely, a class\footnote{Since $K$ is trivial in the homology, $\Sigma$ exists, but it might not be unique. If it is not unique, we should probably sum over possible choices. For simple manifolds like $\RR^4$ and $\RR\times S^3$ that we mostly consider in this paper, this question does not arise.} in the relative homology H$_2(M,K)$. The U$(m)$ bundle is trivialized along the knot $K$, so it makes sense to evaluate its first Chern class on the class $\Sigma$, and to include a factor
\beq
\exp\left(ic\int_\Sigma \tr\, F/2\pi\right)\,\label{magtwist}
\eeq
in the functional integral. Here $c$ is an angular variable, which we conjecture to  equal the twist of the line operator on the electric side.\footnote{Note that one cannot define such twisted operators in the one-sided, purely bosonic theory, because there the gauge bundle is trivialized completely along $y=0$, and not only along the knot.} This proposal can be justified by noting that the insertion (\ref{magtwist}) is essentially an abelian surface operator of type $(0,0,0,\eta^\vee)$, with $\eta^\vee$ valued in the center of the Lie algebra of the magnetic gauge group. After doing the $S$-duality transformation, this becomes an operator of type $(\alpha,0,0,0)$ in the electric theory. The singularity $\alpha\rm{d}\theta$ in the abelian gauge field can be removed by making a gauge transformation around this surface operator. Such a gauge transformation closes only up to the element $\exp(ic)$ of the center, and therefore introduces a twist by $\exp(ic)$ to the boundary hypermultiplets.

Now let us turn to the orthosymplectic Chern-Simons theory. For the OSp$(2m|2n)$ case the magnetic gauge group is SO$(2m)\times\rm{SO}(2n+1)$, and its subgroup which is not broken by the three-dimensional defect is SO$(N)$, where $N=2m$ or $N=2n+1$, depending on $m,n$. As is clear from the electric description of section \ref{twisted}, for the twisted operator to have a non-zero matrix element, the knot $K$ should be trivial in $H_1(M;\ZZ_2)$, that is, we should have $K=\partial\Sigma+2K'$, where $\Sigma$ is a two-cycle in $H_2(M,K)$, and $K'$ is an integral cycle. In the magnetic description we define a twisted operator of quantum-corrected weight $\uplambda=\Lambda+\rho_{\bar{0}}$ by the same irreducible model solution that we would use for an untwisted operator, but we also make an insertion in the functional integral. Namely, when we sum over different bundles, we add an extra minus sign if the SO$(N)$-bundle, restricted to $\Sigma$, cannot be lifted to a Spin$(N)$-bundle. In other words, we add a factor
\beq
(-1)^{\int_{\Sigma}w_2}\,,
\eeq
where $w_2$ is the second Stiefel-Whitney class.\footnote{What we have described about the $S$-duality of twisted line operators is rather similar to the result of \cite{VafaWitten}: choosing a topological type of bundle on one side of the duality translates on the other side to choosing a fugacity in the sum over bundles.}

There is no analog of this for an odd orthosymplectic group $\OSp(2m+1|2n)$.  For example, for $m=n$, the magnetic dual is simply an $\Sp(2n)$ gauge
theory with a fundamental hypermultiplet along the defect.  The existence of this hypermultiplet means that the gauge bundle restricted to $\Sigma$
must be an $\Sp(2n)$ bundle, not a bundle with structure group $\mathrm{PSp}(2n)=\Sp(2n)/\Z_2$.  For $m\not=n$, the model solution has a Nahm pole
valued in $\Sp(|2m-2n|)$, and this is incompatible with a twist defined using the center of $\Sp(2n)$.  The magnetic duals of twisted and untwisted
operators are nonetheless different, but that is because the model solutions used to define them are different, as explained in section
\ref{dunsp}.

\subsubsection{More On The Orthosymplectic Lie Superalgebras}\label{osptheory}
We also need to review some facts about the orthosymplectic Lie superalgebras. We start with the even orthosymplectic superalgebra D$(m,n)\simeq\osp(2m|2n)$. Here we assume that $m>1$, since $m=1$ corresponds to the type I superalgebra C$(n)\simeq\osp(2|2n)$ (the analysis of its line and surface operators is analogous to the $\u(m|n)$ case, which we have discussed in
section \ref{another}).
 We also assume that $n>1$; the case $n=1$ can be treated with minor modifications.

\begin{figure}
 \begin{center}
   \includegraphics[width=150mm]{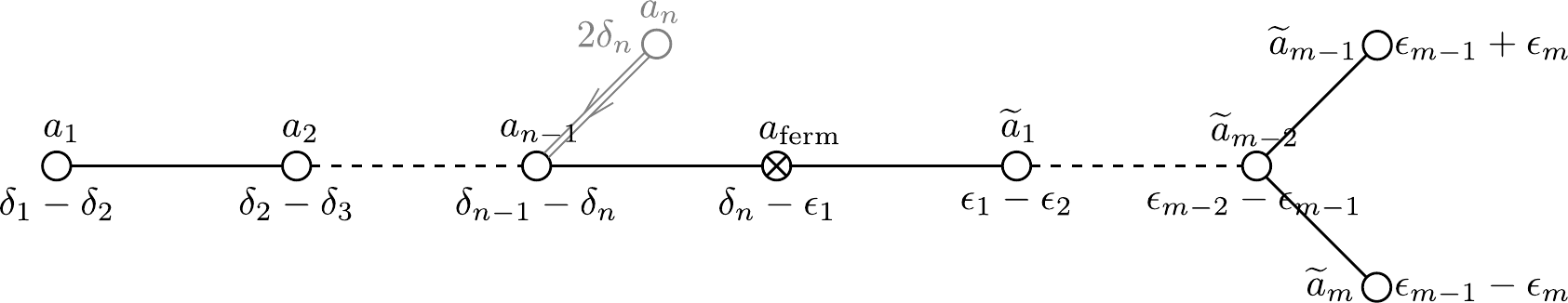}
 \end{center}
\caption{\small Dynkin diagram for the $\mathfrak{osp}(2m|2n)$ superalgebra, $m\ge2$. The subscripts are expressions for the roots in terms of the orthogonal basis $\delta_\bullet$, $\eps_\bullet$. The superscripts represent the Dynkin labels of a weight. The middle root denoted by a cross is fermionic. Roots of the $\mathfrak{sp}(2n)$ and $\mathfrak{so}(2m)$ subalgebras are on the left and on the right of the fermionic root. The site shown in grey and labeled
 $a_n$ is the long simple root of the $\mathfrak{sp}(2n)$ subalgebra, which does not belong to the set of simple roots of the superalgebra.}
 \label{dynkinospeven}
\end{figure}

The Dynkin diagram for D$(m,n)$ is shown on fig. \ref{dynkinospeven}. The positive bosonic and fermionic roots of $\osp(2m|2n)$  are
\beqn
&&\Delta_\bos^+=\bigl\{\delta_i\pm\delta_{i+p},\,2\delta_i,\,\eps_j\pm\eps_{j+p}\bigr\}\,,\nnr
&&\Delta_\ferm^+=\bigl\{\delta_i\pm\eps_j\}\,,
\quad i=1\dots n, \,j=1\dots m,\, p>0\,,\label{ospevenroots}
\eeqn
where the mutually orthogonal basis vectors are normalized as
\beq
\langle \delta_i,\delta_i\rangle=\fr{1}{2}\,,\qquad \langle\eps_i,\eps_i\rangle=-\fr{1}{2},\label{factor2}
\eeq
to ensure that the longest root has length squared 2.
The bosonic and fermionic Weyl vectors are
\beq
\rho_{\bar{0}}=\sum_{i=1}^n\left(n+1-i\right)\delta_i+\sum_{j=1}^m\left(m-j\right)\eps_j\,,\quad \rho_{\bar{1}}=m\sum_{i=1}^n\delta_i\,,\label{evenWeyl}
\eeq
and the superalgebra Weyl vector is $\rho=\rho_{\bar{0}}-\rho_{\bar{1}}$.

 A weight with Dynkin labels\footnote{The Dynkin label of a weight $\Lambda$ for a simple bosonic root $\alpha$ is defined as usual as $a=2\langle\Lambda,\alpha\rangle/
\langle\alpha,\alpha\rangle$. However, the Dynkin labels used in (\ref{evenweight}) are for the simple roots of $\so(2m)\times \sp(2n)$, not for the superalgebra
$\osp(2m|2n)$. In practice, this
means that $a_n$ is the weight for the long root $2\delta_n$ of $\sp(2n)$, and we do not use the label $a_{\mathrm{ferm}}$ associated to the fermionic
root of the superalgebra.} $a_\bullet$, $\tilde{a}_\bullet$ is decomposed in terms of the basis vectors as
\beqn
\Lambda&=&a_1\delta_1+\dots+a_n(\delta_1+\dots+\delta_n)+\tilde{a}_1\eps_1+\dots+\tilde{a}_{m-2}(\eps_1+\dots+\eps_{m-2})\nnr
&+&\fr{1}{2}\tilde{a}_{m-1}(\eps_1+\dots+\eps_{m-1}+\eps_m)+\fr{1}{2}\tilde{a}_{m}(\eps_1+\dots+\eps_{m-1}-\eps_{m})\,.\label{evenweight}
\eeqn
It is a dominant weight of a finite-dimensional representation, if the Dynkin labels are non-negative integers, and also satisfy the following supplementary condition: if $a_n\le m-2$, then no more than the first $a_n$ of the labels $\tilde{a}_\bullet$ can be non-zero; if $a_n=m-1$, then $\tilde{a}_{m-1}=\tilde{a}_{m-2}$; if $a_n\ge m$, there is no constraint. We will call a weight (and the corresponding representation) spinorial  if the number $\tilde{a}_{m-1}+\tilde{a}_m$ is odd. Clearly, a spinorial dominant weight must have $a_n\ge m$. Also, such a weight is always typical.

Now let us turn to the odd orthosymplectic superalgebra B$(m,n)\simeq\osp(2m+1|2n)$. The distinguished Dynkin diagram and the simple roots for $\osp(2m+1|2n)$ and for its bosonic subalgebra $\so(2m+1)\times \sp(2n)$ can be found in fig. \ref{ospDynkin} of section \ref{superrev}. The positive bosonic and fermionic roots of this superalgebra are
\beqn
&&\Delta_\bos^+=\bigl\{\delta_i-\delta_{i+p},\,\delta_i+\delta_{i+p},\,2\delta_i,\,\eps_j-\eps_{j+p},\,\eps_j+\eps_{j+p},\,\eps_j\bigr\}\,,\nnr
&&\Delta_\ferm^+=\bigl\{\delta_i-\eps_j,\,\delta_i+\eps_j,\,\delta_i\bigr\}\,,
\quad i=1\dots n, \,j=1\dots m,\, p>0\,,\label{simpleroots}
\eeqn
where the mutually orthogonal basis vectors are normalized as in (\ref{factor2}).
The bosonic and fermionic Weyl vectors are
\beq
\rho_{\bar{0}}=\sum_{i=1}^n\left(n+1-i\right)\delta_i+\sum_{j=1}^m\left(m+\fr{1}{2}-j\right)\eps_j\,,\quad \rho_{\bar{1}}=\left(m+\fr{1}{2}\right)\sum_{i=1}^n\delta_i\,,\label{rhoodd}
\eeq
and as usual the superalgebra Weyl vector is $\rho=\rho_{\bar{0}}-\rho_{\bar{1}}$.
\begin{figure}
 \begin{center}
   \includegraphics[width=80mm]{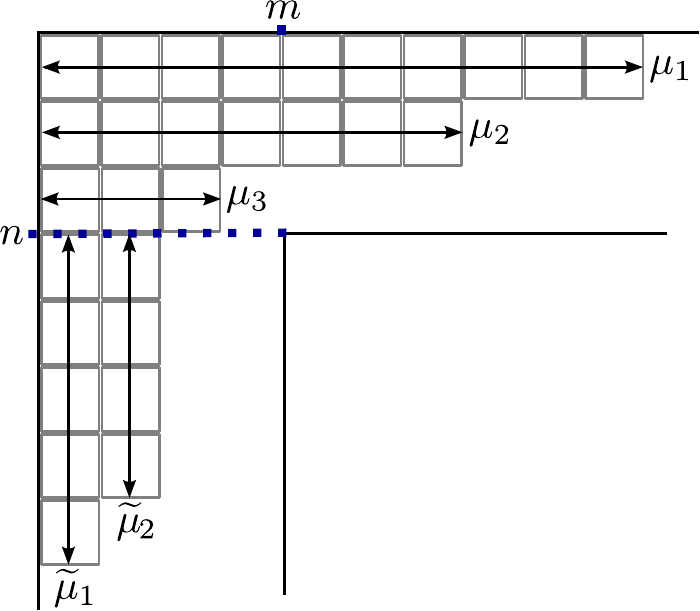}
 \end{center}
\caption{\small Example of a hook partition for $\osp(9|6)$. The labels $\mu_i$, $i=1,\dots n$ and $\tilde{\mu}_j$, $j=1,\dots m$ were defined in (\ref{mus}). Here $\mu_3=3$, and, clearly, no more than the first three $\tilde{\mu}$'s can be non-zero.}
 \label{hook}
\end{figure}

 If we parametrize a weight as 
\beq
\Lambda=\sum_{i=1}^n\mu_i\delta_i+\sum_{i=1}^{m}\tilde{\mu}_i\eps_i\,,\label{rhonext}
\eeq
then, in terms of its Dynkin labels, one has
\beqn\label{pok}
\mu_i=\sum_{j=i}^{n}a_j\,,\quad \tilde{\mu}_i=\sum_{j=i}^{m-1}\ta_j+\fr{1}{2}\ta_m.\label{mus}
\eeqn
A weight $\Lambda$ is a highest weight of a finite-dimensional representation of $\osp(2m+1|2n)$, if its Dynkin labels are non-negative integers, and no more than the first $a_n$ of the $\so(2m+1)$ labels $(\tilde{a}_1,\dots,\tilde{a}_m)$ are non-zero. The last condition is trivial if $a_n\ge m$. We will call an irreducible representation ``large'' if $a_n\ge m$, and ``small'' in the opposite case. An irreducible representation is  spinorial if the Dynkin label $\ta_m$ is odd, and non-spinorial in the opposite case. Clearly, any spinorial representation is ``large.'' It is also easy to see that all the ``small'' representations are atypical, and all the spinorial representations are typical.

Non-spin highest weights can be conveniently encoded in terms of hook partitions \cite{Bars,Farmer,King}. These are simply Young diagrams which are constrained to fit inside a hook with sides of width $n$ and $m$, as shown in fig.~\ref{hook} for $n=3$ and $m=4$. The figure shows how the labels $\mu_\bullet$ and $\tilde{\mu}_\bullet$ parametrizing the weight are read from the diagram. This presentation implements automatically the constraint that only the first  $a_n$ of the $\so(2m+1)$ Dynkin labels can be non-zero. In this notation, the ``small'' representations are those for which the Young diagram does not fill the upper left $n\times m$ rectangle.

Finally, let us note that for typical representations of any superalgebra there exist simple analogs of the Weyl formula to compute characters and supercharacters. For the character of a representation with highest weight $\Lambda$, the formula reads
\beq
{\rm ch}\left(R_\Lambda\right)=L^{-1}\sum_{w\in \mathcal{W}}(-1)^{\ell(w)}\label{WeylChF} \exp\left(w(\Lambda+\rho)\right).
\eeq
Here the sum goes over the elements of the Weyl group $\mathcal{W}$, which, by definition, is generated by reflections along the bosonic roots. The number $\ell(w)$ is the length of the reduced expression for the Weyl group element $w$. The Weyl denominator $L$ is
\beq
L=\frac{\prod_{\alpha\in \Delta^+_{\bar{0}}}(\ex^{\alpha/2}-\ex^{-\alpha/2})}{\prod_{\alpha\in \Delta^+_{\bar{1}}}(\ex^{\alpha/2}+\ex^{-\alpha/2})}.
\eeq

\subsubsection{$\OSp(2m|2n)$: The Mapping Of Line Operators}
To understand the action of the $S^{-1}TS$ duality on the line operators of the D$(m,n)$ Chern-Simons theory, we need to understand the action of the $T$-transformation on their magnetic duals. Since $T$ is just a shift of the theta-angle, it does not change the model solution that is used to define the operator. Therefore one might conclude, as we did for the unitary superalgebra, that line operators are invariant under this transformation. As we now explain, this is indeed true for a subclass of line operators, but not for all of them. 

In section \ref{review} we have defined the instanton number $\mathfrak{N}^\vee$ for the orthogonal group. The action contained a term $i\theta_s\mathfrak{N}^\vee/2$, where $\theta_s$ is the string theory theta-angle. The $2\pi$-periodicity of $\theta_s$ relied on the fact that $\mathfrak{N}^\vee$ takes values in $2\ZZ$. While this assertion is true on $\RR^4$ or $\RR\times S^3$, it is not always true on more general manifolds. We now want to show that it is not true even on simple manifolds like $\RR^4$ in  the presence of some   line operators, and therefore such line operators transform 
non-trivially under the $T$-transformation.

Before explaining the details, let us state clearly  the result. Consider a Wilson-'t~Hooft operator (untwisted or twisted) in the electric theory, located along a knot $K$. We claim that in the presence of its $S$-dual, the instanton number $\mathfrak{N}^\vee$ of the magnetic theory takes values in $2\ZZ$, if the quantum-corrected weight $\uplambda$ of the operator is non-spin, and it takes values\footnote{As we have already explained in footnote \ref{noninteger}, a more precise statement is that the instanton number takes values in $2\ZZ+c$ or $\ZZ+c$ for some constant $c$. Here we are interested only in the difference of instanton numbers for different bundles, so we will ignore the constant shift.} in $\ZZ$, if this weight is spin. Therefore, $T$ acts trivially on the non-spinorial line operators, but not on the spinorial ones. We will show that for spinorial weights the transformation $T$ exchanges twisted and untwisted operators of a given quantum-corrected weight $\uplambda$. In terms of the electric theory, we say that the knot invariants that are obtained from an untwisted spinorial operator in the theory with level $\calK_\osp$ are equal to the invariants obtained from a twisted spinorial operator in the theory with level $\calK_\osp'$, where $\calK_\osp'$ is given by (\ref{Koddtransform}). The mapping of non-spinorial line operators (whether untwisted or twisted) between the Chern-Simons theories with levels $\calK_\osp$ and $\calK_\osp'$ is trivial: the weight is unchanged
and twisted or untwisted operators map to themselves.

Now let us prove our assertions about the instanton number. Assume for simplicity that the four-manifold $M$ is 2-connected (that is, $\pi_1(M)=\pi_2(M)=0$). Our goal is to evaluate the instanton number $\mathfrak{N}^\vee$ for an SO$(2m)\times\SO(2n+1)$ bundle on the knot complement $M\setminus K$ with a fixed trivialization along $K$, which is defined by a model solution of weight $\uplambda$. For now let us assume that $m\le n$, so that the SO$(2m)$ subgroup of the gauge group is left unbroken by the three-dimensional defect at $y=0$.  Let $\Sigma'$ be a two-sphere in $M$ that encircles some point of the knot (this means that the linking number of  $\Sigma'$ with $K$ is 1; for instance, $\Sigma'$ can be the sphere $x^0=0$, 
$r'=$const in the language of fig. \ref{geom} of section 3.3.1), and $\Sigma$ be a surface, bounded by the knot. $\Sigma$ represents the non-trivial cycle in the relative homology $H_2(M,K)$.

We will focus on SO$(2m)$ bundles $\mathcal{V}$ on the knot complement, and ignore what happens in the SO$(2(n-m)+1)$-part of the gauge group, which is broken everywhere at $y=0$ by the boundary condition of the 3d defect. The reason we can do so is that all interesting things will come from different extensions of the SO$(2m)$ bundle from the knot neighborhood $K\times \Sigma'$ to the cycle $\Sigma$, while for the SO$(2(n-m)+1)$ subgroup this extension is uniquely fixed by the boundary condition. This is also the reason that there is no non-trivial analog of this story for the one-sided problem  \cite{5knots}.

So far we have not been precise about the global form of the structure group of our bundle $\mathcal{V}\to M$. In the most general case, the structure group is the projective orthogonal group PSO$(2m)$ (the quotient of $\SO(2m)$ by its center $\{\pm 1\}$), and this structure group 
 might or might not lift to SO$(2m)$ or Spin$(2m)$. If it does lift to SO$(2m)$ or Spin$(2m)$, we say that $\mathcal{V}$ has a vector or a spin structure, respectively. To study obstructions to the existence of a vector or a spin structure (and more generally, obstructions related to $\pi_1(G)$ for $G$-bundles), it is enough to look at the restriction of the bundle to the two-skeleton of the manifold.
Let $\Sigma_0$ be a two-manifold with $G$-bundle $\mathcal V\to \Sigma_0$; we assume that $G$ is  a connected group, and that $\Sigma_0$ 
is closed or that $\mathcal V$ is trivialized on its boundary.  Such a $\mathcal V\to \Sigma_0$ is classified topologically by a characteristic
class $x$ valued in $H^2(\Sigma_0,\pi_1(G))$.  Concretely, $x$ is captured by an element of $\pi_1(G)$ that is used
as a gluing function to construct the bundle $\mathcal V\to\Sigma_0$.  Thus, $x$ associates to $\Sigma_0$ an element $\hat x$ of the center of the universal
cover $\hat G$ of $G$.  A bundle $\mathcal V_R$ associated to $\mathcal V$ in a representation $R$ exists if and only if $\hat x$ acts trivially on $R$.

In our application, $\Sigma_0$ is either $\Sigma$ or $\Sigma'$, and $G=\mathrm{PSO}(2m)$.
 We note that the surface $\Sigma$ can be deformed to lie entirely in the region $y>0$, where the gauge group is SO$(2n+1)$. Since SO$(2m)$ and not PSO$(2m)$ is a subgroup of SO$(2n+1)$, the restriction of $\mathcal{V}$ to $\Sigma$ always has  vector structure.   

Let $\uplambda$ be a non-spinorial weight of the gauge group of the electric theory. This means that $\uplambda$ belongs to the character lattice of SO$(2m)\times \Sp(2n)$, and therefore the parameter of the $S$-dual magnetic operator belongs to the cocharacter lattice of the dual group, which is SO$(2m)\times \SO(2n+1)$. Therefore, the model solution for the line operator defines on $\Sigma'$ a bundle with  vector structure. Together with the facts that we explained a few lines above, this means that $\mathcal{V}$ has vector structure, {\it i.e.} it is an SO$(2m)$ bundle. For its instanton number we can use the formula
\beq
\mathfrak{N}^\vee=\int_M w_2\wedge w_2\,\,{\rm mod}\,\,2\,,
\eeq
where $w_2$ is the second Stiefel-Whitney class, or more precisely an arbitrary lift of it to the integral cohomology. (For a derivation of this formula, see {\it e.g.}  \cite{4dIndex}.) On our manifold we can rewrite\footnote{For a quick explanation, think of $w_2$ in this geometry
as a sum $a+b$, where $a$ is possibly non-trivial on $\Sigma$ but trivial on $\Sigma'$, and $b$ is trivial on $\Sigma$ but possibly
non-trivial on $\Sigma'$.  Then $w_2^2=2ab =0$ mod 2, accounting for the factor of 2 in eqn. (\ref{inst}).} this as 
\beq\
\mathfrak{N}^\vee=2\,\left(\int_\Sigma w_2\right)\,\left(\int_{\Sigma'}w_2\right)\,\,{\rm mod}\,\,2\,,\label{inst}
\eeq
which means that whatever $w_2$ is, the instanton number is even. Therefore, a shift of the theta-angle by $2\pi$ in presence of a non-spinorial line operator is still a symmetry, and such operators are mapped trivially under the $T$-transformation.

Now let the weight $\uplambda$ be spinorial. Then it belongs to the character lattice of $\Spin(2m)\times\Sp(2n)$ (and not to its sublattice corresponding to $\SO(2m)\times \Sp(2n)$), and therefore the parameter of the dual magnetic operator belongs to the cocharacter lattice of PSO$(2m)\times\SO(2n+1)$ (and not to the cocharacter lattice of $\SO(2m)\times\SO(2n+1)$). The bundle that is defined on $\Sigma'$ by such a model solution is a PSO$(2m)$ bundle with no vector structure. What we then expect to get is roughly speaking that the factor $\int_{\Sigma'}w_2$ in (\ref{inst}) now becomes $1/2$, which would give us $\mathfrak{N}^\vee=\int_{\Sigma}w_2\,\,{\rm mod}\,\,2$ for the instanton number. Let us prove this in a more rigorous way.

For that we adapt  arguments used in \cite{4dIndex}, where more detail can be found. The topology of two PSO-bundles that coincide on the two-skeleton can differ only by the embedding of some number of bulk instantons. Therefore the instanton numbers of such bundles can only differ by an even integer. To find $\mathfrak{N}^\vee\,\,{\rm mod}\,\,2$, it is enough to study any convenient bundle with a given behavior on $\Sigma$ and $\Sigma'$. Consider first the case of the group PSO$(6)=\SU(4)/\ZZ_4$. Its fundamental group is $\ZZ_4$. Let $x$ be the $\ZZ_4$-valued characteristic class which defines the topology of the restriction of the bundle to the two-skeleton (i.e., to $\Sigma$ and $\Sigma'$).  Let $\mathcal{L}$ be a line bundle with  first Chern class $c_1=x\,\,{\rm mod}\,\,4$. Let $\cO$ be the trivial line bundle, and consider the bundle
\beq
\mathcal{V}_4=\cL^{1/4}\otimes(\cL^{-1}\oplus\cO\oplus\cO\oplus\cO)\,.
\eeq
It does not exist as an SU$(4)$ bundle, unless $x=0$, but its associated adjoint bundle $3\cL\oplus3\cL^{-1}\oplus9\cO$ does exist; this
bundle has structure group PSO(6).  The associated bundle in the vector representation of $\SO(6)$ is the antisymmetric part of $\mathcal{V}_4\otimes\mathcal{V}_4$; it exists precisely when $x=0$ mod 2, since it contains $\mathcal{L}^{1/2}$. Though $\mathcal{V}_4$ might not exist, we can use the standard formulas to compute its Chern number
\beq
\int_M c_2(\mathcal{V}_4)=-\fr{3}{4}\int_\Sigma c_1(\cL)\int_{\Sigma'}c_1(\cL)=\fr{1}{4}\int_\Sigma x\int_{\Sigma'}x\,\,{\rm mod}\,\,1\,.
\eeq
This Chern number is the instanton number normalized to be $\Z$-valued for an $\SU(4)$ bundle, so it is  $\mathfrak{N}^\vee/2$. Note that, since the bundle on $\Sigma'$ has no vector structure, we have $\int_{\Sigma'}x=\pm 1$. On the contrary, on $\Sigma$ there is  vector structure, and we can write $\int_{\Sigma}x=2\int_\Sigma w_2$ mod 4. We finally get
\beq
\mathfrak{N}^\vee=\int_{\Sigma}w_2\,\,{\rm mod}\,\,2\,.
\eeq
Comparing to the definition of the magnetic duals of the twisted operators in section \ref{mtws}, we conclude that the $T$-transformation, besides shifting the theta-angle by $2\pi$, ialso interchanges the twisted and untwisted spinorial line operators. One can easily extend these arguments to the even orthogonal groups other than SO$(6)$. The relevant facts are explained in \cite{4dIndex} in a similar context, and will not be repeated here.

In our discussion, we have assumed that the ranks of the two gauge groups satisfy $m\le n$. One can extend the arguments to the case $n>m$ with some technical modifications. Rather than explaining this, we will now give an alternative argument, which uses the language of surface operators, and does not depend on the rank difference $n-m$. 

\subsubsection{$\OSp(2m|2n)$: The Mapping Of Surface Operators}
Our discussion will be analogous to what we have said about the case of the unitary superalgebra in section \ref{another}. The $S^{-1}TS$ duality transformation acts on the half-BPS surface operators in the following way,
\beq
\label{longosp}\begin{pmatrix}\alpha\cr \eta\end{pmatrix}\xrightarrow{S} \begin{pmatrix}\eta\cr -\alpha\end{pmatrix}\xrightarrow{T}
\begin{pmatrix}\eta\cr \eta^{*{\mathrm so}}-\alpha\end{pmatrix}\xrightarrow{S^{-1}}\begin{pmatrix}\alpha-\eta^{*{\mathrm so}}\cr \eta\end{pmatrix} .
\eeq
Here the $T$-transformation acts in the magnetic description of the theory. Therefore, its definition involves taking the dual of $\eta$ with respect to the canonically-normalized Killing form of the orthogonal Lie group, which is the gauge group in the magnetic description. To emphasize this fact, we have denoted this dual by $\eta^{*{\mathrm so}}$.

Recall that the action in the electric theory was defined using the canonically-normalized Killing form of the superalgebra, whose bosonic part, according to (\ref{ospnorm}), is $\kappa_\osp=\kappa_\sp-2\kappa_\so$, where $\kappa_\so$ and $\kappa_\sp$ are the canonically-normalized Killing forms for the corresponding bosonic Lie algebras. Let us consider the positive-definite form $\kappa_\sp+2\kappa_\so$, and denote the dual with respect to this form by a star. (In fact, this notation has already been defined in footnote \ref{starcircle}.) The equation (\ref{longosp}) in this notation is equivalent to
\beq
\begin{pmatrix}\alpha\cr \eta\end{pmatrix}\xrightarrow{ { S^{-1}TS } }\begin{pmatrix}\alpha-2\eta^{*}\cr \eta\end{pmatrix} .\label{ospsts}
\eeq
For the $\so(2m)$ part of the parameters, the factor of two in this formula simply follows from the analogous factor in front of $\kappa_\so$ in $\kappa_\osp$. For the $\sp(2n)$ part of the parameters, one needs to compare the canonically-normalized Killing forms of $\sp(2n)$ and $\so(2n+1)$ on $\mathfrak{t}^*_\sp\simeq\mathfrak{t}_\so$. The $S$-duality maps the root lattice in $\mathfrak{t}^*_\sp$ to the coroot lattice in $\mathfrak{t}_\so$. Comparing these lattices, one finds that in $\mathfrak{t}^*_\sp\simeq\mathfrak{t}_\so$ the $S$-duality identifies $\delta_i$ with $\eps_i$, in the notations of section \ref{osptheory}. The canonically-normalized forms for $\sp(2n)$ and $\so(2n+1)$ give respectively\footnote{Note that the canonical normalization of the Killing form for $\so(2n+1)$ is different from the superalgebra normalization (\ref{factor2}).} $\langle \delta_i,\delta_j\rangle_{\sp}=\delta_{ij}/2$ and $\langle\eps_i,\eps_j\rangle_\so=\delta_{ij}$, and their ratio gives the factor of two in (\ref{ospsts}).

The equation (\ref{surfweight}), which defines the relation between the weight and the parameters of a surface operator in the electric theory, continues to hold for the orthosymplectic Chern-Simons theory, if one replaces the level $\calK$ in that equation by $\calK_\osp$. Using this, and also the transformation laws (\ref{Koddtransform}) and (\ref{ospsts}), we conclude that the $S^{-1}TS$ duality transforms the weights according to
\beq
\fr{\uplambda'}{\calK_\osp'}=\fr{\uplambda}{\calK_\osp}\,.\label{ospsts1}
\eeq
Again, the procedure of section \ref{gettingpolynomials} for computing knot invariants using surface operators is obviously invariant under this transformation.

Let us compare the surface operator and the line operator approaches in the case that the weight $\uplambda$ is integral. The equation (\ref{ospsts1}) can alternatively be written as
\beq
\uplambda'=\uplambda+2\calK_\osp'\uplambda\,.\label{ospsts2}
\eeq
First let us look at the part $\uplambda_r$ of the weight, which corresponds to the symplectic Lie subalgebra. In the action (\ref{ospnorm}), the level $\calK_\osp$ multiplies the Chern-Simons term for the $\sp(2n)$ subalgebra, which is defined using the canonically-normalized $\sp(2n)$ Killing form. Therefore the knot invariants computed using the surface operators are unchanged when the weight $\uplambda_r$ is shifted by $\calK_\osp$ times an integral coroot of the $\sp(2n)$ subalgebra. If $\uplambda_r$ is an integral weight, then $2\uplambda_r$ is an integral coroot, and therefore the difference between $\uplambda_r'$ and $\uplambda_r$ in (\ref{ospsts2}) is inessential for computing the knot invariants.

For the part $\uplambda_\ell$ of the weight, which corresponds to the orthogonal subalgebra, the situation is more complicated. The canonically-normalized Chern-Simons term for the orthogonal subalgebra in the action (\ref{ospnorm}) is multiplied by $2\calK_\osp$. For this reason, the knot invariants computed using the surface operators are invariant under the shift of $\uplambda_\ell$ by $2\calK_\osp$ times an integral coroot of the $\mathfrak{so}(2m)$ subalgebra. Therefore, the shift of $\uplambda_\ell$ in the equation (\ref{ospsts2}) is trivial from the point of view of the knot observables if and only if the integral weight $\uplambda_\ell$ is actually a coroot. What if it is not? Since the $\so(2m)$ Lie algebra is simply-laced, any integral weight is also an element of the dual root lattice $\Gamma_r^*$. Therefore the group element $\exp(2\pi\uplambda_\ell)$ actually belongs to the center of the orthogonal group. Let us make a singular gauge transformation in the electric theory around the surface operator on the left side of the three-dimensional defect, using the group element $\exp(\theta\uplambda_\ell)$, where $\theta$ is the azimuthal
angle in the plane normal to the surface operator. This transformation maps a surface operator corresponding to the weight $\uplambda_\ell'$ back to a surface operator with weight $\uplambda_\ell$. Since our  gauge transformation is closed only up to the central element $\exp(2\pi\uplambda_\ell)$, it also introduces a twist of the boundary hypermultiplets by this group element. In the fundamental representation of SO$(2m)$, to which the hypermultiplets belong, the element $\exp(2\pi\uplambda_\ell)$ acts trivially if the weight $\uplambda$ is non-spinorial, and it acts by $-1$ if $\uplambda$ is spinorial.
We have reproduced the result that was derived in the previous section in the language of line operators:  $S^{-1}TS$ acts trivially on Chern-Simons line observables labeled by non-spinorial representations, but exchanges the twisted and the untwisted operators for a spinorial weight.

\subsubsection{$\OSp(2m|2n)$: Comparing The Representations}\label{comparing}
We would like to look closer at the mapping of spinorial line operators. 
Consider a line operator, labeled by a supergroup representation of spinorial
highest weight $\Lambda=\uplambda-\rho$, and an $S^{-1}TS$-dual twisted operator, which is labeled by a representation of the bosonic subgroup with highest weight $\Lambda'=\uplambda-\rho_{\bar{0}}$. Note that the Weyl vectors $\rho$ and $\rho_{\bar{0}}$, which can be found from (\ref{evenWeyl}), are non-spinorial integral weights, and therefore the property of being spinorial/non-spinorial is the same for the weights and for the quantum-corrected weights of OSp$(2m|2n)$.

We would like to see more explicitly how the duality mapping acts in terms of  representations. We have $\uplambda=\Lambda+\rho=\Lambda'+\rho_{\bar{0}}$, or equivalently, $\Lambda'=\Lambda-\rho_{\bar{1}}$. Using the formulas (\ref{evenWeyl}) and (\ref{evenweight}), this can be translated into a mapping of Dynkin labels,
\begin{align}
\tilde{a}_j'&=\tilde{a}_j,\quad j=1,\dots,m\,,\nnr
a_i'&=a_i, \quad i=1,\dots,n-1\,,\nnr
a_n'&=a_n-m\,.
\end{align}
As was noted in section \ref{osptheory}, for a spinorial superalgebra representation one has\footnote{As we have mentioned in a similar context in section \ref{various},
we do not really know why the supplementary condition should be imposed in the present discussion, since it is not a general condition on 1/2-BPS line operators.
Nonetheless, imposing this condition works nicely, as we have just seen. 
This shows once again that our understanding of line operators in the theory is incomplete. We will find something similar for odd $\OSp$ supergroups.} $a_n\ge m$. Therefore, the mapping of Dynkin labels written above is a one-to-one correspondence between the irreducible spinorial representations of the $D(m,n)$ superalgebra and its bosonic subalgebra.

We can make an additional test of the duality by comparing the local framing anomalies of the line operators. Recall that the knot polynomials in Chern-Simons theory are invariants of framed knots. If the framing
of a knot is shifted by one unit via a $2\pi$ twist, the knot polynomial is multiplied by a factor
\beq\label{hello}
\exp(2\pi i \Delta_O)\,,
\eeq
where $\Delta_O$ is the dimension of the conformal primary $O$ that corresponds in the WZW model\footnote{As we explain in Appendix
\ref{currel}, there actually is not a straightforward relation between 3d Chern-Simons theory and 2d current algebra in the case of a supergroup.
Nonetheless, some results work nicely and the one we are stating here seems to be one. } 
to the given Wilson line. For a Wilson line in representation $R$,
this framing factor is
\beq
\exp\left(i\pi \,\frac{c_2(R)}{k+h\,\,\sign(k)}\right)=q^{c_2(R)}\,,\label{ospfram}
\eeq
where $c_2(R)=\langle\uplambda,\uplambda\rangle-\langle\rho,\rho\rangle$ is the value of the quadratic Casimir in the representation $R$. The variable $q$ was defined for the D$(m,n)$ superalgebra in (\ref{qeven}). In the bosonic, one-sided case these formulas have been derived in \cite{5knots} from the magnetic description of the theory. It would be desirable to give such a derivation for the two-sided case, but we will not attempt to do it here.

To compare the framing factors for our dual operators, we need to derive a formula for the framing anomaly of a twisted operator. The energy-momentum tensor of the conformal field theory is given by the Sugawara construction
\beq
T(z)=\fr{\hat{\kappa}_{nm} :\hspace{-1mm}J^m(z)J^n(z)\colon}{2(k+h)}\,,\label{Sugawara}
\eeq
where $\hat{\kappa}=\kappa\oplus\omega$ is the superinvariant bilinear form\footnote{Here we slightly depart from our usual notation, and use indices $m,n,\dots$ both for bosonic and fermionic generators of the superalgebra. Also, note that the inverse tensor is defined by $\hat{\kappa}^{mn}\hat{\kappa}_{pn}=\delta^m_p$, hence the unusual order of indices in the Sugawara formula.} on the superalgebra, and $J^m(z)$ is the holomorphic current with the usual OPE
\beq
J^m(z)J^n(w)\sim \fr{k\,\hat{\kappa}^{mn}}{(z-w)^2}+\fr{f^{mn}_p J^p(w)}{z-w}.\label{currentalg}
\eeq
One can easily verify that for a simple superalgebra the formula (\ref{Sugawara}) gives the energy-momentum tensor with a correct OPE.

Normally, the current $J^m(z)$ is expanded in integer modes. The eigenvalue of the Virasoro generator $L_0$, acting on a primary field, is determined by the action of the zero-modes of the current, which give the quadratic Casimir, as stated in eqn. (\ref{ospfram}). However, for a primary field corresponding to a twisted operator in Chern-Simons, one naturally expects the fermionic components of the current $J^m(z)$ to be antiperiodic. In that case, the bosonic part of the current gives the usual contribution to the conformal dimension, which for a weight $\Lambda$ is proportional to the bosonic quadratic Casimir $\langle \Lambda+2\rho_\bos,\Lambda\rangle$. The fermionic part of the current in the twisted sector has no zero-modes, and its contribution to the $L_0$ eigenvalue is just a normal-ordering constant, independent of the weight $\Lambda$. One can evaluate this constant from (\ref{Sugawara}), (\ref{currentalg}), and get for the dimension of the operator 
\beq
\Delta_O^{tw}=\fr{\langle\Lambda+2\rho_\bos,\Lambda\rangle-k\,{\rm dim}(\mathfrak{g}_{\ferm})/8}{2(k+h)}.
\eeq
Using the identity $\langle \rho_\bos,\rho_\bos\rangle=\langle\rho,\rho\rangle+h\,\,{\rm dim}(\mathfrak{g}_\ferm)/8$, which actually is valid for any of our superalgebras, one obtains an expression for the framing factor 
\beq
\exp\left(i\pi\,\frac{\langle \uplambda,\uplambda\rangle -\langle\rho,\rho\rangle}{k+h\,\,\sign(k)}\right)\exp\left(-i\pi{\rm dim}(\mathfrak{g}_\ferm)\sign(k)/8\right).\label{ospframtw}
\eeq
Here we have restored the dependence on the sign of the level $k$, and used our definition of $\uplambda$ for the twisted operators. The second factor in this formula does not map correctly under the duality, but that is what one could have expected, since this factor is non-analytic in $\calK=k+h\,\sign(k)$ (compare to the discussion of the global framing anomalies in sections \ref{franom} and \ref{ospfranom}). The first factor is analytic in $\calK$, and it is clear from comparison to eq. (\ref{ospfram}) that it does map correctly under the duality.

\subsubsection{Duality For The Odd Orthosymplectic Superalgebra}\label{dunsp}
Let us turn to the case of the odd orthosymplectic superalgebra. As was already noted in section \ref{linerev}, the definition of line operators in this theory has some peculiarities. As follows from the equation (\ref{rhoodd}), for B$(m,n)$ the bosonic Weyl vector $\rho_{\bar{0}}$ is an integral spinorial weight, while the superalgebra Weyl vector $\rho$ is not an integral weight: it has a half-integral Dynkin label with respect to the short coroot of the $\sp(2n)$ subalgebra. This means that the quantum-corrected weight $\uplambda=\Lambda+\rho$ for an untwisted operator is not an integral weight, and therefore a Wilson-'t~Hooft operator, as defined in section (\ref{defmonodr}), is not gauge-invariant classically. The resolution of this puzzle should come from another peculiarity of the B$(m,n)$ Chern-Simons theory. The definition of the path-integral of this theory includes an $\eta$-invariant (\ref{orez}), which comes from the one-loop determinant (or rather the Pfaffian) of the hypermultiplet fermions. In the presence of a monodromy operator, one should carefully define this fermionic determinant, and we expect an anomaly that will cancel the problem that exists at the classical level. We will not attempt to explain the details of this in the present paper.

Unlike the case $\OSp(2m|2n)$, a magnetic line operator of  $\OSp(2m+1|2n)$ is completely determined\footnote{Here we ignore the issues related to the atypical representations. We will say a little more on this later in this section.} 
 by its weight $\uplambda$, as explained at the end of section
\ref{mtws}.  However, the quantum-corrected weights for  twisted and   untwisted operators belong to different lattices, due to the different properties of $\rho$ and $\rho_{\bar{0}}$, mentioned above.  So the magnetic duals of twisted and untwisted electric line operators are simply described by
different model solutions.
 Since the $T$-transformation preserves the model solution, the $S^{-1}TS$ duality should preserve the quantum-corrected weight.
 
We need to introduce some further notation. In the orientifold construction, we took the Killing form to be positive on the $\sp$ part of B$(m,n)$. In the dual theory, it will be positive on the $\so$ part, and for this reason we denote the superalgebra of the dual theory by B$'(n,m)$. The basis vectors in the dual ${\mathfrak{t}^*}'$ of the Cartan subalgebra  of B$'(n,m)$ will be denoted by $\delta_j'$, $j=1,\dots,m$, and $\eps_i'$, $i=1,\dots,n$, and their scalar products are defined to have opposite sign relative to (\ref{factor2}). The Dynkin labels for the representations of B$'(n,m)$ will be denoted as $a_j'$, $j=1,\dots,m$, and $\tilde{a}_i'$, $i=1,\dots,n$. To make precise sense of the statement that the $S^{-1}TS$ duality preserves the quantum-corrected weight, it is necessary to specify how one identifies $\mathfrak{t}^*$ and ${\mathfrak{t}^*}'$. We use the mapping which identifies $\eps_i'$ with $\delta_i$ and $\delta_j'$ with $\eps_j$. This linear map preserves the scalar product. In principle, one could derive this identification from the $S$-duality transformations of surface operators, but we will simply take it as a conjecture and show that it passes some non-trivial tests.

We can make one such test before we go into the details of the operator mapping. According to the equations (\ref{ospfram}), (\ref{ospframtw}) and the definition (\ref{qodd}) of the variable $q$, the framing anomaly factor in the B$(m,n)$ theory for an operator of quantum-corrected weight $\uplambda$ is equal to $q^{2c_2}$, where $c_2=\langle \uplambda,\uplambda\rangle-\langle\rho,\rho\rangle$. (This formula is true for both  twisted and  untwisted operators, modulo non-analytic terms.) From this we can see that our map does preserve the framing anomaly.\footnote{To be precise, there is actually a little mismatch for the spinorial operators. In that case the quadratic Casimir $c_2$ can be non-integral, and therefore there is a difference by a root of unity due to the fact that $q$ is mapped to $-q$. Hopefully, this discrepancy can be cured in a more accurate treatment.} Indeed, it preserves $\uplambda$ and the scalar product, and although the Weyl vectors $\rho$ and $\rho'$ for the two dual superalgebras B$(m,n)$ and B$'(n,m)$ are different, their lengths happen to coincide, as one can verify from the explicit formula (\ref{rhoodd}). 

In the rest of this section we will examine the mapping 
\beq
\uplambda=\uplambda'\label{lambdadual}
\eeq 
in more detail. We will see that it gives a correspondence between the untwisted non-spinorial operators of the two theories, maps the twisted non-spinorial operators to the untwisted spinorial operators, and finally indentifies the twisted spinorial operators of one theory with the twisted spinorial operators of the other one. To put it shortly, it exchanges the spin and the twist. It is important to note that one might need to refine the mapping (\ref{lambdadual}) for  atypical weights. We will indeed encounter an ambiguity in interpreting (\ref{lambdadual}) for the ``small'' atypical weights.

First let us focus on the non-spinorial untwisted line operators, for which the duality should give a correspondence between the non-spinorial representations of the two superalgebras. The map (\ref{lambdadual}) of the dominant weights is  already known in the literature for the special case of $m=0$. In fact, a remarkable correspondence between  finite-dimensional representations of 
$\osp(1|2n)$ and non-spinorial finite-dimensional representations of $\so(2n+1)$ was established in \cite{Rittenberg}. It preserves the full set of Casimirs, 
including the quadratic one.  
For $n=1$, the map is so elementary that one can describe it by hand.  This will make our later discussion more concrete. The spin $s$ representation of $\so(3)$, for non-negative integer $s$, is mapped to the trivial representation of $\osp(1|2)$ for $s=0$, and otherwise to the
representation of $\osp(1|2)$ that is a direct sum of bosonic states of spin $s/2$ (under $\sp(2)\cong\su(2)$) and fermionic states of spin $(s-1)/2$. Note
that if we ignore the statistics of the states, the given $\so(3)$ and $\osp(1|2)$ representations both have dimension $2s+1$.  This
is a special case of a correspondence between characters found in \cite{Rittenberg}. 

An equivalent explanation is that  a representation of $\so(3)$ whose highest weight is $s$ is mapped, if $s$ is an integer,
 to a representation of $\osp(1|2)$ whose highest weight is $s$ times the smallest
strictly positive weight of this algebra.  
The spinorial representations of $\so(3)$ -- the representations with half-integral $s$ -- do not participate in this correspondence,
since there is no representation of $\osp(1|2)$ whose highest weight is a half-integral multiple of the smallest positive weight.  The spinorial representations of $\so(3)$ have a dual in terms of twisted line operators, but not in terms of representations. 

This correspondence between $\so(s)$ and $\osp(1|2)$ maps tensor products of $\so(3)$ representations to tensor
products of $\osp(1|2)$ representations if one ignores whether the highest weight of an $\osp(1|2)$ representation is bosonic or fermionic. To illustrate this correspondence, let $\s$ denote an irreducible $\so(3)$ representation of spin $s$.  Let $\s'$ and $\t\s'$ denote irreducible
$\osp(1|2)$ representations whose highest weight is $s$ times
the smallest positive weight, with the highest weight vector being bosonic or fermionic, respectively.   Then one has, for example,  
\begin{equation}\label{zerb}
\begin{cases}
\1\otimes\1\cong \2\oplus \1\oplus \0 &  \mbox{for} ~\so(3)\cr \1'\otimes\1'\cong \2'\oplus \t\1'\oplus \0' &  \mbox{for}~ \osp(1|2).\end{cases}\end{equation}  There is an obvious matching,
if we ignore the reversed statistics of $\t\1'$ on the $\osp(1|2)$ side.  We interpret  this matching to reflect the fact that the duality between 
$\so(3)$ and $\osp(1|2)$ preserves the operator production expansion for Wilson line operators. 
(In Chern-Simons theory, for generic $q$ the OPE of line operators is given by the classical tensor product, so we can compare such OPE's
by comparing classical tensor products.)   However, we do not know the interpretation of the reversed statistics of $\t\1'$.  Perhaps
it somehow involves the fact that the quantum duality changes the sign of $q$.
In \cite{Rittenberg}, it is shown that an analogous matching of tensor products holds in general.

Additional relevant results are in  \cite{qOSp}. Let $\mathcal U_q(\osp(1|2n))$ and $\mathcal U_{q'}(\so(2n+1))$ be the quantum deformations of the universal enveloping algebras of the corresponding Lie (super)algebras. It has been shown in \cite{qOSp} that there exists a natural map between the representations of these two quantum groups if one takes $q'=-q$, and restricts to non-spinorial representations of the latter.  One would expect such a result from our duality,  assuming that Chern-Simons theory of a supergroup is related to a corresponding quantum group in the manner
that is familiar in the bosonic world.  (While this is a plausible hope, some attempts to derive such a statement will not work, as we see in Appendix \ref{currel}.)

Now we return to our mapping $\uplambda=\uplambda'$ (eqn. (\ref{lambdadual})),  which extends the known results described above to general $m$ and $n$. It has several nice properties. As follows from our discussion of the framing anomaly, it preserves the quadratic Casimir. From the Harish-Chandra isomorphism, it follows that, for non-spinorial weights, (\ref{lambdadual}) gives a natural mapping not only of the quadratic Casimir, but of the higher Casimirs as well. It would be interesting to find an explanation of this directly from the quantum field theory. The map also preserves the atypicality conditions (\ref{acondition}). Next, let us look at the Weyl character formula (\ref{WeylChF}), assuming that the weights are typical. The Weyl groups for the two superalgebras are equivalent and act in the same way on $\mathfrak{t}^*\simeq {\mathfrak{t}^*}'$; therefore, with the mapping (\ref{lambdadual}), the numerators of the character formula coincide for the dual representations. The denominators are also equal, as one can easily check, using the list of simple roots (\ref{simpleroots}). However, the supercharacters are not mapped in any simple way. In particular, the duality preserves the dimensions of typical representations, but not the superdimensions.\footnote{Of course, for $m,n\ne 0$ the superdimensions of typical representations on both sides of the duality are simply zero. But for $m$ or $n$ equal to 0, they are non-zero and do not agree.}

Let us actually see what the equation (\ref{lambdadual}) says about the map of representations. Writing it as $\Lambda'=\Lambda+\rho-\rho'$ and using equations (\ref{rhoodd}), (\ref{rhonext}), one gets that the labels $\mu_\bullet$ and $\tilde{\mu}_\bullet$, defined in those equations, transform into $\mu_j'=\tilde{\mu}_j+n$, $\tilde{\mu}_i'=\mu_i-m$. According to the equation (\ref{mus}), this gives a mapping for the Dynkin labels,
\begin{align}
\tilde{a}_i'&=a_i,\quad i=1,\dots,n-1\,,\nnr
a_j'&=\tilde{a}_j, \quad j=1,\dots,m-1\,,\nnr
\tilde{a}_n'&=2(a_n-m)\,,\nnr
a_m'&=\fr12\tilde{a}_m+n.\label{weightmap}
\end{align}
If we restrict to ``large'' non-spin dominant weights ($a_n\geq m$), then this formula gives a one-to-one correspondence.
The non-spin condition means that $\tilde a_m$ is even, so that the mapping (\ref{weightmap}) is well-defined, and the ``large'' condition
$a_n\geq m$ ensures that $\t a_n'\geq 0$.

It is not immediately obvious what to say
 for ``small'' representations, since for them the dual Dynkin label $\tilde{a}_n'$ comes out negative. Note that all the ``small'' representations are atypical, and in general we have less control over them by methods of this paper. There can be different possible conjectures as to how to make sense of our map for them. First of all, we can still treat (\ref{lambdadual}) as a correspondence between monodromy operators. Then to understand to which representation a given operator corresponds, we should make a Weyl transformation on $\uplambda'$, to bring it to a positive Weyl chamber. This is one possible way to understand the map (\ref{lambdadual}) for the ``small'' representations.  (For an atypical weight, there can be several different ways to conjugate it to the positive Weyl chamber; these give different weights, though belonging to the same atypical block.)

There is another very elegant possibility.
If we simply transpose the hook diagram for a B$(m,n)$ weight, we will get some weight of B$'(n,m)$. It is a curious observation that for the ``large'' representations, this operation reproduces our duality (\ref{weightmap}). Moreover, one can prove that even for the ``small'' representations this flip preserves the quadratic Casimir operator and therefore the framing anomaly, and can be a candidate for the generalization of our map to the ``small'' highest weights. Unfortunately, this is merely a possible guess.

In short, we have found a natural 1-1 mapping between non-spinorial representations of $\OSp(2m+1|2n)$ and $\OSp(2n+1|2m)$.  Now let us turn to spinorial ones. The mapping (\ref{lambdadual}) sends spinorial line operators to twisted operators. Here is a simple consistency check of this statement.  In the electric theory, consider a Wilson-'t~Hooft operator in a spinorial representation $R$ that is supported on a knot $K$ in a three-manifold $W$. If the class of $K$ in $H_1(W;\Z_2)$ is nonzero, then the expectation of the operator vanishes because it is odd under a certain ``large'' gauge transformation that
is single-valued in $\SO(2m+1)$ but not if lifted to $\Spin(2m+1)$. (The gauge transformations along a Wilson-'t~Hooft operator are constrained to lie in the maximal torus, but there is no problem in choosing such an abelian ``large'' gauge transformation.) The dual of such a Wilson-'t~Hooft operator under the $S^{-1}TS$ duality
should have the same property. Indeed, a twisted operator, as described in section \ref{twisted}, does have this property (in this case because the definition of the twisted operator involves picking a $\Z_2$ bundle with monodromy around $K$).

Let $\Lambda$ be a spinorial dominant weight of the B$(m,n)$ superalgebra, and let $\Lambda'$ be a non-spinorial weight of the bosonic algebra $\so(2n+1)\times
\sp(2m)$ that we use in defining a twisted line operator. The
mapping (\ref{lambdadual}) would then be  $\Lambda'+\rho_\bos'=\Lambda+\rho$.  The bosonic Weyl vector that is used here can be obtained from (\ref{rhoodd}) by exchanging  $\eps_\bullet$ with $\delta_\bullet$ and $m$ with $n$. 
From this one finds  that the coefficients in the expansion of the weights in the $\delta_\bullet$, $\eps_\bullet$ basis transform as $\tilde{\mu}_i'=\mu_i-m$, $\mu_j'=\tilde{\mu}_j-1/2$. Therefore, according to (\ref{mus}), the Dynkin labels of the weights are related as
\begin{align}
\tilde{a}_i'&=a_i,\quad i=1,\dots,n-1\,,\nnr
a_j'&=\tilde{a}_j, \quad j=1,\dots,m-1\,,\nnr
\tilde{a}_n'&=2(a_n-m)\,,\nnr
a_m'&=\fr12(\tilde{a}_m-1).\label{weightmap1}
\end{align}
This gives a one-to-one correspondence between the spinorial supergroup representations and the non-spinorial weights of the bosonic algebra $\so(2n+1)\times\sp(2m)$. In fact, for a spinorial representation of $\osp(2m+1|2n)$, $\tilde a_m$ is odd, ensuring that $a'_m$ is an integer.  On the other hand, $\t a_n'$ is always
even, so the twisted line operator with Dynkin labels $a'_i, \t a'_j$ is always associated to a non-spinorial representation of the bosonic subalgebra of $\OSp(2n+1|2m)$. Moreover, the supplementary condition guarantees that $a_n-m$ is non-negative for a spinorial superalgebra representation.

The twisted operators for spinorial representations of the bosonic subgroup should be mapped into similar twisted spinorial operators. The mapping (\ref{lambdadual}) reduces in this case to $\Lambda+\rho_\bos=\Lambda'+\rho_\bos'$. This gives $\tilde{\mu}_i'=\mu_i+1/2$, $\mu_j'=\tilde{\mu}_j-1/2$, or, in terms of the Dynkin labels, 
\begin{align}
\tilde{a}_i'&=a_i,\quad i=1,\dots,n-1\,,\nnr
a_j'&=\tilde{a}_j, \quad j=1,\dots,m-1\,,\nnr
\tilde{a}_n'&=2a_n+1\,,\nnr
a_m'&=\fr12(\tilde{a}_m-1)\,,\label{weightmap2}
\end{align}
which is indeed a one-to-one correspondence between the spinorial representations of the bosonic subgroups.  In other words,
the weights $a'_i$ and $ \t a'_j$ are integers if the $a_i $ and $\t a_j$ are integers and  $\t a_m$ is odd, and moreover in that case $\t a'_n$ is odd.

\section{Symmetry Breaking}\label{symbr}
\subsection{Detaching D3-Branes}\label{detachment}
The bosonic fields of $\N=4$ super Yang-Mills theory are the gauge field $A_i$ and scalars $\vec X$ and $\vec Y$.  From the point of view
of the brane construction that was reviewed in the introduction, $\vec X$ describes motion of D3-branes along a fivebrane, while $\vec Y$ describes
motion of the D3-branes normal to the fivebrane. In the present
section, let us assume that the fivebrane is an NS5-brane, leading to a Chern-Simons construction in which
$A$ and $\vec X$ are combined to a complex gauge field $\AA$.  

In the one-sided problem, related to Chern-Simons theory of a purely bosonic gauge group,
it is possible to break some of the gauge symmetry by giving an expectation value to $\vec X$.  (What one learns this way is summarized
at the end of section \ref{symbrane}.) But in the two-sided
case, there is an additional possibility of gauge-symmetry breaking by giving an expectation value\footnote{\label{whyonly} In discussing
symmetry breaking, we want to assume that the three-manifold $W$ is such that attempting to specify the vacuum expectation value
of a massless scalar field on $W\times \RR$ or $W\times \RR_+$ does not lead to infrared divergences.  This condition actually forces
us to assume that $W$ has at least 2 noncompact dimensions.  In practice, we will consider only the case $W=\RR^3$.}  to $\vec Y$.
Let us explore this case.

As usual, we  write $\vec Y_\ell$ and $\vec Y_r$ for scalar fields on D3-branes to the left or right of the fivebrane, respectively.
If some D3-branes are to be displaced normal to the fivebrane -- so that they no longer meet it -- then for those D3-branes, 
 $\vec Y_\ell$ and $\vec Y_r$ must combine to a field $\vec Y$ that is continuous at $x_3=0$.
  Concretely, when we give expectation values to $\vec Y_{\ell}$ and $\vec Y_r$, we also
have to give an expectation value to the scalar fields $Q_{\dot A}^I$ in the defect hypermultiplet, because of the boundary condition of eqn. (\ref{Ybc1}).  
This boundary condition forces the nonzero eigenvalues of $\vec Y_\ell$  and $\vec Y_r$ to be equal at $x_3=0$.  (For the component $\sigma$
of $\vec Y$ defined in eqn. (\ref{tork}), this is  demonstrated  in eqn. (\ref{mostr}).)   Thus, this symmetry breaking mechanism consists of giving 
expectation values to $\vec Y_\ell$, $\vec Y_r$, and $Q_{\dot A}^I$ in such a way that $r$ semi-infinite branes at $x_3<0$ 
(for some integer $r>0$) and $r$  semi-infinite branes at $x_3>0$
combine together into $r$ D3-branes whose support spans the whole range $-\infty<x_3<\infty$, and which are located at $r$ nonzero values\footnote{As usual,
$\N=4$ super Yang-Mills theory contains a potential energy term $-\sum_{\dot a,\dot b}\Tr\,[Y_{\dot a},Y_{\dot b}]^2$.  To minimize the energy, the components
of $\vec Y$ must commute, and the common eigenvalues of $\vec Y$ are the positions of the D3-branes.  The fact that the components of $\vec Y$
have to commute is actually an ingredient in deducing from (\ref{mostr}) that $\vec Y$ is continuous at $x_3=0$ up to a unitary gauge transformation.}  of $\vec Y$.
The values of $\vec Y$ for these D3-branes can be varied independently, preserving the full supersymmetry of the system and in particular the $\Qb$
symmetry that is used in the relation to supergroup  Chern-Simons theory.

What is special about this method of symmetry breaking is that none of the fields $\vec Y_\ell$, $\vec Y_r$, $Q_{\dot A}^I$ that receive
expectation values appear in the Chern-Simons action.    We recall that the basis for the relation between $\N=4$ super Yang-Mills theory and Chern-Simons
theory is that the action $I$ of the topologically twisted $\N=4$ theory differs from the Chern-Simons action $\CS(\AA)$ by a $\Qb$-exact term:
\begin{equation}\label{copyact} I=i\calK \CS(\AA)+\{\Qb,\dots\}.  \end{equation}
In the method of symmetry breaking that we have described, the fields that receive expectation values appear only inside $\{\Qb,\dots\}$.
Let us discuss in very general terms the implications of this fact.

We consider an arbitrary theory with fields $\Phi$ and a classical action $I(\Phi)$ on a non-compact spacetime $M$.  A standard problem is
to perform the Feynman path integral with $\Phi$ required to vanish at infinity.  Now, let us consider symmetry breaking.  Symmetry breaking
means that we pick a classical solution $\Phi_0$ that does not vanish at infinity, write $\Phi=\Phi_0+\Phi'$, and perform the path integral over fields
$\Phi'$ that are required to vanish at infinity.  The action in the path integral over $\Phi'$ is
\begin{equation}\label{opyact}\h I(\Phi')=I(\Phi_0+\Phi'). \end{equation}
In general, the function $\h I$ does not coincide with the original action function $I$, and that is why typically symmetry breaking affects the physics.

Now consider a theory with a fermionic symmetry $\Qb$, obeying $\Qb^2=0$, and suppose that we are only interested in $\Qb$-invariant observables.
Suppose furthermore that the action function has a decomposition $I=I_0+\{\Qb,\dots\}$ (where in our case $I_0$ is the Chern-Simons action), and that
the classical solution $\Phi_0$ is $\Qb$-invariant and is such that  $I_0(\Phi_0+\Phi')=I_0(\Phi')$.  Then
\begin{equation}\label{pycat}\h I(\Phi')=I(\Phi')+\{\Qb,\dots\}. \end{equation}
The $\Qb$ invariance ensures that the $\Qb$-exact terms decouple,\footnote{The assumption that $\Phi_0$ is $\Qb$-invariant is necessary here;
otherwise, the
proof of the decoupling of $\Qb$-exact terms will fail because of a Goldstone fermion contribution.}  and therefore symmetry breaking under these
assumptions does not affect $\Qb$-invariant observables.  

Symmetry breaking by giving expectation values to $Y_\ell$, $Y_r$, and $Q_{\dot A}^I$ (and no other fields) satisfies these conditions, and therefore
will be an invariance of the supergroup Chern-Simons theory that is derived from the brane construction.  We consider the implications next.  

\subsection{Symmetry Breaking In The Brane Construction}\label{symbrane}

Let us consider the usual brane construction with $m$ D3-branes ending on an NS5-brane on one side
and $n$ on the other.  The gauge symmetry is $\U(m)\times \U(n)$, and along the NS5-brane is a $\U(m|n)$ Chern-Simons theory. 
If we remove a  D3-brane to $\vec Y\not=0$, then this brane supports a $\U(1)$ gauge symmetry which we will call $\U^*(1)$.
The original $\U(m)\times \U(n)$ gauge symmetry is broken 
 to $\U(m-1)\times \U(n-1)\times \U^*(1)$, and the Chern-Simons theory now has the gauge group $\U(m-1|n-1)$.  
 
 In this construction, fields with nonzero $\U^*(1)$ charge acquire masses proportional to $|\langle \vec Y\rangle|$.  The masses
 mean that these fields decouple in computing Wilson loop expectation values.  Indeed, topological invariance of the supergroup Chern-Simons
 theory means that the knots whose expectation values we may wish to compute can always be scaled up to a very large size, and then
 massive fields clearly cannot affect their expectation values.  So the knot invariants of $\U(m|n)$ can be computed in a $\U(m-1|n-1)\times \U^*(1)$ theory.
 But in fact the $\U^*(1)$ factor completely decouples, and the computation can be done simply in $\U(m-1|n-1)$ Chern-Simons theory.  This is almost
 obvious from the fact that a D3-brane at $\vec Y\not=0$ does not intersect the NS5-brane, so the $\U^*(1)$ gauge field has no Chern-Simons coupling
 and hence can scarcely contribute to knot invariants.  Momentarily, we give a more detailed explanation using some elementary supergroup theory.

First we state the claim precisely. For a reason explained in footnote \ref{whyonly}, we limit ourselves to the case of knot invariants
in $\RR^3$ (it may be possible to replace $\RR^3$ with $\RR^2\times S^1$).
 Let $K$ be a knot and $R$ a representation of $\U(m|n)$, and let $\W_R(K)$ be the corresponding
Wilson loop operator.   By restricting $R$ to a representation of $\U(m-1|n-1)$, we can view $\W_R(K)$ as an observable in $\U(m-1|n-1)$
Chern-Simons theory.  (Even if $R$ is indecomposable as a representation of $\U(m|n)$, it is typically decomposable as a representation of $\U(m-1|n-1)$,
so as an observable of $\U(m-1|n-1)$ Chern-Simons theory, $\W_R(K)$ may be a non-trivial sum of Wilson operators for indecomposable representations.)

\def\F{{\mathcal F}}
Let us study the representation $R$ as a representation of $\U(m-1|n-1)\times \U^*(1)$.  Actually, $\U(m|n)$ contains a sub-supergroup
$\U(m-1|n-1)\times \U(1|1)$, and $\U^*(1)$ is the center of $\U(1|1)$.  (Since $\U^*(1) $ was left unbroken when we gave an expectation value
to the hypermultiplet fields $Q_{\dot A}^I$, it commutes with the fermionic generators of $\U(1|1)$ and thus is central.)  
We recall that if we view $\U(1|1)$ as a matrix supergroup with odd generators
\begin{equation}\label{multo}\F=\begin{pmatrix} 0 & 1\cr 0 & 0 \end{pmatrix},~~~\t \F=\begin{pmatrix} 0 & 0\cr 1& 0 \end{pmatrix},\end{equation}
then the center of $\U(1|1)$ is generated by
\begin{equation}\label{ulto} E=\{\F,\t \F\}.\end{equation}
An indecomposable representation of $\U(1|1)$ with $E\not=0$ is two-dimensional, with one bosonic state and one fermionic one; in such a representation,
the supertrace of any element of $\U^*(1)$ vanishes.  More generally, this is so in any representation of $\U(1|1)$ with $E\not=0$.  Hence,
when we decompose $R$ under $\U(m-1|n-1)\times \U(1|1)$, the subspace with $E\not=0$ will not contribute to the expectation value of
$\W_R(K)$ (or of  any product of such operators): once we reduce to a computation in $\U(m-1|n-1)\times \U^*(1)$,
the contribution of the subspace with $E\not=0$ vanishes because of the supertrace
that is part of the definition of $\W_R(K)$.  The subspace of $R$ with $E=0$ does contribute in the evaluation of $\langle \W_R(K)\rangle$,
but since $\U^*(1)$ acts trivially on this subspace, the computation reduces to a computation in $\U(m-1|n-1)$ Chern-Simons theory.

Obviously, this argument can be iterated to show that knot invariants on $\RR^3$ of the supergroup $\U(m|n)$ can be replaced by knot invariants
of the ordinary group $\U(|n-m|)$.  In this reduction, for example, the fundamental representation of $\U(m|n)$ is replaced by the fundamental representation
of $\U(|n-m|)$.  

We can analyze in a similar way the orthosymplectic brane construction studied in section \ref{ortho}.  In this case, removing a D3-brane to $\vec Y\not=0$
reduces the gauge symmetry $\O(2m+1)\times \Sp(2n)$ to $\O(2m-1)\times \Sp(2n-2)\times \U^*(1)$, 
where again $\U^*(1)$ is supported on the D3-brane at nonzero $\vec Y$.  The
 supergroup $\OSp(2m+1|2n)$ is reduced to $\OSp(2m-1|2n-2)\times U^*(1)$. If $R$ is a representation of $\OSp(2m+1|2n)$ that we restrict to $\OSp(2m-1|2n-2)$, and $K$ is a knot in $\RR^3$,
then the expectation value $\langle \W_R(K)\rangle$ is unchanged in replacing $\OSp(2m+1|2n)$ by $\OSp(2m-1|2n-2)$.  One way to show
the decoupling of $\U^*(1)$ is to observe that $\OSp(2m+1|2n)$ contains a subgroup $\OSp(2m-1|2n-2)\times \OSp(2|2)$.  Moreover,
$\OSp(2|2)$ has a subgroup $\U(1|1)$ whose center is $\U^*(1)$.  The decoupling of $\U^*(1)$ then follows from the same elementary facts about $\U(1|1)$
that were used  before.  By repeating this symmetry breaking process, we can reduce knot invariants of $\OSp(2m+1|2n)$ to those of
$\O(2(m-n)+1)$ if $m\geq n$, or of $\OSp(1|2(n-m))$ if $n\geq m$.  Of course, in section \ref{another}, we have found a duality between orthogonal and
orthosymplectic Chern-Simons theories with gauge groups of these types that also changes the sign of the quantum parameter $q$.  

An interesting fact about this construction is that we can {\it not} make a similar argument for the symmetry breaking mechanism that involves
giving an expectation value to $\vec X$ rather than $\vec Y$.  This mechanism was mentioned at the beginning of section \ref{detachment},
and was explored in \cite{GaiW} in the one-sided case.  What goes wrong is that if $\langle \vec X\rangle\not=0$ but $\langle\vec Y\rangle=0$,
it is not true that all charged modes -- that is, modes that do not commute with $\langle \vec X\rangle$ -- acquire mass.  Such modes are massive in bulk,
but can remain massless along the NS5-brane.  Indeed, the boundary condition obeyed by $\vec X$ at $x_3=0$ is the condition
$\partial \vec X/\partial x_3+\vec X\times \vec X=0$ related to Nahm's equation.  In expanding around a symmetry-breaking solution
characterized by a nonzero constant value of $\vec X$, the boundary condition and the equations of motion are satisfied by certain
modes proportional to $\exp(-|x_3\langle \vec X\rangle|)$.  These modes are localized near the NS5-brane, and remain massless
in the three-dimensional sense.  As a result, for example, symmetry breaking from $\mathrm{SU}(2)$ to $\U(1)$ by taking $\langle \vec X\rangle\not=0$
cannot be used to reduce $\mathrm{SU}(2)$ Chern-Simons theory to $\U(1)$ Chern-Simons theory. No simple reduction of that sort exists.
Such symmetry breaking can, however, be used to deduce a vertex
model for knot invariants in $\RR^3$ in $\mathrm{SU}(2)$ Chern-Simons theory.  See \cite{GaiW}, especially section 6.7.4.  We expect that this derivation of a vertex model 
has an analog for supergroups.

\subsection{Analog In Pure Chern-Simons Theory}\label{toldu}

In Chern-Simons theory of a supergroup $SG$, viewed
as a purely three-dimensional theory, without the four-dimensional interpretation, one can reach a similar result by
adapting a argument that has been given in  \cite{CCMS} for two-dimensional sigma-models with supergroup symmetry.  We
will be brief, leaving details to the reader. (As we explain in Appendix \ref{Anomalous}, some fundamental
issues about Chern-Simons theory of a supergroup as a purely three-dimensional theory are not entirely clear.  We do not expect those
issues to affect the following discussion.)  

  In Chern-Simons theory on $\RR^3$, with all fields required to vanish at infinity, we can regard
$SG$ (acting by gauge transformations that are constant at infinity) as a group of global symmetries.  Now pick\footnote{Such a subgroup
exists for all supergroups of the class we consider except $\OSp(1|2n)$, which is the case in which symmetry breaking by $\vec Y$ is not
possible.} a subgroup $\mathrm F\subset SG$ of dimension
$0|1$; its Lie algebra is generated by an element $\mathcal C\subset \frak{sg}$ that obeys $\{\mathcal C,\mathcal C\}=0$.    Let us write $\Qb_{\mathcal C}$ for the symmetry of $SG$ Chern-Simons theory generated by $\mathcal C$.
Thus  $\Qb_{\mathcal C}^2=0$, and we can analyze the path integral of $SG$ Chern-Simons theory by localizing on the
space of fields invariant under the action of $\Qb_{\mathcal C}$.  This localization with respect to this symmetry can be used to reduce
Chern-Simons theory of the supergroup $SG$ to Chern-Simons theory of the supergroup $SH$ whose Lie algebra is the cohomology
of $\mathcal C$.  For example, if $SG=\U(m|n)$ and $\mathcal C$ is the fermionic raising operator $\F$ of a $\U(1|1)$ subgroup,
then $SH=\U(m-1|n-1)$.

Notice that in section \ref{symbrane}, writing $C$ for the fermion number one component of the hypermultiplet fields $Q^I_{\dot A}$,
the expectation value $\mathcal C=\langle C\rangle$ was not required to obey $\{\mathcal C,\mathcal C\}=0$.   This condition
is satisfied precisely if the branes are separated in $x_7$ but not in $x_8$ or $x_9$.

\section{Lift to A D4-D6 System And to M-Theory}\label{lifting}

\subsection{Review}\label{kreview}
So far in this paper, we have considered an ``electric'' description via a D3-NS5 system (with a variant of this in section
\ref{another}), and a dual ``magnetic'' description via a D3-D5 system.
It is straightforward to ``lift'' the D3-D5 system to a D4-D6 system and in turn to relate this to a description via M5-branes.
In the one-sided case, this procedure has been described in
  \cite{5knots}, starting in section 4.   (Some matters are more briefly explained in \cite{WittenK}, \cite{TwoLectures}.)
Here we will provide a mini-review, which can be regarded as a continuation of the introduction to this paper. 
In section \ref{goingon}, we describe
the two-sided generalization.  For the most part, this generalization is straightforward, but  in section \ref{orthagain} we describe some
interesting and novel details for the orthosymplectic case.

The lift from D3-D5 to D4-D6 is a standard $T$-duality.  We compactify on a circle one of the directions normal to the D3-D5 system.
In the notation of section \ref{nstype}, these are the $x_7,x_8,x_9$ directions.  Since we have interpreted the rotation of the $x_8-x_9$ plane
as the ghost number or fermion number symmetry $F$ and we wish to preserve this symmetry, we compactify the $x_7$ direction.  Then we perform a conventional
$T$-duality, converting the D3-D5 system to a D4-D6 system, which one studies by similar methods to those that one uses for the D3-D5 case.  

The $T$-duality converts D3-branes supported on  $\RR^3\times \RR_+$ to D4-branes supported on $S^1\times \RR^3\times \RR_+$.
However, after this $T$-duality, one can decompactify the $S^1$.  In the one-sided case, the D4-branes then 
live on $\RR^4\times\RR_+$.
Part of the reason that the lift from D3-D5 to D4-D6 is useful is that
the condition of $\Qb$-invariance turns out to give a system of elliptic\footnote{Roughly speaking, elliptic differential
equations have properties similar to  familiar, physical sensible differential equations in Euclidean signature such as the Laplace or Dirac equation.
Given suitable boundary conditions and some technical conditions (which have not yet been rigorously proved in the present context), 
the solutions of an elliptic differential equation can be counted in a sensible way.  The ability to make such a counting
is assumed below.  Were we to make a further $T$-duality to a D5-D7 system, we would lose ellipticity
and some of the related special properties that hold in the D4-D6 case.} differential equations for the pair $A,B$, and moreover
these equations are invariant under rotations of $\RR^4$, as a result of which $\RR^4$ can be replaced by a more general oriented
four-manifold.
The bosonic fields on $\RR^4\times \RR_+$ that have ghost number zero are the gauge field $A$ and an adjoint-valued
 three-component field $B$ that
transforms under rotations of $\RR^4$ as a self-dual two-form.     The equations can be written
\begin{align}\label{helpful}\notag F^+-\frac{1}{4}B\times B -\frac{1}{2}D_y B & = 0 \\
                                            F_{y\mu}+\,D^\nu B_{\nu\mu} &= 0, \end{align}
 (For a fuller explanation, see \cite{5knots}, section 5.2.)                                           
These equations reduce
to the equations (\ref{localization})  of the D3-brane gauge theory if one considers solutions that are ``time''-independent, in other words  invariant under
translations in one direction in $\RR^4$.  In this reduction, which effectively undoes the results of the $T$-duality, the `'time'' component
of $A$ combines with the three-component field $B$ to make the four-component field that was called $\phi$ in the four-dimensional
localization equations (\ref{localization}).

In describing the $T$-duality in gauge theory language,
the Nahm pole boundary condition of the D3-D5 system naturally lifts to a Nahm pole boundary condition of the D4-D6 system.
Similarly, the  line operators of the D3-D5 system, which are described by magnetic impurities,
 lift to surface operators of the D4-D6 system.  One can think of the support of such a surface operator as the worldvolume of a
 knot or string propagating in four dimensions.  It makes sense to consider a surface operator supported on a general two-manifold
 $\Sigma$ in the four-dimensional D4-brane boundary; this gives a  framework for understanding the
 ``morphisms'' of Khovanov homology.
 But we will here focus on the time-independent case that the D4-brane world-volume
is  $S^1\times \RR^3\times \RR_+$ (or its cover $\RR^4\times \RR_+$) and $\Sigma=S^1\times K$ (or its cover $\RR\times K$). 

\def\Nb{\pxitfont N}
\def\nb{\pxitfont n}
\def\mb{\pxitfont m}
In general, in a supersymmetric theory, a supersymmetric path integral on $S^1\times X$, for any $X$, with supersymmetry-preserving
boundary conditions, is equal to an ``index,'' computed in the space of supersymmetric states on $X$.    So for our case, with
$X=\RR^3\times \RR_+$,
and a magnetic impurity supported on $K\times \{0\}$ (where $K\subset \RR^3$ is a knot and $\{0\}$ is the endpoint of $\RR_+$), let
us discuss what are the supersymmetric states.   In the classical approximation, a supersymmetric state of this 
system on $\RR^3\times \RR_+$ is represented by a time-independent solution of the equations (\ref{helpful}).  
Such a solution has two conserved charges, the fermion number $F$,
and the instanton number $\Nb$, defined in eqn. (\ref{elba}). In this approximation, there is a space $\H_{0;K,R}$ of
approximate supersymmetric states; $\H_{0;K,R}$ has a basis consisting of a basis vector $|i\rangle$ for every time-independent
classical solution $i$.   $F$ and $\Nb$ are diagonal in this basis: $F|i\rangle=f_i|i\rangle$, $\Nb|i\rangle =\nb_i|i\rangle$.
In the space $\H_{0;K,R}$, one defines a differential $\h\Qb$ by evaluating the matrix elements of $\Qb$ from $|i\rangle$
to $|j\rangle$ (and conjugating, in a manner explained in \cite{WittenMorse,GMW} or in section 10 of \cite{Hori}, so that these matrix elements become
integers).  The matrix element of $\h\Qb$ from $|i\rangle$ to $|j\rangle$ (which vanishes unless $f_j=f_i+1$ and $\nb_j=\nb_i$,
since $\Qb$ increases $F$ by 1 and commutes with $\Nb$) is
\begin{equation}\label{mosto}\h\Qb|i\rangle=\sum_{j|f_j=f_i+1}\mb_{ij}|j\rangle.\end{equation}
Here $\mb_{ij}$ is computed by ``counting,'' with signs, 
 time-dependent classical 
solutions that are asymptotic to the time-independent solution $i$ in the past and to $j$ in the future.  If $S_{ij}$ is the set of
such solutions, and for $x\in S_{ij}$, $(-1)^{w(x)}$ is the sign of the fermion determinant that arises in expanding around the 
solution $x$, then 
\begin{equation}\label{osto}\mb_{ij}=\sum_{x\in S_{ij}}(-1)^{w(x)}. \end{equation}
One has $\h\Qb^2=0$, and the space $\H_{K,R}$ of exact supersymmetric states in the presence of the knot is the cohomology
of $\h\Qb$.  The conserved charges $F$ and $\Nb$ act on $\H_{K,R}$, as they do on the classical approximation
$\H_{0;K,R}$, generating an action of\footnote{We have to subtract a constant from $\Nb$ to make its eigenvalues integers
if we want the group to be precisely $\U(1)\times \U(1)$.  Topologically, one needs to pick a framing of $K$ to define $\Nb$
so that its eigenvalues are integers.  Note that $\H_{K,R}$ is defined over $\Z$ (since $\h\Qb$ has integer matrix elements);
to emphasize this, one may prefer to speak of a $\Z\times \Z$ grading by the eigenvalues of $F$ and $\Nb$ rather than an 
action of $\U(1)\times \U(1)$.} $\U(1)\times \U(1)$.  The contribution
of a state of given charges $F$ and $\Nb$ to the index -- the path integral on $S^1\times \RR^3\times \RR_+$ -- 
is\footnote{The reason for the factor of $q^\Nb$ is the same as it was in 
our discussion of the D3-brane system in section \ref{magpre}.  Just like the action (\ref{Imagnetic}) of  the D3-brane gauge theory,
the action of the D4-brane gauge theory is the sum of a $\Qb$-exact term and a multiple of the instanton number $\Nb$.  Now $\Nb$
is understood as a conserved charge, integrated over an initial value surface $\RR^3\times \RR_+\subset S^1\times \RR_3\times \RR_+$. } 
$(-1)^F q^\Nb$. The sum over all states gives $\Tr\, (-1)^F q^\Nb$.  Here, because of the invariance of the index 
under $\Qb$-invariant perturbations, it does not matter if the trace is taken
in the space $\H_{0;K,R}$ generated by all time-independent classical solutions or the space $\H_{K,R}$ of exact quantum ground states.
If we take the trace in the space $\H_{0;K,R}$, we simply recover the prescription that was reviewed in section \ref{magpre} for computing
the quantum knot invariant $Z_{K,R}(q)$ corresponding to the choice of $K$ and $R$.   However, we get a more interesting
formula if we take the trace in the space $\H_{K,R}$ of exact quantum ground states:
\begin{equation}\label{meloxic} Z_{K,R}(q)=\Tr_{\H_{K,R}}\,(-1)^F q^\Nb. \end{equation}

 This version of the formula  is the significant one because standard arguments using the fact that the stress tensor of the system
 is $\Qb$-exact
show that, in contrast to $\H_{0;K,R}$, $\H_{K,R}$ is a topological invariant: it depends  only on the knot $K$ and 
representation $R$ but not on the embedding of $K$ in $\RR^3$.   $\H_{K,R}$ is a candidate for the Khovanov homology \cite{Kh} associated
to the given knot and representation.  $\H_{K,R}$ determines the usual quantum knot invariants $Z_{K,R}(q)$ via
the formula (\ref{meloxic}), but actually $\H_{K,R}$ contains more information (because the eigenvalues of $F$ are integers, while  $Z_{K,R}(q)$ depends on
these integers only mod 2,
and because by considering surface operators with time-dependent support, one can define 
certain natural operators that act on $\H_{K,R}$ but which cannot be formulated in terms of  $Z_{K,R}(q)$).
The candidate $\H_{K,R}$ for Khovanov homology  is a close cousin of a previous proposal by Gukov, Schwarz, and Vafa
\cite{GSV}, which
in turn was based on  much earlier work on BPS states in the presence of a knot by Ooguri and Vafa \cite{OV}.
A straightforward elaboration of this procedure to the two-sided case gives -- as we discuss in section
(\ref{goingon}) -- a candidate for an analog of Khovanov homology for a supergroup.

\def\TN{{\mathrm {TN}}}
One important further refinement of this construction involves a lift from Type IIA superstring theory to M-theory.  We recall that Type IIA
superstring theory on $\RR^{10}$ lifts to M-theory on $\RR^{10}\times \h S^1$ (here $\h S^1$ is a circle that should be distinguished from
 the circles that entered the discussion of $T$-duality).  Type IIA on $\RR^{10}$ with a D6-brane supported on a copy of $\RR^7\subset \RR^{10}$ lifts
to M-theory on $\RR^7\times \TN$, where $\TN$ is a Taub-NUT space --  a hyper-Kahler manifold that asymptotically looks
like a $\h S^1$ bundle over $\RR^3$, rather than a simple product $\RR^3\times \h S^1$.  $\TN$ has a $\U(1)$ symmetry that rotates
$\h S^1$, and the hyper-Kahler moment map for this symmetry gives a map
\begin{equation}\label{molk}\pi:\TN\to \RR^3\end{equation}
that is a fibration except over a single point $\{0\}\in \RR^3$ where the fiber $\h S^1$ collapses to a point
(the point $\{0\}$ corresponds in the Type IIA description to the position of the D6-brane). 
  A system of $N$ D4-branes that end on the D6-brane
(and thus have world-volume $\RR^4\times \RR_+$) lifts in M-theory to a system of $N$ M5-branes supported on $\RR^4\times C$, where $C$
is a semi-infinite cigar (fig. \ref{cigar}(a)).   (Here $C=\pi^{-1}(\RR_+)$, where $\RR_+\subset \RR^3$ is a half-line ending at the origin.)
In this description, the instanton number $\Nb$ becomes a geometrical symmetry, associated to the rotation of the cigar around its tip.  
By taking a low energy limit, the description in terms of M5-branes can be understood as  a description via a certain six-dimensional
superconformal field theory formulated on $\RR^4\times C$, with a topological twisting that preserves some supersymmetry.  
The six-dimensional theory in question has been sometimes modestly called the $(0,2)$ model, and sometimes more gloriously
called Theory X.
\begin{figure}
 \begin{center}
   \includegraphics[width=170mm]{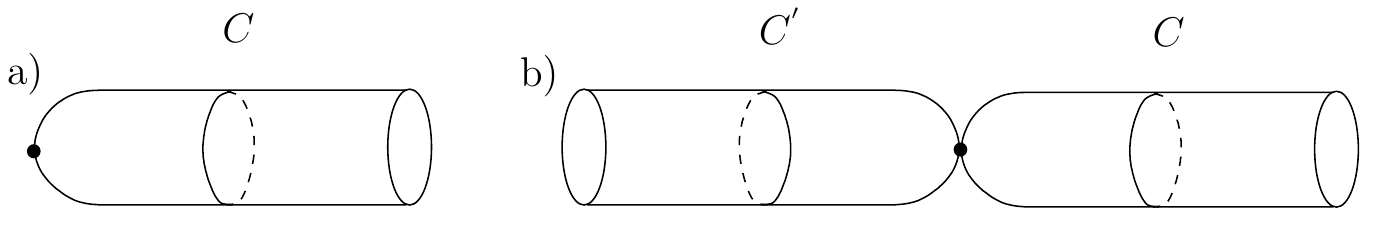}
 \end{center}
\caption{\small (a) A semi-infinite cigar $C$.  (b) A pair of semi-infinite cigars $C$ and $C'$ in $\TN$, meeting at their common tip.
$C$ and $C'$ intersect transversely, though this is difficult to draw since it is not possible in three dimensions. }
 \label{cigar}
\end{figure}

\subsection{Two-Sided Analog}\label{goingon}

In most respects, the two-sided generalization of what we have just explained is straightforward.

The $T$-dual of a configuration with D3-branes on both sides of a D5-brane is a configuration with D4-branes
on both sides of a D6-brane.  Changing our notation slightly, we will parametrize the D4-brane worldvolume
by $x_0,\dots, x_4$, and assume that the D6-brane is located at $x_4=0$.  We further assume that there are $m$ D4-branes,
supporting a $\U(m)$ gauge symmetry, for $x_4<0$, and $n$ D4-branes, supporting a $\U(n)$ gauge symmetry, for $x_4>0$.  

   For $x_4<0$ or $x_4>0$,  the condition for $\Qb$-invariance is precisely (\ref{helpful}), with the fields $A$ and $B$
   being now $\frak{u}(m)$-valued or $\frak{u}(n)$-valued.  To complete the description, of course,
one needs to describe what happens at $x_4=0$.  The answer depends on the value of $n-m$, and is quite similar to the
description of the D3-D5 system in section \ref{magnetic}.

The simplest case is $m=n$. In this
case, we are simply considering a system of $n$ D4-branes intersecting a D6-brane at $x_4=0$, and this system can be weakly
coupled (on the worldsheet and in spacetime).   From the point of view
of the $\U(n)$ gauge theory on the D4-branes, the intersection with the D6-branes produces a hypermultiplet in the fundamental
representation of $\U(n)$ that is supported on the defect at $x_4=0$.  Just as in section \ref{eqrank}, in the twisted theory, the scalar fields $Z^A$
in the hypermultiplet become spinors.   But now these spinors live in four dimensions instead of three, and we have to distinguish
the two spinor representations of the group $\Spin(4)$.  In the context of the D4-D6 system,
$Z^A$ is a spinor of definite chirality; which chirality it has depends on how the theory is
twisted.  If our convention is such that the self-dual two-form $B$ of eqn. (\ref{helpful}) has spin $(1,0)$ under $\Spin(4)$, then
$Z^A$ has spin $(1/2,0)$.  This being so, the hyper-Kahler moment map, which is still defined by eqn. (\ref{momentmap}),
has spin $(1,0)$, and hence it is possible to write an analog of eqn. (\ref{discX}) for the discontinuity in $B$ across $x_4=0$:
\begin{equation}\label{toldo}\left.B^a\right|^+_-=\frac{1}{2}\mu^a. \end{equation}
To complete the description of the system, we also need an equation of motion for $Z$.  $T$-duality converts the field called $\phi_3$
in eqn. (\ref{bono}) into the covariant derivative with respect to a new coordinate, so after the $T$-duality, 
the equation for $Z$ becomes
a simple 4d Dirac equation
\begin{equation}\label{poldo}\slashed{D}Z=0.\end{equation}
The equations (\ref{helpful}) and (\ref{poldo}) and the discontinuity condition (\ref{toldo}) are the conditions of $\Qb$-invariance
of the fields $A,B,Z$.  The corresponding condition on the remaining bosonic field $\sigma$ is the obvious generalization of
eqn. (\ref{melb}):
\begin{equation}\label{oldo} D_\mu\sigma=[B,\sigma]=[\bar \sigma,\sigma]=0.\end{equation}
This equation says that $\sigma$ generates a symmetry of the whole configuration (so in the case of an irreducible solution,
$\sigma$ vanishes).

The analog of this for $n\not=m$ is quite similar to what was described in section \ref{unequalrk}.  For $n\not=m$,
there are no fields supported on the defect at $x_4=0$.  We may as well assume that $n>m$.  If $n=m+1$, then the fields $A$ and $B$ 
are $m\times m$ matrices for $x_4<0$ but have an extra row and column for $x_4>0$.  
Just as in section \ref{unequalrk}, the $m\times m$ matrices are continuous at $x_4=0$, and the extra row and column of $A$ and $B$
that exist for $x_4>0$
obey simple  boundary conditions at $x_4=0$.  (In the context of the first order equations (\ref{helpful}), as opposed to the second
order equations of the full physical theory, the extra row and column of 
$A$ vanish at $x_4=0$ and there is no restriction on the extra row and column of $B$.)  

For $n-m>1$, the picture is precisely analogous to what is described in section \ref{unequalrk}, with a Nahm pole in $B$
of rank $n-m$, and other details lifted in an obvious way from the D3-D5 system.

This construction leads to an analog of Khovanov homology for the supergroup $\U(m|n)$.  A classical approximation 
$\H_{0;K,R}$ to the space of supersymmetric states has a basis corresponding to  time-independent classical solutions
that satisfy the appropriate boundary conditions; the corresponding exact quantum space $\H_{K,R}$ is found in the usual
way by taking time-dependent solutions into account.  

A lift to $M$-theory is also possible.
As we explained in section \ref{review}, $n$ D4-branes supported on $\RR^4\times \RR_+$ lift in $M$-theory to $n$ M5-branes
supported on $\RR^4\times C$, where $C=\pi^{-1}(\RR_+)$ is a semi-infinite cigar; here  $\RR_+$ is a ray emanating
from the origin  $0\in\RR^3=\pi(\TN)$.  In the two-sided problem, we have $n$ D4-branes on $\RR^4\times \RR_+$ and
$m$ D4-branes on $\RR^4\times \RR_+'$, where $\RR'_+$ is another copy of $\RR_+$.  We can think of $\RR_+$ and $\RR'_+$
as rays that that emanate in opposite directions from the same endpoint $0\in \RR^3$.  The $M$-theory lift then involves $n$
M5-branes on $\RR^4\times C$ and $m$ M5-branes on $\RR^4\times C'$.  Here $C$ and $C'$ are two semi-infinite cigars
in $\TN$ that meet transversely at their tips (fig. \ref{cigar}(b)).

One interesting question here is to compare Khovanov homology of $\U(m|n)$ to Khovanov homology of $\U(m-1|n-1)$.
We can try to make this comparison along lines described in section
\ref{detachment}, by displacing a D4-brane normal to the D6-brane (and thus in the $x_8-x_9$ directions,
explicitly breaking the fermion number symmetry).  However, now that we are aiming to describe a space $\H_{K,R}$ of quantum states, rather than a function
$Z_{K,R}(q)$, the necessary reasoning is more delicate and somewhat beyond the scope of the present paper.   From direct study of Khovanov homology for supergroups \cite{GGS}, it appears that in the $\U(m|n)$ Khovanov homology of a knot
$K$ and representation $R$, it is possible to define a differential whose homology is the $\U(m-1|n-1)$ Khovanov homology for
the same knot and with a corresponding representation of $\U(m-1|n-1)$.

\subsection{$\OSp(2m+1|2n)$ And Khovanov Homology}\label{orthagain}

\subsubsection{Orthosymplectic Theory In Five Dimensions}\label{orthlift}
In section \ref{another}, we explored a duality between Chern-Simons theory of $\OSp(2m+1|2n)$ with coupling
parameter $q$ and Chern-Simons theory of $\OSp(2n+1|2m)$ with parameter $-q$.  On the electric side, this duality
is rather interesting and surprising.  However, the magnetic counterpart was less interesting.  It involved a comparison
between a description based on $\Sp(2m)\times \Sp'(2n)$, in which 
one has to count certain classical solutions  with a sign factor
\begin{equation}\label{tifox} (-1)^Fq^{\Nb^\vee}(-1)^{\nu_r} \end{equation}
and a description based on $\Sp'(2m)\times \Sp(2n)$, in which one has to count the same classical solutions with the sign
factor
\begin{equation}\label{nifox}(-1)^F (-q)^{\Nb^\vee}(-1)^{\nu_\ell}. \end{equation}
These factors are equal, because of the relatively elementary identity $\nu_\ell+\nu_r=\nu=\Nb^\vee$.

However, after $T$-duality to a D4-D6 system, the relation between the two magnetic descriptions becomes much more interesting
and involves some topological subtleties.  To understand these subtleties, 
we set $m=n$, so that the gauge group is everywhere $\Sp(2n)$.  In the D3-D5 system,
the $\Sp(2n)$ gauge theory is defined on a four-manifold $M=W\times\RR$, where $\RR$ is parametrized by $y$; it interacts with a hypermultiplet in the fundamental
representation that is supported on $W\times \{y=0\}$, which we just call $W$.  $T$-duality
replaces $M$ with the five-manifold $Y=M\times S^1$ and $W$ by the four-manifold $V=W\times S^1$.  We write $Y_\ell$ or
$Y_r$ for the portions of $Y$ with $y<0$ or $y>0$, respectively.

A purely four-dimensional $\Sp(2n)$ gauge theory on a four-manifold  
coupled to a single fundamental hypermultiplet is inconsistent.
It suffers from a global anomaly that was described in \cite{GlobalAnom}.  We will have to understand this anomaly
and its relation to certain phenomena in $\Sp(2n)$ gauge theory in five dimensions.

\def\R{{\mathcal R}}

\subsubsection{An Anomaly And A Discrete Theta-Angle}\label{anad}
The starting point is the fact that $\pi_4(\Sp(2n))=\Z_2$.  This homotopy group is associated with a discrete $\theta$-like
angle for an $\Sp(2n)$ bundle on a five-dimensional spin manifold $U_5$.  In odd dimensions, there is no distinction between
fermions of positive or negative chirality, so the familiar notion of the integer-valued index of the Dirac operator does not apply.
However, in certain cases in odd dimensions, one can define a mod 2 index of the Dirac operator.   $\Sp(2n)$ gauge
theory in five dimensions is an example.  In five dimensions, spinors are pseudoreal, and as the fundamental representation $\R$ of
$\Sp(2n)$  is also pseudoreal, $\R$-valued spinors are real.  The five-dimensional Dirac operator $\slashed{D}_5$  acting
on $\R$-valued spinors is a real, skew-symmetric operator.  The number of its zero-modes is a topological invariant mod 2,
and this quantity, which we will call $\zeta$, is known as the mod 2 index of the Dirac operator.   It is possible to introduce
a sort of discrete $\theta$ angle in $\Sp(2n)$ gauge theory on a five-dimensional spin 
manifold $U_5$ without boundary by weighting the contribution
to the path integral of every $\Sp(2n)$ bundle over $U_5$ by a factor of $(-1)^\zeta$.  We will denote the $\Sp(2n)$ gauge
theory with this discrete parameter turned on as $\Sp^*(2n)$.  

Unlike the more familiar integer-valued index of the Dirac operator in even dimensions, $\zeta$ is not described by any integral
formula and in fact it is unusual to have a convenient way to calculate it.  (There is a mod 2 version of the Atiyah-Singer index
theorem, but what it says is a little abstract.) However,  there is one situation in which there
is a simple formula for $\zeta$.  Let $V_4$ be a four-dimensional spin manifold, and 
suppose that $U_5=V_4\times S^1$ with a product metric and a periodic (unbounding) spin
structure on $S^1$.  Consider on $U_5$ an $\Sp(2n)$ bundle that is a pullback from $V_4$.  The zero-modes of the five-dimensional Dirac operator $\slashed{D}_5$ are pullbacks from $V_4$ 
(since a fermion mode with non-zero momentum around the $S^1$ would not be a zero-mode)
so $\zeta$ can be expressed in terms of the kernel of the four-dimensional Dirac operator $\slashed{D}_4$.  In four dimensions,
fermion zero-modes can have either positive or negative chirality; let $c_+$ and $c_-$ be the number of zero-modes
of $\slashed{D}_4$ of the indicated chirality.  So the index of $\slashed{D}_4$ is $\nu=c_+-c_-$.  On the other hand,
all zero-modes on $V_4$ of either chirality pull back to fermion zero-modes on $U_5$, so the total number of zero-modes of
$\slashed{D}_5$ is $c_++c_-$.  So $\zeta=c_++c_-$ mod 2, or equivalently $\zeta=\nu$ mod 2:
\begin{equation}\label{mexto}(-1)^\zeta=(-1)^\nu. \end{equation}

We need one more general comment as preparation.
  Let $I$ be a unit interval and let $\pi:V_4\times I\to V_4$ be
the natural projection.   Suppose that we are given an $\Sp(2n)$ bundle $E\to V_4$ and a gauge transformation $g$ of this
bundle.   We pull back $E$ to a bundle $E'=\pi^*(E)\to V_4\times I$ and use $g$ as gluing data in gluing together the
fibers of $E'$ over the two ends of $V_4\times I$ to build an $\Sp(2n)$ bundle  $E''\to V_4\times S^1$.  We say that $g$ 
detects $\pi_4$ if $\zeta(E'')\not=0$.  For every $V_4$ and $E$, there exists a gauge transformation that detects $\pi_4$ in this
sense.  (One can choose $g$ to be 1 outside a small ball in $V_4$ and construct $g$ inside the ball using a homotopically non-trivial map
$S^4\to \Sp(2n)$.  In general saying that $g$ detects $\pi_4$
is not  equivalent to saying that $g$ is topologically non-trivial, since the topological classification
of $\Sp(2n)$ gauge transformations on a four-manifold also involves $\pi_3(\Sp(2n))\cong\Z$.)

Now we can explain the global anomaly in four dimensions and also its relation to five dimensions.  A theory on a four-dimensional
spin manifold $V_4$ with a single multiplet of fermions in the fundamental representation of $\Sp(2n)$ (or equivalently a single
hypermultiplet in that representation) is inconsistent because one cannot consistently define the sign of $\Pf(\slashed{D}_4)$,
the Pfaffian of the four-dimensional Dirac operator $\slashed{D}_4$.  The sign of this Pfaffian is reversed by a gauge transformation if and only
if this gauge transformation detects $\pi_4$.    However, if $V_4$ is the boundary of a five-dimensional spin manifold $U_5$,
and the $\Sp(2n)$ gauge field and (therefore) its gauge transformations extend over $U_5$, then there is no inconsistency.   One
way to explain this is to observe that a gauge transformation on $V_4$ that detects $\pi_4$ cannot be extended\footnote{If it
could be so extended, the bundle $E''\to V_4\times S^1$ constructed in the last paragraph could be extended over $U_5\times
S^1$, contradicting the cobordism invariance of the mod 2 index of the Dirac operator.} over $U_5$, so the problem does not
arise.  A more complete explanation is as follows.  Consider the Dirac operator $\slashed{D}_5$ on $U_5$ with APS boundary
conditions along $V_4=\partial U_5$.    Let $\zeta$ be the mod 2 index of this operator.  The product
\begin{equation}\label{modo}|\Pf(\slashed{D}_4)|(-1)^\zeta \end{equation}
is completely gauge-invariant and is a good and physically sensible substitute for the ill-defined object $\Pf(\slashed{D}_4)$.
The basis for this last statement is that whenever the gauge field $A$ is varied so that an eigenvalue of $\slashed{D}_4$ changes
sign (more precisely so that a pair of eigenvalues change sign in opposite directions, so that the fermion path integral
 $\Pf(\slashed{D}_4)$ should
change sign), the mod 2 index $\zeta$
 in five dimensions computed with APS boundary conditions also jumps, so that the product $|\Pf(\slashed{D}_4)|(-1)^\zeta$
 does have the desired change in sign.
 
 Hopefully it is clear that all this has a converse.  In gauge theory on a five-dimensional spin manifold $U_5$ without
 boundary, it makes sense to include a factor $(-1)^\zeta$ in the functional integral.  If however $U_5$ has a boundary $V_4$,
 with APS boundary conditions, but without $\R$-valued fermions on the boundary, then it does not make sense to include
 the factor $(-1)^\zeta$, because the jumps that this factor can have when the gauge field is varied would be unphysical.  
 The relation that we have explained between the anomaly in four dimensions and the discrete theta-angle in five dimensions
 can be regarded as a discrete version of anomaly inflow \cite{CH}.    There is also an analogy with topological insulators,
 along the lines that we already discussed in one dimension less in section \ref{anomfour}.

In our application, our four-manifolds and five-manifolds are typically   
not compact, but have ends of simple types.  If one allows only square-integrable solutions
of the Dirac equation and gauge transformations that are trivial at infinity, all the above statements apply.  

\subsubsection{Two Lifts To Five Dimensions}\label{twotheories}

Now we return to the situation considered in section \ref{orthlift}.  A five-manifold $Y$ that supports an $\Sp(2n)$ gauge
group is divided by a four-manifold $V$ into portions $Y_\ell $ and $Y_r$.  Along $V$, there is a hypermultiplet in the fundamental
representation. Since this is anomalous by itself, there must be a discrete theta-angle either on $Y_\ell$ or $Y_r$, but not both.
In other words, if $\zeta_\ell$ and $\zeta_r$ are the mod 2 indices of the $\R$-valued Dirac operator on $Y_\ell$ and $Y_r$
respectively (in each case with APS boundary conditions along $V$), then the path integral must have a factor of either
$(-1)^{\zeta_\ell}$ or $(-1)^{\zeta_r}$.  

Differently put, when a system of D4-branes supporting a symplectic gauge group crosses a D6-brane, the discrete
theta-angle jumps and the theory is $\Sp(2n)$ on one side and $\Sp^*(2n)$ on the other side, or vice-versa.  (This possibility
has been suggested in the past; see section 3.1.1 of \cite{HananyOrientifolds}.)  Though we have made these arguments
for the case that the gauge group does not jump in crossing the D6-brane, a consideration of what happens when one displaces
some D4-branes away from the D6-brane on one side or the other shows that this restriction is not essential.  A D4-D6 system
that naively describes $\Sp(2m)$ gauge theory on the left of a D6-brane
 and $\Sp(2n)$ on the right is really $\Sp(2m)\times \Sp^*(2n)$ or
$\Sp^*(2m)\times \Sp(2n)$.  

These two variants of the D4-D6 system are simply the lifts to five dimensions of the two magnetic duals
of $\OSp(2m+1|2n)$ Chern-Simons that were described on the right half of fig. \ref{ospodd} in section \ref{prelims}.  Now let us explore
the analogs of Khovanov homology.  We consider a ``time''-independent
case in which the five-manifold $Y$ of the D4-D6 system is $\RR\times M$ (or $S^1\times M$), where $M=W\times \RR$ is the four-volume of the D3-D5 system that we studied in section \ref{ortho}.  A classical approximation $\H_{0;K,R}$ to the space
of supersymmetric states has a basis given by solutions of the localization equations that obey the appropriate boundary conditions,
just as in sections \ref{kreview} or \ref{goingon}.  These equations do not know about the discrete theta-angle, which also does not
affect the definition of the fermion number $F$ or the instanton number $\Nb^\vee$.  So $\H_{0;K,R}$ is the same in the two
cases.  

When we compute the space of {\it exact} supersymmetric
 states, rather than a classical approximation to it, we do see the discrete theta-angles.  The exact supersymmetric states
 are the cohomology of a ``differential'' $\h\Qb$ acting on $\H_{0;K,R}$.  Here $\h\Qb$ is defined by a formula (\ref{mosto})
 where as in (\ref{osto}), the coefficients $\mb_{ij}$ are computed by counting, with signs, certain time-dependent solutions.  It is here that the discrete theta-angles enter:
 the appropriate factor of $(-1)^{\zeta_\ell}$ or $(-1)^{\zeta_r}$ must be included in eqn. (\ref{osto}) for the coefficients of the 
 differential, which becomes\footnote{In writing the formula this way, we assume that $m=n$.  The two factors on the right hand side of (\ref{tono})
 have the same origin: $(-1)^{w(x)}$ is the sign of the path integral of the bulk fermions, and $(-1)^{\zeta_\ell(x)}$ or $(-1)^{\zeta_r(x)}$ is the sign
 of the path integral of the fermions that live on the defect, defined with the help of $\zeta_\ell$ or $\zeta_r$ to cancel the anomaly.  
 If deform to $m\not=n$ by moving some D4-branes on one side or the other away from the D6-brane,
 then, as discussed in section \ref{thedual} for the D3-D5 case, Higgsing occurs  and we can no longer distinguish the two kinds of fermion.  The
 contribution to $\mb_{ij}$ of a time-dependent solution $x$ is always the sign of the fermion path integral in linearizing around $x$.}
 \begin{equation}\label{tono}\mb_{ij}=\sum_{x\in S_{ij}}(-1)^{w(x)}(-1)^{\zeta_\ell(x)} ,\end{equation}
 or the same formula with $\zeta_\ell\to \zeta_r$.  Thus, we can construct two different differentials -- say $\h\Q'$ if we use $\zeta_\ell$ and $\h\Q''$ if we use
 $\zeta_r$.  Their cohomologies give two different spaces of physical states, say $\H'_{K,R}$ and $\H''_{K,R}$.  
 
 In general, there is nothing simple that we can say about either $\zeta_\ell(x)$ or $\zeta_r(x)$ (except that they have a behavior
 under gluing that ensures that $(\Qb')^2=(\Qb'')^2=0$).  The only simple relation between $\Qb'$ and $\Qb''$ is that they
 are congruent mod 2 (this statement makes sense because their matrix elements are integers).  Similarly, $\H'_{K,R}$ and
 $\H''_{K,R}$ (which are defined over $\Z$ because $\Qb'$ and $\Qb''$ are) are equivalent if reduced mod 2.  But otherwise
 $\H'_{K,R}$ and $\H''_{K,R}$ may be quite different; for example, one might be $\Z^p\times \Z_2^q$ and the other
 $\Z^q\times \Z_2^p$, as these are congruent mod 2.   
 
 Simplicity, however, is recovered by taking a trace or more precisely an index.  For this, we replace $Y$ by $S^1\times M$ (with
 supersymmetric boundary conditions around $S^1$).
 As always, a path integral on $S^1\times M$ computes an index. In either of the two constructions, 
  the index will equal the quantum knot
 invariant $Z_{K,R}(q)$ that can be studied  via the D3-D5 system on $M$, without the $T$-duality
 to $Y=S^1\times M$; taking the index effectively undoes the $T$-duality.
  In the $\Sp(2m)\times \Sp^*(2n)$ construction, the index can be represented, as in section \ref{review},
   as a trace in either 
 $\H_{0;K,R}$ or $\H''_{K,R}$, and in the $\Sp^*(2m)\times \Sp(2n)$ construction, it can be similarly represented as a trace in either
 $\H_{0;K,R}$ or $\H'_{K,R}$.  However, the interesting formulas are the ones that involve traces in $\H'_{K,R}$ or $\H''_{K,R}$,
 because the standard arguments show that these spaces are topological invariants of the knot $K$.
 
 To express the quantum knot invariant as  a trace in $\H''_{K,R}$, we have to weight each contribution by
  a factor $(-1)^{\zeta_r}$, but in a time-independent
 situation, this reduces to $(-1)^{\nu_r}$, as we learned in section \ref{anad}.
 The formula we get is then
 \begin{equation}\label{modz}Z_{K,R}(q)=\Tr_{\H''_{K,R}}\,(-1)^F q^\Nb (-1)^{\nu_r}. \end{equation}
 To get the analogous formula for $Z_{K,R}(q)$ as a trace in $\H'_{K,R}$, we must replace $\nu_r$ by $\nu_\ell$,
 but we must also remember to replace $q$ by $-q$:
  \begin{equation}\label{modzi}Z_{K,R}(q)=\Tr_{\H'_{K,R}}\,(-1)^F (-q)^\Nb (-1)^{\nu_\ell}. \end{equation}
  These formulas reduce to what we had in section \ref{ortho} if we replace $\H'_{K,R}$ and $\H''_{K,R}$ by
  their classical approximation $\H_{0;K,R}$, but the formulas in terms of  $\H'_{K,R}$ and $\H''_{K,R}$ are much more
  significant, since those spaces are topological invariants.
    
  We have presented this construction for the orthosymplectic group $\OSp(2m+1|2n)$ for general $m,n$, but it is interesting
  to set $n=0$ and specialize to the purely bosonic group $\O(2m+1)$. 
  When $n=0$, $\nu_r=0$ in eqn. (\ref{modz}), and $\nu_\ell=\Nb$ in eqn. (\ref{modzi}), so these formulas reduce\footnote{For
  $n,m>0$, we cannot eliminate the factors $(-1)^{\zeta_\ell}$, $(-1)^{\zeta_r}$ in this way.  Perhaps we should interpret
  these factors as representing an extra $\Z_2$ grading of Khovanov homology for the orthosymplectic group $\OSp(2m+1|2n)$,
  $m,n>0$.} to
  \begin{equation}\label{mof}Z_{K,R}(q)=\Tr_{\H'_{K,R}}\,(-1)^Fq^\Nb=\Tr_{\H''_{K,R}}\,(-1)^Fq^\Nb. \end{equation}
Thus, we have identical formulas expressing the quantum knot invariants in terms of two
 Khovanov-like theories, congruent to each other mod 2, that can be used
  to study quantum knot invariants of $\O(2m+1)$.  
  
  Although the brane construction that has guided us in much of this
  paper leads to $\O(2m+1)$, from a gauge theory point of view one can replace this with another Lie group with the same
  Lie algebra, such as $\Spin(2m+1)$.  In the particular case $m=1$, we have the exceptional isomorphism $\Spin(3)\cong \SU(2)$.
  It is known \cite{ORS} that at least for the two-dimensional representation of $\SU(2)$, Khovanov homology has a cousin --
  known as odd Khovanov homology -- that gives a second way to write the quantum knot invariant as a trace in a space of
  physical states.  Odd Khovanov homology is congruent mod 2 to Khovanov homology, just as $\H'_{K,R}$ is congruent
  mod 2 to $\H''_{K,R}$.
  
  Moreover, it has been discovered recently \cite{Lauda} that odd Khovanov homology of $\SU(2)$ has a close relation to the
  orthosymplectic group $\OSp(1|2)$.  
  This was one of the main clues leading to our analysis in section 
  \ref{another} and also in the present section. 
  The relation of odd Khovanov homology to $\OSp(1|2)$ and
   the mod 2 congruences between the two pairs of theories make us suspect that our two theories correspond
   to even and odd Khovanov homology.  Our construction
  does not make completely clear which is which of $\H'_{K,R}$ and $\H''_{K,R}$, but the
  relation of quantum knot invariants to $\H''_{K,R}$  generalizes more directly to an arbitrary simple Lie group $G$, so we surmise
  that one should identify
  $\H''_{K,R}$ with ordinary Khovanov homology and $\H'_{K,R}$  with the more special odd theory.  

\vskip.7cm
\noindent{\bf Acknowledgments}  Research of EW is partly supported by  NSF Grant PHY-1314311. We thank N.~Dedushenko, S.~Pufu, D. Spielman, N. Snyder, and R. Zhang for useful discussions, and  A. Lauda for pointing out the relation of odd Khovanov homology to  the supergroup $\OSp(1|2)$.

\appendix
\section{Conventions And Supersymmetry Transformations}\label{N4SUSY}
We mostly follow the notation of \cite{Janus, 5knots}, with some minor differences. Euclidean signature is used, except in section \ref{nstype} and the beginning of section \ref{magnetic}. The Lorentz vector indices are denoted by Greek letters $\mu,\,\nu,\dots$ in four dimensions and by Latin $i,\,j,\,k$ in three dimensions. The defect is at $x^3=0$, and $x^3$ is assumed to be the normal coordinate such that $\partial_3$ is the unit normal vector at the defect. The 3d spinor indices are denoted by $\alpha,\,\beta,\,\dots$. When the indices are not shown explicitly, they are contracted as $v^\alpha w_\alpha$. They are raised and lowered with epsilon symbols,
\beqn
&&\epsilon^{12}=\epsilon_{12}=1\,,\nnr
&&v^{\alpha}=\epsilon^{\alpha\beta} v_\beta.
\eeqn
Vector and spinor notation are related by sigma-matrices,
\beqn
&&V_{\alpha\beta}=\sigma_{\alpha\beta}^i V_i=\left(\bea{cc}
-iV_2+V_3& iV_1\\
iV_1& iV_2+V_3
\eea\right).\label{sigmas}
\eeqn
With this definition, the product of the sigma-matrices is
\beq
\sigma^{i\alpha\beta}\sigma^j_{\beta\gamma}=\delta^{ij}\delta^\alpha_\gamma+\eps^{ijk}\sigma^{\alpha}_{k\gamma}.
\eeq
The boundary conditions are invariant under 3d supersymmetry, with $R$-symmetry group $\SU(2)_X\times \SU(2)_Y$. The spinor indices for these two groups are denoted by $A, B,\dots$ and $\dA, \dB,\dots$, respectively, and the vector indices are denoted by $a, b, c$ and $\dot{a},\dot{b},\dot{c}$. Conventions for the $R$-symmetry indices are the same as for the Lorentz indices. In particular, the $R$-symmetry  sigma-matrices are as in \ref{sigmas}. 

Fields that take values in the adjoint representation are understood as anti-hermitian matrices.

The three-dimensional $\cN=4$ supersymmetry acts on the fields in the following way:
\beqn
&&\delta A_i=-\fr{1}{\sqrt{2}}\veps^\alpha_{A\dB}\left(\Psi_1^{A\dB\beta}\sin\ang+\Psi_2^{A\dB\beta}\cos\ang\right)\sigma_{i\alpha\beta}\,,\nnr
&&\delta A_3=-\fr{i}{\sqrt{2}}\veps^\alpha_{A\dB}\left(-\Psi_{1\alpha}^{A\dB}\cos\ang+\Psi_{2\alpha}^{A\dB}\sin\ang\right)\,,\nnr
&&\delta X^a=-\fr{i}{\sqrt{2}}\veps^A_\dB\Psi_1^{B\dB}\sigma^a_{AB}\,,\nnr
&&\delta Y^a=\fr{i}{\sqrt{2}}\veps_A^{\dA}\Psi_2^{A\dB}\sigma^a_{\dA\dB}\,,\nonumber
\eeqn

\beqn
&&\sqrt{2}\delta\Psi_{1\alpha}^{A\dB}=\veps^{\beta B\dB}\left(-\slashed{D}_{\alpha\beta}X^A_B-\fr{i}{2}\eps_{\alpha\beta}\sin\ang[X^{AC},X_{BC}]\right)\nnr
&&~~~~~~~~~~~~~~-\veps^{A\dA}_\alpha\left(i D_3Y^{\dB}_{\dA}+\fr{i}{2}\sin\ang[Y_{\dA\dC},Y^{\dB\dC}]\right)\nnr
&&~~~~~~~~~~~~~~+i\cos\ang\veps^{C\dC}_\alpha[X^A_C,Y^\dB_\dC]+\veps^{\beta A\dB}\left(\fr{i}{2}\sin\ang\eps_{ijk}F^{ij}+\cos\ang F_{k3}\right)\sigma_{k\alpha\beta}\,,\nnr
&&\sqrt{2}\delta\Psi_{2\alpha}^{A\dB}=\veps^{\beta A\dA}\left(\slashed{D}_{\alpha\beta}Y^\dB_\dA-\fr{i}{2}\eps_{\alpha\beta}\cos\ang[Y^{\dB\dC},Y_{\dA\dC}]\right)\nnr
&&~~~~~~~~~~~~~~-\veps^{B\dB}_\alpha\left(i D_3X^{A}_{B}-\left[\fr{i}{2}\mu^A_B\delta\bigl(x^3\bigr)\right]-\fr{i}{2}\cos\ang[X^{AC},X_{BC}]\right)\nnr
&&~~~~~~~~~~~~~~-i\sin\ang\veps^{C\dC}_\alpha[X^A_C,Y^\dB_\dC]-\veps^{\beta A\dB}\left(-\fr{i}{2}\cos\ang\eps_{ijk}F^{ij}+\sin\ang F_{k3}\right)\sigma_{k\alpha\beta}\,,\nonumber
\eeqn
\beqn
&&\delta \Q ^I_\dA=-\veps^A_\dA\lambda^I_A\,,\nnr
&&\delta\lambda^I_{\alpha A}=\veps^{\beta\dA}_A i\slashed{D}_{\alpha\beta}\Q ^I_\dA-\veps_{\alpha A\dA}\omega^{IJ}\fr{\partial W_4}{\partial \Q ^J_\dA}+\veps^{B\dB}_\alpha \sin\ang X^m_{AB}T^I_{mJ}\Q ^J_\dB\,,\nnr
&&\delta Z^A=-\veps^A_\dA\zeta^\dA\,,\nnr
&&\delta \bar{Z}^A=-\veps^A_\dA\bar{\zeta}^\dA\,,\nnr
&&\delta\zeta^\dA=\veps^\dA_A i\slashed{D}{Z}^A-\veps_{B\dB}Y^{\dA\dB}Z^B\,,\nnr
&&\delta\bar{\zeta}^\dA=\veps^\dA_A i\slashed{D}\bar{Z}^A+\veps_{B\dB}\bar{Z}^BY^{\dA\dB}.\label{N4formulas}
\eeqn
The term with the moment map $\mu^A_B$ in the transformation of the  $\Psi_2$ fermion is present only for the magnetic theory. In the language of $\cN=1$ three-dimensional superfields, it comes from the $\delta(x_3)$ term in the auxiliary field $F_Y$ (see eqn.
(\ref{zumbolt})  for more details). This term propagates in all equations in combination with $D_3X^a$, canceling the delta-contribution from the discontinuity of the field $X^a$.

\section{Details On The Action And The Twisting}\label{technical1}
\subsection{Constructing The Action From $\cN=1$ Superfields}
In this section, we review the construction \cite{Janus} of the action for the D3-NS5 system. One of the reasons for discussing this in some detail is that we will need parts of it  to write out the action for the magnetic theory.

Here we work in Euclidean signature. In \cite{Janus}, the D3-NS5 action was constructed by writing an $\cN=1$ 3d supersymmetric action with a global $\SU(2)$ symmetry, and then adjusting the couplings to extend this symmetry to a product $\SU(2)_X\times \SU(2)_Y$. This group does not commute with the supersymmetry generators, and therefore extends the $\cN=1$ supersymmetry to $\cN=4$. The $\cN=1$ multiplets in the bulk are a vector multiplet\footnote{The subscript $A$ in $\fsigma_A$ is not an $R$-symmetry index.} $(A_i, \fsigma_A)$ and three chiral multiplets $(X^a,\rho^a_1, F^a_X)$, $(Y^a,\rho_2^a,F_Y^a)$ and $(A_3, \fsigma_3, F_3)$, where $X^a$ and $Y^a$ are the six scalars of the $\cN=4$ SYM\footnote{In non-$R$-symmetrized expressions, where only the diagonal subgroup of the $\SU(2)_X\times \SU(2)_Y$ is explicitly visible, it does not make sense to distinguish $\SU(2)_X$ and $\SU(2)_Y$ indices.}, and $A_3$ is a component of the gauge field. The fermionic fields can be packed into two $\cN=4$ SUSY covariant combinations
\beqn
&&\sqrt{2}\Psi_1^{A\dB}=-i\rho_1^{(A\dB)}+\eps^{A\dB}(-\sin\ang\,\fsigma_A+\cos\ang\,\fsigma_3)\,,\nnr
&&\sqrt{2}\Psi_2^{A\dB}=-i\rho_2^{(A\dB)}+\eps^{A\dB}(-\cos\ang\,\fsigma_A-\sin\ang\,\fsigma_3).\label{Psis}
\eeqn
The action of the bulk $\cN=4$ super Yang-Mills, rephrased in three-dimensional notation, has the following form,
\beqn
&&-\fr{1}{g_{\rm YM}^2}\int \rmd^4x\,\tr\left(\fr12 F_{\mu\nu}^2+(D_i X^a)^2+(D_i Y^a)^2\right.\nnr
&&+i\Psi_{1\alpha}^{A\dB}D^\alpha_\beta\Psi_{1A\dB}^\beta+i\Psi_{2\alpha}^{A\dB}D^\alpha_\beta\Psi_{2A\dB}^\beta+2\Psi_{2\alpha}^{A\dB}D_3\Psi_{1A\dB}^\alpha\nnr
&&+X^{A}_B\left(-\sin\ang([\Psi_{1}^{B\dC\alpha},\Psi_{1A\dC\alpha}]-[\Psi_2^{B\dC\alpha},\Psi_{2A\dC\alpha}])-2\cos\ang[\Psi_2^{B\dC\alpha},\Psi_{1A\dC\alpha}]\right)\nnr
&&+Y^{\dC}_{\dD}\left(-\cos\ang ([\Psi_{1}^{A\dD\alpha},\Psi_{1A\dC\alpha}]-[\Psi_{2}^{A\dD\alpha},\Psi_{2A\dC\alpha}])+2\sin\ang[\Psi_{2}^{A\dD\alpha},\Psi_{1A\dC\alpha}]\right)\nnr
&&-F_X^2-F_Y^2-F_3^2+2D_3\left(F_X Y\right)-2F_3[X,Y]\nnr
&&+F_X^a\left(-2D_3Y_a-\sin\ang\eps_{abc}([X^b,X^c]-[Y^b,Y^c])-2\cos\ang\eps_{abc}[X^b,Y^c]\right)\nnr
&&\left.+F_Y^a\left(2D_3X_a-\cos\ang\eps_{abc}([X^b,X^c]-[Y^b,Y^c])+2\sin\ang\eps_{abc}[X^b,Y^c]\right)\right)\nnr
&&+\fr{i\theta_{\rm YM}}{8\pi^2}\int \tr\left(F\wedge F\right)\nnr
&&+\int\rmd^4x\,\tr\left(\fr{\theta_{\rm YM}}{8\pi^2}\pt_3\left(\fsigma_A^2\right)-\fr{1}{g_{\rm YM}^2}\pt_3\left((\fsigma_A^2-\fsigma_3^2)\sin\ang\cos\ang-2\fsigma_3\fsigma_A\cos^2\ang\right)\right).\label{N4}
\eeqn
Here the first four lines are the usual kinetic and Yukawa terms. The next three lines contain the auxiliary fields, after eliminating which these terms will give the usual quartic $\cN=4$ super Yang-Mills potential, but they will also give some total $\pt_3$ derivatives, which we cannot drop if we want to couple the theory to the defect in a supersymmetric way. Next, there is also a theta-term, and finally in the last line there are some total derivatives of the non-$R$-symmetric combinations of fermions, which appear from rearranging the fermionic kinetic terms and from the theta-term.

For the NS5-type defect we can use (\ref{canonical}) to reduce the last line in (\ref{N4}) to
\beq
\fr{\cot\ang}{g_{\rm YM}^2}\int\rmd^4x\,\partial_3\tr\left(\fsigma_A\cos\ang+\fsigma_3\sin\ang\right)^2.\label{24}
\eeq
This term is important for $R$-symmetrizing the fermionic couplings on the boundary.

On the three-dimensional defect live chiral multiplets $(\Q^A,\lambda^A,F_\Q^A)$. In $\cN=1$ notation, the action on the defect includes a usual kinetic term for the $Q$-multiplet, a quartic superpotential~$\fr{\calK}{4\pi}{\mathcal W}_4(\cQ)$ with
\beqn
&&\mathcal{W}_4=\fr{1}{12}t_{IJ;KS}\eps^{AB}\eps^{CD}\cQ^I_A\cQ^J_B\cQ^K_C\cQ^S_D\,,\nnr
&&t_{IJ;KS}=\fr{1}{4}{\kappa}^{mn}\left(\tau_{mIK}\tau_{nJS}-\tau_{mIS}\tau_{nJK}\right)\,,
\eeqn
and a superpotential that couples the four-dimensional scalar $X^a$ to the defect theory,
\beq
\mathcal{W}_{\Q X\Q}=-\fr{\calK}{4\pi}\sin\ang \cQ^{IA} \mathcal{X}^m_{AB}\tau_{mIJ}\cQ^{JB}.
\eeq
This choice of the superpotential corresponds to the case when the $NS5$-brane is stretched in directions $456$. Indeed, the bifundamental fields will have a mass term proportional to $X^2$, {\it i.e.} their mass is proportional to the displacement in these directions. 

The boundary conditions of the theory form a current multiplet of three-dimensional $\cN=4$ supersymmetry,
\beqn
&&Y^m_{\dA\dB}=-\fr{1}{2\cos\ang} \tau^m_{IJ}\Q^I_{\dA}\Q^J_{\dB}\,,\nnr
&&\sqrt{2}\Psi^m_{2\alpha A\dB}=\fr{i}{\cos\ang} \tau^m_{IJ}\lambda^I_{\alpha A}\Q^J_{\dB}\,,\nnr
&&\sin\ang F^m_{k3}-\fr{i}{2}\cos\ang\eps_{ijk}F^m_{ij}=-\fr{2\pi}{\cos\ang}\kappa^{mn} J_{nk}\,,\nnr
&&D_3X^m_{a}-\fr12\cos\ang\eps_{abc} f^m_{np} X^{bn}X^{cp}=\fr12\tan\ang\,\omega_{IJ}\eps_{\dA\dB}X^n_{a}T^{mI}_{K}T^{J}_{nS}\Q^{K\dA}\Q^{S\dB}\nnr
&&\hspace{60mm}-\fr{1}{4\cos\ang}\lambda^I_A \sigma^{AB}_a\tau^m_{IJ}\lambda^J_B\,,\label{bc}
\eeqn
where $J_{mk}$ is the current
\beq
J_{mi}=\fr{\delta I_\Q }{\delta A^{im}}=\fr{1}{4\pi}\tau_{mIJ}\left(\eps^{\dA\dB}\Q_\dA^ID_i\Q^J_\dB+\eps^{AB}\fr{i}{2}\lambda^I_{A}\sigma_i\lambda^J_{B}\right).\label{current}
\eeq
The first of the boundary conditions has the following origin. At stationary points of the action the auxiliary field $F^a_X$ has a contribution from the boundary, proportional to the delta function. Then the term $F_X^2$ would produce a square of the delta function. To avoid this and to make sense of the action, the boundary contribution to $F^a_X$ should be set to zero, and this gives the boundary condition for the field $Y^\da$. The other three boundary conditions can be obtained in a usual way from the variation of the action, after eliminating the auxiliary fields.

The complete action after eliminating the auxiliary fields is
\beqn
I_{\rm electric}&=&I_{\rm SYM}+\fr{i\theta_{\rm YM}}{2\pi}\CS(A)+\calK I_\Q (A)\nnr
&&+\fr{\calK}{4\pi}\int\rmd^3x\left(\fr{1}{2}\sin^2\ang\omega_{IJ}\eps_{\dA\dB}X^{ma}X^{na}T^{I}_{mK}T^{J}_{nS}\Q^{K\dA}\Q^{S\dB}-\frac{1}{2}\sin\ang\lambda^I_A X^{mAB}\tau_{mIJ}\lambda^J_B\right)\nnr
&&+\fr{1}{g_{\rm YM}^2}\int\rmd^3x\,\Tr\left(-\fr{2}{3}\eps_{abc}\cos\ang X^aX^bX^c-\fr{2}{3}\eps_{abc}\sin\ang Y^aY^bY^c+2\Psi_1^{A\dB}\Psi_{2A\dB}\right)\,,\label{actionform}
\eeqn
where
\beqn
&&I_\Q (A)=\fr{1}{4\pi}\int\rmd^3x\left(\fr{1}{2}\eps^{\dA\dB}\omega_{IJ}D_i\Q^I_\dA D^i\Q^J_\dB-\fr{i}{2}\eps^{AB}\omega_{IJ}\lambda^{I}_A\slashed{D}\lambda^J_{B}\right.\nnr
&&\left.+\fr{1}{4} {\kappa}^{mn}\tau_{mIJ}\tau_{nKS}\Q^{I\dA}\Q^K_\dA\lambda^{JC}\lambda^S_{C}+\fr{1}{2}\eps_{\dA\dB}\omega^{IJ}\fr{\partial W_4}{\partial \Q^I_\dA}\fr{\partial W_4}{\partial \Q^J_\dB}\right).\label{SQ}
\eeqn
is the $\cN=4$ super Chern-Simons action with the CS term omitted. 

Before proceeding to twisting, it is useful to remove the term $\lambda X\lambda$ in the action, using the last line in the boundary conditions\footnote{One might be worried that after this transformation the action no longer gives the same boundary conditions from the boundary variation. In section ~\ref{bcsection} we will make our argument more accurate.} (\ref{bc}). Then the action is
\beqn
I_{\rm electric}&=&I_{\rm SYM}+\fr{i\theta_{\rm YM}}{2\pi}\CS(A)+\calK I_\Q (A)\nnr
&&+\fr{\calK}{4\pi}\int\rmd^3x\left(-\fr{1}{2}\sin^2\ang\omega_{IJ}\eps_{\dA\dB}X^{ma}X^{na}T^{I}_{mK}T^{J}_{nS}\Q^{K\dA}\Q^{S\dB}\right)\nnr
&&+\fr{1}{g_{\rm YM}^2}\int\rmd^3x\,\Tr\left(-\fr{2}{3}\eps_{abc}\cos\ang X^aX^bX^c-\fr{2}{3}\eps_{abc}\sin\ang Y^aY^bY^c+2\Psi_1^{A\dB}\Psi_{2A\dB}\right)\nnr
&&-\fr{2}{g_{\rm YM}^2}\int\rmd^3x\,\Tr\left(X^aD_3X_a-\cos\ang\eps_{abc}X^aX^bX^c\right).\label{action2}
\eeqn
The supersymmetry transformations for this theory can be found by $R$-symmetrization of the $\cN=1$ supersymmetry transformations, or, for the bulk super Yang-Mills fields, by reduction from the $\cN=4$ formulas in four dimensions. The result can be found in Appendix \ref{N4SUSY}.

\subsection{Twisted Action}\label{tbcast}
Now we would like to twist the theory and to couple it to the metric. Let us recall, what is the set of fields of our topological theory. The four scalars $X^a$ and $Y^1$ of the bulk super Yang-Mills become components of a 1-form $\phi$, and the other two scalars are combined as $\sigma=\fr{Y_2-iY_3}{\sqrt{2}}$ and $\bar{\sigma}=\fr{Y_2+iY_3}{\sqrt{2}}$. The fermions of the twisted bulk theory are \cite{Langlands} two scalars $\eta$ and $\tilde{\eta}$, two one-forms $\psi$ and $\tilde{\psi}$, and a 2-form $\chi$. The selfdual and anti-selfdual parts of the two forms are denoted by $\pm$ superscripts. These fermions are related to the fields of the physical theory as follows,
\beqn
2\sqrt{2}\Psi_1^{\alpha A\dA}&=&(\tilde{\eta}-t^{-1}\eta)\eps^{\alpha A}v^\dA+(-\tilde{\psi}-t\psi)_3\eps^{\alpha A}u^\dA+\nnr
&&+2(t^{-1}\chi^++\chi^-)_{i3}\sigma_i^{\alpha A}v^\dA+(\tilde{\psi}-t\psi)_i\sigma_i^{\alpha A}u^{\dA}\,,\nnr
-2\sqrt{2}i\Psi_2^{\alpha A\dA}&=&(-\tilde{\eta}-t^{-1}\eta)\eps^{\alpha A}v^\dA+(-\tilde{\psi}+t\psi)_3\eps^{\alpha A}u^\dA+\nnr
&&+2(t^{-1}\chi^+-\chi^-)_{i3}\sigma_i^{\alpha A}v^\dA+(\tilde{\psi}+t\psi)_i\sigma_i^{\alpha A}u^\dA.\label{fermcomp}
\eeqn
Here is a summary of $\Qb$-transformations of the bulk fields, as derived in \cite{Langlands},
\begin{alignat}{2}
&\delta A=it\tilde{\psi}+i\psi\,,&\qquad&\delta\phi=-i\tilde{\psi}+it\psi\,,\nnr
&\delta\sigma=0\,,&\quad&\delta\bar{\sigma}=it\tilde{\eta}+i\eta\,,\nnr
&\delta\eta=tP+[\bar{\sigma},\sigma]\,,&\quad&\delta\tilde{\eta}=-P+t[\bar{\sigma},\sigma]\,,\nnr
&\delta\psi=D\sigma+t[\phi,\sigma]\,,&\quad&\delta\tilde{\psi}=tD\sigma-[\phi,\sigma]\,,\nnr
&\delta\chi=H\,,
\end{alignat}
where on-shell
\beq
P=D^\mu\phi_\mu\,,\quad H^+=\mathcal{V}^+(t)\,,\quad H^-=t\mathcal{V}^-(t)
\eeq
and
\beqn
&&\mathcal{V}^+(t)=\left(F-\phi\wedge\phi+tD\phi\right)^+\,,\nnr
&&\mathcal{V}^-(t)=\left(F-\phi\wedge\phi-t^{-1}D\phi\right)^-.
\eeqn

 As it was described in \cite{Langlands}, the manifestly $\Qb$-invariant topological action for the bulk super Yang-Mills theory contains a topological term and a $\Qb$-variation of a fermionic expression (see section 3.4 of that paper). In our case the theory is defined on the two half-spaces with the defect~W between them, and therefore the equations have to be completed with some boundary terms:
\beqn
I_{\rm SYM}&=&\bigl\{\Qb,\dots\bigr\}-\fr{t-t^{-1}}{t+t^{-1}}\fr{4\pi}{g_{\rm YM}^2}\,\CS(A)\nnr
&+&\fr{1}{g_{\rm YM}^2}\int_W\Tr\left(\fr{4}{t+t^{-1}}\left(F\wedge\phi-\fr{1}{3}\phi\wedge\phi\wedge\phi\right)+\fr{t-t^{-1}}{t+t^{-1}} \phi\wedge D\phi\right)\nnr
&+&\fr{1}{g_{\rm YM}^2}\int_W\rmd^3x\sqrt{\gw}\,\Tr\left(2\bar{\sigma}D_3\sigma+\gw^{ij}\phi_iD_j\phi_3-\gw^{ij}\phi_3 D_j\phi_i\right).\label{QSYM}
\eeqn
Let us give some explanations on this formula. Recall that in our notation, $I_{\mathrm{SYM}}$ is the part of the bulk super Yang-Mills action, which is proportional to $1/g_{\rm SYM}^2$, -- that is, with the $\theta_{\mathrm{YM}}$-part omitted. Here and in what follows we ignore expressions on $W$ bilinear in the bulk fermions, because in the end they have to cancel by supersymmetry, anyway. As usual, the Chern-Simons form $CS(A)$ is just a notation for the bulk topological term. By $\gw$ we denote the induced metric on W. The third component of various bulk tensors on the boundary is defined as a contraction of these tensors with a unit vector field $n^\mu$, normal to the defect. For example, $D_j\phi_3$ means a pullback to W of a one-form $n^\nu D_\mu\phi_\nu$. 

The first line in the expression above is the formula that was used in \cite{Langlands}. The coefficient of the topological term in this expression adds with the usual theta parameter $\theta_{\rm{YM}}$ to become the canonical parameter, which we called $\calK$. The second line in this formula is what appeared in the purely bosonic Chern-Simons case \cite{5knots}. Finally, the last line was dropped in that paper as a consequence of the boundary conditions, but in our case it is non-zero.

A useful transformation is to integrate by parts in the last line of (\ref{QSYM}) to change $-\phi_3D^i\phi_i$ into another $\phi^iD_i\phi_3$, but in doing so we have to remember that the metric connection in the covariant derivatives is four-dimensional. Because of this, the integration by parts produces a curvature term
\beq
\fr{1}{g_{\rm YM}^2}\int_W\rmd^3x\sqrt{\gw}\,\Tr\left(-s_{ij}\phi^i\phi^j+s^i_i\phi_3\phi_3\right)\,,
\eeq
where $s_{ij}$ is the second fundamental form of the hypersurface W. This curvature term should be canceled by adding a curvature coupling to the last line in (\ref{action2}).

We will substitute what we have just learned about $I_{\mathrm{SYM}}$ into the action (\ref{action2}) of the theory, but first let us make some transformations of the action (\ref{action2}). We would like to complexify the gauge field in the hypermultiplet action $I_\Q (A)$. The seagull term for $(D\Q)^2$ comes from $XX\Q\Q $ in the second line of (\ref{action2}). To change the terms linear in the gauge field we need to add and subtract $i\sin\ang X$ times the boundary current~(\ref{current}). Using the third of the boundary conditions (\ref{bc}), the current can be rewritten as a combination of gauge field strengths. After these manipulations, a twisted version of (\ref{action2}) will look like
\beqn
I_{\rm electric}&=&I_{\rm SYM}+\fr{i\theta_{\rm YM}}{2\pi}\CS(A)+\calK I_\Q (\cAb)+\fr{1}{g_{\rm YM}^2}\int\rmd^3x\,\Tr\left(-\fr{2}{3}\cos\ang \phi\wedge\phi\wedge\phi\right)\nnr
&&+\fr{1}{g_{\rm YM}^2}\int\rmd^3x\,\sqrt{\gw}\,\Tr\left(-i\sin\ang \phi_3[\bar{\sigma},\sigma]-2\phi^iD_3\phi_i-2i\sin\ang\,\phi^i F_{i3}+s_{ij}\phi^i\phi^j-s^i_i\phi_3\phi_3\right)\nnr
&&+\fr{2\cos\ang}{g_{\rm YM}^2}\int\rmd^3x\,\Tr\left(\phi\wedge\phi\wedge\phi-\phi\wedge F\right).\label{action3}
\eeqn

Now we substitute here the expression (\ref{QSYM}) for the super Yang-Mills action. The Chern-Simons term in (\ref{QSYM}) changes the coefficient in front of the Chern-Simons term in (\ref{action3}) from $\theta_{\rm YM}/{2\pi}$ to $\calK$. Expression in the second line in (\ref{QSYM}) and the term with $\phi\wedge\phi\wedge\phi$ in the first line of (\ref{action3}) combine with the Chern-Simons term, changing the gauge field in it from $A$ into complexified gauge field $\cAb$, as shown in \cite{5knots}. We are left with the following action,
\beqn
&&I_{\rm electric}=\bigl\{\Qb,\dots\bigr\}+i\calK\CS(\cAb)+\calK I_\Q (\cAb)\nnr
&&+\fr{1}{g_{\rm YM}^2}\int \rmd^3 x\,\sqrt{\gw}\,\Tr\left(-i\sin\ang \phi_3[\bar{\sigma},\sigma]+2\bar{\sigma}D_3\sigma\right)\nnr
&&+\fr{1}{g_{\rm YM}^2}\int \rmd^3 x\,\sqrt{\gw}\,\Tr\left(-2\phi^iD_3\phi_i+2\phi^iD_i\phi_3-2i\sin\ang\,\phi^i F_{i3}\right)\nnr
&&+\fr{2\cos\ang}{g_{\rm YM}^2}\int\Tr\left(-\phi\wedge F+\phi\wedge\phi\wedge\phi\right).
\eeqn
We are almost done. All we need to show is that the last three lines here are $\Qb$-exact. This is indeed so (again, we ignore the fermion bilinears):
\beqn
&&\int\rmd^3x\,\sqrt{\gw}\,\Tr\left(\bar{\sigma}D_3\sigma\right)=-\fr{1}{2\cos\ang}\left\{\Qb,\int\rmd^3x\,\sqrt{\gw}\,\Tr\left(\bar{\sigma}(t^{-1}\psi_3+\tilde{\psi}_3)\right)\right\}\,,\nnr
&&\int\rmd^3x\,\sqrt{\gw}\,\Tr\left(\phi_3[\bar{\sigma},\sigma]\right)=-\fr{1}{2\cos\ang}\left\{\Qb,\int\rmd^3x\,\sqrt{\gw}\,\Tr\left(\bar{\sigma}(t^{-1}\tilde{\psi}_3-\psi_3)\right)\right\}\,,\label{Wexact}\\
&&\int\Tr\left(\phi\wedge\left(\star D\phi-i\sin\ang\,\star F-\cos\ang (F-\phi\wedge\phi)\right)\right)=\left\{\Qb,\int\Tr\left(\phi\wedge(t^{-1}\chi^++\chi^-)\right)\right\}.\nonumber
\eeqn

Up to $\Qb$-exact terms, our action is the sum of the Chern-Simons term and the twisted action~$I_\Q(\cAb)$. This combination is just the (twisted) action of the $\cN=4$ Chern-Simons theory. Let us see, how it is related~\cite{KapustinSaulina} to the Chern-Simons theory with a supergroup. We define the fields of the twisted theory as
\beqn
&&\Q^\dA = iv^\dA \bar{C}  + \fr{1}{2} u^\dA C\,,\nnr
&&\lambda^{\alpha A}=-\fr{i}{2}\epsilon^{\alpha A} B +i\sigma^{i\alpha A} \cAf_{i}.\label{twistedQ}
\eeqn
Substituting this into the action and using the explicit form (\ref{SQ}) of $I_\Q(A)$, one finds,
\beq
i\calK\,\CS(\cAb)+\calK\, I_\Q (\cAb) = i\calK\, \CS(\cA)+i\calK I_{\rm g.f.}\,,
\eeq
where $\cA=\cAb+\cAf$ is the complexified superconnection. The $\Qb$-exact gauge fixing term $I_{\rm g.f.}=\left\{\Qb,V_{\rm g.f.}\right\}$ for the fermionic part of the superalgebra is
\beqn
I_{\rm g.f.}&=&\int\rmd^3x\,\sqrt{\gw}\, \Str\left(-\mathcal{D}_b^iB\cAf_{i} + \mathcal{D}_b^i\bar{C}\mathcal{D}_{bi}C+\{\cAf,\bar{C}\}\{\cAf,C\}\right.\nnr
&+&\left.\fr{1}{4}\{\bar{C},B\}\{C,B\}+\fr{1}{16}[C,\{\bar{C},\bar{C}\}][\bar{C},\{C,C\}]\right)\,,\label{CSgfix}\\
V_{\rm g.f.}&=&\int \rmd^3x\,\sqrt{\gw}\,\Str\left(-\mathcal{D}^i_b\bar{C}\cAf_{i}+\fr{1}{8}\{\bar{C},\bar{C}\}\{C,B\}\right).\nonumber
\eeqn

\subsection{Boundary Conditions}\label{bcsection}
Let us rewrite the boundary conditions (\ref{bc}) in terms of fields of the twisted theory. The first line of that formula gives
\beq
\sigma=\fr{i}{2}\fr{1}{1+t^2}\{C,C\}\,,\quad \bar\sigma=\fr{i}{1+t^{-2}}\{\bar{C},\bar{C}\}\,,\quad \phi_3=-\fr{1}{t+t^{-1}}\{C,\bar{C}\}.\label{bcb1}
\eeq
These three formulas are related to one another by $\SU(2)_Y$ rotations. The boundary condition for the fermion in (\ref{bc}) gives one new relation
\beq
t^{-1}\chi^+_{i3}-\chi^-_{i3}=\frac{2}{t+t^{-1}}\{\cAf_{i},\bar{C}\}\,,\label{bcf1}
\eeq
two relations, that can be obtained from (\ref{bcb1}) by $\Qb$-transformations
\beq
\tilde{\eta}+t^{-1}\eta=\frac{2}{t+t^{-1}}\{B,\bar{C}\}\,,\quad -\tilde{\psi}_3+t\psi_3=\frac{i}{t+t^{-1}}\{B,C\}\,,
\eeq
and one relation which comes from the bulk and boundary $\Qb$-variation of the gauge field $\cAb$, which we have already discussed,
\beq
 \tilde{\psi}_i+t\psi_i=-\frac{2i}{t+t^{-1}}\{\cAf_{i},C\}.
\eeq
The third line in (\ref{bc}) gives boundary condition for the gauge field,
\beq
\cos\ang\,\imath^*\left(i\sin\ang\star F+\cos\ang F\right)=-\cAf\wedge\cAf+\fr{1}{2}\star_3\left(\{C,D\bar{C}\}-\{\bar{C},DC\}+[B,\cAf]\right).\label{bcg1}
\eeq
The twisted version of the last line in (\ref{bc}) is a long expression with a contribution from the curvature coupling. It can be somewhat simplified by subtracting a $D_i$ derivative of the boundary condition (\ref{bcb1}) for $\phi_3$. The result is the following,
\beq
\cos\ang\imath^*\left(\star D\phi+\cos\ang\,\phi\wedge\phi\right)=-\cAf\wedge\cAf+
\fr{1}{2}\star_3\left(D\{C,\bar{C}\}+i\sin\ang\left(\{\bar{C},[\phi,C]\}-\{C,[\phi,\bar{C}]\}\right)-[B,\cAf]\right).\label{bcb2}
\eeq
If we subtract (\ref{bcb2}) and (\ref{bcg1}), we get just a $\Qb$-variation of the fermionic boundary condition (\ref{bcf1}). A new relation results, if we add these two:
\beq
\cFb+\cAf\wedge\cAf=\star_3\{C,D\bar{C}-i\sin\ang[\phi,\bar{C}]\}-\left\{\Qb,\chit\right\}\,,
\eeq
where we defined 
\beq
\chit=\fr{t^{-2}-3}{4}\chi^++\fr{t^2-3}{4}\chi^-.
\eeq
$\Qb^2$ acts as a gauge transformation with parameter $-i(1+t^2)\sigma$ in the bulk and with parameter $\{C,C\}/2$ on the defect (\ref{Q1}), (\ref{Q2}). This agrees with the boundary conditions.

The $\Qb$-transformations of the set of boundary ghosts $\bar{C}$, $C$ and $B$ were given in ($\ref{Q2}$). To fix the residual gauge symmetry in perturbation theory, we introduce the usual ghosts $c$, $\bar{c}$ and the Lagrange multiplier field $b$, and the BRST-differential $\Qb_{\rm bos}$, associated to this gauge fixing. This differential acts on all fields in the usual fashion. The topological differential $\Qb$ acts trivially on $b$ and $\bar{c}$, but generates the following transformation, when acting on $c$:
\beq
\delta c=i(1+t^2)\sigma.
\eeq
On the boundary, this corresponds to \cite{KapustinSaulina}
\beq
\delta c=-\fr{1}{2}\{C,C\}.
\eeq
The full BRST differential in the gauge fixed theory is the sum $\Qb+\Qb_{\rm bos}$. This operator squares to zero,  and in the boundary theory it corresponds to the usual gauge fixing for the full supergroup gauge symmetry.

Finally, let us comment on the fact that we used the boundary conditions to transform the action (to pass from (\ref{SQ}) to (\ref{action2}), and then to get (\ref{action3})). We did it to exploit more directly the relation to the $\cN=4$ Chern-Simons theory, but that transformation was not really necessary. Indeed, the terms that came from using the boundary conditions gave essentially the last line in the list (\ref{Wexact}) of $\Qb$-exact expressions.  The combination of the boundary conditions that we used was just a $\Qb$-variation of the boundary condition for the $\chi$ fermion (\ref{bcf1}). (More precisely, this combination differs by a derivative of (\ref{bcb1}), but this is fine, since the boundary condition (\ref{bcb1}) is Dirichlet.) So we could equally well keep the expressions that involved the hypermultiplet fields, instead of transforming them into the bulk fields, and this would give $\Qb$-exact expressions as well.

\section{Details On The Magnetic Theory}\label{tech2}
\subsection{Action Of The Physical Theory}
Here we would like to give some details on the derivation of the action and the boundary conditions for the D3-D5 system, with equal numbers if the D3-branes in the two sides of the D5-brane. This action has been constructed in \cite{Ooguri}, but our treatment of the boundary conditions is slightly different.

As in the electric theory, we write the action in the three-dimensional $\cN=1$ formalism. The bulk super Yang-Mills part of the action has been given in (\ref{N4}) (one should set $\ang$ to $\pi$ in that formula). On the defect there is a fundamental hypermultiplet $(Z^A,\zeta^A,F^A)$, where the first two fields have already appeared in our story, and $F^A$ is the auxiliary field. Besides the usual kinetic term, the boundary action contains a superpotenial that couples the bulk and the boundary fields,
\beq
{\mathcal W}_{\bar{Z}YZ}=-\bar{\mathcal{Z}}_A \mathcal{Y}^A_B\mathcal{Z}^B.\label{SdualW}
\eeq
This superpotential has been chosen in such a way as to make the boundary interactions invariant\footnote{As we have said, we choose $t^\vee=1$. For $t^\vee=-1$ the sign of the superpotential would be the opposite.} under the full $SO(3)_X\times SO(3)_Y$ $R$-symmetry group. Specifically, the boundary action contains Yukawa couplings $-i\bar{\zeta}_A\fsigma_AZ^A+i\bar{Z}_A\fsigma_A\zeta^A$ coming from the kinetic term, and $\bar{Z}_A\rho^{AB}_{2}\zeta^B+\bar{\zeta}_A\rho^{AB}_{2}Z_B$ from the superpotential. They can be packed into $R$-symmetric couplings
\beq
i\sqrt{2}\left(\bar{Z}_A\Psi_2^{A\dB}\zeta_\dB+\bar{\zeta}_\dB\Psi^{A\dB}_2Z_A\right)\,,
\eeq
where the $\cN=4$ fermion $\Psi^{A\dB}_{2}$ was defined in (\ref{Psis}). 

The superpotential contains a coupling of the auxiliary field $F_Y$ to the moment map $\mu_m^a$, which was defined in (\ref{momentmap}). This coupling will add a delta-function contribution to the equation for the auxiliary field,
\beq
F_Y^{am}=D_3X^{am}+\fr{1}{2}\eps^{abc}([X_b,X_c]-[Y_b,Y_c])^m-\fr{1}{2}\mu^{am}\delta(x^3).\label{zumbolt}
\eeq
The square of the auxiliary field in the Yang-Mills action would produce a term with a square of this delta-function. To make this contribution finite, we require the scalars $X^a$ to have a discontinuity across the defect. This discontinuity equation extends via the supersymmetry to a set of equations for two three-dimensional current multiplets,
\beqn
&&X^{am}\bigr|^+_-=\fr{1}{2}\mu^{am}\,,\nnr
&&\sqrt{2}\Psi_{1m}^{A\dB}\bigr|^+_-={i}\left(\bar\zeta^\dB T_m Z^A+\bar{Z}^A T_m \zeta^\dB\right)\,,\nnr
&&F_{3i}^m\bigr|^+_-=\fr{1}{2}J^m_i\,,\nnr
&&D_3Y^{a}_m\bigr|^+_-=\fr{1}{2}\left(\bar{Z}_A\{Y^a,T_m\}Z^A-\bar{\zeta}^\dA T_m\zeta^\dB\sigma^a_{\dA\dB}\right)\,,\label{gluing}
\eeqn
where the current is
\beq
J_{mi}=\fr{\delta I^\vee_{\mathrm{hyp}}}{\delta A^{mi}}=-\bar{Z}_A T_mD_iZ^A+D_i\bar{Z}_AT_mZ^A-i\bar{\zeta}^\alpha_\dA T_m\sigma^{\beta}_{i\alpha} \zeta^\dA_\beta.
\eeq
Next we have to substitute expressions for all the auxiliary fields into the Lagrangian, and make it manifestly $R$-symmetric. Also, we would like to rearrange the action in such a way that the squares of the delta-function would not appear. In the Yang-Mills action (\ref{N4}) there is a potentially dangerous term $F_Y^2$, but with the gluing conditions (\ref{gluing}) it is non-singular and produces no finite contribution at $x_3=0$. Then for this term we can replace the $x_3$-integral over $\RR$ by an integral over $x_3<0$ and $x_3>0$. The term $F_X^2$ is also non-singular, so we delete the plane $x^3=0$ in the same way. There is a singular term $D_3(F_XY)$, but in can be dropped as a total derivative. The total $\pt_3$ derivative of the non-$R$-symmetric fermion combination in (\ref{N4}) can be dropped in the same way. There is also a delta-function contribution from the $D_3$ part of the fermionic kinetic term. Collecting all the boundary terms in the integrals with $x_3=0$ deleted, we get a simple action
\beqn
I_{\rm magnetic}&=&I_{\rm SYM}+\fr{i\theta_\YM^\vee}{8\pi^2}\int\tr\left(F\wedge F\right)\nnr
&+&\fr{1}{(g^\vee_\YM)^2}\int\rmd^3x\left(D_i\bar{Z}_AD^iZ^A-i\bar{\zeta}_\dA\slashed{D}\zeta^\dA-\bar{\zeta}_\dA Y^\dA_\dB\zeta^\dB-\bar{Z}_AY^aY_aZ^A\right)\nnr
&+&\fr{1}{(g^\vee_\YM)^2}\int \rmd^3 x\,\fr{2}{3}\tr\left(\eps_{abc}(X^aX^bX^c)\bigr|^+_-\right).\label{SDaction}
\eeqn
Here in $I_{\rm SYM}$ the usual super Yang-Mills Lagrangian in the bulk is integrated over the two half-spaces $x_3<0$ and $x_3>0$, with the hyperplane $x_3=0$ deleted. On the defect the $\bar{Z}YYZ$ terms from the superpotential combined with the $XYY$ term from the bulk action into an $R$-symmetric coupling. The Yukawa terms $\bar{\zeta}\Psi_2Z+\bar{Z}\Psi_2\zeta$ canceled with the delta-contribution from the bulk fermionic kinetic energy.

\subsection{Action Of The Twisted Theory}\label{acttwist}
From the action of supersymmetry (\ref{N4formulas}) one finds the following $\Qb$-transformations for the boundary fields of the twisted theory,
\beqn
&&\delta Z=-2i\zeta_u\,,\nnr
&&\delta\bar{Z}=-2i\bzeta_u\,,\nnr
&&\delta\zeta_u=\sigma Z\,,\nnr
&&\delta\bzeta_u=-\bar{Z}\sigma.
\eeqn
The two other fermions transform as $\delta \zeta_v=f$ and $\delta\bzeta_v=\bar{f}$, where
\beqn
&&f=\slashed{D}Z+\phi_3 Z\,,\nnr
&&\bar{f}=\slashed{D}\bar{Z}-\bar{Z}\phi_3\,,
\eeqn
but with these transformation rules the algebra does not close off-shell. For this reason we introduce two auxiliary bosonic spinor fields $F$ and $\bar{F}$, for which the equations of motion should impose $F=f$ and $\bar{F}=\bar{f}$. The topological transformations are then
\beqn
&&\delta\zeta_v=F\,,\nnr
&&\delta\bzeta_v=\bar{F}\,,\nnr
&&\delta F=-2i\sigma\zeta_v\,,\nnr
&&\delta \bar{F}=2i\bzeta_v\sigma.
\eeqn
The transformation rules for the auxiliary fields were chosen in a way to ensure that the square of the topological supercharge acts by the same gauge transformation, by which it acts on the other fields.

Now we would like to prove our claim that the action of the magnetic theory is $\Qb$-exact (\ref{Imagnetic}), up to the topological term. The first step is to notice that the following identity holds, up to terms bilinear in the bulk fermions,
\beqn
&&\int\rmd^3x\sqrt{\gamma}\left(D_i\bar{Z}_\alpha D^iZ^\alpha-i\bar{\zeta}_\dA\slashed{D}\zeta^\dA+\bar{\zeta}_\dA Y^\dA_\dB\zeta^\dB+\bar{Z}_\alpha\left(-\phi^2_3-\{\bar{\sigma},\sigma\}+\fr{1}{4}R\right)Z^\alpha\right)\nnr
&&=\left\{\Qb,\int\rmd^3x\sqrt{\gamma}\left(\left(\fr{1}{2}\bar{F}-\bar{f}\right)\zeta_v+\bzeta_v\left(\fr{1}{2}F-f\right)+\bar{Z}\bar{\sigma}\zeta_u-\bzeta_u\bar{\sigma}Z\right)\right\}\nnr
&&+\int\rmd^3x\sqrt{\gamma}\,\tr\left(\phi_3D_i\mu^{i}\right)-\int\rmd^3x\,\tr\left(F\wedge\mu\right).\label{lichnerowicz}
\eeqn
In the first line $R$ is the scalar curvature of the three-dimensional metric $\gamma_{ij}$, which appears in this equation from the Lichnerowicz identity.

We can apply this formula to the action (\ref{SDaction}) of the theory, after adding appropriate curvature couplings. We see that there are several unwanted terms, which are not $\Qb$-exact. They come from the last line in the identity (\ref{lichnerowicz}), from the boundary terms in the Yang-Mills action (\ref{QSYM}), and, finally, there is a cubic $XXX$ term in (\ref{SDaction}). Using the Dirichlet boundary condition (\ref{discX}), we see that most of these terms cancel. What is left is the $\tr(\bar{\sigma}D_3\sigma|^\pm)$ term from the super Yang-Mills action (\ref{QSYM}), but this term is $\Qb$-exact (after adding appropriate fermion bilinear), as we noted in (\ref{Wexact}). So the only non-trivial term in the action of the magnetic theory is the topological term. This is, of course, what one would expect, since in the electric theory we are integrating the fourth descendant of the scalar BRST-closed observable $\tr\sigma^2$. In the $S$-dual picture this should map to the fourth descendant of the analogous scalar operator, which gives precisely the topological term.

Let us comment on the role of the discontinuity equations (\ref{gluing}) in the localization computations. In fact, only the first condition in (\ref{gluing}) should be explicitly imposed on the solutions of the localization equations. Indeed, one can show with some algebra that the last two conditions in that formula follow from the first one automatically, if the localization equations $\{\Qb,\lambda\}=0$ for every fermion are satisfied.

\section{Local Observables}\label{finicky}
In a topological theory of cohomological type (see \cite{coho} for an introduction), 
there generally are interesting local observables.  In fact, typically there are $\Qb$-invariant zero-form
observables (local operators that are inserted at points) and also $p$-form
observables which must be integrated over $p$-cycles to  achieve $\Qb$ invariance. They are derived from the local observables by a  ``descent'' procedure.

We will describe here the local observables in our problem and the descent procedure.  In the magnetic description, everything
proceeds in a rather standard way, so we have little to say. 
The action of electric-magnetic duality on local observables is also straightforward.  The zero-form operators of the electric theory
are gauge-invariant
polynomials in $\sigma$, as we discuss below, and duality maps them to the corresponding gauge-invariant polynomials in $\sigma^\vee$;
the duality mapping of $k$-form observables is then determined by applying the descent procedure on both sides of the duality.
We focus here on the peculiarities of the electric description that reflect the fact that there are two different
gauge groups on the two sides of a defect. 

First we recall what happens in bulk, away from the defect.  The theory has a complex adjoint-valued scalar $\sigma$ (defined in eqn. (\ref{tork}))
that has ghost number 2  (that is, charge 2 under $\U(1)_F$).  This ensures that $\{\Qb,\sigma\}=0$, as super Yang-Mills theory has no field
of dimension $3/2$ and ghost number 3 (the elementary fermions have ghost number $\pm 1$).  The gauge-invariant and $\Qb$-invariant
local operators are simply the gauge-invariant polynomials in $\sigma$.  For a semisimple Lie group of rank $r$, it is a polynomial ring with $r$ generators.
To be concrete, we consider gauge group $\U(n)$, in which the generators are  $\O_k =\fr{1}{k}\tr\,\sigma^k$, $k=1,\dots,n$.  These are the basic
$\Qb$-invariant local observables.
  
In a topological field theory, one would expect that it does not matter at what point in spacetime the operator  $\O_k$ is inserted.
This follows from the identity
\begin{equation}\d \O_p=\left\{\Qb,\fr{1}{2}\tr\,\left(\sigma^{p-1}(t^{-1}\tilde{\psi}+\psi)\right)\right\}, \label{zodoc}\end{equation}
where $\d=\sum \d x^\mu\partial_\mu$ is the exterior derivative, and $\psi$ and $\tilde{\psi}$ are fermionic one-forms. (See Appendix \ref{tbcast} for a list of fields of the bulk theory and their $\Qb$-transformations.) This identity, which says that the derivative of $\O_k$ is $\Qb$-exact,
is actually the first in a hierarchy \cite{WitDon}.  If we rename $\O_k$
as $\O_k^{(0)}$ to emphasize the fact that it is a zero-form valued operator, then for each $k$, 
there is a hierarchy of $s$-form valued operators $\O_k^{(s)}$, $s=0,\dots,4$, obeying
\begin{equation}\label{ocon}\d \O_k^{(s)}=[\Qb,\O_k^{(s+1)}\}.   \end{equation}
Construction of this hierarchy is sometimes called the descent procedure.
This formula can be read in two ways.  If $\Sigma_s$ is a closed, oriented $s$-manifold in $W$, then 
$I_{k,\Sigma_s}=\int_{\Sigma_s}\O_k^{(s)}$ is a $\Qb$-invariant
observable, since
\begin{equation} \left[\Qb,\int_{\Sigma_s}\O_k^{(s)}   \right\}=\int_{\Sigma_s}\d\O_k^{(s-1)}=0.  \label{telf}\end{equation}
And $I_{k,\Sigma_s}$ only depends, modulo $[\Qb,\dots\}$,
 on the homology class of $\Sigma_s$, since if $\Sigma_s$ is the boundary of some $\Sigma_{s+1}$, then
\begin{equation}\label{zilo}\int_{\Sigma_s}\O_k^{(s)}=\int_{\Sigma_{s+1}}\d \O_k^{(s)}=\left[\Qb,\int_{\Sigma_{s+1}}\O_k^{(s+1)}\right\}. \end{equation}
For $s=0$, $\Sigma_s$ is just a point $p$ , and $\int_p\O_k^{(0)}$ is just the evaluation of $\O_k=\O_k^{(0)}$ at $p$; the statement that
$\int_{\Sigma_s}\O_k^{(s)}$ only depends on the homology class of $\Sigma_s$ means that it is independent of $p$, as we explained already above via
eqn. (\ref{zodoc}).

In the magnetic description, we simply carry out this procedure as just described.  However, in the electric theory, it is not immediately obvious how much of this standard picture survives when a four-manifold $M$ is divided into two halves $M_\ell$ and $M_r$ by
a defect $W$.  Starting with zero-forms, to begin with we can define separate observables $\O_{k,\ell}=\fr{1}{k}\tr_\ell \,\sigma^k$ and $\O_{k,r}=\fr{1}{k}\tr_r\,\sigma^k$
in $M_\ell$ and $M_r$ respectively.  $\O_{k,\ell}$ is constant mod $\{\Qb,\dots\}$ in $M_\ell$, and similarly $\O_{k,r}$ is constant mod
$\{\Qb,\dots\}$ in $M_r$.  But is there any relation between these two observables?   Such a relation follows from
 boundary condition (\ref{Ybc1}), which tells us that on the boundary
\beq
\sigma=\fr{i}{2}\fr{1}{1+t^2}\{C,C\}.
\eeq
(This concise formula, when restricted to the Lie algebras of $G_\ell$ or of $G_r$, expresses the boundary value of $\sigma$ on $M_\ell$
or on $M_r$ in terms of the same boundary field $C$.)
Hence the invariance of the supertrace implies that  $\Str\,\sigma^k=0$ along $W$, or in other words that 
\begin{equation}\label{mostr} \tr_\ell\,\sigma^k=\tr_r\sigma^k \end{equation}
when restricted to the boundary between $M_\ell$ and $M_r$, where it makes sense to compare these two operators.

Now let us reconsider the descent procedure in this context.  We will try to construct an observable by integration on a closed one-cycle $\Sigma_1=\Sigma_{1\ell}\cup\Sigma_{1r}$, which lies partly in $M_\ell$ and partly in $M_r$,
\beq
\int_{\Sigma_1}\O_{k}^{(1)} \equiv\int_{\Sigma_{1\ell}}\O_{k,\ell}^{(1)}+\int_{\Sigma_{1r}}\O_{k,r}^{(1)}.
\eeq
Given that $\O_{k,\ell}^{(0)}=\O_{k,r}^{(0)}$ along $M_\ell\cap M_r=W$, and in particular on $C_0=\Sigma_1\cap W$, our observable is $\Qb$-closed,
\beq
\left[\Qb,\int_{\Sigma_1}\O_{k}^{(1)} \right\}=\int_{C_0} \left(\O_{k,r}^{(0)}-\O_{k,\ell}^{(0)}\right)=0.
\eeq
The relative minus sign comes in here, because $\Sigma_{1\ell}$ and $\Sigma_{1r}$ end on $C_1$ with opposite orientations.

Next we would like to go one step further and define an analogous 2-observable. To check $\Qb$-invariance of such an observable, analogously to the case just considered, we would need a relation between $\O_{k,\ell}^{(1)}$ and $\O_{k,r}^{(1)}$. From the relations $\d\O_{k,\ell}^{(0)}=[\Qb,\O_{k,\ell}^{(1)}\}$ in $M_\ell$, $\d\O_{k,r}^{(0)}=[\Qb,\O_{k,r}^{(1)}\}$ in $M_r$, it follows that, if $\imath:\,W\hookrightarrow M$ is the natural embedding, then
\begin{equation} \left[\Qb,\imath^*(\O_{k,\ell}^{(1)}-\O_{k,r}^{(1)})\right\}=0.              \label{welg}\end{equation}
In topological theory, a $\Qb$-closed unintegrated one-form should be $\Qb$-exact, so there should exist some operator $\tilde{\O}_{k}^{(1)}$, such that
\beq
\imath^*(\O_{k,\ell}^{(1)}-\O_{k,r}^{(1)})=\left[\Qb,\tilde{\O}_{k}^{(1)}\right\},.\label{io1}
\eeq
Then for a closed 2-cycle $\Sigma_2=\Sigma_{2\ell}\cup\Sigma_{2r}$ that intersects $W$ along some $C_1$ we can define an observable
\beq
\int_{\Sigma_2}\O_{k}^{(2)} \equiv\int_{\Sigma_{2\ell}}\O_{k,\ell}^{(2)}+\int_{\Sigma_{2r}}\O_{k,r}^{(2)}+\int_{C_1}\tilde{\O}_{k}^{(1)}.\label{O2}
\eeq
This observable is $\Qb$-closed.

Let us see how to define the next descendant. From the definition of $\O^{(2)}$ and from (\ref{io1}) we have
\beq
\left[\Qb,\imath^*(\O_{k,\ell}^{(2)}-\O_{k,r}^{(2)})\right\}=\left[\Qb,{\rm d}\tilde{O}^{(1)}_k\right\}\,,
\eeq
therefore, there exists $\tilde{\O}_{k}^{(2)}$ such that
\beq
\imath^*(\O_{k,\ell}^{(2)}-\O_{k,r}^{(2)})={\rm d}\tilde{O}^{(1)}_k+\left[\Qb,\tilde{\O}_{k}^{(2)}\right\}.
\eeq
Continuing in the same way, we find $\tilde{\O}_{k}^{(n)}$ such that
\beq
\imath^*(\O_{k,\ell}^{(n)}-\O_{k,r}^{(n)})={\rm d}\tilde{O}^{(n-1)}_k+\left[\Qb,\tilde{\O}_{k}^{(n)}\right\}\,,
\eeq
and the $\Qb$-invariant $(n)$-observable can be defined as
\beq
\int_{\Sigma_n}\O_{k}^{(n)} \equiv\int_{\Sigma_{n\ell}}\O_{k,\ell}^{(n)}+\int_{\Sigma_{nr}}\O_{k,r}^{(n)}+\int_{C_{n-1}}\tilde{\O}_{k}^{(n-1)}.\label{On}
\eeq

Let us find explicit representatives for all these operators in our case. A formula for $\O^{(1)}_{k}$ was already given in the right hand side of (\ref{zodoc}):
\beq
\O_{k,\ell,r}^{(1)}=\tr_{\ell,r}\left(\sigma^{k-1}\psit\right)\,,\label{1obs}
\eeq
where we now defined
\beq
\psit=\fr{1}{2}(t^{-1}\tilde{\psi}+\psi).
\eeq
This field has useful properties
\beq
\{\Qb,\psit\}=\cDb\sigma\,,\quad [\Qb,\cFb]=i(1+t^2)\cDb\psit\,,
\eeq
and satisfies the boundary condition
\beq
\imath^*(\psit)=-\fr{i}{1+t^2}\{\cAf,C\}.\label{psibc}
\eeq
Therefore on the defect
\beq
\imath^*\left(\O_{k,\ell}^{(1)}-\O_{k,r}^{(1)}\right)\sim\Str\left(\{C,C\}^{k-1}\{C,\cAf\}\right)=0.\label{1zero}
\eeq
Since this is zero, $\tilde{\O}^{(1)}_k$ vanishes, and the 2-observable can be defined without a boundary contribution. A representative for the 2-observable is
\beq
\O_{k,\ell,r}^{(2)}=\tr_{\ell,r}\left(\fr{1}{2}\sum_{k-2}\sigma^{\mathrm{j}_1}\,\psit\wedge \sigma^{\mathrm{j}_2}\,\psit-\fr{i}{1+t^2}\,\sigma^{k-1}\cFb\right)\,,
\eeq
where $\cFb$ is the field strength for the complexified gauge field $\cAb$. Here and in what follows we use the notation $\sum_m$ for a sum where the set of indices $\mj_1,\mj_2,\dots$ runs over partitions of $m$.

Using the boundary condition (\ref{psibc}) and invariance of the supertrace, one finds on the boundary,
\beq
\imath^*\left(\O_{k,\ell}^{(2)}-\O_{k,r}^{(2)}\right)=\fr{i}{1+t^2}\,\Str\left(\sigma^{k-1}\mathcal{F}'\right)\,,\label{O2boundary}
\eeq
where $\mathcal{F}'=\cFb+\cAf\wedge\cAf$ is the part of the super gauge field strength that lies in the bosonic subalgebra. The expression under the supertrace is non-zero, but we know that it should be $\Qb$-exact. Indeed, one finds that this is a $\Qb$-variation of
\beq
\tilde{\mathcal{O}}^{(2)}_k=\fr{1}{2}\left(\fr{i}{1+t^2}\right)^k\Str\left(C^{2k-3}\cDb\cAf\right).
\eeq

Proceeding further with the descent procedure, we can find the 3-descendant,
\beq
\mathcal{O}^{(3)}_{k,\ell,r}=\tr_{\ell,r}\left(\fr{1}{3}\sum_{k-3}\sigma^{\mathrm{j}_1}\,\psit\wedge\sigma^{\mathrm{j}_2}\,\psit\wedge\sigma^{\mathrm{j}_3}\,\psit
-\fr{i}{1+t^2}\sum_{k-2}\sigma^{\mathrm{j}_1}\,{\cFb}\wedge\sigma^{\mathrm{j}_2}\,\psit\right).
\eeq
On the boundary after some computation we find
\beq
\tilde{\mathcal{O}}^{(3)}_k=\fr{1}{2}\left(\fr{i}{1+t^2}\right)^{k}\Str\left(\sum_{2k-4}C^{\mathrm{j}_1}\cAf C^{\mathrm{j}_2}\cDb\cAf\right).\label{tO3}
\eeq
The bulk part of the four-observable has a representative
\beqn
\O^{(4)}_{k,\ell,r}&=&\tr_{\ell,r}\left(\fr{1}{4}\sum_{k-4}\sigma^{\mj_1}\psit\wedge\sigma^{\mj_2}\psit\wedge\sigma^{\mj_3}\psit\wedge\sigma^{\mj_4}\psit -\fr{i}{1+t^2}\sum_{k-3}\sigma^{\mj_1}\cFb\wedge\sigma^{\mj_2}\psit\wedge\sigma^{\mj_3}\psit\right.\nnr
&-&\left.\fr{1}{2(1+t^2)^2}\sum_{k-2}\sigma^{\mj_1}\cFb\wedge\sigma^{\mj_2}\cFb\right).\label{O4}
\eeqn
The four-observable, which is formed from (\ref{tO3}) and (\ref{O4}), has ghost number zero for $k=2$. In this case, of course, it reduces just to our super Chern-Simons action.

One might wonder how unique this procedure is. Clearly, for the $n^{th}$ descendant of $\O^{(0)}_k$, we can try to modify it by adding a suitable $(n-1)$-observable with ghost number $(2k-n)$, integrated over $C_{n-1}=\Sigma_n\cap W$. Since  $C_{n-1}$ is a boundary in the bulk (it is the boundary of $\Sigma_n\cal M_\ell$, for example), such a modification would be non-trivial only if the observable that we add cannot be extended into the bulk. One possible example is adding a Wilson loop to $\tilde{\O}^{(1)}_1$ in the 2-descendant of the operator $\tr\,\sigma$. What other boundary observables  might one consider? If we denote the bosonic subgroup of the supergroup by $\SG_{\bar{0}}\cong G_{\ell}\times G_r$, the $\Qb$-invariant scalar observables on the defect correspond to the $SG_{\bar{0}}$-invariant polynomials of the ghost field $C$. However, one can check that for the basic classical Lie superalgebras all such polynomials come\footnote{See, e.g., a list of these polynomials in \cite{BoeKujawa}.} from the invariant polynomials in $\sigma\sim\{C,C\}$, and therefore the corresponding observables are bulk observables.

\section{Puzzles In Three Dimensions}\label{Anomalous}

Naively, one might expect that familiar facts about three-dimensional Chern-Simons 
theory with a bosonic  gauge group\footnote{In stating some of these facts, for
 simplicity, we will assume $G$ (and similarly the maximal bosonic subgroup of a supergroup $SG$) to be connected and simply-connected,
to make the definition of the Chern-Simons action $\CS(A)$ straightforward.  Also, to avoid additional subtleties,
we always pick the compact form
for the maximal bosonic subgroup of a supergroup $SG$.  Some questions we discuss
below have rough analogs for bosonic Chern-Simons theory of a group such as $SL(2,\RR)$ whose Lie algebra admits an invariant,
non-degenerate quadratic form, but not one that is positive-definite.  Some of the questions have been discussed in that
context in \cite{BarNatan}.}  $G$ would
have direct analogs with $G$ replaced by a supergroup $SG$.  Actually,
one runs immediately into a variety of puzzles, some of which we describe here.\footnote{It will perhaps not be a surprise that 
$\OSp(1|2n)$ is an exception to most of
our statements.}
The puzzles mostly have their roots in the fact that the invariant quadratic form on the bosonic part of the superalgebra $\frak{sg}$
is not positive-definite.   We have not had to grapple with these puzzles in this paper, because the paper 
is really not devoted to an abstract three-dimensional  Chern-Simons theory
but to a four-dimensional construction that in some sense gives an analytically-continued version of that theory.  The analytically
continued version of the supergroup theory seems to present no puzzles analogous to those that we discuss here.

\subsection{Relation To Current Algebra}\label{currel}

To see that some standard statements in the bosonic world are not likely to  have simple superanalogs, let $W$ be an oriented
three-manifold with boundary $\Sigma$.  The orientation of $W$ determines an orientation of $\Sigma$ so that (once a conformal structure is picked) it becomes
a complex Riemann surface  and we know what
is ``holomorphic'' or ``antiholomorphic.'' Also, on an oriented three-manifold, one can define the Chern-Simons function $\CS(A)$
for a gauge field $A$, and the corresponding action\footnote{
In a more intrinsic approach to Chern-Simons theory, one does not
pick an orientation on $M$, and instead of interpreting $k$ as an integer, one interprets it as an element of $H^3(W;\Z)$ (which is
non-canonically isomorphic to $\Z$, with an isomorphism given by a choice of orientation of $W$).  We have chosen not to use that language here, because it is probably unfamiliar and would
detract from our main point about supergroups.} $I=ik\CS(A)$, with $k\in\Z$.  $\CS(A)$ is defined using a positive-definite quadratic
form on the Lie algebra $\frak g$; one has to specify this so as to know what is meant by $k$ rather than $-k$.
 If $k$ is positive, one relates Chern-Simons theory on
$W$ to a holomorphic current algebra  on $\Sigma$; if $k$ is negative, one relates it to an antiholomorphic current algebra on 
$\Sigma$.
This relationship  is the basis for claiming that Chern-Simons theory of $G$ is exactly soluble, so certainly
we would like to try to generalize it to supergroups.

In the case of a supergroup $SG$, there is still an invariant quadratic form $\Str$ on the superalgebra $\frak{sg}$.  But (with
the usual exception of $\OSp(1|2n)$) it is not
possible to pick this invariant quadratic form to be positive-definite when restricted to a maximal bosonic subalgebra of $\frak{sg}$.
Hence, there is no natural notion of what it means for $k$ to be ``positive'' or ``negative.''  Having picked a quadratic form,
we can decompose the maximal bosonic subgroup of $SG$ in the usual way as $G_\ell\times G_r$ so that the quadratic
form is positive on the Lie algebra of $G_\ell$ and negative on that of $G_r$.  Then on $\Sigma=\partial W$, we would expect
to see holomorphic current algebra of $G_\ell$ and antiholomorphic current algebra of $G_r$.  There is no obvious way to
extend this to an $SG$-invariant story.  So it is hard to see what could be a superanalog of the usual statement for a bosonic
simple Lie group.

To understand this better, we 
 now take a step back and recall {\it why}  Chern-Simons theory of a compact bosonic group $G$ can be related to a
holomorphic current algebra on the boundary.  The action $I=ik\CS(A)$ is gauge-invariant 
(mod $2\pi$) on a three-manifold $W$ without
boundary, but on a three-manifold with boundary, this action has an anomaly at level $k$, supported on the boundary.  To
cancel the anomaly, one can couple to boundary degrees of freedom that have an equal and opposite anomaly.  In general,
there are many ways to do this and any of them may be of interest.  However, there is one canonical procedure that 
leads to the relation between Chern-Simons theory and current algebra.
 One takes the boundary theory to be a WZW model with target space $G$ and level $k$.  
 This theory
has both holomorphic and antiholomorphic current algebras, associated  respectively to the left and right action of $G$ on itself.
One can cancel the anomaly of the bulk Chern-Simons theory by coupling the bulk gauge fields to the holomorphic or antiholomorphic
currents of the WZW model; the choice of which currents should be gauged depends on the sign of $k$.  After gauging one
set of currents, the currents of the other set survive as global symmetry generators.  For $k>0$ or $k<0$, the  
surviving currents are a holomorphic current algebra
at level $k$, or an antiholomorphic one at level $-k$.  This gives the basic relation of bulk Chern-Simons theory to a boundary
current algebra in the case of a bosonic Lie group $G$.    For some elaboration on this picture, see \cite{OnHol}.

To imitate this for a supergroup $SG$, we have to pick a boundary theory with $SG$ global symmetry that is anomalous when gauged.
One may be tempted to use the WZW model with target $SG$, but here we run into the fact that the WZW model with supergroup
target is much more problematical than Chern-Simons theory of the same supergroup.  The reason for this is that the path integral
of bosonic Chern-Simons theory is, in the leading approximation, an oscillatory Gaussian integral that makes sense for either
sign of the quadratic form. The basic one-dimensional oscillatory Gaussian integral was already defined in eqn. (\ref{oblo}):
\begin{equation}\label{noblo}\int_{-\infty}^\infty\frac{\d x}{\sqrt\pi}\exp(i\lambda x^2) =\frac{\exp(i(\pi/4)\sgn\lambda)}{|\lambda|},\end{equation}
It makes sense for either sign of $\lambda$, so in Chern-Simons theory, there is no immediate problem if the quadratic
form on the bosonic part of $\sg$ is indefinite.  However, the WZW model is another story.  
In addition to the Wess-Zumino term, which is imaginary (when the WZW model is formulated in Euclidean signature),
the action of the WZW model of a group $G$ contains a real term
proportional to  the metric on $G$.  As a result,
the WZW path integral is only convergent if the metric  of $G$ is positive definite.  In the case of a supergroup $SG$,
the metric should be positive on a maximal bosonic subgroup of $SG$.  But this is precisely the condition that is not satisfied
(except for $\OSp(1|2n)$), so generically there is no straightforward notion of a WZW model of a supergroup. One could try to define something by analytic continuation of some kind, but one should not expect a simple answer.  In fact,
sigma-models with homogeneous supermanifolds as targets are rather tricky even when this particular problem does not
arise \cite{Supersigma}.

Going back to $SG$ Chern-Simons theory on a manifold with boundary, in general we certainly may be able to cancel the
anomaly by coupling to some other boundary theory with $SG$ symmetry, rather than a WZW model.  But then we do not get a 
framework for relating $SG$ Chern-Simons theory in bulk to $SG$ current algebra on the boundary.  What we do get depends on
what boundary theory we pick.

This conclusion is not  entirely satisfactory, because at a  couple of points in the present paper  we obtained reasonable results assuming formulas
 that would follow from a relation between $SG$ Chern-Simons theory in three dimensions and $SG$ current algebra in 
 two dimensions.\footnote{One example is the discussion of the framing anomaly  in section \ref{comparing}.  Another is the relation
 (see section \ref{dunsp})
 between the duality between orthosymplectic theories with parameters $q$ and $-q$ and a corresponding duality of quantum groups
 in \cite{qOSp}.} One would like to understand why 
 some such statements are valid, even though others probably are not.  It may be that the explanation has to be given in the framework of analytically
 continued 3d Chern-Simons theory, formulated as in this paper in terms of 4d super Yang-Mills theory, and not in a relation between 3d and 2d theories.  
  
\subsection{The One-Loop Shift}\label{olshift}

Now we will discuss a purely three-dimensional question -- not involving an attempt to compare to a two-dimensional theory -- in
which familiar facts in the bosonic world do not have a straightforward extension to a supergroup. 

Does there exist an abstract 3d Chern-Simons theory of a supergroup $SG$, in which the coupling $k$ has to be an integer for
topological reasons and whose path integrals  give invariants of arbitrary framed three-manifolds (possibly containing
framed links)?  The answer to this question is apparently ``yes'' \cite{ZhThree,Bluman} for $\OSp(1|2n)$.
For other supergroups, one cannot hope for such a strong result; some observables that depend inversely on the volume of $SG$
will diverge, as explained in eqn. (\ref{infinite}).    However, good evidence will be presented elsewhere \cite{VM} that a partial Chern-Simons theory of $SG$ -- with many but not all amplitudes being well-defined -- does exist.

 Here we will try to understand the large $k$ behavior of the path integral in such a theory.  As we will see, some familiar ideas
 for the case of a purely bosonic compact group $G$ do not carry over well for supergroups.

 In Chern-Simons theory
of a simple compact Lie group $G$ on a three-manifold $W$, one may construct a perturbative expansion in powers of $1/k$ around any classical
solution.   A classical solution is a flat connection $A_0$, corresponding to a homomorphism $\varrho:\pi_1(W)\to G$.
For simplicity, let us assume that $A_0$ is isolated and irreducible (meaning that after gauge-fixing, neither ghosts nor matter
fields have zero-modes in expanding around $A_0$).  This assumption makes it straightforward to construct perturbation theory
around the given solution.  The semiclassical approximation is the exponential of minus the classical action 
times a product of bosonic and fermionic
Gaussian integrals.  (The fermions in questions are the ghosts and antighosts that arise in gauge-fixing.)  
The fermionic Gaussian integrals are real but (simply because the classical action $I=ik \CS(A)$ is imaginary), the 
bosonic Gaussian integrals are oscillatory.  As usual in a bosonic Gaussian integral, one can pick a basis $x_j$ of
integration variables such that the action is a diagonal expression $-i\sum_j\lambda_j x_j^2$.  The integral over each $x_j$ produces
a phase $\exp(i(\pi/4)\sgn\lambda_j)$, as stated in eqn. (\ref{noblo}).  With $\zeta$-function regularization, the product of all
these phases gives an Atiyah-Patodi-Singer $\eta$-invariant, so that the phase of the path integral in the semiclassical
approximation is $\exp(ik\CS(A_0))\exp(i\pi \eta(A_0)/4)$, where the first factor is the exponential of minus the classical action
and the second factor comes from the one-loop correction.

Modulo a constant that does not depend on $A_0$ (see \cite{FreedGompf} for details), $\pi\eta(A_0)/4$ is equivalent
to $h\,\sgn\, k \,\,\CS(A)$, where $h$ is the dual Coxeter number of $G$.  (This statement is often formulated only for $k>0$,
in which case $\sgn\,k=1$. For $k<0$, the $\lambda_j$ all change sign and the sign of the $\eta$-invariant is reversed, explaining the factor of $\sgn \,k$.)  
Thus the phase of the semiclassical approximation to the path integral expanded around $A_0$, 
apart from a constant factor that is independent of $A_0$, is 
\begin{equation}\label{zordo}\exp(i(k+h\,\sgn\,k)\CS(A_0)). \end{equation}
This is often summarized by saying that the one-loop correction induces a shift
\begin{equation}\label{mordo} k\to k+h\,\sgn\,k  \end{equation}
in the effective action.   

Now let us ask what the analog may be for a supergroup $SG$.  First of all, the basic question does have an analog.
We expand around a flat connection $A_0$ that is isolated and irreducible.  A flat $\sg$-valued connection that has no odd or even
moduli has a structure group that reduces to a maximal bosonic subgroup of $SG$, so we will assume that $A_0$ has
this property.
We ask, in the semiclassical approximation,
what is the phase of its contribution to the effective action.   Naively, we expect by analogy with (\ref{zordo}) that the 
answer will be
\begin{equation}\label{pordo}\exp(i(k+h_\sg\sgn\,k)\CS(A_0)) \end{equation}
(apart from a constant factor independent of $A_0$).  However, we should expect trouble here (with the usual exception
of $\OSp(1|2n)$) because as the quadratic
form on $\sg$ is indefinite, it is not natural to say that $k$ is positive or negative so it is not clear what could be meant by $\sgn\,k$.

To make the problem more concrete, let us consider the case of the supergroup $\SU(m|n)$. (It should be obvious that the discussion
has an analog for any $SG$.)   The bosonic part of the Lie algebra is
$\frak g_{\overline 0}= \frak{su}(m)\oplus \frak{su}(n)\oplus \frak{u}(1)$.  We pick a quadratic
form that is positive on $\frak{su}(n)$ and negative on $\frak{su}(m)$.   It is positive on $\frak{u}(1)$ if $n<m$ and negative if $n>m$
(we must take $n\not=m$, since  $\SU(n|n)$ does not have an invariant nondegenerate 
quadratic form).  The dual Coxeter number is $h=n-m$,
so naively we expect a shift
\begin{equation}\label{mexor}k\to k+(n-m)\sgn(k). \end{equation}

Let us see what actually happens, assuming that we can treat the one-loop approximation to the path integral as an oscillatory integral.
 Because of our assumption that the structure group of $A_0$ reduces to a purely bosonic subgroup,
we can consider separately the parts of the Gaussian integrals over fields valued in $\frak{g}_{\bar 0}$ and $\frak{g}_{\bar 1}$
(we recall that these are the even and odd parts of $\sg$).  The $\frak{g}_{\bar 0}$-valued part is the same Gaussian integral
that would give the one-loop correction to $\frak g_{\bar 0}$-valued gauge fields.  Since the dual
Coxeter numbers for $\frak{su}(n)$, $\frak{su}(m)$ and $\frak{u}(1)$ are $n, $ $m$, and 0, and  the quadratic form on $\sg$
is positive on $\su(n)$ and negative on $\su(m)$, the one-loop shifts for the parts of $A_0$ that are valued in $\su(n)$, $\su(m)$,
and $\u(1)$ are respectively
\begin{equation}\label{gremo}k\to \begin{cases} k+n\,\sgn(k) &\su(n) \cr
                                                                             k+m\,\sgn(k) & \su(m)\cr 
                                                                               k & \u(1). \end{cases}\end{equation}
To explain the first two statements, the $\su(n)$ theory is at level $k$, so its shift is $k\to k+n\,\sgn(k)$, but the $\frak{su}(m)$
theory is at level $k'=-k$, so its shift is $k'\to k'+m\,\sgn(k')=k'-m\,\sgn(k)$; this is equivalent to $k\to k+m\,\sgn(k)$.  

What about the shift due to the odd part of $\sg$?  From the point of view of $\frak{su}(n)$, $\frak g_{\bar 1}$ consists of $m$
copies of the fundamental plus antifundamental representation; from the point of view of $\frak{su}(m)$, it consists of $n$
copies of the fundamental plus antifundamental.   So if $\alpha$ is the one-loop shift due to a single copy of the fundamental
plus antifundamental representation, then  the $\su(n)$ shift due to $\frak g_{\bar 1}$ is 
$k\to k+m\alpha$.  Likewise the $\su(m)$ shift is $k'\to k'+n\alpha$ or $k\to k-n\alpha$.
When we combine this with (\ref{gremo}), we see that the overall one-loop shifts
are $k\to k+n\,\sgn(k) +m\alpha$ for $\su(n)$ and $k\to k+m\,\sgn(k)-n\alpha$ for $\su(m)$.  There is no way to reconcile
these statements  with the naive expectation of eqn. (\ref{mexor}), regardless of what we assume for $\alpha$.  (It does not even help to
use different values of $\alpha$ for $\su(n)$ and $\su(m)$, or to change the sign of $\sgn\,k$ in eqn. (\ref{mexor}).)

But what is the natural value of the shift due to $\frak g_{\overline 1}$?  Our tentative inclination is to claim (for any supergroup)
that the natural value is 0.  In other
words, the claim is that in the most
natural interpretation of the three-dimensional fermionic path integral, the
integration over $\frak g_{\overline 1}$-valued fields produces no shift in the effective values of the Chern-Simons couplings.   In fact,
the contribution of $\frak g_{\bar 0}$ to the shift ultimately derives from the fact that the bosonic oscillatory Gaussian integral
(\ref{noblo}) is not analytic in $\lambda$.  The product over all modes of this nonanalytic factor gives the shift in $k$.
An analogous fermionic ``oscillatory'' Gaussian integral is completely analytic:
\begin{equation}\label{urtzu}\int\d\psi_1\,\d\psi_2\,\exp(i\lambda\psi_1\psi_2)=i\lambda. \end{equation}
There is no factor dependent on $\sgn\,\lambda$ that might produce an $\eta$-invariant. 
 
 The motivation for our suggestion is that the path integral over $\frak g_{\overline 1}$-valued fields is naturally real,
 although not naturally positive.
To explain this, consider performing the $\frak g_{\overline 1}$-valued part of the one-loop path integral in the background
of an arbitrary $\frak g_{\bar 0}$-valued gauge field  $A_0$ (not necessarily a classical solution).  After putting the part of the action quadratic
in $\frak g_{\bar 1}$-valued fields in a canonical form, we formally have to consider a product of many integrals like (\ref{urtzu}):
\begin{equation}\label{wurtz}   \prod_k \int\d\psi_{1,k}\,\d\psi_{2,k}\,\exp(i\lambda_k\psi_{1,k}\psi_{2,k})=
\prod_k (i\lambda_k). \end{equation}  (There are also $\frak g_{\bar 1}$-valued bosonic ghosts and antighosts, but their
path integral is real and so not relevant to possible shifts in $k$.)
Although there are infinitely many factors of $i$ in (\ref{wurtz}), 
they do not contribute, because the number of modes is independent of $A_0$.    
We explain this better in a moment.    After dropping the factors of $i$, the product
in (\ref{wurtz}) is naturally real.  However, it is not naturally positive; as one varies $A_0$, one of the $\lambda_k$ might
pass through 0 and then the sign of the $\frak g_{\bar 1}$ path integral should change.  We conclude that the one-loop path
integral over $\frak g_{\bar 1}$-valued fields, in a background whose structure group reduces to the maximal bosonic subgroup
of $SG$, is naturally real but not naturally positive.  

To   explain the statement that the factors of $i$ are not relevant, we use $\zeta$-function regularization.  We let $\DD$ be the  
kinetic operator for the $\frak g_{\bar 1}$-valued fermions (it is imaginary and self-adoint).  The $\zeta$ function of this
operator is defined as $\zeta_\DD(s)=\sum_k|\lambda_k|^{-s}$.  It converges for $\mathrm{Re}\,s$ sufficiently large
and can be analytically continued to $s=0$.  In $\zeta$-function regularization, the regularized version of the total number of
modes of $\DD$ is $\zeta_\DD(0)$. But $\zeta_\DD(0)=0$ (this is a special case of a more general statement
about elliptic differential operators in odd dimenisons),  justifying our claim.  The difference for the bosons
is that because the argument of the oscillatory Gaussian integral (\ref{noblo}) has a non-analytic contribution $(\pi/4)\sgn\,\lambda$,
to compute the phase of the one-loop integral for $\frak g_{\bar 0}$-valued fields, we need a regularized version of
$\sum_k \sgn\,\lambda_k$.  Here the analog of $\zeta$-function regularization leads to an $\eta$-invariant.

We are thus led to propose that the $\frak g_{\bar 1}$-valued part of the path integral produces no one-loop shift, while the
$\frak g_{\bar 0}$-valued part produces the shifts summarized in eqn. (\ref{gremo}).  Then the semiclassical approximation to
the $SG$ Chern-Simons path integral, expanded around a classical solution $A_0$ valued in a maximal bosonic subgroup,
is
\begin{equation}\label{turkey}\exp\left(i(k+n\,\sgn(k))
\CS(A_0^{\su(n)})+i(k+m\,\sgn(k))\CS(A_0^{\su(m)})+ik\CS(A_0^{\u(1)})\right), \end{equation}                                                      
where $A_0^{\su(n)}$, $A_0^{\su(m)}$, and $A_0^{\u(1)}$ are the projections of $A_0$ to $\su(n)$, $\su(m)$, and $\u(1)$,
respectively.

The reader may object that this formula does not look gauge-invariant, but actually there is no simple objection along these
lines.  Since we have assumed that $A_0$ is valued in a maximal bosonic subgroup, which one may denote as ${\mathrm{S}}(\U(m)\times \U(n))$,
there is no problem in defining mod $2\pi$ the expression                                                                  
$(k+n\,\sgn(k))
\CS(A_0^{\su(n)})+(k+m\,\sgn(k))\CS(A_0^{\su(m)})+k\CS(A_0^{\u(1)})$ that appears in eqn. (\ref{turkey}).  It is a linear combination
of the three independent Chern-Simons invariants of a gauge field with structure group ${\mathrm S}(\U(m)\times \U(n))$.  Nor is the claim that
(\ref{turkey}) is well-defined 
an artifact of our assuming that $A_0$ is valued in a maximal bosonic subgroup of $SG$.  
Any connected supermanifold $\mathcal F$ that parametrizes a family of $SG$-valued flat connections
has a connected reduced space $\mathcal F_{\mathrm{red}}$ that parametrizes flat connections whose structure group does reduce to a purely bosonic
subgroup.  The expression in (\ref{turkey}) can be evaluated for a flat connection corresponding to  a point 
$q\in \mathcal F_{\mathrm{red}}$, giving a result that is independent of the choice of $q$, and this gives the appropriate value
 for the whole family $\mathcal F$.  
 
 Thus there is no immediate contradiction in our suggestion
  that (\ref{turkey}) captures the appropriate one-loop shifts for the supergroup
 $\SU(m|n)$.   Moreover, this proposal has an obvious analog for any supergroup $SG$.  
  
 But it is hard to see how an abstract 3d theory in which what we have described is the right answer
 can be related to the analytically-continued
 theory that can be derived from $\N=4$ super Yang-Mills theory in four dimensions.  That theory
 is described in terms of an effective coupling $\calK$ that is the same for $G_\ell$ and $G_r$. Moreover, it is not clear whether our
 proposal for interpreting the $\frak g_{\bar 1}$-valued part of the path integral is compatible with actual results for $\OSp(1|2n)$, for
 which an abstract 3d theory does exist \cite{ZhThree,Bluman}.  We have little to offer here and
  can only lamely conclude that if there is an abstract 3d version of supergroup Chern-Simons theory  that is supposed
  to be related to the 4d picture,
then what we have said is not a good approach to its path integral.  Perhaps the reader can explain a better point of view.

\enddocument